\documentclass{article}

 \usepackage[preprint]{neurips_2026}

\usepackage{amsmath,amsfonts,bm,amssymb}
\usepackage{mathtools}
\usepackage{amsthm}

\def\eqref#1{equation~\ref{#1}}

\def\1{\bm{1}}

\DeclareMathAlphabet{\mathsfit}{\encodingdefault}{\sfdefault}{m}{sl}
\SetMathAlphabet{\mathsfit}{bold}{\encodingdefault}{\sfdefault}{bx}{n}

\usepackage{graphicx} 

\usepackage{xcolor}

\usepackage{colortbl}

\usepackage[most]{tcolorbox}

\usepackage{booktabs}

\usepackage{multirow}

\usepackage{adjustbox}

\usepackage{makecell}

\usepackage{tabularx}

\usepackage{longtable}

\usepackage{array}

\usepackage{subcaption}

\usepackage{wrapfig} 

\usepackage[utf8]{inputenc} 

\usepackage[patch=none]{microtype}

\usepackage{stfloats}

\usepackage{lineno}

\usepackage{textcomp}

\usepackage[OT1]{fontenc} 

\usepackage{balance}

\usepackage{enumitem} 

\usepackage{nicefrac}       

\usepackage{amsmath}
\usepackage{amsfonts}
\usepackage{amssymb}
\usepackage{bm}       

\usepackage{amsthm}

\usepackage{pifont}

\usepackage{listings}

\usepackage{verbatim}

\usepackage{hyperref}

\usepackage[textsize=tiny]{todonotes}

\usepackage[capitalize,noabbrev]{cleveref}

\def\HiLi{\leavevmode\rlap{\hbox to \hsize{\color{gray!20}\leaders\hrule height .8\baselineskip depth .5ex\hfill}}}

\definecolor{darkblue}{rgb}{0, 0, 0.5}

\hypersetup{
  colorlinks=true,   
  citecolor=darkblue, 
  linkcolor=darkblue, 
  urlcolor=darkblue   
}

\theoremstyle{plain}

\theoremstyle{definition}

\theoremstyle{remark}

\title{When Does Multi-Agent Collaboration Help? An Entropy Perspective}

\author{%
  Yuxuan Zhao\textsuperscript{1,2} \quad
  Sijia Chen\textsuperscript{2,\thanks{Corresponding author: sijiachen@hkust-gz.edu.cn}} \quad
  Ningxin Su\textsuperscript{2} \\[0.3em]
  \textsuperscript{1}Yantai Research Institute of Harbin Engineering University \\
  \textsuperscript{2}The Hong Kong University of Science and Technology (Guangzhou) \\[0.5em]
  Project page: \href{https://multiagent-entropy.github.io/}{\textcolor[HTML]{DA3D8F}{https://multiagent-entropy.github.io/}}
}

\begin{document}

\maketitle

\begin{abstract}
Multi-agent systems (MAS) have emerged as a prominent paradigm for leveraging large language models (LLMs) to tackle complex tasks. However, the mechanisms governing the effectiveness of MAS built upon publicly available LLMs, specifically the underlying rationales for their success or failure, remain largely unexplored. In this paper, we revisit MAS through the perspective of \textit{entropy}, considering both intra- and inter-agent dynamics by investigating entropy transitions during problem-solving across various topologies, six reasoning benchmarks, and two agentic tasks. By analyzing 245 features spanning token-, agent-, and round-level entropy, we counterintuitively find that a single agent outperforms MAS in approximately 43.3\% of cases, and that entropy dynamics are largely determined during the first round of interaction. Furthermore, we provide three key observations: 1) \textit{Certainty Preference}: peak entropy directly harms and stable entropy directly benefits MAS correctness; 2) \textit{Base Entropy}: base models with lower entropy during problem-solving causally drive MAS performance; and 3) \textit{Task Awareness}: entropy dynamics of MAS play varying roles across different tasks. Building on these insights, we introduce a simple yet effective algorithm, the \textit{Entropy Judger}, to select solutions from MAS's pass@$k$ results, leading to consistent accuracy improvements across all MAS configurations and tasks. Our source code is available at \href{https://github.com/AgenticFinLab/multiagent-entropy}{this https URL}.
\end{abstract}

\addtocontents{toc}{\protect\setcounter{tocdepth}{-1}}

\section{Introduction}
\label{sec:intro}

Multi-agent systems (MAS), with each agent built upon large language models (LLMs), are broadly applied in diverse domains~\citep{mas-debate-icml23,metagpt-iclr23,agentbench-iclr24,balrog-iclr25,mas-collaboration-neurips25} and even regarded as the only choice for problem-solving~\citep{agent-forest-arxiv24}. However, it remains largely unexplored whether MAS, particularly those built upon open-source LLMs, can outperform their single-agent counterparts, and what underlies their effectiveness.

Existing work has observed that single-agent systems (SAS) can match or even surpass MAS on certain tasks~\citep{sas-or-mas-arxiv25,sas-outperform-mas-token-budget-arxiv26}, and the failures of MAS often stem from communication breakdowns, inter-agent misalignment, and insufficient verification~\citep{why-mas-fail-neurips25}. In particular, recent work establishes scaling principles for MAS through quantitative analysis~\citep{science-of-scaling-mas-arxiv25}. However, these studies are primarily conditioned on simple metrics such as accuracy, latency, and cost without providing a deeper understanding of the underlying mechanisms.

Entropy has emerged as a key perspective for understanding LLM reasoning, ranging from reinforcement learning (RL) analysis~\citep{entropy-mechanism-neurips25} to training algorithms that balance exploration and exploitation~\citep{entro-duction-acl25,eas-aaai25}. Notably, even individual observations of correlations between entropy and accuracy have spurred further research in regularization, advantage shaping, and token updates~\citep{clip-entropy-neurips25workshop,entropy-adv-shaping-aaai25,80/20rule-neurips25}.

Therefore, for LLM-based MAS, which are inherently complex and exhibit uncertainty at the token, agent, and other levels, it is crucial to build a comprehensive relationship between entropy and reasoning reliability. Although recent works approach it from an information-theoretic perspective~\citep{info-theory-mas-arxiv25,sas-outperform-mas-token-budget-arxiv26}, they are limited to specific architectures or token-budget comparisons.

In this paper, we revisit MAS by investigating the entire lifecycle of entropy across diverse levels, steps, and phases involved in reasoning, under varying MAS topologies and tasks. Specifically, by mining the fine-grained relationship between large-scale information entropy, derived from both intra-agent and inter-agent interactions, and MAS performance, we demonstrate that MAS effectiveness is largely determined by early-round entropy dynamics, with peak entropy universally harmful. We further validate through causal inference that these entropy-performance associations reflect genuine causal mechanisms rather than mere correlations. In summary, our key contributions are:
\begin{itemize}[leftmargin=*,topsep=0pt,itemsep=-0.2em]
\item A systematic study of entropy dynamics across six reasoning tasks and two agentic tasks, under four MAS topologies, analyzing 245 features at token, agent, and interaction-round levels;
\item The counterintuitive finding that SAS outperform MAS in approximately 43.3\% of cases, underscoring the trade-off between system complexity and performance;
\item Three insights on entropy behavior: Certainty Preference (peak entropy directly harms and stable low entropy directly benefits MAS correctness), Base Entropy (lower base model entropy causally drives MAS performance), and Task Awareness (entropy patterns differ across tasks); these findings generalize to agentic settings, where tool-call entropy and first-round inter-agent dispersion jointly constrain MAS correctness;
\item The \textit{Entropy Judger}, which selects high-quality outputs from MAS pass@$k$ results and consistently boosts accuracy across all configurations and tasks.
\end{itemize}

\section{Related Work}
\label{sec:related-work}

\textbf{Large Language Model-based Multi-Agent Systems (MAS)} decompose complex tasks into specialized agents that interact through diverse coordination topologies~\citep{agentverse-iclr23,chain-agent-neurips24,mas-collaboration-neurips25,metaagent-icml25}, thereby enhancing problem-solving capabilities. Beyond improving accuracy and efficiency of MAS, several studies examine the conditions under which MAS are effective. Scaling analyses show that gains from adding agents or varying coordination structures are often offset by communication overhead and inter-agent misalignment, yielding diminishing or even negative returns~\citep{science-of-scaling-mas-arxiv25}. Recent work identifies key failure modes in MAS, such as insufficient verification and communication breakdowns~\citep{why-mas-fail-neurips25}. More recently, SAS equipped with rich skill libraries have been shown to match or surpass MAS in both accuracy and efficiency~\citep{single-agent-skills-arxiv26}. While these works inform MAS practice, they rely on simple metrics that fail to capture system complexity and therefore cannot uncover the underlying rationales governing MAS effectiveness. Recent information-theoretic analysis~\citep{info-theory-mas-arxiv25,sas-outperform-mas-token-budget-arxiv26} offers deeper insight but is limited to specific architectures or token budget comparison.

However, analyzing \textbf{entropy in LLM reasoning} has deepened our understanding of the underlying reasoning process~\citep{entropy-mechanism-neurips25,eas-aaai25}. In RL training, reducing entropy sharpens output distribution and improve accuracy~\citep{entropymin-arxiv25,reasoning-with-sampling-iclr26}, but this benefit is highly dependent on the base model capabilities, and aggressive entropy reduction may trigger \textit{entropy collapse}, wherein the model becomes overconfident and converges to suboptimal policies~\citep{rl-incentivize-reasoning-neurips25,nofreelunch-arxiv25}. To mitigate this, recent work proposes entropy-intervention methods, including entropy regularization~\citep{aer-arxiv25,clip-entropy-neurips25workshop}, entropy-based advantage shaping~\citep{entropy-adv-shaping-aaai25}, and entropy-guided token updates~\citep{80/20rule-neurips25}. At test time, entropy also signals uncertainty to dynamically adjust reasoning depth and direction~\citep{entropy-gated-branching-eacl26,min-tt-intervention-acl26}. While these studies enhance single-agent reasoning by probing entropy, they largely overlook how entropy propagates across multiple interacting agents. To address this gap, we analyze entropy dynamics in LLM-based MAS to explain when and why they succeed.
\section{Preliminaries}
\label{sec:preliminaries}

Let \(M_{\text{base}}\) be an LLM with parameters \(\theta\), defining a distribution \(\pi_\theta(v \mid s)\) over tokens \(v \in \mathcal{V}\) given state \(s\).
We formalize a Multi-Agent System (MAS) as \(\mathsf{M} = (A, G, R)\), where \(A\) is a set of agents each instantiated from \(M_{\text{base}}\), \(G = (A, E)\) is a directed interaction graph, and \(R\) is the number of interaction rounds.
In each round \(r\), agent \(a\) generates a reasoning trajectory \(\tau_a^{(r)} \sim \pi_\theta(\cdot \mid \mathcal{H}_a^{(r)})\), where the context \(\mathcal{H}_a^{(r)}\) is determined by \(G\) as follows, and the system outputs a final prediction \(\hat{y}\).

\textbf{Single Agent System (SAS).}
\(A = \{a\}\), \(G = (\{a\}, \emptyset)\), \(\mathcal{H}_a^{(r)} = \{\tau_a^{(r')}\}_{r' < r}\), \(\hat{y} = \tau_a^{(R)}\).

\textbf{Sequential.}
\(A = \{a_1, \ldots, a_N\}\), \(G\) is a directed path \(a_1 \to \cdots \to a_N\):
\(\mathcal{H}_{a_j}^{(r)} = \{\tau_a^{(r')}\}_{a \in A,\, r' < r}\) if \(j = 1\), else \(\mathcal{H}_{a_j}^{(r)} = \{\tau_{a_{j-1}}^{(r)}\}\); \(\hat{y} = \tau_{a_N}^{(R)}\).

\textbf{Centralized.}
\(A = A_{\text{work}} \cup \{a_o\}\), \(G\) is a star centered at orchestrator \(a_o\):
\(\mathcal{H}_{w_i}^{(r)} = \{\tau_{a_o}^{(r-1)}\}\), \(\mathcal{H}_{a_o}^{(r)} = \{\tau_{w_i}^{(r)}\}_{w_i \in A_{\text{work}}}\), \(\hat{y} = \tau_{a_o}^{(R)}\).

\textbf{Debate.}
\(A = \{a_1, \ldots, a_N\}\), \(G\) is a directed acyclic graph over \(A\) with total order \(a_1 \prec \cdots \prec a_N\), where each \(a_j\) observes all predecessors \(a_{j'}\) (\(j' < j\)) within the current round and all agents from prior rounds:
\(\mathcal{H}_{a_j}^{(r)} = \{\tau_{a_{j'}}^{(r)}\}_{j' < j} \cup \{\tau_a^{(r')}\}_{a \in A,\, r' < r}\); \(\hat{y}\) is determined by majority voting over \(\{\tau_a^{(R)}\}_{a \in A}\).

\textbf{Hybrid.}
\(A = A_{\text{work}} \cup \{a_o\}\) as in Centralized, with expanded worker context:
\(\mathcal{H}_{w_i}^{(r)} = \{\tau_{a_o}^{(r-1)}\} \cup \{\tau_w^{(r-1)}\}_{w \in A_{\text{work}}} \cup \{\tau_{w_j}^{(r)}\}_{j < i}\), \(\mathcal{H}_{a_o}^{(r)} = \{\tau_w^{(r)}\}_{w \in A_{\text{work}}}\), \(\hat{y} = \tau_{a_o}^{(R)}\).

\paragraph{Entropy Metrics.} Building on the established link between entropy dynamics and LLM reasoning~\citep{entro-duction-acl25,entropy-mechanism-neurips25}, we extend these insights to MAS through hierarchical entropy metrics. Specifically, let \(s_t^{(i,a,r)}\) denote the decoding state preceding the \(t\)-th token of agent \(a\) in round \(r\) for sample \(i\). The token-level entropy is \(H_t^{(i,a,r)} = -\sum_{v \in \mathcal{V}} \pi_\theta(v \mid s_t^{(i,a,r)}) \log \pi_\theta(v \mid s_t^{(i,a,r)})\). During experiments, we log \(H_t^{(i,a,r)}\), latency, and token costs for every agent to analyze hierarchical entropy dynamics.
\section{Exploring MAS with Entropy Dynamics}
\label{sec:exploring_uncertainty_dynamics}

While prior work investigates failure modes of MAS~\citep{why-mas-fail-neurips25,science-of-scaling-mas-arxiv25}, it relies on proprietary models and aggregate metrics, such as accuracy, latency, and cost, overlooking how entropy evolves within and across agents. We address this gap by analyzing entropy dynamics in open-source LLM-based MAS across diverse topologies and tasks.

\subsection{Evaluation Protocol}

Small, open-source LLMs enable cost-effective multi-agent collaboration through specialized task allocation and coordination~\citep{small-llm-tool-learner-emnlp24}. Their open access to token-level probabilities enables direct entropy computation, which is critical for agent decision-making. More discussion can be found in Appendix~\ref{app:model_rationale}. To this end, in contrast to prior work which evaluates only proprietary API-based models~\citep{science-of-scaling-mas-arxiv25}, our study focuses on publicly available LLMs, including the LLaMA series (3.1-8B-Instruct, 3.2-3B-Instruct)~\citep{llama3-arxiv24} and the Qwen3 series (0.6B, 4B, 8B)~\citep{qwen3-arxiv25}. We evaluate across six benchmarks: \texttt{GSM8K}~\citep{gsm8k-arxiv21}, \texttt{MATH500}~\citep{math500-neurips21}, \texttt{AIME2024}, and \texttt{AIME2025} for mathematics; \texttt{HumanEval}~\citep{humaneval-arxiv21} for code generation; and \texttt{MMLU}~\citep{mmlu-iclr21} for knowledge question-answering (Q\&A). All systems use the same $M_{\text{base}}$ with $R=2$ interaction rounds. Additional experimental details are provided in Appendix~\ref{app:experimental_details}. We further extend the evaluation to two agentic benchmarks: \texttt{GAIA}~\citep{gaia-iclr23} and \texttt{FinanceAgent}~\citep{FinanceAgentBench-arxiv25} in Appendix~\ref{app:agentic_task}.

\subsection{Measuring Reasoning with Entropy and Beyond}
\label{sec:entropy_features}

Entropy has proven effective for analyzing single-agent reasoning~\citep{eas-aaai25,entro-duction-acl25}. Recent work has even leveraged entropy-based features to train simple machine learning models that predict LLM correctness~\citep{entropy-sentinel-arxiv26}. However, this approach is limited to a single LLM and computes entropy only from the top-20 token probabilities, yielding just 11 entropy-related features. We extend entropy analysis to MAS by designing hierarchical entropy features that capture how entropy evolves across agents and rounds. Specifically, for each sample, we extract 254 features by aggregating the logged traces across all agents and rounds:

\textbf{Entropy features} (\(\mathcal{F}_{E}\), 239 features) measure entropy across hierarchical levels: 
\begin{itemize}[leftmargin=*,topsep=0pt,itemsep=-0.2em]
\item \textit{Agent-level statistics} capture per-agent reasoning trajectories, including their statistical properties, variations across rounds, and inter-agent entropy divergence;  
\item \textit{Round-level dynamics} track aggregate entropy metrics for each round and their relative changes;  
\item \textit{Sample-level statistics} aggregate entropy measures across all agents for a given sample;  
\item \textit{System-level aggregation} provides global entropy measures across different topologies;  
\item \textit{Base-model entropy} (\(\mathcal{F}_{\text{base-E}}\), 17 features) additionally captures entropy characteristics of \(M_{\text{base}}\) and quantifies shifts in entropy between \(M_{\text{base}}\) and MAS;
\end{itemize}

\textbf{Computational metrics} (\(\mathcal{F}_{C}\), 15 features) capture non-entropy-related quantities, including reasoning time, token usage, inference counts, and \(M_{\text{base}}\) correctness (\(\mathcal{F}_{\text{base-C}}\)), at the same hierarchical levels.

Excluding 9 experimental identifier columns yields 245 trainable features. We define three feature groups to isolate the influence of \(M_{\text{base}}\) on MAS performance: (1) \textbf{MAS only} (\(\mathcal{G}_{\text{MAS}}\), \(d = 224\)): \(\mathcal{G}_{\text{MAS}} = (\mathcal{F}_{E} \cup \mathcal{F}_{C}) \setminus (\mathcal{F}_{\text{base-E}} \cup \mathcal{F}_{\text{base-C}})\), capturing entropy dynamics intrinsic to multi-agent interaction; (2) \textbf{Base entropy} (\(\mathcal{G}_{\text{base-H}}\), \(d = 241\)): \(\mathcal{G}_{\text{base-H}} = \mathcal{G}_{\text{MAS}} \cup \mathcal{F}_{\text{base-E}}\), evaluating the impact of \(M_{\text{base}}\) entropy on MAS performance; (3) \textbf{Base full} (\(\mathcal{G}_{\text{base-full}}\), \(d = 245\)): \(\mathcal{G}_{\text{base-full}} = \mathcal{G}_{\text{base-H}} \cup \mathcal{F}_{\text{base-C}}\), directly measuring how \(M_{\text{base}}\)'s reasoning capability conditions or limits MAS effectiveness. Detailed feature descriptions, together with a discussion of feature redundancy, are provided in Appendix~\ref{app:entropy_features}.

\subsection{Mining Effectiveness of MAS}

MAS built on LLMs inherently exhibit uncertainty during individual reasoning, and their interactions among agents compound this effect~\citep{why-mas-fail-neurips25}. Prior work largely relies on simple metrics that fail to capture internal entropy dynamics or identify the factors underlying MAS failure.

To address this, we leverage the hierarchical entropy features defined in Section~\ref{sec:entropy_features} to reformulate MAS evaluation as a supervised learning problem: predicting per-sample correctness $y_i \in \{0,1\}$ from entropy traces $\mathbf{x}_i \in \mathbb{R}^d$, where $\mathbf{x}_i$ is the feature vector extracted from sample $i$'s MAS execution logs according to one of the three feature groups. We train an ensemble classifier, the \textit{Entropy Judger}, by averaging the predicted probabilities of XGBoost~\citep{xgboost-2016} and LightGBM~\citep{lightgbm-2017}, yielding a final prediction $f(\mathbf{x}_i) \in [0,1]$. This formulation enables data-driven discovery of fine-grained factors governing MAS effectiveness. Beyond binary prediction, the \textit{Entropy Judger} supports label-free pass@$k$ selection by choosing the candidate with the highest predicted correctness; details are in Appendix~\ref{app:entropy_judger}. To interpret learned patterns, we perform SHAP analysis~\citep{shap-neurips17} on both XGBoost and LightGBM. For each feature $j$, we extract two metrics: (1) \textbf{mean feature importance} $\bar{I}_j$, the average min-max-normalized importance across both models; and (2) \textbf{SHAP correlation} $\rho_j$, the average Pearson correlation between feature values and their SHAP attributions. A positive $\rho_j$ indicates that higher feature values increase predicted correctness. Formal definitions are in Appendix~\ref{app:shap_analysis}. To distinguish correlation from causation, we further conduct a causal inference analysis in Section~\ref{sec:causal_analysis} to validate these findings.

\begin{figure*}[t]
    \centering
    \includegraphics[width=\textwidth]{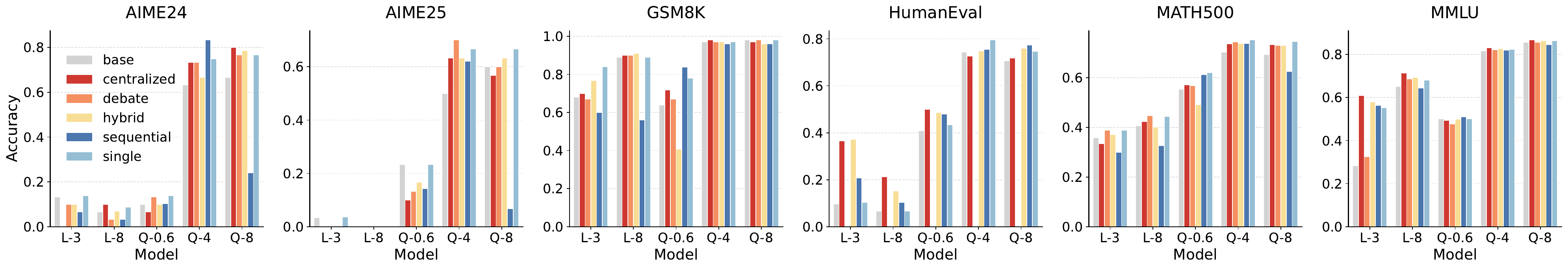}
    \caption{Accuracy comparison of SAS and MAS across models and datasets. For brevity, LLaMA-3.2-3B-Instruct and LLaMA-3.1-8B-Instruct are denoted as L-3 and L-8, while Qwen3-0.6B/4B/8B are denoted as Q-0.6/Q-4/Q-8. The \texttt{base} denotes the accuracy of a single $M_{\text{base}}$.}
    \label{fig:acc-comparison}
    \vspace{-1em}
\end{figure*}

\subsection{Examining Entropy Impacts on MAS}
\label{sec:examining_uncertainty_impacts}

\paragraph{MAS does not always outperform SAS.} Conventional assumptions suggest that more agents improve MAS performance~\citep{agent-forest-arxiv24}. However, consistent with recent findings~\citep{science-of-scaling-mas-arxiv25,sas-outperform-mas-token-budget-arxiv26}, we show that MAS does not universally surpass SAS, and we substantially extend this observation to open-source LLMs across a broader range of tasks. As shown in Figure~\ref{fig:acc-comparison}, SAS matches or exceeds the performance of at least one MAS topology in 26/30 cases across 5 models and 6 datasets. Specifically, SAS achieves the highest accuracy in 13 cases (43.3\%), surpassing the average MAS accuracy by 6.28\%, particularly on math tasks and with smaller models.

\begin{figure*}[t]
    \centering
    \includegraphics[width=\textwidth]{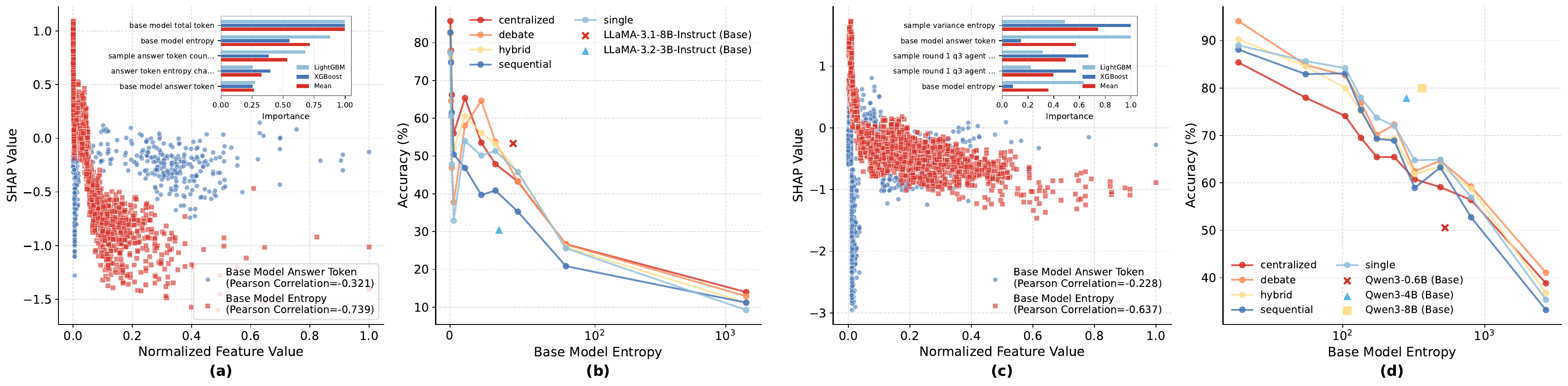}
    \caption{Base-model entropy limits MAS effectiveness. The left two subfigures show results for LLaMA; the right two for Qwen. (a) Relationship between feature values and SHAP values for the most important entropy features on \(\mathcal{G}_{\text{base-H}}\), sorted by \(\bar{I}_j\) and annotated with \(\rho_j\). (b) MAS performance across deciles of \(M_{\text{base}}\) entropy: \(M_{\text{base}}\) entropy is partitioned into ten equal-sized bins, and average MAS accuracy, aggregated over datasets and model sizes, is computed per bin. Additionally, the average \(M_{\text{base}}\) entropy and accuracy across all datasets are overlaid as markers.}
    \label{fig:base-model-analysis}
    \vspace{-1em}
\end{figure*}

\paragraph{Base model entropy limits MAS effectiveness.} Prior work shows MAS performance depends on \(M_{\text{base}}\) capability~\citep{metaagent-icml25}, a trend we also observe in Figure~\ref{fig:acc-comparison}; moreover, we find that \(M_{\text{base}}\) entropy further constrains MAS effectiveness. On \(\mathcal{G}_{\text{base-H}}\), the top predictors are \(M_{\text{base}}\)'s total token-level entropy and answer length: for LLaMA, total token count (\(\rho \approx -0.47\), \(\bar{I} = 1.0\)) and entropy (\(\rho \approx -0.73\), \(\bar{I} \approx 0.72\)); for Qwen3, answer token count (\(\rho \approx -0.18\), \(\bar{I} \approx 0.58\)) and entropy (\(\rho \approx -0.64\), \(\bar{I} \approx 0.36\)). Critically, higher \(M_{\text{base}}\) entropy consistently reduces MAS accuracy across both families, as shown in Figure~\ref{fig:base-model-analysis}b,d, with performance dropping sharply when entropy exceeds 100; additional results appear in Appendix~\ref{app:base_model_entropy}. Notably, despite this shared trend, entropy scales differ: LLaMA operates in low-entropy ranges (0-100) but achieves lower accuracy, while Qwen uses higher entropy (100-1,000) yet performs better. This reflects divergent reasoning styles, as Qwen verifies and refines its answers before finalizing, generating self-correcting trajectories that suppress error propagation in MAS despite higher entropy, whereas LLaMA tends to reuse others' answers without verification, leading to uncontrolled error propagation. We use an example question to illustrate this contrast in Appendix~\ref{app:model-comparison}. We treat entropy primarily as a predictive feature for MAS correctness rather than a universal measure of uncertainty; detailed calibration analyses showing its dependence on model family and task difficulty are provided in Appendix~\ref{app:calibration}. We further examine how RL fine-tuning reshapes this entropy-performance relationship in Appendix~\ref{app:rl_finetuned_base}.

\begin{figure*}[t]
    \centering
    \includegraphics[width=\textwidth]{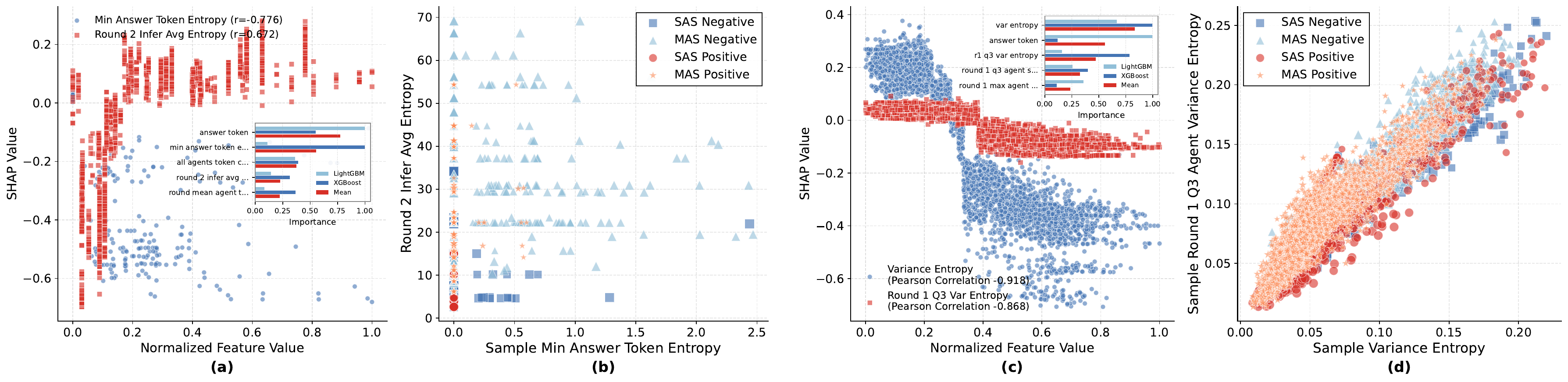}
    \caption{MAS mainly fails on inter-agent misalignment. The left two subfigures show results for LLaMA; the right two for Qwen. (a) Same as Figure~\ref{fig:base-model-analysis}a, but for entropy features in \(\mathcal{G}_{\text{MAS}}\). (b) Impact of these features on sample predicted correctness: we plot feature values against the average predicted probability of correctness from XGBoost and LightGBM.}
    \label{fig:mas-analysis}
    \vspace{-1em}
\end{figure*}

\paragraph{MAS mainly fails on inter-agent misalignment.} Prior work identifies inter-agent misalignment as a key cause of MAS failure~\citep{why-mas-fail-neurips25}. We deepen this understanding by analyzing fine-grained entropy dynamics within MAS. On \(\mathcal{G}_{\text{MAS}}\), as shown in Figure~\ref{fig:mas-analysis}, we examine the top entropy-related features ranked by \(\bar{I}\). For Qwen, failure is driven by high entropy variance across agents during problem solving (\(\rho \approx -0.92\), \(\bar{I} \approx 0.83\)) and strong agent disagreement in round 1 (\(\rho \approx -0.87\), \(\bar{I} \approx 0.47\)), indicating that early divergence leads to increasingly incompatible reasoning trajectories. Figure~\ref{fig:mas-analysis}d further shows that correctly solved MAS samples (\textit{MAS Positive}) cluster at low sample-level entropy variance and low round-1 per-agent entropy variance, whereas SAS solves samples correctly even at higher entropy variance, indicating that \textbf{MAS imposes a stricter entropy constraint than SAS for successful problem-solving}. In contrast, LLaMA failures are characterized by verbose and uncertain final answers, with answer-token count (\(\rho \approx -0.63\), \(\bar{I} \approx 0.78\)) and minimum answer-token entropy (\(\rho \approx -0.78\), \(\bar{I} \approx 0.56\)) as dominant predictors. Notably, LLaMA round-2 entropy positively correlates with correctness, unlike Qwen, which relies on early convergence. More analysis can be found in the Appendix~\ref{app:mas_failure}.

\section{Deep Analysis}
\label{sec:deep_analysis}

\subsection{Effective MAS Requires Stable Deliberation}

We categorize mathematical reasoning tasks by average MAS accuracy into easy (\texttt{GSM8K}), medium (\texttt{Math500}), and hard (\texttt{AIME24/25}). We then analyze entropy features in \(\mathcal{G}_{\text{MAS}}\) across these levels to understand how entropy shapes MAS effectiveness under varying task difficulty.

\textbf{Stable and Low Entropy for Simple Problems.} On \texttt{GSM8K}, the top features include final answer length and dispersion of early-round agent entropy (with \(\bar{I} \in [0.51, 0.56]\) and \(|\rho| \leq 0.15\)), indicating only a mild influence on predicted correctness. In contrast, overall round-1 entropy (\(\bar{I} \approx 0.47\), \(\rho \approx -0.64\)) and the stability index which measures consistency of entropy across agents (\(\bar{I} \approx 0.44\), \(\rho \approx -0.79\)), are both highly predictive and strongly negatively correlated with success. This suggests that \textbf{simple problems are solved when agents converge quickly to low-entropy, stable answers.}

\textbf{Balanced Exploration for Medium Problems.} On \texttt{Math500}, high average per-agent reasoning entropy (\(\bar{I} \approx 0.77, \rho \approx 0.63\)) and longer reasoning time in round 1 (\(\bar{I} \approx 0.15, \rho \approx 0.71\)) correlate positively with MAS success, suggesting that medium-difficulty problems benefit from sustained exploration with moderate entropy. Conversely, excessive early entropy, measured by maximum total entropy across agents in round 1 (\(\rho \approx -0.73\)), and verbose final answers (\(\rho \approx -0.48\)) strongly predict failure. Notably, median round-1 entropy also shows a negative correlation (\(\rho \approx -0.36\)), indicating that while some entropy aids discovery, uncontrolled divergence hinders consensus. The divergence between mean and median correlations suggests a right-skewed entropy distribution: moderate sustained entropy benefits MAS, whereas rare extreme spikes dominate the median signal and predict failure. Together, these findings show that \textbf{on medium-difficulty problems, MAS succeeds when agents explore long enough with moderate entropy, but avoid excessive early divergence, and converge on a concise answer.}

\textbf{Structured Deliberation for Hard Problems.} On \texttt{AIME24/25}, round-1 total reasoning time is the top predictor (\(\bar{I} = 1.0\), \(\rho \approx 0.73\)), confirming that olympiad-level problems require substantial early effort. Crucially, entropy-related features reveal a sharp trade-off: excessive early entropy harms performance, evident in strong negative correlations for round-1 max entropy (\(\rho \le -0.70\)) and per-token entropy (\(\rho \approx -0.37\)), while moderate average output entropy shows a positive correlation (\(\rho \ge 0.37\)). Entropy dispersion in later rounds also degrades accuracy, with high inter-agent variance in round-2 strongly predicting failure (\(|\rho| \ge 0.66\)). Together, these results show that \textbf{success on hard problems demands not just long reasoning, but controlled entropy: early exploration must be bounded, and inter-agent reasoning must remain aligned.}

\textbf{Stable Entropy for Code Generation.} On \texttt{HumanEval}, total reasoning time is the strongest predictor, and moderate average entropy improves performance, while both overconfident and erratic entropy profiles degrade it. This echoes the principle observed \textbf{in mathematical reasoning: deliberation benefits MAS only when entropy remains structured and stable.}

\textbf{Inter-Agent Agreement for Knowledge Q\&A.} On \texttt{MMLU}, MAS success hinges on inter-agent agreement rather than extended deliberation, unlike in math and code tasks, where longer reasoning improves performance. High entropy variance harms accuracy, and larger teams worsen consensus, confirming that \textbf{consensus, not duration, drives success in knowledge-intensive Q\&A.}

\begin{figure*}[t]
    \centering
    \includegraphics[width=\textwidth]{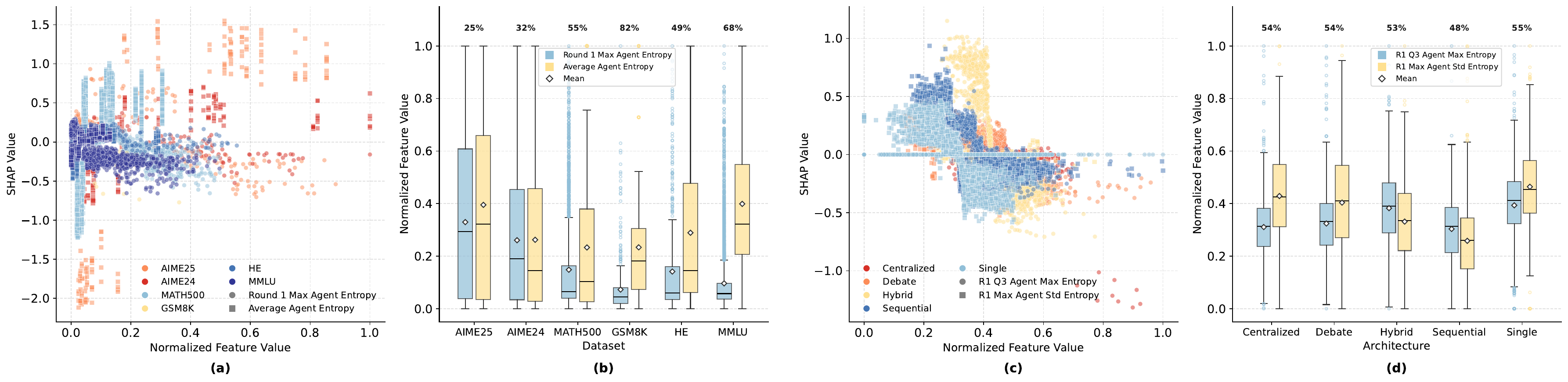}
    \caption{Entropy dynamics in MAS exerts distinct effects depending on task difficulty and the coordination architecture. (a, c) Feature-SHAP relationships for top entropy features in \(\mathcal{G}_{\text{MAS}}\), grouped by dataset (a) and architecture (c). (b, d) Corresponding box plots across all models, annotated with average MAS correctness per dataset (b) or per architecture (d).}
    \label{fig:dataset-arch-analysis}
    \vspace{-1em}
\end{figure*}

Overall, \textbf{entropy-performance relationships are task-aware}: simple tasks demand rapid convergence to low, stable entropy, whereas medium and hard tasks benefit from moderate average entropy but are harmed by peak or dispersed entropy. This reveals that harder problems require more exploratory reasoning, yet uncontrolled entropy spikes remain universally detrimental. Additionally, \textbf{extended reasoning consistently improves MAS performance on hard tasks}, with trajectory length strongly predictive of success. This aligns with findings in LLM reinforcement learning, where longer reasoning chains reflect stronger underlying reasoning capabilities~\citep{longcot-iclr25}. 

Beyond per-dataset analyses, we examine the two most important entropy features across all models and datasets: round-1 maximum agent entropy and average entropy of agents' outputs during problem solving. Figure~\ref{fig:dataset-arch-analysis}a shows that harder tasks exhibit wider SHAP value distributions and benefit from moderate average entropy, whereas high early entropy consistently harms performance. In contrast, easier datasets gain little from higher entropy. Figure~\ref{fig:dataset-arch-analysis}b further reveals that as dataset accuracy declines, from 82\% on \texttt{GSM8K} to 25\% on \texttt{AIME25}, both entropy features increase in magnitude and dispersion, with round-1 max entropy showing the strongest sensitivity to task difficulty. This suggests that difficult problems not only induce higher entropy, but also amplify inter-agent disagreement, making early entropy control increasingly critical as task complexity grows. Finally, these results show that \textbf{MAS succeeds by reasoning more while keeping entropy low and consistent across agents}. We provide further analysis in Appendix~\ref{app:diff_tasks}, and visualize token-level entropy trajectories across different model families and datasets in Appendix~\ref{app:case_study}. These patterns generalize to agentic tasks with external tools: tool-call entropy negatively predicts accuracy (mirroring sample-level entropy in reasoning), while step-level features show that the \emph{first reasoning step} dominates the round-1 predictive signal, thus underscoring the critical role of early-stage entropy control (Appendix~\ref{app:agentic_task}).

\subsection{Peak Entropy Is Universally Harmful in MAS}
\label{sec:arch_entropy_analysis}

We analyze how entropy features influence MAS effectiveness across five architectures on \(\mathcal{G}_{\text{MAS}}\). \textbf{Centralized} systems are highly sensitive to early-round entropy: peak agent entropy in round 1 and peak answer entropy strongly predict failure, as the orchestrator's single-context aggregation allows any erratic agent to contaminate the entire reasoning process. \textbf{Debate} architectures depend critically on early consensus: high initial agent divergence reflects how initial divergence amplifies across rounds, preventing convergence; yet once aligned, cumulative entropy becomes beneficial, indicating productive exploration. \textbf{Hybrid} systems balance robustness and depth: early peak entropy remains harmful, but dual feedback from peers and orchestrator enables recovery through extended deliberation. \textbf{Sequential} systems are most fragile: answer-level entropy dominates the top predictors, reflecting error propagation through strict role chaining with no cross-checking. \textbf{Single (SAS)} prioritize brevity and penalize both high answer entropy and entropy variance, indicating that internal consistency is critical for success.

\textbf{Architecture determines which entropy matters.} Aggregation-based systems (centralized, debate) fail on inter-agent dispersion, as noisy inputs corrupt shared contexts. Sequential systems fail on answer-level entropy, where specialized roles propagate errors without cross-verification. Hybrid systems are most robust, reconciling conflicts through dual feedback. \textbf{Universally, peak entropy harms, while cumulative entropy helps}, showing that MAS succeeds by shaping entropy distribution, not eliminating entropy. This mirrors recent findings in multi-turn agent training~\citep{epo-arxiv25}, where bounding policy entropy variance within historical averages, rather than maximizing absolute entropy, is what stabilizes learning and prevents collapse, suggesting the principle that \emph{entropy stability}, not absolute entropy level, governs effective deliberation, which we extend from single-agent training to multi-agent inference. 

Beyond per-architecture analysis, we compare two key features across all architectures: upper-quartile agent peak entropy and maximum entropy dispersion across agents in round 1. Figure~\ref{fig:dataset-arch-analysis}c shows that both negatively predict correctness within each architecture. Surprisingly, Figure~\ref{fig:dataset-arch-analysis}d reveals an inverse trend across architectures: sequential (lowest-accuracy) shows the lowest feature averages, while single (highest-accuracy) shows the highest. This indicates that \textbf{the relationship between entropy and performance depends on architectural capacity to control entropy, not merely on minimizing entropy}. Further details are provided in Appendix~\ref{app:diff_tasks}. Causal discovery in Section~\ref{sec:causal_analysis} subsequently confirms that maximum answer-token entropy is a consensus direct cause of MAS correctness, elevating the peak-entropy harm finding from correlation to causation.

\subsection{More Rounds Are Not Always Better}
\label{sec:round}

\begin{figure*}[t]
    \centering
    \includegraphics[width=\textwidth]{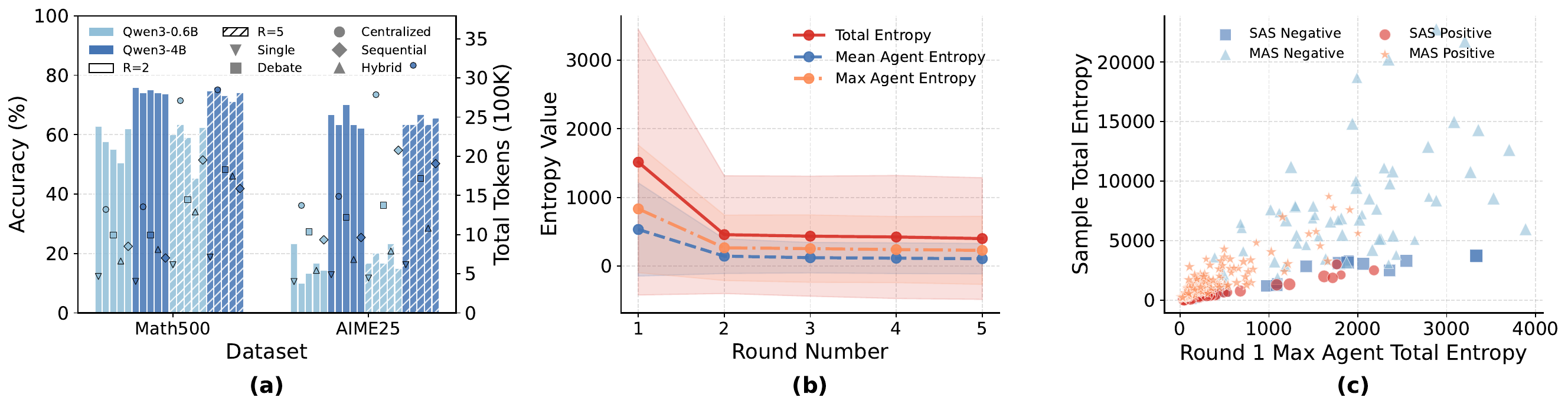}
    \caption{More rounds do not necessarily improve MAS performance. (a) Accuracy and token consumption for different MAS architectures with $R=2$ and $R=5$ on two benchmarks. (b) Evolution of three key entropy metrics across rounds. (c) The impact of two prominent entropy features, notable for their high importance (\(\bar{I}\)) and strong correlation (\(|\rho|\)) with sample correctness.}
    \label{fig:round-analysis}
    \vspace{-1em}
\end{figure*}

To investigate whether additional rounds improve performance, we extend our analysis from fixed $R=2$ to $R=5$ using Qwen3-0.6/4B on \texttt{MATH500} and \texttt{AIME2025}, expanding the feature space from 224 to 494 dimensions. Comparing \(R=2\) and \(R=5\), we find that \textbf{extending deliberation rarely improves performance and often harms it}, even at the cost of higher token consumption, as shown in Figure~\ref{fig:round-analysis}a. On challenging benchmarks like \texttt{AIME25} and \texttt{MATH500}, most architectures, including debate and hybrid, exhibit performance degradation with additional rounds, especially for smaller models. The only consistent gains occur in centralized systems, where strong orchestration enables effective aggregation over longer trajectories. In contrast, peer-based architectures (debate, hybrid) appear to suffer from prolonged disagreement, as repeated interactions without convergence amplify noise rather than refine reasoning. This is further supported by entropy dynamics: as Figure~\ref{fig:round-analysis}b shows, key entropy metrics, maximum, mean, and total entropy, drop sharply from round 1 to round 2, but remain nearly flat from round 2 to round 5, indicating that agents largely stabilize after the second round. Together, these results demonstrate that simply increasing the number of rounds does not enhance MAS performance; instead, \textbf{the benefit of extended deliberation depends critically on an architecture's ability to align agents and stabilize entropy early in the process}.

Despite the expanded feature space with \(R=5\), early-round entropy remains the dominant failure mode, as shown in Figure~\ref{fig:round-analysis}c. Round-1 features dominate the top predictors: peak cumulative agent entropy in round 1 ranks second ($\bar{I} \approx 0.60$, $\rho \approx -0.91$), underscoring the critical role of early consensus. Even cumulative sample-level entropy is strongly harmful ($\rho \approx -0.73$), reinforcing that uncontrolled entropy, not just its timing, is detrimental. These results confirm a central principle: \textbf{MAS effectiveness is largely determined in the first round, and additional deliberation cannot reliably recover from initial misalignment}. See Appendix~\ref{app:more_rounds} for details. Causal discovery in Section~\ref{sec:causal_analysis} directly supports this: round-1 total entropy is a consensus direct cause of MAS correctness across both PC and FCI algorithms, confirming that first-round entropy has a genuine causal footprint on the outcome rather than being merely predictive. In addition, controlled experiments in Appendix~\ref{app:sas-mas-comparison} confirm that inter-agent interaction rarely yields genuine accuracy improvements, further validating that MAS outcomes are predominantly fixed by round-1 dynamics rather than subsequent deliberation.

\subsection{Entropy Causally Drives MAS Correctness}
\label{sec:causal_analysis}

\begin{figure*}[t]
    \centering
    \begin{subfigure}[t]{0.34\textwidth}
        \centering
        \includegraphics[width=\textwidth]{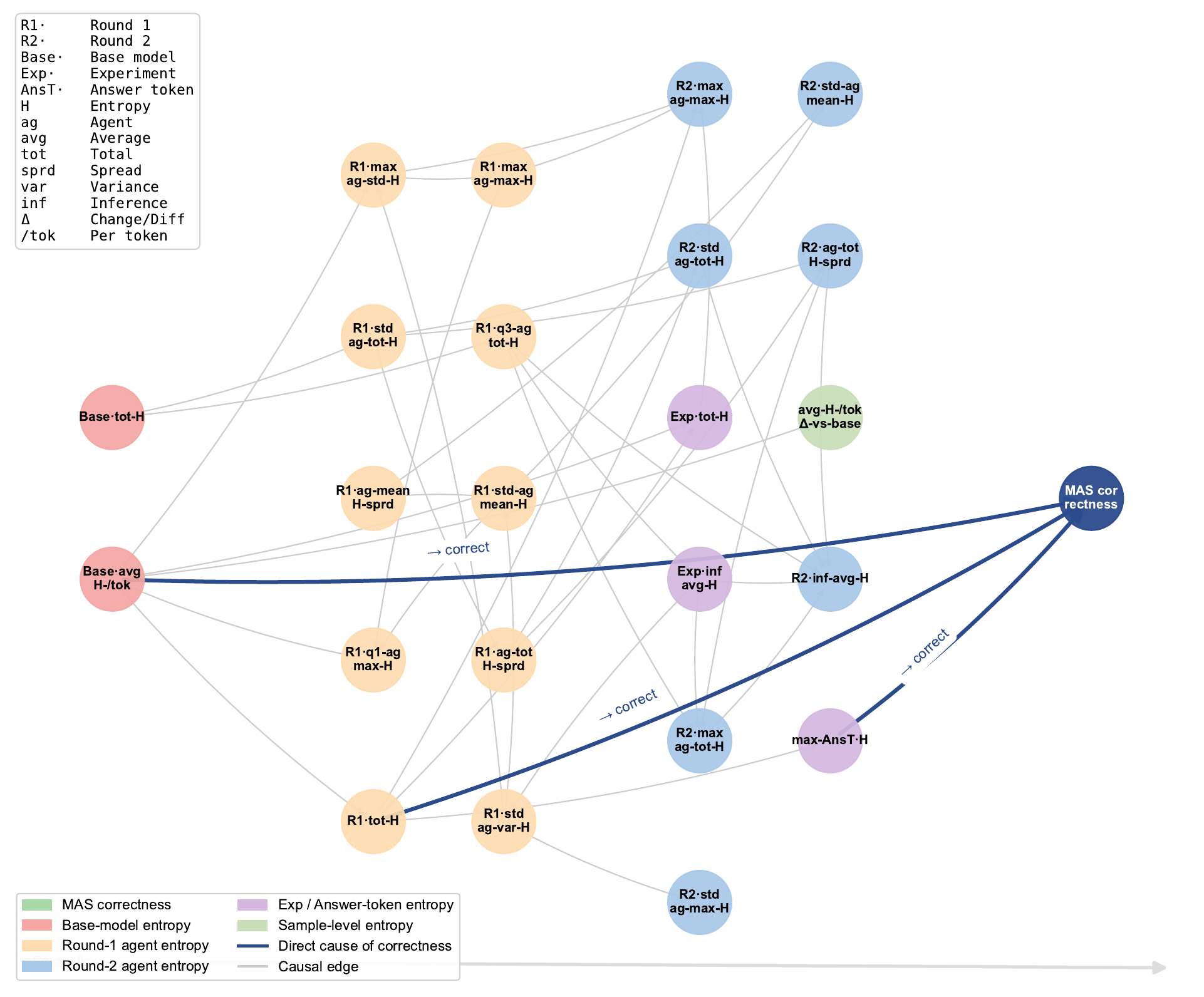}
        \caption{}
        \label{fig:causal-dag-main}
    \end{subfigure}
    \hfill
    \begin{subfigure}[t]{0.32\textwidth}
        \centering
        \includegraphics[width=\textwidth]{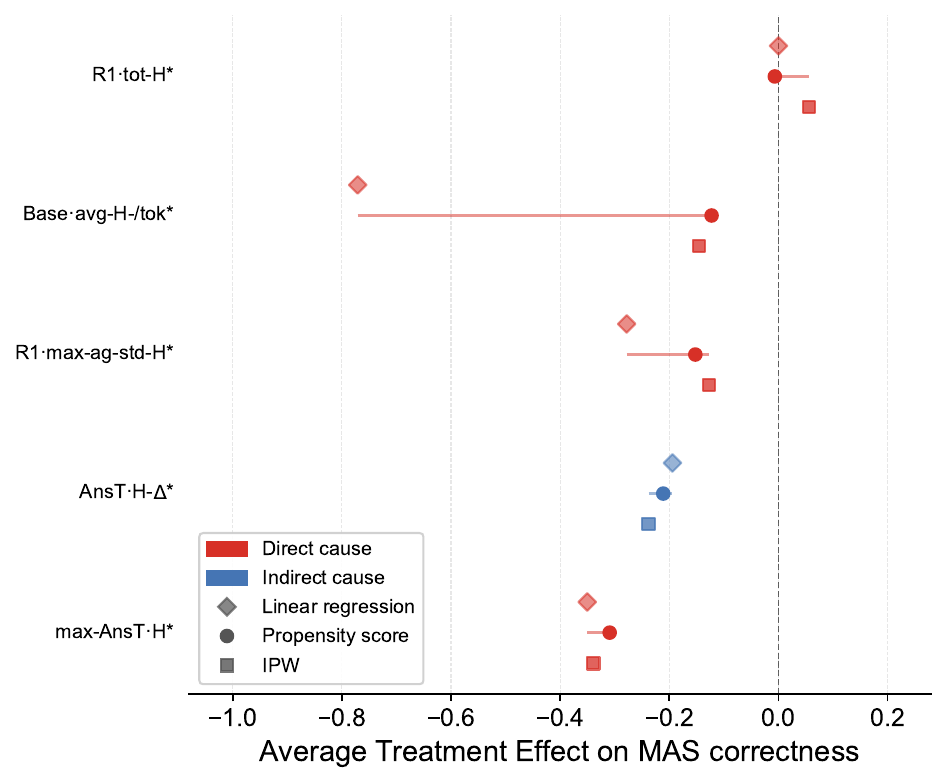}
        \caption{}
        \label{fig:causal-ate-forest}
    \end{subfigure}
    \hfill
    \begin{subfigure}[t]{0.32\textwidth}
        \centering
        \includegraphics[width=\textwidth]{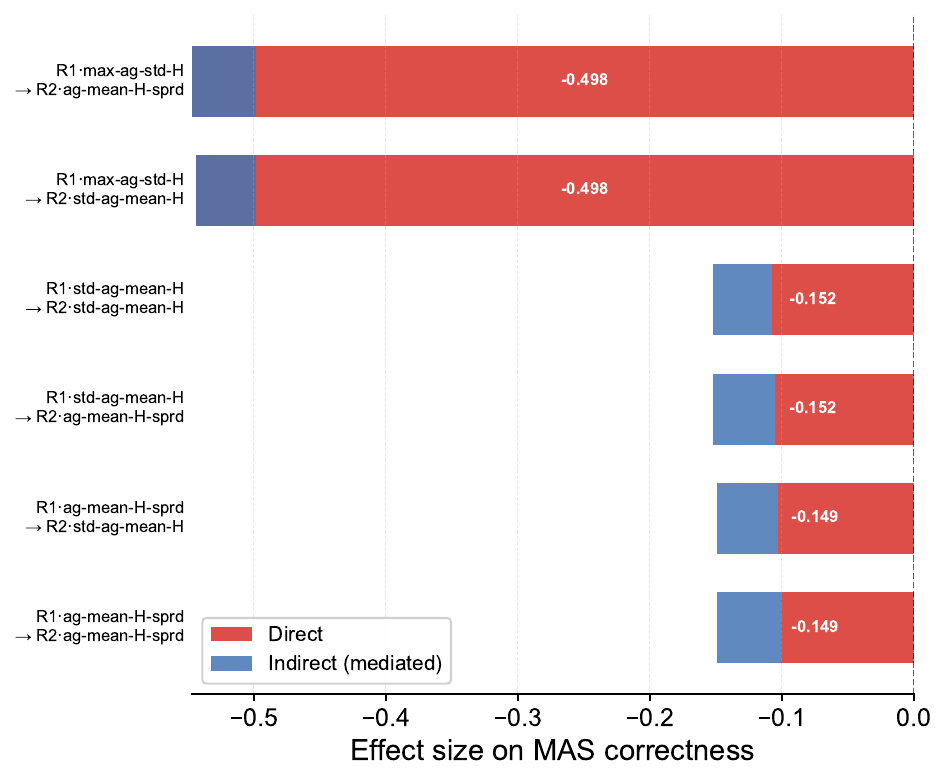}
        \caption{}
        \label{fig:causal-mediation-main}
    \end{subfigure}
    \caption{Causal analysis of entropy and MAS correctness. (a) Consensus causal DAG (PC $\cap$ FCI): base-model entropy, round-1 total entropy, and maximum answer-token entropy are the three consensus direct causes of MAS correctness; other entropy signals act through a layered cascade from round-1 to round-2 agent entropy. (b) ATE forest plot (LR / PS / IPW, sorted by PS-ATE): consensus direct causes (red) show tighter estimator spread than indirect causes (blue). (c) Mediation decomposition of significant round-1 paths (direct $c'$ in red, indirect $a{\times}b$ in blue).}
    \label{fig:causal-analysis-main}
    \vspace{-1em}
\end{figure*}

To elevate SHAP-based correlational findings to causal mechanisms, we employ PC and FCI algorithms with temporal constraints, followed by DoWhy effect estimation. \textbf{Both algorithms identify three consensus direct causal factors of MAS correctness}: base-model average per-token entropy ($\text{ATE}_\text{PS} = -0.12$, $p < 10^{-21}$), round-1 total entropy ($p < 10^{-19}$), and maximum answer-token entropy ($\text{ATE}_\text{PS} = -0.31$, $p < 10^{-28}$), all refutation tests pass; the consensus DAG is shown in Figure~\ref{fig:causal-dag-main}. Round-1 maximum agent entropy dispersion is additionally a direct cause under PC but not FCI, suggesting a possible latent confounder (e.g., problem difficulty). Propensity-based estimators (PS/IPW) serve as the primary quantitative evidence; for base-model entropy they yield $-0.12$ to $-0.15$, and for maximum answer-token entropy $-0.31$ to $-0.34$, confirming consistent negative causal effects. Round-1 total entropy shows near-zero propensity-based effects (PS $= -0.007$, IPW $= +0.056$), consistent with it acting primarily through downstream entropy nodes rather than as an independent linear driver. \textbf{Across all estimators, no treatment variable shows a robustly positive effect} (Figure~\ref{fig:causal-ate-forest}), confirming that higher entropy does not causally benefit MAS correctness.

Mediation analysis further reveals that \textbf{round-1 inter-agent entropy dispersion transmits 30--33\% of its causal effect on correctness through round-2 entropy} (Figure~\ref{fig:causal-mediation-main}), causally supporting the first-round dominance finding: early misalignment compounds into the subsequent round rather than self-correcting. Base-model entropy exerts its causal effect primarily through direct pathways rather than through sample-level mediators, indicating that base-model uncertainty directly shapes the multi-agent output distribution. Together, these confirm that \textbf{entropy is a causal driver of MAS performance, operating through hierarchical, multi-round mechanisms}. Full details in Appendix~\ref{app:causal-discovery}. Beyond this global analysis, per-finding causal validations are provided in Appendix~\ref{app:experimental_results}.

\subsection{Entropy Predicts MAS Correctness}

The \textit{Entropy Judger} achieves high cross-validation accuracy using only MAS-derived entropy features: 72.6\% / 79.1\% (LLaMA / Qwen) on $\mathcal{G}_{\text{MAS}}$, rising to 74.5\% / 80.7\% when base-model entropy is added ($\mathcal{G}_{\text{base-H}}$), and 81.2\% / 91.6\% with base-model correctness included ($\mathcal{G}_{\text{base-full}}$). This demonstrates that entropy dynamics alone are highly predictive of correctness, and that $M_{\text{base}}$ further influence MAS effectiveness. Beyond binary prediction, the \textit{Entropy Judger} enables label-free pass@$k$ selection and consistently improves accuracy across all MAS configurations; see details in Appendix~\ref{app:entropy_judger}.
\section{Conclusion}
\label{sec:conclusion}

This study presents a comprehensive analysis of entropy dynamics in LLM-based MAS, examining 245 entropy features across six reasoning tasks, two agentic tasks, and four MAS topologies. Our findings challenge prevailing assumptions: single agents outperform MAS in 43.3\% of cases, and MAS effectiveness is largely determined by first-round entropy dynamics. We identify three principles governing MAS performance: (1) \textit{Certainty Preference}, where peak entropy directly harms and stable low entropy directly benefits MAS correctness; (2) \textit{Base Entropy}, where lower base model entropy is a direct causal driver of MAS performance; and (3) \textit{Task Awareness}, where optimal entropy profiles vary by task difficulty. These findings extend to agentic settings: tool-call entropy and first-round inter-agent dispersion jointly constrain MAS performance. Building on these insights, the \textit{Entropy Judger} leverages learned entropy patterns to select high-quality outputs from MAS pass@$k$ candidates, achieving consistent accuracy gains without ground-truth labels.

\bibliography{reference}
\bibliographystyle{iclr2026_conference}

\newpage
\appendix
\onecolumn
\addtocontents{toc}{\protect\setcounter{tocdepth}{2}}
\renewcommand{\contentsname}{Appendix Contents}
\tableofcontents
\newpage

\section{Rationale for Open-Source Small LLMs}
\label{app:model_rationale}

Our study exclusively employs open-source small LLMs from the LLaMA and Qwen3 series, rather than proprietary API-based models. This choice is driven not merely by practical considerations but by the necessity of accessing internal probability distributions for entropy-based analysis.

\paragraph{Full Probability Access.}
Proprietary APIs typically return only generated text or at most top-$k$ logprobs with $k \leq 20$~\citep{entropy-sentinel-arxiv26}. This truncation prevents accurate entropy computation, which requires the complete token-level probability distribution $P(x_t \mid x_{<t})$ over the full vocabulary. Open-weight models provide full access to these probabilities, enabling the 245-dimensional hierarchical entropy features that span token, agent, and round levels. Without this access, our core entropy analysis would be impossible.

\paragraph{Reproducibility and API Instability.}
Proprietary models often undergo silent updates that change their behavior without notice, making longitudinal studies unreliable. In contrast, open-source models with fixed weights guarantee that our experimental protocol can be exactly replicated. This stability is essential for scientific validity, as entropy dynamics measured today must match those observed in future reproductions.

\paragraph{Cost-Effective Multi-Agent Scaling.}
Exploring diverse MAS configurations across five architectures, five models, six datasets, and multiple rounds requires thousands of LLM calls. Using proprietary APIs would incur prohibitive costs, whereas local inference with small models makes comprehensive evaluation feasible. Furthermore, multiple specialized small agents can match or even exceed the performance of a single large model through task decomposition~\citep{small-llm-tool-learner-emnlp24}.

\paragraph{Complementing Prior Work.}
Existing MAS studies rely on proprietary models and report only aggregate metrics such as accuracy, latency, or cost, thereby overlooking internal entropy dynamics. Our focus on open-source models reveals how entropy evolves within and across agents, providing mechanistic insights that remain inaccessible under API-only evaluation. This approach complements rather than duplicates prior findings.

\section{Experimental Details}
\label{app:experimental_details}

\subsection{Evaluation}
\label{app:evaluation}

\paragraph{Models and Configurations.}
We evaluate five open-source LLMs on reasoning benchmarks: LLaMA-3.1-8B-Instruct, LLaMA-3.2-3B-Instruct~\citep{llama3-arxiv24}, and Qwen3-0.6B, Qwen3-4B, Qwen3-8B~\citep{qwen3-arxiv25}. All systems are built using the same $M_{\text{base}}$, with the number of interaction rounds fixed to $R=2$, temperature set to 0.6, top-$p$ to 0.95, and maximum sequence length to 8,192 tokens. On the challenging \texttt{AIME24/25} benchmarks, the maximum sequence length is set to 16,384 tokens. For Qwen3 models, thinking mode is enabled by default (\texttt{enable\_thinking=True}), which activates extended chain-of-thought reasoning prior to the final response; the thinking tokens are included in the entropy computation. We validate the robustness of our findings across varying temperatures (Appendix~\ref{app:temperature-ablation}) and verify their generalizability to 14B-parameter models (Appendices~\ref{app:model-size} and \ref{app:gaia}).

\paragraph{Benchmarks.}
We evaluate on six reasoning benchmarks across three task types: for mathematical reasoning, \texttt{GSM8K}~\citep{gsm8k-arxiv21}, \texttt{MATH500}~\citep{math500-neurips21}, \texttt{AIME2024}, and \texttt{AIME2025}; for code generation, \texttt{HumanEval}~\citep{humaneval-arxiv21}; and for knowledge question-answering (Q\&A), \texttt{MMLU}~\citep{mmlu-iclr21}. Each question is accompanied by a chain-of-thought prompt. We further evaluate on two agentic benchmarks: \texttt{GAIA}~\citep{gaia-iclr23} (165 questions, full validation split) and \texttt{FinanceAgent}~\citep{FinanceAgentBench-arxiv25}; see Appendix~\ref{app:agentic_task} for details.

\paragraph{Evaluation and Infrastructure.}
For mathematics and \texttt{MMLU}, models enclose final answers in \texttt{\textbackslash boxed\{\}}, and correctness is verified using the \textit{Math Verify} tool~\cite{math-verify} or exact string matching. For \texttt{HumanEval}, code blocks are extracted from markdown output and validated by executing the provided test cases with a 10-second timeout per sample. We exclude debate architecture on \texttt{HumanEval}, as majority voting is generally ineffective for code generation tasks. For agentic benchmarks, each agent operates in a ReAct loop; \texttt{GAIA} uses the official answer-matching protocol, and \texttt{FinanceAgent} uses exact answer matching against ground-truth financial metrics. All experiments are conducted on four RTX 5090 GPUs.

\subsection{MAS Architecture Details}
\label{app:mas_details}

All architectures use LangGraph for workflow orchestration. Each architecture runs for $R=2$ rounds by default. Table~\ref{tab:mas_comparison} summarizes the key differences.

\paragraph{Single Agent (SAS).}
A single \texttt{SingleSolver} agent processes the input and iteratively refines its answer across rounds. Each round receives the accumulated history from previous rounds. LLM calls: $R \times 1$.

\paragraph{Sequential.}
Four specialized agents form a pipeline: \texttt{Planner} $\rightarrow$ \texttt{Solver} $\rightarrow$ \texttt{Critic} $\rightarrow$ \texttt{Judger}. The planner generates step-by-step instructions (no calculations), the solver executes the plan, the critic reviews and identifies errors, and the judger produces the final answer. The planner (first agent) receives the accumulated outputs of all agents from prior rounds; each subsequent agent receives only its immediate predecessor's output within the current round. LLM calls: $R \times 4$.

\paragraph{Centralized.}
Three domain experts (\texttt{MathAgent}, \texttt{ScienceAgent}, \texttt{CodeAgent}) execute in parallel, and an \texttt{OrchestratorAgent} aggregates their outputs. In rounds $r < R$, the orchestrator provides feedback to all workers; in round $R$, it produces the final answer. LLM calls: $R \times 3 + R \times 1 = R \times 4$.

\paragraph{Debate.}
Three debate agents (\texttt{Agent1}, \texttt{Agent2}, \texttt{Agent3}) execute sequentially. Each agent observes all prior agents' outputs from both current and previous rounds. The final answer is determined by majority voting over \texttt{\textbackslash boxed\{\}} extractions, without additional LLM inference. LLM calls: $R \times 3$.

\paragraph{Hybrid.}
Combines centralized and debate structures. Workers receive both orchestrator feedback and peer outputs, enabling dual feedback channels. The orchestrator provides guidance while workers can observe how peers interpret that guidance. LLM calls: $R \times 3 + R \times 1 = R \times 4$.

\begin{table*}[ht]
\centering
\caption{Comparison of MAS architectures. $N$: number of worker agents; $R$: communication rounds.}
\label{tab:mas_comparison}
\footnotesize
\begin{tabular}{@{}lcccp{2.2cm}p{4.5cm}@{}}
\toprule
\textbf{Architecture} & \textbf{$N$} & \textbf{LLM Calls} & \textbf{Orchestrator} & \textbf{Decision Rule} & \textbf{Agent Roles \& Functions} \\
\midrule
Single (SAS) & 1 & $R$ & None & Last round output &
SingleSolver: solves and iteratively refines answer \\
\midrule
Sequential & 4 & $4R$ & None & Judger output &
Planner: generates plans; Solver: executes plans; Critic: reviews solutions; Judger: outputs final answer \\
\midrule
Centralized & 3+1 & $4R$ & LLM-based & Orchestrator &
Math / Science / CodeAgent: domain-specific reasoning; Orchestrator: aggregates feedback and outputs final answer \\
\midrule
Debate & 3 & $3R$ & Voting & Majority vote &
Agent 1--3: independent solvers observing all prior outputs \\
\midrule
Hybrid & 3+1 & $4R$ & LLM-based & Orchestrator &
Math / Science / CodeAgent: domain-specific reasoning; Orchestrator: aggregates + peer feedback \\
\bottomrule
\end{tabular}
\end{table*}

\subsection{Agent Prompts}

The prompts for the Sequential architecture are detailed in Appendix~\ref{app:model-comparison}. Here we present the prompts for the Centralized architecture on mathematical reasoning tasks. Prompts for code generation and knowledge Q\&A follow similar patterns and can be found in the source code.

\paragraph{First-Layer Expert Agents.} Each expert agent receives a system prompt defining its role and a user prompt containing the question. All agents may receive orchestrator feedback from previous rounds.

\textbf{MathAgent:}
\begin{itemize}[leftmargin=*,topsep=2pt,itemsep=0pt]
    \item \textit{System}: ``You are the MathAgent. Solve the given question with clear steps. Your input may include feedback from the Orchestrator from the previous round.''
    \item \textit{User}: ``Question: \{question\} Provide a concise mathematical solution, showing key steps.''
\end{itemize}

\textbf{ScienceAgent:}
\begin{itemize}[leftmargin=*,topsep=2pt,itemsep=0pt]
    \item \textit{System}: ``You are the ScienceAgent. Analyze and solve the given question with scientific reasoning. Your input may include feedback from the Orchestrator from the previous round.''
    \item \textit{User}: ``Question: \{question\} Explain your scientific reasoning and provide a final result.''
\end{itemize}

\textbf{CodeAgent:}
\begin{itemize}[leftmargin=*,topsep=2pt,itemsep=0pt]
    \item \textit{System}: ``You are the CodeAgent. Provide a self-contained Python function that solves the problem. Your input may include feedback from the Orchestrator from the previous round.''
    \item \textit{User}: ``Question: \{question\} Write a single self-contained Python function in a markdown code block that solves the problem.''
\end{itemize}

\paragraph{Orchestrator Agent.} The orchestrator operates in two modes:

\textbf{Feedback Mode (intermediate rounds):}
\begin{itemize}[leftmargin=*,topsep=2pt,itemsep=0pt]
    \item \textit{System}: ``You are the Orchestrator Agent. Your task is to review the solutions provided by the first-layer agents in the current round. Analyze the provided solutions, identify any issues or areas for improvement, and provide constructive feedback. You may rewrite content, provide specific feedback, and offer improvement suggestions as needed. Your feedback will be used by the agents in the next round to improve their solutions.''
    \item \textit{User}: ``Question: \{question\} Here are the solutions from the expert agents in the current round: === Solutions === \{block\} === Solutions === Review these solutions and provide feedback for the next round. If corrections are needed, specify the issues and suggest improvements. If the solutions are satisfactory, acknowledge them and provide guidance for further refinement.''
\end{itemize}

\textbf{Aggregation Mode (final round):}
\begin{itemize}[leftmargin=*,topsep=2pt,itemsep=0pt]
    \item \textit{System}: ``You are the Orchestrator Agent. Your task is to aggregate the solutions provided by the first-layer agents and produce a final answer wrapped in \textbackslash boxed\{\}.''
    \item \textit{User}: ``Question: \{question\} Here are the solutions from the expert agents: === Solutions === \{block\} === Solutions === Based on these inputs, provide the final answer wrapped in \textbackslash boxed\{\}.''
\end{itemize}

\subsection{Data Mining}
\label{app:shap_analysis}

We employ an ensemble of XGBoost and LightGBM rather than a single model to obtain more robust and stable feature importance estimates. These two algorithms use different tree construction strategies: XGBoost grows trees level-wise, whereas LightGBM grows trees leaf-wise. By averaging their attributions, we capture a broader range of feature interactions and reduce variance due to correlated features or random data splits, yielding rankings that better reflect true relevance.

For each feature $j$, we compute two key metrics. The \textbf{mean feature importance} $\bar{I}_j$ is obtained by: (1) extracting raw feature importances $I_j^{(m)}$ from each model $m \in \{\text{XGB}, \text{LGB}\}$; (2) applying min-max normalization to obtain $\tilde{I}_j^{(m)} = (I_j^{(m)} - \min_k I_k^{(m)}) / (\max_k I_k^{(m)} - \min_k I_k^{(m)}) \in [0,1]$; and (3) averaging: $\bar{I}_j = \frac{1}{2}(\tilde{I}_j^{\text{XGB}} + \tilde{I}_j^{\text{LGB}})$. The \textbf{SHAP correlation} $\rho_j$ is computed as the average Pearson correlation between feature values $\mathbf{x}_j = (x_{1j}, \ldots, x_{nj})$ and their SHAP attributions $\boldsymbol{\phi}_j^{(m)} = (\phi_{1j}^{(m)}, \ldots, \phi_{nj}^{(m)})$ across both models: $\rho_j = \frac{1}{2}\sum_{m \in \mathcal{M}} \text{corr}(\mathbf{x}_j, \boldsymbol{\phi}_j^{(m)})$. The magnitude $|\rho_j|$ quantifies the strength of feature $j$'s influence on predicted correctness, while $\text{sign}(\rho_j)$ indicates its direction.

\section{Entropy Features}
\label{app:entropy_features}

We design a hierarchical feature set to capture entropy dynamics across agents and rounds in MAS. This section provides formal definitions for all 254 features used in our analysis.

\subsection{Feature Hierarchy}

Our features are organized into four hierarchical levels, reflecting the nested structure of MAS execution:

\paragraph{Token Level.}
For each generated token $t$ in agent $a$'s output, we compute Shannon entropy from the softmax distribution over vocabulary $\mathcal{V}$: \(
    H_t = -\sum_{v \in \mathcal{V}} p(v|x_{<t}) \log p(v|x_{<t})
\).

\paragraph{Agent Level.}
For each agent $a \in A$ in round $r$, we aggregate token-level entropy into summary statistics: total entropy $H_a^{(r)} = \sum_t H_t$, and distributional measures (mean, max, min, std, variance, median, Q1, Q3).

\paragraph{Round Level.}
For each round $r \in \{1, \ldots, R\}$, we aggregate agent-level statistics across all agents: total round entropy $\tilde{H}^{(r)} = \sum_{a \in A} H_a^{(r)}$ and mean round entropy $\bar{H}^{(r)} = \tilde{H}^{(r)} / |A|$.

\paragraph{Sample Level.}
For each input sample $i$, we aggregate round-level statistics and compute cross-round dynamics.

\subsection{Feature Groups}

We organize features into semantically coherent groups for analysis.

\paragraph{Entropy Features ($\mathcal{F}_E$, 239 features).} Hierarchical structure:

\paragraph{Agent-level statistics (156 features)} capture per-agent reasoning trajectories:
\begin{itemize}[leftmargin=*,topsep=0pt,itemsep=-0.2em]
    \item Per-round agent entropy: For each round $r$ and statistic $s \in \{\text{max}, \text{mean}, \text{std}, \ldots\}$, we compute $s(\{H_a^{(r)}\}_{a \in A})$, yielding features like \texttt{sample\_round\_1\_max\_agent\_total\_entropy}.
    \item Inter-agent divergence: Variance and coefficient of variation across agents within each round.
\end{itemize}

\paragraph{Round-level dynamics (27 features)} track temporal evolution:
\begin{itemize}[leftmargin=*,topsep=0pt,itemsep=-0.2em]
    \item Round totals: Total entropy and token count per round.
    \item Cross-round changes: First-to-last difference $\Delta H = \tilde{H}^{(R)} - \tilde{H}^{(1)}$, ratio $\tilde{H}^{(R)}/\tilde{H}^{(1)}$, and slope per round.
    \item Volatility: Standard deviation of entropy across rounds.
\end{itemize}

\paragraph{Sample-level statistics (29 features)} aggregate across the full MAS execution:
\begin{itemize}[leftmargin=*,topsep=0pt,itemsep=-0.2em]
    \item Basic statistics: $\sum_r \sum_a H_a^{(r)}$, mean, max, min, std, variance, quartiles.
    \item Distribution shape: Range, IQR, Bowley skewness $(Q_3 + Q_1 - 2 \cdot \text{median}) / \text{IQR}$, coefficient of variation $\sigma/\mu$, tail weight $(\max - Q_3)/\text{IQR}$.
    \item Stability index: $1 - \sigma_H / \mu_H$, measuring consistency across agents.
    \item Answer token entropy: Statistics computed over tokens in the final \texttt{\textbackslash boxed\{\}} output.
\end{itemize}

\paragraph{System-level aggregation (10 features)} provides global measures:
\begin{itemize}[leftmargin=*,topsep=0pt,itemsep=-0.2em]
    \item Architecture-specific: Number of agents, total inference count.
    \item Experiment totals: Aggregate entropy, average entropy per inference.
\end{itemize}

\paragraph{Base-model entropy ($\mathcal{F}_{\text{base-E}}$, 17 features)} captures the single-agent baseline:
\begin{itemize}[leftmargin=*,topsep=0pt,itemsep=-0.2em]
    \item Base model statistics: Total token-level entropy $H_{\text{base}}$, token count, average entropy per token of $M_{\text{base}}$.
    \item Comparison between MAS and base model: Entropy ratio $H_{\text{MAS}}/H_{\text{base}}$, reduction $H_{\text{base}} - H_{\text{MAS}}$.
    \item Answer entropy shift: Difference and ratio between base model and MAS final answer entropy.
\end{itemize}

\paragraph{Computational Metrics ($\mathcal{F}_C$, 15 features).}
Non-entropy quantities at the same hierarchical levels:
\begin{itemize}[leftmargin=*,topsep=0pt,itemsep=-0.2em]
    \item Timing: Total reasoning time, per-round time.
    \item Token usage: Total tokens generated, per-agent token count, answer length.
    \item Inference counts: Number of LLM calls per round and total.
    \item Base-model correctness ($\mathcal{F}_{\text{base-C}}$, 4 features): Whether $M_{\text{base}}$ answered correctly, format compliance.
\end{itemize}

\subsection{Key Feature Definitions}

Table~\ref{tab:key_features} lists representative features from each category with their formal definitions.

\begin{table}[ht]
\centering
\caption{Representative entropy features and their definitions.}
\label{tab:key_features}
\footnotesize
\begin{tabular}{@{}lll@{}}
\toprule
\textbf{Feature} & \textbf{Level} & \textbf{Definition} \\
\midrule
\small
\texttt{sample\_total\_entropy} & Sample & $\sum_{r,a} H_a^{(r)}$ \\
\texttt{sample\_entropy\_stability\_index} & Sample & $1 - \sigma_H / \mu_H$ \\
\texttt{sample\_entropy\_cv} & Sample & $\sigma_H / \mu_H$ (coefficient of variation) \\
\texttt{sample\_entropy\_bowley\_skewness} & Sample & $(Q_3 + Q_1 - 2 \cdot \text{med}) / \text{IQR}$ \\
\texttt{sample\_max\_answer\_token\_entropy} & Sample & $\max_t H_t$ for $t \in$ answer tokens \\
\midrule
\texttt{round\_1\_total\_entropy} & Round & $\sum_{a \in A} H_a^{(1)}$ \\
\texttt{round\_1\_2\_change\_entropy} & Round & $\sum_a H_a^{(2)} - \sum_a H_a^{(1)}$ \\
\midrule
\texttt{sample\_round\_1\_max\_agent\_max\_entropy} & Agent & $\max_{a \in A} (\max_t H_t^{(a,1)})$ \\
\texttt{sample\_round\_1\_mean\_agent\_std\_entropy} & Agent & $\frac{1}{|A|} \sum_{a} \sigma(H_t^{(a,1)})$ \\
\texttt{sample\_round\_1\_variance\_agent\_total\_entropy} & Agent & $\text{Var}(\{H_a^{(1)}\}_{a \in A})$ \\
\midrule
\texttt{base\_sample\_total\_entropy} & Base & $H_{\text{base}}$, trajectory-level entropy of $M_{\text{base}}$ \\
\texttt{sample\_entropy\_ratio\_vs\_base\_total} & Base & Entropy ratio $H_{\text{MAS}} / H_{\text{base}}$ \\
\bottomrule
\end{tabular}
\end{table}

\subsection{Feature Redundancy Discussion}
\label{app:feature-redundancy}

Our goal is to comprehensively explore which aspects of entropy influence MAS effectiveness, rather than to identify a minimal predictor set. Achieving broad coverage requires summarizing the same token-level entropy signal through related statistics (mean and median, std and variance, quartiles and skewness) and pairing each base-model feature with its MAS counterpart, which inevitably introduces correlation among features. We acknowledge this cost because different statistics expose different regimes: the median is robust to outlier tokens while the maximum captures peak entropy, and running SHAP over the full set reveals which one matters in which context.

To verify that the resulting redundancy does not compromise our findings, we characterize it with four methods: pairwise correlation, principal component analysis, recursive feature elimination, and cross-method importance validation. The analyses below support two complementary conclusions. First, the redundancy does not undermine our predictive or interpretive results. The 183:1 sample-to-feature ratio (44{,}780 to 245), tree-ensemble regularization, and consistent top features across four independent importance methods together show that the conclusions in Section~\ref{sec:examining_uncertainty_impacts} are stable under correlated inputs. Second, redundancy does matter for causal inference, where collinearity biases structure learning and confounds effect estimation. We therefore prune redundancy in our causal pipeline, applying Borda-fusion feature selection followed by a $|\rho| > 0.85$ threshold to obtain 28 non-redundant features for PC and FCI graph discovery and DoWhy ATE estimation (Appendix~\ref{app:causal-discovery}). The exploratory and causal feature sets thus play complementary roles: we keep redundancy where it aids exploration and remove it where it would distort causal claims.

\subsubsection{Pairwise Correlation Analysis}


To characterize feature redundancy, we analyze the pairwise Pearson correlation matrix of $\mathcal{G}_{\text{MAS}}$. Consistent with expectations, highly correlated features form distinct blocks (e.g., \texttt{sample\_mean\_*} with \texttt{sample\_median\_*}; \texttt{std\_*} with \texttt{variance\_*}). Rather than discarding these as redundant, we leverage this structure to observe which specific statistic SHAP prioritizes within each block, thereby offering finer-grained interpretability than selecting a single representative feature. More notably, base-model entropy features show strong associations with MAS entropy dynamics. For example, \texttt{base\_model\_min\_answer\_token\_entropy} is correlated with \texttt{answer\_token\_entropy\_change\_direction}, and \texttt{base\_model\_is\_finally\_correct} is correlated with \texttt{is\_finally\_correct}. These cross-tier correlations directly reinforce our finding in Section~\ref{sec:examining_uncertainty_impacts} that $M_{\text{base}}$ entropy and correctness condition MAS effectiveness. In addition, \texttt{architecture} exhibits strong correlations with numerous entropy features ($|\rho| > 0.5$), confirming that different MAS topologies induce distinct entropy dynamics, consistent with Section~\ref{sec:arch_entropy_analysis}. The correlation map also informs the $|\rho| > 0.85$ pruning threshold subsequently applied in our causal pipeline (Appendix~\ref{app:causal-discovery}).

\subsubsection{PCA Analysis}

\begin{figure}[t]
\centering
\begin{subfigure}{0.62\columnwidth}
\centering
\includegraphics[width=\textwidth]{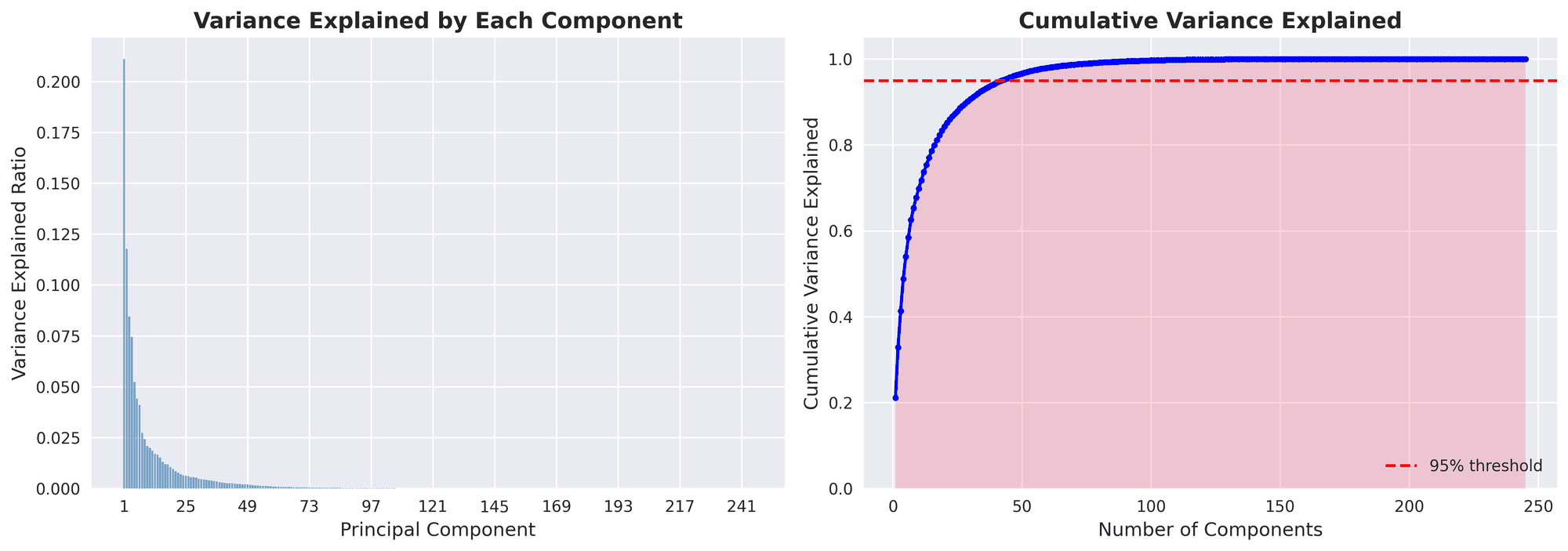}
\caption{}
\label{fig:pca-analysis}
\end{subfigure}
\hfill
\begin{subfigure}{0.36\columnwidth}
\centering
\includegraphics[width=\textwidth]{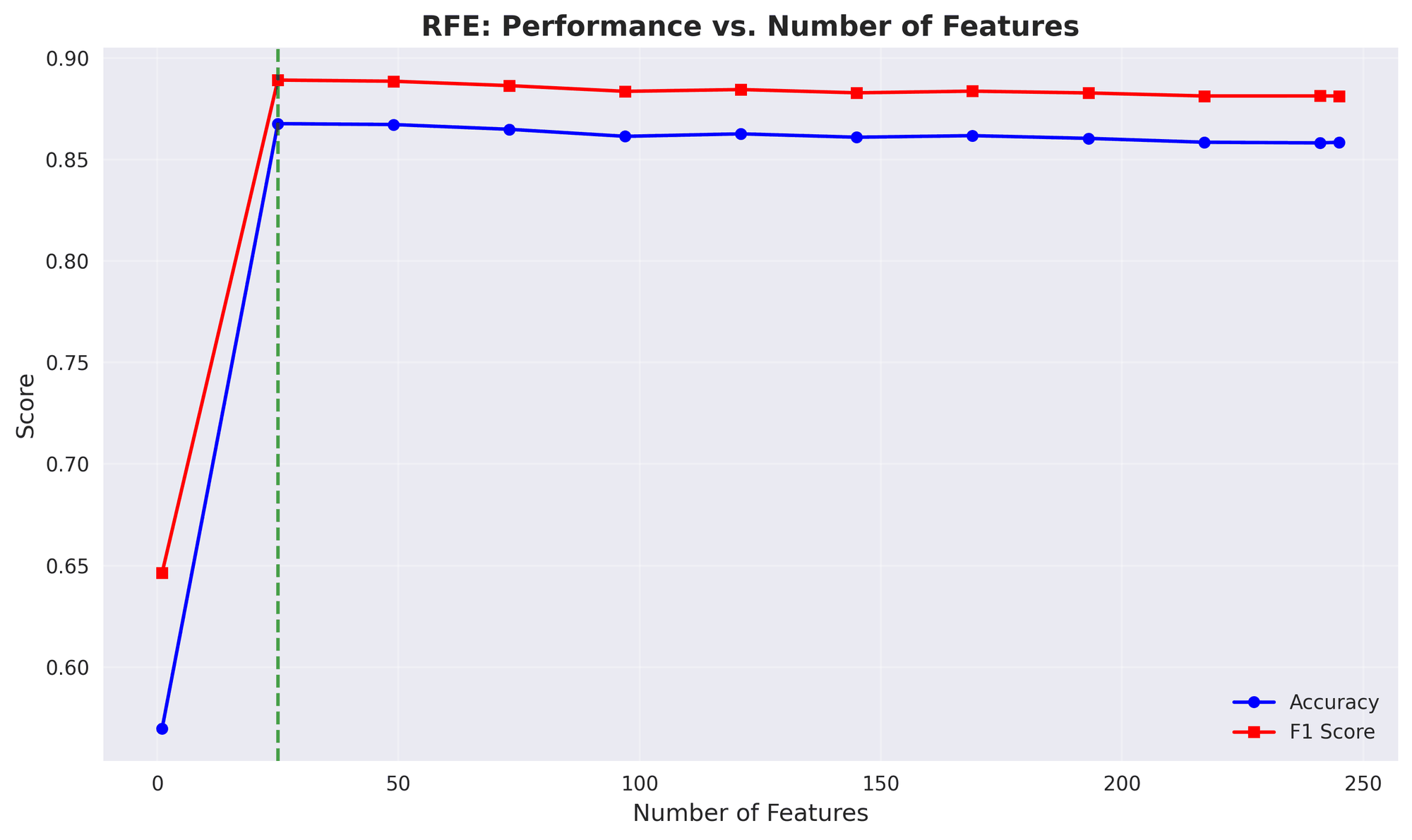}
\caption{}
\label{fig:feature-ablation}
\end{subfigure}
\caption{PCA analysis and feature ablation study. (a) PCA variance explained for the 245-dimensional entropy feature space. The curve shows cumulative explained variance as a function of the number of principal components. Achieving 95\% explained variance requires 43 components, indicating that information is distributed across many dimensions rather than concentrated in a few dominant directions. (b) Recursive feature elimination (RFE) performance curve. The optimal subset of 25 features achieves the highest accuracy (86.77\%) and F1 (88.92\%). Performance plateaus and slightly decreases as more features are added, indicating that the model does not overfit to the high-dimensional space.}
\end{figure}

We apply principal component analysis to the standardized 245-dimensional feature space computed over 44{,}780 samples. As shown in Figure~\ref{fig:pca-analysis}, the eigenvalue spectrum reveals a distributed variance structure. The first principal component explains 21.1\% of total variance, the top three components account for 41.3\%, the top five for 54.0\%, and 43 components are required to reach 95\% cumulative explained variance, an 82.4\% dimensionality reduction.

We draw two operational implications, neither of which contradicts the existence of definitional redundancy. First, the absence of a small dominant subspace (the top three components explain only about 41\%) indicates that the underlying entropy signal is spread across many orthogonal directions. Even where individual features overlap, the feature set as a whole captures diverse aspects of MAS entropy rather than a single dominant axis. Second, although 43 components suffice for 95\% variance, replacing the raw features with this PCA representation incurs a non-trivial classification cost. XGBoost F1 drops from 89.16\% to 86.13\% ($\Delta$F1 = $-$3.03\%), LightGBM from 89.10\% to 86.03\% ($\Delta$F1 = $-$3.07\%), and Random Forest from 88.13\% to 86.06\% ($\Delta$F1 = $-$2.07\%). The raw features therefore carry complementary nonlinear information that a linear projection erases. This is why we retain the full set for the Entropy Judger and reserve targeted redundancy removal for the causal analysis, where collinearity is genuinely harmful.

\subsubsection{Feature Ablation Study}

To assess how much unique predictive content the redundant feature set carries, we perform recursive feature elimination (RFE) using Random Forest as the base estimator with 5-fold cross-validation. Figure~\ref{fig:feature-ablation} presents the performance trajectory as features are progressively removed.

The results reveal a characteristic plateau pattern. A compact subset of 25 features achieves near-optimal performance (accuracy 86.77\%, F1 88.92\%), and adding the remaining 220 features only marginally changes performance (full-set accuracy 85.85\%, F1 88.13\%). Performance does not continuously increase with more features. It slightly decreases beyond the optimal 25, consistent with a regularized model that absorbs correlated inputs without overfitting. This plateau is consistent with the redundancy we explicitly designed into the feature set. A small core of features carries most of the predictive signal, while the remainder provides interpretive coverage across statistics and hierarchical levels at no cost to predictive performance. The same observation directly motivates the targeted feature selection used in our causal pipeline (Appendix~\ref{app:causal-discovery}), where keeping the redundant tail would distort structure learning rather than aid interpretation.

We further examine individual feature contributions through leave-one-out ablation from the full 245-feature set. Removing \texttt{base\_sample\_token\_count} causes the largest single-feature accuracy drop ($\Delta$Acc = $-$0.0163, $\Delta$F1 = $-$0.0142), followed by \texttt{sample\_answer\_token\_count} ($\Delta$Acc = $-$0.0151, $\Delta$F1 = $-$0.0125). Even within a high-dimensional space with correlated features, removing specific features still produces measurable degradation, indicating that the redundancy is partial rather than total. Each feature retains some unique residual signal beyond its correlated neighbors.

The favorable sample-to-feature ratio of our dataset (44{,}780 samples and 245 features, about 183 to 1) substantially exceeds conventional overfitting thresholds (typically 10:1 to 20:1), and the 5-fold cross-validation standard deviation remains below 0.02 across all settings. Together with the plateau behavior above, this confirms that retaining the full redundant feature set does not destabilize the predictive conclusions reported in Section~\ref{sec:examining_uncertainty_impacts}.

\subsubsection{Cross-Method Feature Importance Validation}

A potential concern with SHAP-based feature importance is that correlated features may cause importance to be distributed unreliably across redundant groups, yielding rankings that depend on the choice of estimator. To check whether our interpretive conclusions are robust to this effect, we compare feature importance rankings produced by four independent methods with fundamentally different assumptions: tree-based importance (Random Forest), logistic regression coefficients, chi-square statistical association, and mutual information.

\begin{figure}[t]
\centering
\includegraphics[width=0.9\linewidth]{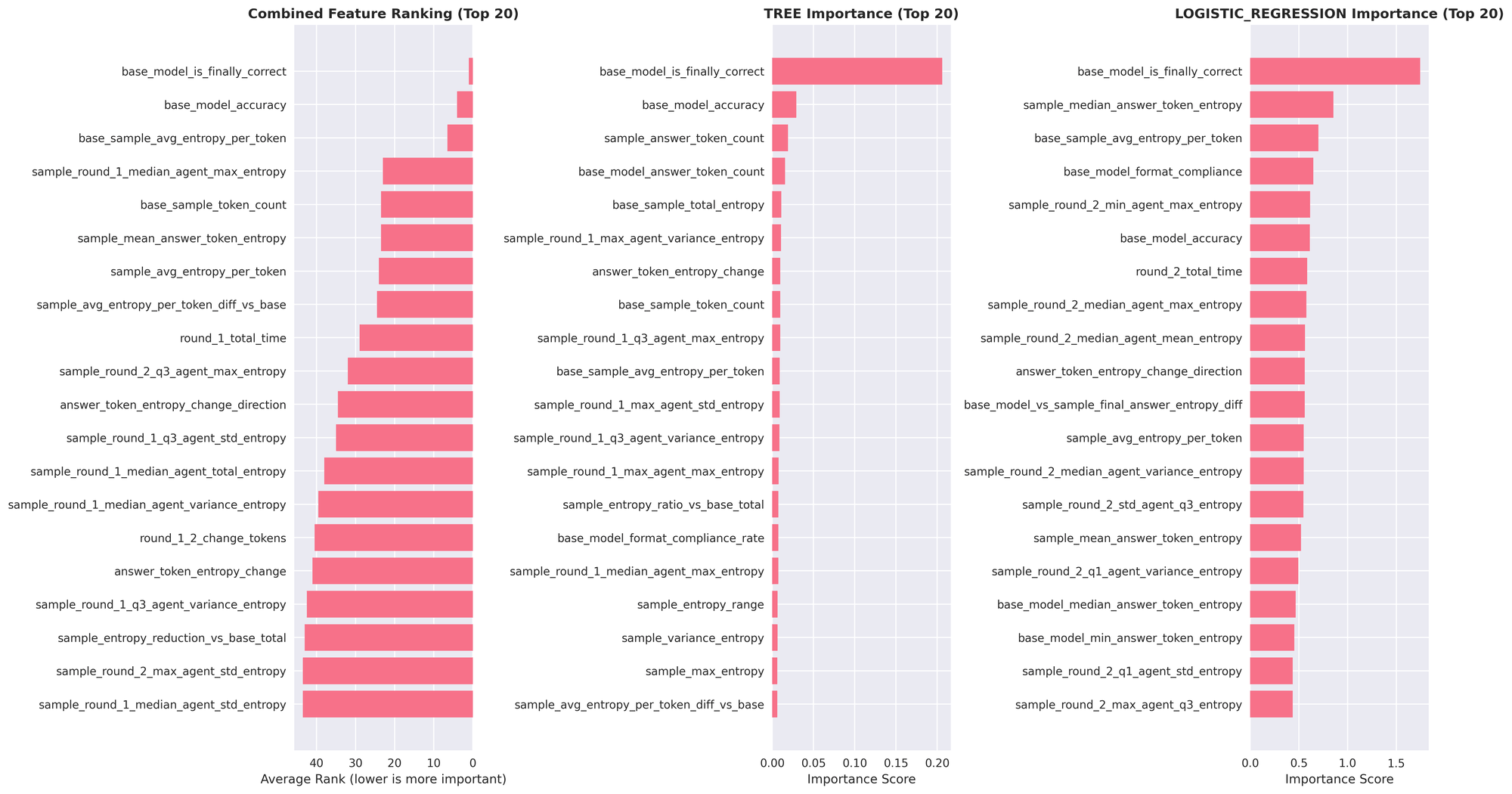}
\caption{Cross-method feature importance comparison across tree-based (Random Forest), logistic regression, chi-square, mutual information, and F-statistic methods. Despite fundamentally different mechanisms, the methods produce consistent top-feature rankings, validating the robustness of our feature importance findings.}
\label{fig:cross-method-importance}
\end{figure}

As shown in Figure~\ref{fig:cross-method-importance}, the methods produce remarkably consistent top-feature rankings despite their distinct mechanisms. The feature \texttt{base\_model\_is\_finally\_correct} ranks first in tree-based importance, logistic regression, chi-square, and F-statistic methods, and third in mutual information. Averaging tree-based and logistic regression ranks gives a top-5 of \texttt{base\_model\_is\_finally\_correct} (avg.\ rank 1.0), \texttt{base\_model\_accuracy} (4.0), \texttt{base\_sample\_avg\_entropy\_per\_token} (6.5), \texttt{sample\_round\_1\_median\_agent\_max\_entropy} (23.0), and \texttt{base\_sample\_token\_count} (23.5).

These top-5 features span three distinct categories: base-model correctness indicators (ranks 1 and 2), base-model entropy characteristics (ranks 3 and 5), and round-level agent entropy statistics (rank 4). This category coverage matters for our argument. Since the same high-importance signals appear across base-model quality, single-agent entropy, and multi-agent coordination, the conclusions in Section~\ref{sec:examining_uncertainty_impacts} cannot be artifacts of one correlated cluster dominating a single ranker.

The methods do reveal complementary secondary emphases. Tree models prioritize correctness indicators and answer token counts, logistic regression highlights answer entropy and format compliance, chi-square emphasizes quartile entropy statistics, and mutual information favors total entropy and round-level features. This divergence is exactly what the redundant feature design is meant to enable. When several correlated features encode overlapping signals, different importance methods surface different members of each block, and retaining the full set lets us read off these complementary perspectives instead of committing to a single representative chosen up front.

\section{Causal Discovery and Effect Estimation}
\label{app:causal-discovery}

Appendix~\ref{app:entropy_features} characterized the redundancy structure of the 245-dimensional entropy feature set and noted that, while this redundancy is benign for SHAP-based exploration, it would distort causal structure learning and effect estimation. This appendix takes the next step: we apply Borda-fusion feature selection with a $|\rho| > 0.85$ pruning threshold to obtain a 28-feature non-redundant subset, then run constraint-based causal discovery (PC and FCI) and the DoWhy framework on the resulting variables to identify direct causal factors and mediation pathways for MAS correctness across 44{,}780 samples. The pipeline is the global counterpart of the per-finding causal validations summarized in Appendix~\ref{app:experimental_results}, and it is complemented from a different angle by the controlled three-way SAS/MAS-Round~1/MAS-Round~2 experiment in Appendix~\ref{app:sas-mas-comparison}, which isolates the role-assignment and inter-agent-interaction contributions to entropy dynamics rather than estimating the effect of entropy on correctness.

\subsection{Feature Selection Pipeline}
\label{app:causal-discovery-feature}

The original feature space comprises 245 token-level entropy features. To ensure identifiability and interpretability for causal discovery, we apply a multi-method fusion pipeline that reduces dimensionality by 88.6\% while preserving coverage of all hierarchical feature levels.

\paragraph{Ranking Methods.}
Four complementary methods contribute to a weighted reciprocal-rank (Borda Count) fusion:
\begin{itemize}[nosep,leftmargin=1.5em]
    \item \textbf{Combined Tree + Logistic Regression importance} (weight $= 3.0$): captures both non-linear and linear predictive relevance.
    \item \textbf{Mutual Information} (weight $= 1.5$): measures general statistical dependence with the outcome.
    \item \textbf{Chi-squared test} (weight $= 1.0$): assesses categorical association strength.
    \item \textbf{ANOVA F-test} (weight $= 1.0$): evaluates between-group variance for continuous features.
\end{itemize}

\begin{figure}[h]
\centering
\begin{minipage}[c]{0.48\linewidth}
\centering
\includegraphics[width=\linewidth]{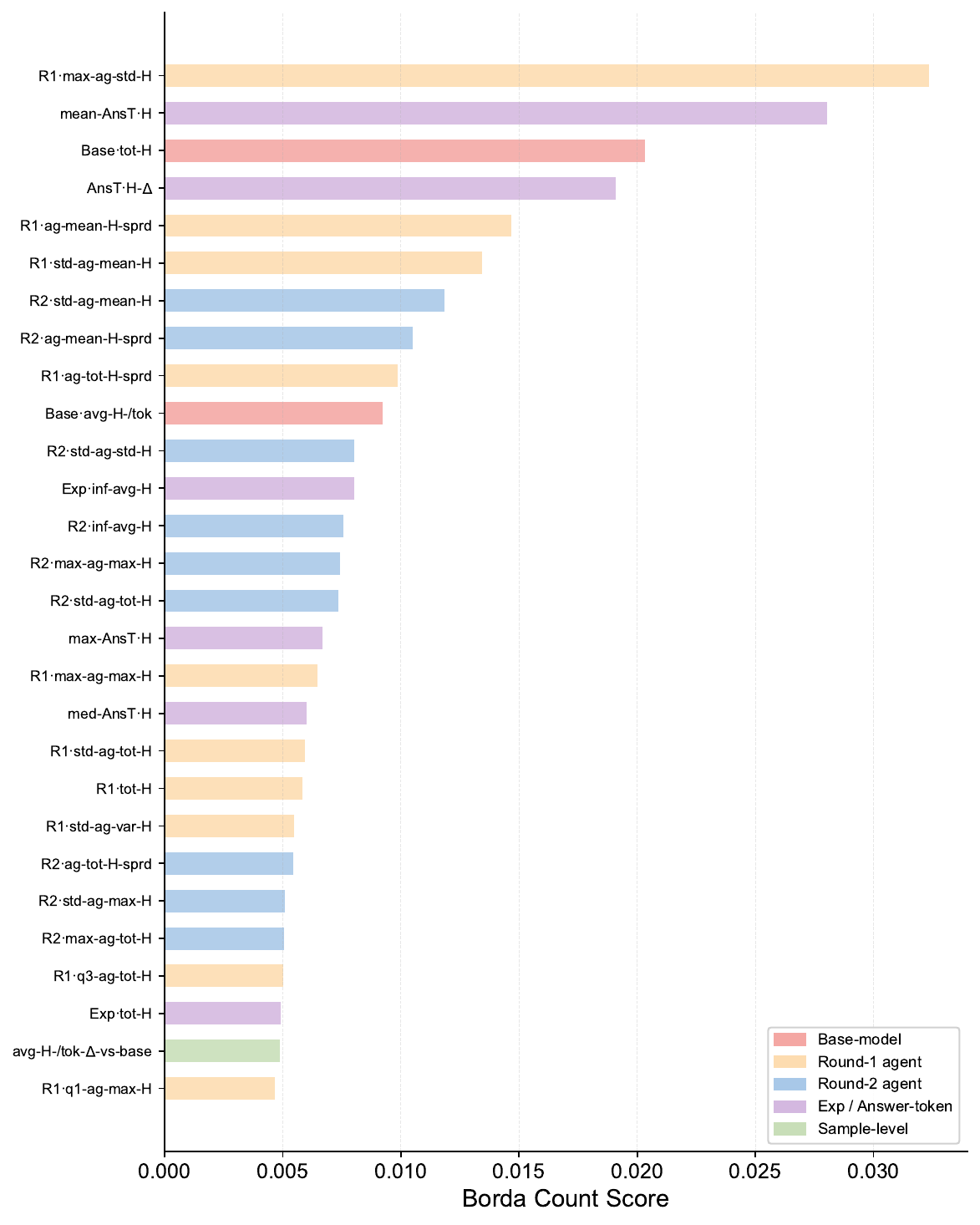}
\captionof{figure}{Borda fusion scores for the 28 selected features, colored by hierarchical tier. Features are sorted by score; abbreviated names follow the conventions used in the main text.}
\label{fig:causal-borda-scores}
\end{minipage}
\hfill
\begin{minipage}[c]{0.48\linewidth}
\centering
\captionof{table}{Selected features ranked by Borda score, grouped by hierarchical feature level.}
\label{tab:causal-features}
\vspace{2pt}
\tiny
\begin{tabularx}{\linewidth}{@{}r X r@{}}
\toprule
\textbf{Rank} & \textbf{Feature} & \textbf{Score} \\
\midrule
\multicolumn{3}{@{}l}{\textit{Base-Model Entropy $\mathcal{F}_{\text{base-E}}$ (2 features)}} \\
3  & \texttt{base\_sample\_total\_entropy}                    & 0.0203 \\
10 & \texttt{base\_sample\_avg\_entropy\_per\_token}          & 0.0092 \\
\midrule
\multicolumn{3}{@{}l}{\textit{Agent-Level Statistics, Round~1 (11 features)}} \\
1  & \texttt{sample\_round\_1\_max\_agent\_std\_entropy}      & 0.0323 \\
5  & \texttt{sample\_round\_1\_agent\_mean\_entropy\_spread}  & 0.0147 \\
6  & \texttt{sample\_round\_1\_std\_agent\_mean\_entropy}     & 0.0134 \\
9  & \texttt{sample\_round\_1\_agent\_total\_entropy\_spread} & 0.0099 \\
17 & \texttt{sample\_round\_1\_max\_agent\_max\_entropy}      & 0.0065 \\
19 & \texttt{sample\_round\_1\_std\_agent\_total\_entropy}    & 0.0059 \\
21 & \texttt{sample\_round\_1\_std\_agent\_variance\_entropy} & 0.0055 \\
25 & \texttt{sample\_round\_1\_q3\_agent\_total\_entropy}     & 0.0050 \\
20 & \texttt{round\_1\_total\_entropy}                        & 0.0058 \\
12 & \texttt{exp\_infer\_average\_entropy}                    & 0.0080 \\
28 & \texttt{sample\_round\_1\_q1\_agent\_max\_entropy}       & 0.0047 \\
\midrule
\multicolumn{3}{@{}l}{\textit{Round~2 \& Sample-Level (15 features)}} \\
2  & \texttt{sample\_mean\_answer\_token\_entropy}            & 0.0280 \\
4  & \texttt{answer\_token\_entropy\_change}                  & 0.0191 \\
7  & \texttt{sample\_round\_2\_std\_agent\_mean\_entropy}     & 0.0118 \\
8  & \texttt{sample\_round\_2\_agent\_mean\_entropy\_spread}  & 0.0105 \\
11 & \texttt{sample\_round\_2\_std\_agent\_std\_entropy}      & 0.0080 \\
13 & \texttt{round\_2\_infer\_avg\_entropy}                   & 0.0076 \\
14 & \texttt{sample\_round\_2\_max\_agent\_max\_entropy}      & 0.0074 \\
15 & \texttt{sample\_round\_2\_std\_agent\_total\_entropy}    & 0.0074 \\
16 & \texttt{sample\_max\_answer\_token\_entropy}             & 0.0067 \\
18 & \texttt{sample\_median\_answer\_token\_entropy}          & 0.0060 \\
22 & \texttt{sample\_round\_2\_agent\_total\_entropy\_spread} & 0.0054 \\
23 & \texttt{sample\_round\_2\_std\_agent\_max\_entropy}      & 0.0051 \\
24 & \texttt{sample\_round\_2\_max\_agent\_total\_entropy}    & 0.0051 \\
26 & \texttt{exp\_total\_entropy}                             & 0.0049 \\
27 & \texttt{sample\_avg\_entropy\_per\_token\_diff\_vs\_base} & 0.0049 \\
\bottomrule
\end{tabularx}
\end{minipage}
\end{figure}

\begin{figure}[h]
\centering
\includegraphics[width=0.8\linewidth]{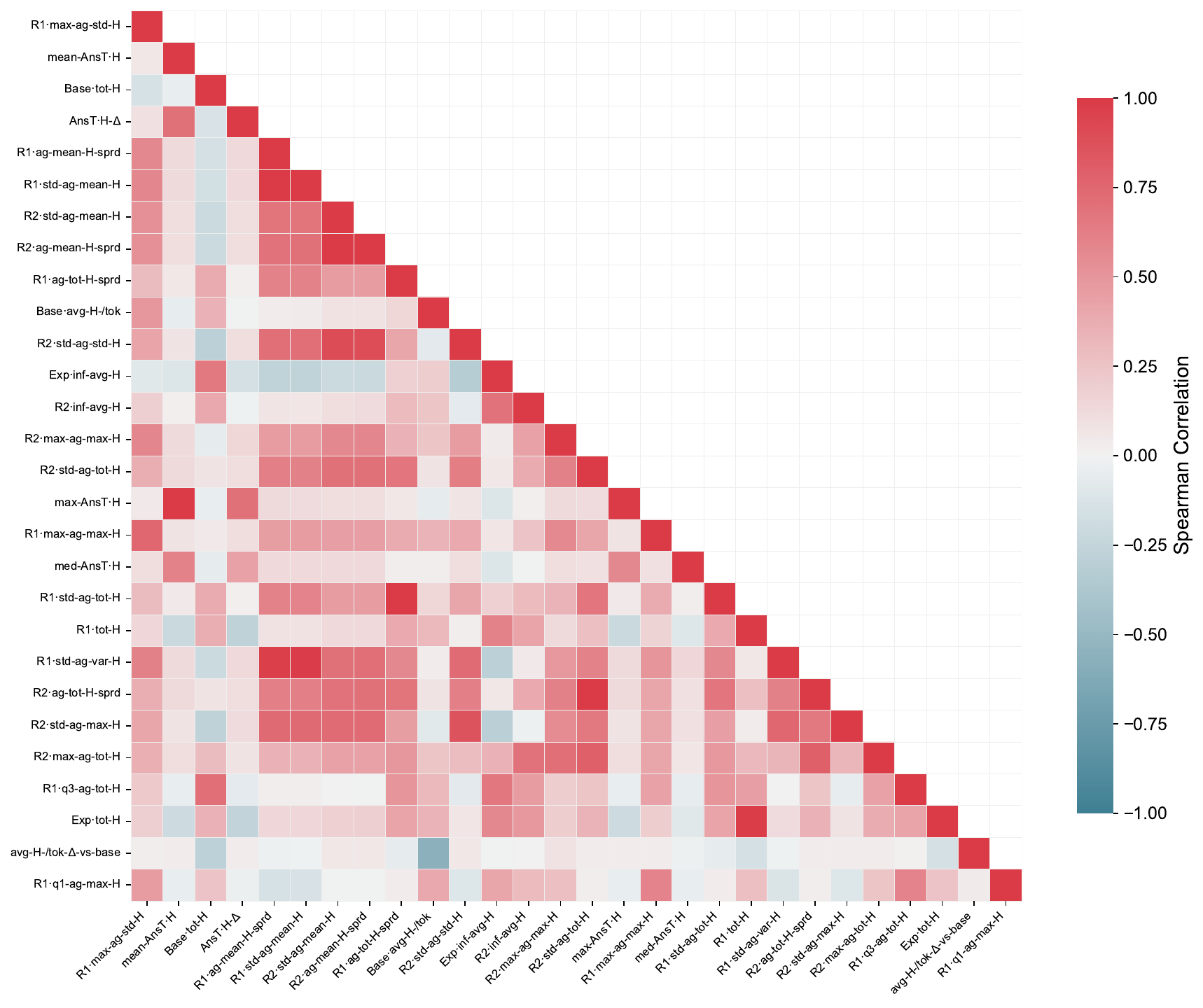}
\caption{Pairwise Spearman correlation matrix of the 28 selected features. The upper-triangular heatmap confirms that no remaining pair exceeds the $|\rho| > 0.85$ redundancy threshold, validating that the selected subset is non-redundant while spanning all hierarchical feature levels.}
\label{fig:causal-correlation-matrix}
\end{figure}

\paragraph{Redundancy Removal and Hierarchical Coverage.}
After ranking, we remove redundant features whose pairwise Spearman correlation exceeds $|\rho| > 0.85$, retaining the higher-scored feature in each correlated pair. A coverage constraint then ensures that the final subset contains at least one representative from each hierarchical level defined in Appendix~\ref{app:entropy_features}: agent-level statistics, round-level dynamics, sample-level statistics, system-level aggregation, base-model entropy ($\mathcal{F}_{\text{base-E}}$), and computational metrics ($\mathcal{F}_C$). The final subset contains \textbf{28 features}; Figure~\ref{fig:causal-borda-scores} visualizes the Borda fusion scores colored by hierarchical tier, Table~\ref{tab:causal-features} lists all selected features grouped by hierarchical level, and Figure~\ref{fig:causal-correlation-matrix} shows the pairwise Spearman correlation matrix of the 28 selected features, confirming that no remaining pair exceeds the $|\rho| > 0.85$ redundancy threshold.

\subsection{Causal Structure Discovery}
\label{app:causal-discovery-structure}

We apply two constraint-based algorithms to learn the causal graph over the 28 selected features plus the binary outcome \texttt{is\_finally\_correct} (29 variables total).

\begin{figure}[t]
\centering
\begin{subfigure}{0.48\columnwidth}
\centering
\includegraphics[width=\textwidth]{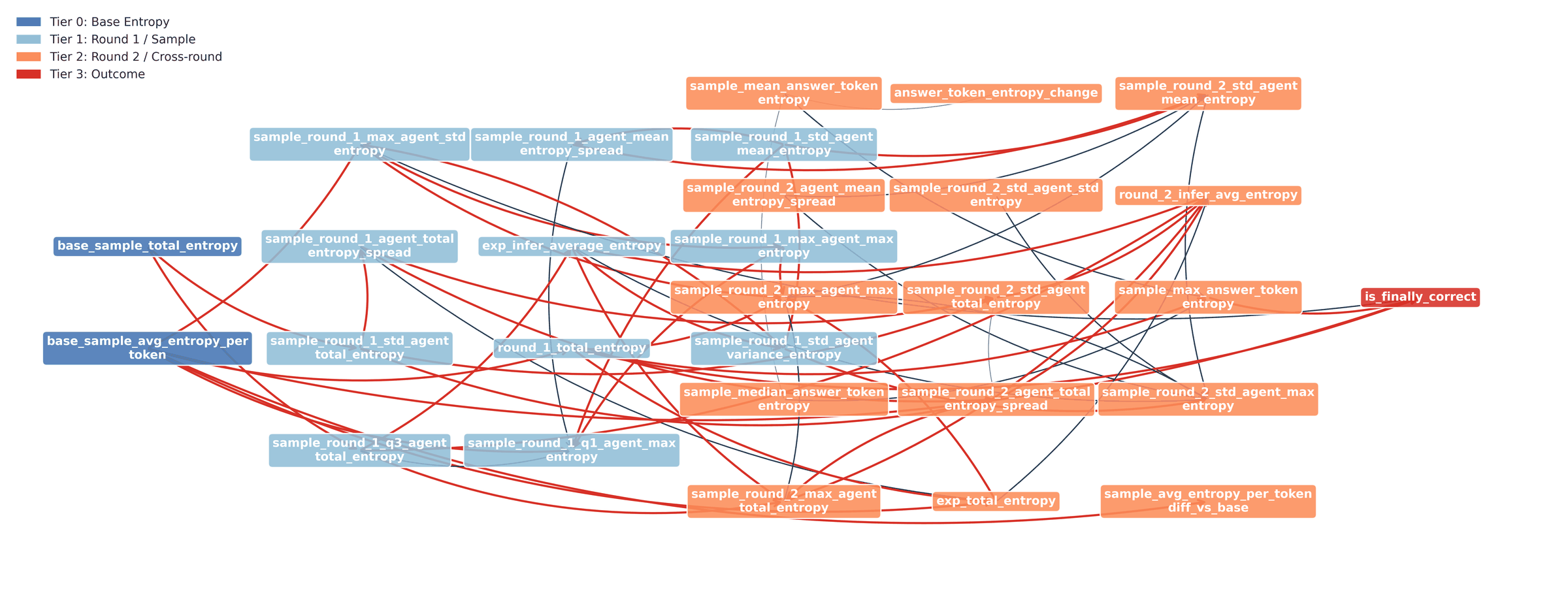}
\caption{PC algorithm (55 directed + 3 undirected edges).}
\label{fig:causal-graph-pc}
\end{subfigure}
\hfill
\begin{subfigure}{0.48\columnwidth}
\centering
\includegraphics[width=\textwidth]{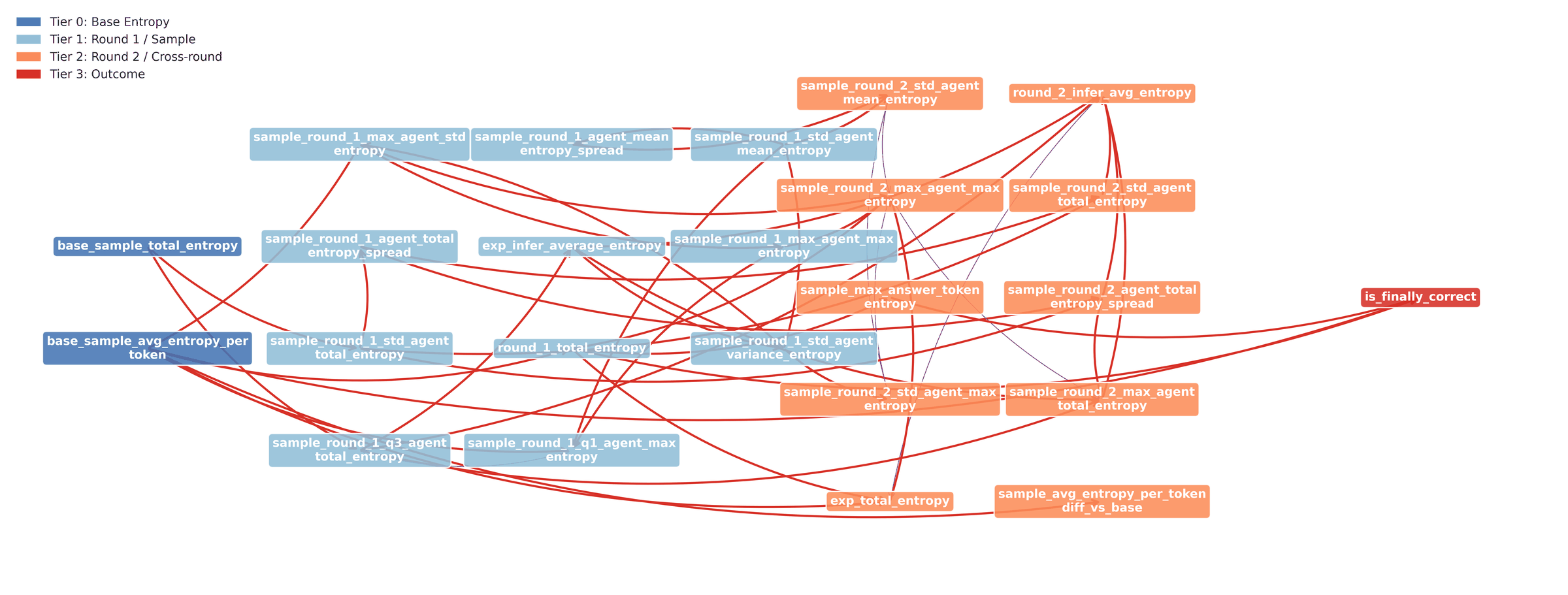}
\caption{FCI algorithm (40 directed and 6 bidirected edges).}
\label{fig:causal-graph-fci}
\end{subfigure}
\caption{Individual causal graphs from constraint-based discovery. The PC graph (a) assumes causal sufficiency and produces undirected edges where orientation is undetermined. The FCI graph (b) relaxes causal sufficiency, introducing bidirected edges ($\leftrightarrow$) to indicate possible latent confounding between connected variables.}
\label{fig:causal-graphs-individual}
\end{figure}

\paragraph{Algorithm Configuration.}
Both the PC algorithm and the FCI algorithm use the Fisher-Z conditional independence test at significance level $\alpha = 0.01$. We encode temporal background knowledge as tier constraints to enforce temporal consistency:
\begin{itemize}[nosep,leftmargin=1.5em]
    \item \textbf{Tier~0} (2 variables): Base model properties - exogenous, cannot be caused by downstream features.
    \item \textbf{Tier~1} (11 variables): Round~1 entropy features - cannot cause Tier~0 variables.
    \item \textbf{Tier~2} (15 variables): Round~2, cross-round, and sample-level aggregation features - cannot cause Tier~0 or Tier~1 variables.
    \item \textbf{Tier~3} (1 variable): Outcome (\texttt{is\_finally\_correct}) - no outgoing edges permitted.
\end{itemize}

\begin{figure}[t]
\centering
\includegraphics[width=0.9\columnwidth]{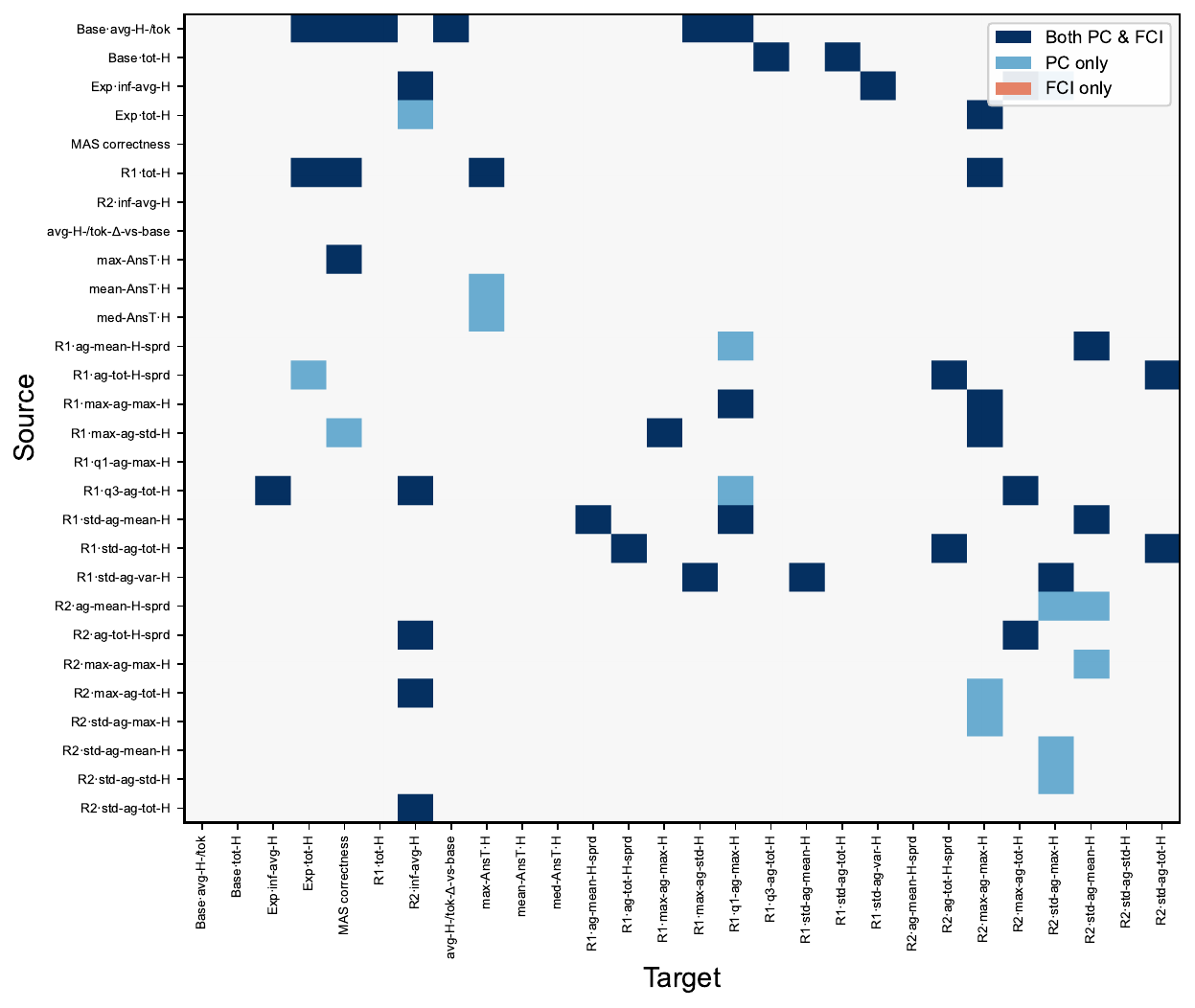}
\caption{PC vs.\ FCI edge agreement heatmap over 29 nodes. Dark blue = both algorithms agree (80 edges); light blue = PC only (15 edges); red = FCI only (0 edges). High overlap confirms that the consensus DAG is stable across Markov-boundary assumptions.}
\label{fig:causal-edge-agreement}
\end{figure}

\paragraph{Results.}
The PC algorithm discovers 55 directed and 3 undirected edges; FCI discovers 40 directed and 6 bidirected edges (the latter indicating possible latent confounding). A total of 80 edges are identified by both algorithms (consensus edges), out of 95 unique directed edges in the union. Figure~\ref{fig:causal-graphs-individual} presents the full causal graphs from each algorithm separately, and Figure~\ref{fig:causal-dag-main} (in Section~\ref{sec:causal_analysis}) shows the consensus graph. Figure~\ref{fig:causal-edge-agreement} confirms the stability of the consensus DAG across both algorithms.

\paragraph{Direct Causes of Correctness.}
The PC algorithm identifies four direct causes of MAS correctness: base-model average per-token entropy, round-1 total entropy, maximum answer-token entropy, and round-1 maximum agent entropy dispersion. FCI, which accounts for latent confounders, identifies three of these: base-model average per-token entropy, round-1 total entropy, and maximum answer-token entropy. \textbf{The consensus direct causes across both algorithms are thus base-model average per-token entropy, round-1 total entropy, and maximum answer-token entropy, indicating that these three entropy signals exert the most robust direct causal influence on MAS correctness.} Round-1 maximum agent entropy dispersion is identified as a direct cause by PC only, suggesting it may share an unobserved common cause with correctness that FCI identifies as a latent confounder.

\subsection{Causal Effect Estimation}
\label{app:causal-discovery-effect}

We estimate the Average Treatment Effect (ATE) of five candidate treatment variables on \texttt{is\_finally\_correct} using the DoWhy framework with 15 confounders identified from the learned causal graph.

\paragraph{Estimation Methods.}
Three estimators are applied for each treatment: Linear Regression (LR), Propensity Score Stratification (PS), and Inverse Probability Weighting (IPW). Robustness is assessed via three refutation tests: (1)~\emph{Random Common Cause}: adding a random confounder should not alter the estimate; (2)~\emph{Placebo Treatment}: replacing the treatment with a random variable should eliminate the effect; (3)~\emph{Data Subset}: re-estimating on a random 80\% subset should yield a consistent estimate.

\begin{table}[h]
\centering
\caption{Causal effect estimates and refutation results. ``Direct" indicates whether the variable is a consensus direct cause from PC/FCI. {*}Direct cause in PC only (not in FCI). RCC = Random Common Cause; all refutation tests pass.}
\label{tab:causal-ate}
\vspace{2pt}
\scriptsize
\begin{tabular}{@{}lrrrclccc@{}}
\toprule
\textbf{Treatment} & \textbf{ATE\textsubscript{LR}} & \textbf{ATE\textsubscript{PS}} & \textbf{ATE\textsubscript{IPW}} & \textbf{$p$-value} & \textbf{Direct} & \textbf{RCC} & \textbf{Placebo} & \textbf{Subset} \\
\midrule
base\_sample\_avg\_ent\_per\_tok    & $-0.771$ & $-0.123$ & $-0.146$ & $1.3\text{e-}21$ & Yes & \ding{51} & \ding{51} & \ding{51} \\
round\_1\_total\_entropy            & $+0.000$ & $-0.007$ & $+0.056$ & $1.3\text{e-}19$ & Yes & \ding{51} & \ding{51} & \ding{51} \\
sample\_max\_ans\_tok\_ent          & $-0.350$ & $-0.309$ & $-0.339$ & $6.8\text{e-}29$ & Yes & \ding{51} & \ding{51} & \ding{51} \\
sample\_r1\_max\_agent\_std\_ent    & $-0.278$ & $-0.152$ & $-0.127$ & $3.6\text{e-}3$  & Yes\textsuperscript{*} & \ding{51} & \ding{51} & \ding{51} \\
ans\_tok\_ent\_change               & $-0.194$ & $-0.211$ & $-0.238$ & $5.8\text{e-}10$ & No  & \ding{51} & \ding{51} & \ding{51} \\
\bottomrule
\end{tabular}
\par\vspace{2pt}
\begin{minipage}{\linewidth}
\end{minipage}
\end{table}

\begin{figure}[t]
\centering
\includegraphics[width=0.8\columnwidth]{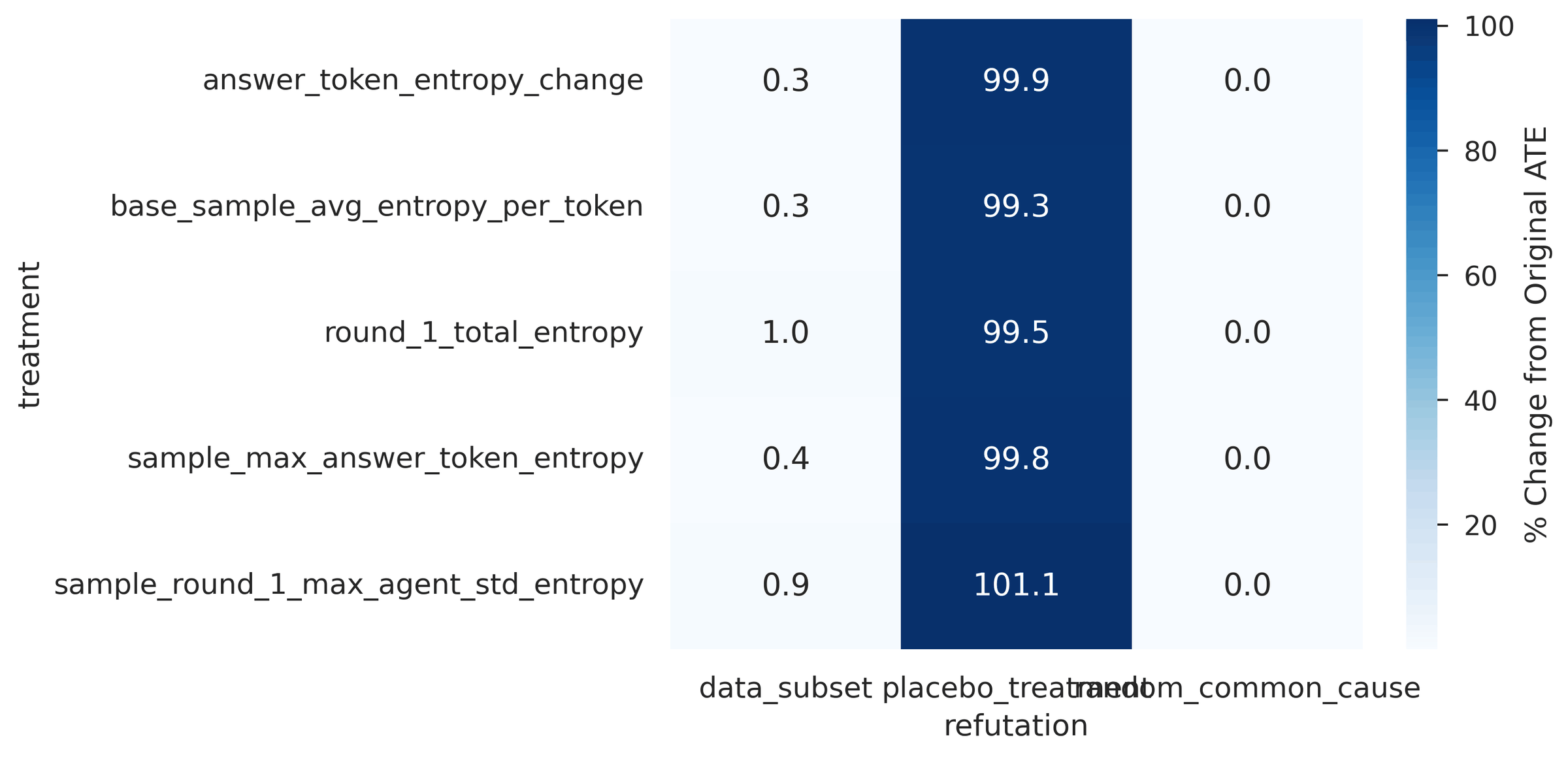}
\caption{Refutation test results for all five treatment variables. All treatments pass all three tests (Random Common Cause, Placebo Treatment, Data Subset), confirming robustness of the causal effect estimates against unmeasured confounding, spurious associations, and sampling variability.}
\label{fig:causal-refutation}
\end{figure}

Figure~\ref{fig:causal-ate-forest} (in Section~\ref{sec:causal_analysis}) compares the ATE estimates across treatment variables and estimation methods, and Figure~\ref{fig:causal-refutation} summarizes the refutation test results.

\paragraph{Interpretation.}
All five treatment variables exhibit statistically significant effects ($p < 0.01$). The strongest and most consistent effect belongs to \texttt{base\_sample\_avg\_entropy\_per\_token} ($p < 10^{-21}$): higher base model entropy strongly reduces MAS correctness probability. The propensity-based estimators, PS ($-0.123$) and IPW ($-0.146$), constitute the primary quantitative evidence for the effect magnitude, as they reweight rather than linearly project and are robust to model misspecification. \texttt{round\_1\_total\_entropy} carries a near-zero LR estimate ($0.000$) and small PS/IPW effects ($-0.007$ and $+0.056$), reflecting that its direct causal influence is largely mediated through downstream entropy nodes rather than acting as an independent linear driver; the highly significant $p < 10^{-19}$ reflects its strong structural role in the learned causal graph rather than a large marginal effect size. All five treatment variables pass all three refutation tests, confirming that the estimated effects are robust to unmeasured confounding, spurious treatment assignment, and sampling variability.

\subsection{Mediation Analysis}
\label{app:causal-discovery-mediation}

To understand the mechanistic pathways through which entropy features influence MAS correctness, we conduct mediation analysis using the Baron-Kenny framework with bootstrap confidence intervals ($n = 1{,}000$ resamples). Of 7 candidate mediation paths, 6 exhibit significant indirect effects (bootstrap 95\% CI excluding zero).

\begin{table}[h]
\centering
\caption{Significant mediation paths ordered by $|\text{indirect effect}|$. CI denotes the bootstrap 95\% confidence interval. Proportion mediated $= \text{Indirect} / \text{Total}$ (Baron-Kenny), where $\text{Total} = \text{Direct} + \text{Indirect}$.}
\label{tab:causal-mediation}
\vspace{2pt}
\scriptsize
\begin{tabular}{@{}llrrrr@{}}
\toprule
\textbf{Treatment} & \textbf{Mediator} & \textbf{Indirect} & \textbf{Total} & \textbf{Prop.} & \textbf{95\% CI} \\
\midrule
\multicolumn{6}{@{}l}{\textit{Round~1 Inter-Agent Dispersion $\to$ Round~2 Entropy $\to$ Correctness}} \\
r1\_max\_agent\_std\_ent   & r2\_agent\_mean\_ent\_spread  & $+0.049$ & $-0.496$ & $-9.8$\%  & [$+0.033$, $+0.066$] \\
r1\_agent\_mean\_ent\_spread & r2\_agent\_mean\_ent\_spread & $-0.048$ & $-0.149$ & 32.5\%   & [$-0.066$, $-0.030$] \\
r1\_std\_agent\_mean\_ent  & r2\_agent\_mean\_ent\_spread  & $-0.047$ & $-0.152$ & 30.9\%   & [$-0.066$, $-0.028$] \\
r1\_agent\_mean\_ent\_spread & r2\_std\_agent\_mean\_ent   & $-0.046$ & $-0.149$ & 30.8\%   & [$-0.064$, $-0.028$] \\
r1\_max\_agent\_std\_ent   & r2\_std\_agent\_mean\_ent     & $+0.046$ & $-0.496$ & $-9.2$\%  & [$+0.030$, $+0.062$] \\
r1\_std\_agent\_mean\_ent  & r2\_std\_agent\_mean\_ent     & $-0.045$ & $-0.152$ & 29.4\%   & [$-0.061$, $-0.027$] \\
\bottomrule
\end{tabular}
\par\vspace{2pt}
\begin{minipage}{\linewidth}
\end{minipage}
\end{table}

\begin{figure}[t]
\centering
\includegraphics[width=0.6\columnwidth]{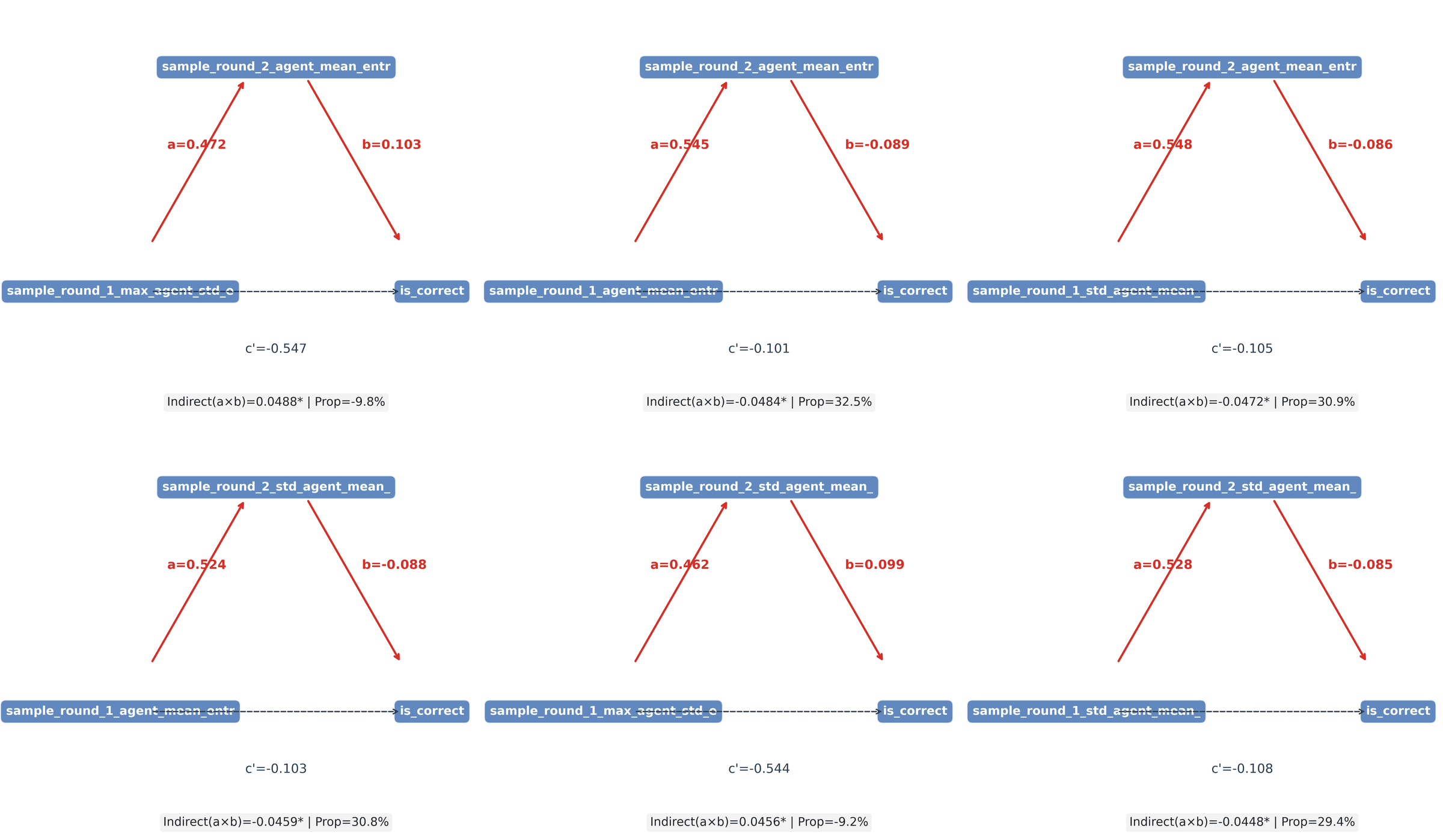}
\caption{Schematic path diagrams for the cross-round causal mediation chain: Round~1 inter-agent entropy dispersion $\to$ Round~2 entropy $\to$ correctness. Arrows represent directed causal links identified from the learned causal graph.}
\label{fig:causal-path-diagrams}
\end{figure}

\begin{figure}[t]
\centering
\includegraphics[width=0.8\columnwidth]{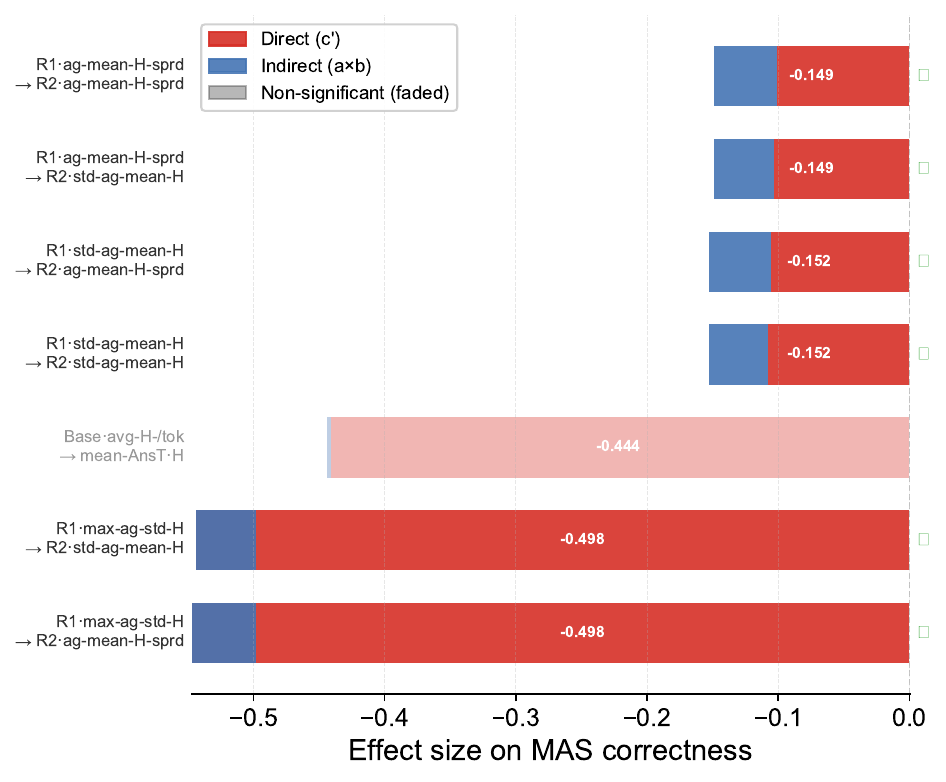}
\caption{Full mediation decomposition for the 7 candidate paths by $|\text{total effect}|$. Stacked bars show direct effect (red, $c'$) and mediated indirect effect (blue, $a \times b$); faded bars indicate non-significant indirect effects. Paths are sorted by total effect magnitude. The 6 paths with significant indirect effects match those reported in Table~\ref{tab:causal-mediation}.}
\label{fig:causal-mediation-full}
\end{figure}

Figure~\ref{fig:causal-path-diagrams} illustrates the mediation pathway, and Figure~\ref{fig:causal-mediation-full} presents the full mediation decomposition of direct and indirect effects for all 7 candidate paths.

\paragraph{Cross-Round Entropy Propagation.}
Round~1 agent-level entropy dispersion features transmit 29--33\% of their causal effects on correctness through Round~2 entropy, specifically through the spread of agent mean entropy in round 2. This confirms that initial entropy heterogeneity among agents causally shapes the post-interaction entropy landscape: early misalignment compounds into the subsequent round rather than self-correcting. The two paths originating from \texttt{sample\_round\_1\_max\_agent\_std\_entropy} show negative indirect-to-total ratios ($-9.8\%$ and $-9.2\%$), reflecting partial suppression: this high-dispersion signal propagates positively into round-2 spread features whose direct effect on correctness is itself negative, creating a suppressive chain that partially offsets the strong direct harm of round-1 maximum dispersion on correctness.

\section{Entropy Patterns Generalize to Agentic Tasks}
\label{app:agentic_task}

This appendix validates that the entropy-performance findings from reasoning benchmarks extend to agentic settings where agents must orchestrate external tools. We present full results on the general agentic task \texttt{GAIA}~\citep{gaia-iclr23} and the financial domain agentic task \texttt{FinanceAgent}~\citep{FinanceAgentBench-arxiv25}.

\subsection{GAIA: First-Round Dispersion Mediates Tool Execution Failure}
\label{app:gaia}

\subsubsection{Experimental Setup}
\label{app:gaia-setup}

\paragraph{Benchmark and Models.} We evaluate on the full \texttt{GAIA} validation split (165 questions across Levels 1, 2, and 3), using six base models spanning two families and four parameter scales: Llama-3.1-8B-Instruct, Llama-3.2-3B-Instruct, Qwen3-0.6B, Qwen3-4B, Qwen3-8B, and Qwen3-14B. We refer to these as L-8, L-3, Q-0.6, Q-4, Q-8, and Q-14 in figures.

\paragraph{Architectures and configuration.} We evaluate the same five architectures (Single, Sequential, Centralized, Hybrid, Debate) at $R{=}2$ interaction rounds, plus the base-model baseline. Each agent operates in a ReAct loop with up to 8192 new tokens per step. We use the official \texttt{GAIA} prompt template and require the model to place its final answer after the \texttt{FINAL ANSWER:} marker.

\paragraph{Tool suite.} The agent has access to five tools: web search via SerperAPI, a sandboxed math-only calculator, a file reader supporting PDF, Excel, CSV, DOCX, and PPTX, a sub-process Python executor, and a multimodal viewer. Since all our base models are text-only LLMs, we expose the multimodal viewer as a tool so the agent can delegate questions that require visual understanding to Doubao-Seed-2.0-Lite. Each tool runs under an asynchronous timeout (calculator and file reader 30s, web search 90s, multimodal viewer 120s, Python executor 150s). In addition, when the model issues the same tool call with identical arguments three times in a row, we force it to emit a final answer in the next loop iteration to avoid unproductive cycles.

\paragraph{Entropy capture.} Unlike the single-pass reasoning benchmarks, \texttt{GAIA} requires multiple rounds of tool invocation and reasoning before a final answer can be produced. Each agent therefore runs a ReAct loop of up to ten steps, in which the LLM first reasons, decides which tool to invoke, receives the tool's output, and then reasons again on the next step. We record token-level entropy at every such step, which inserts an additional \emph{step} granularity between the existing token and agent levels. The resulting hierarchy (token, step, agent, sample, round) captures the model's uncertainty at each tool-calling decision rather than only at the round level used in the main analysis.

\subsubsection{Main Results}
\label{app:gaia-results}

\begin{figure*}[t]
    \centering
    \begin{minipage}{0.95\textwidth}
    \begin{subfigure}{0.48\textwidth}
        \centering
        \includegraphics[width=\linewidth]{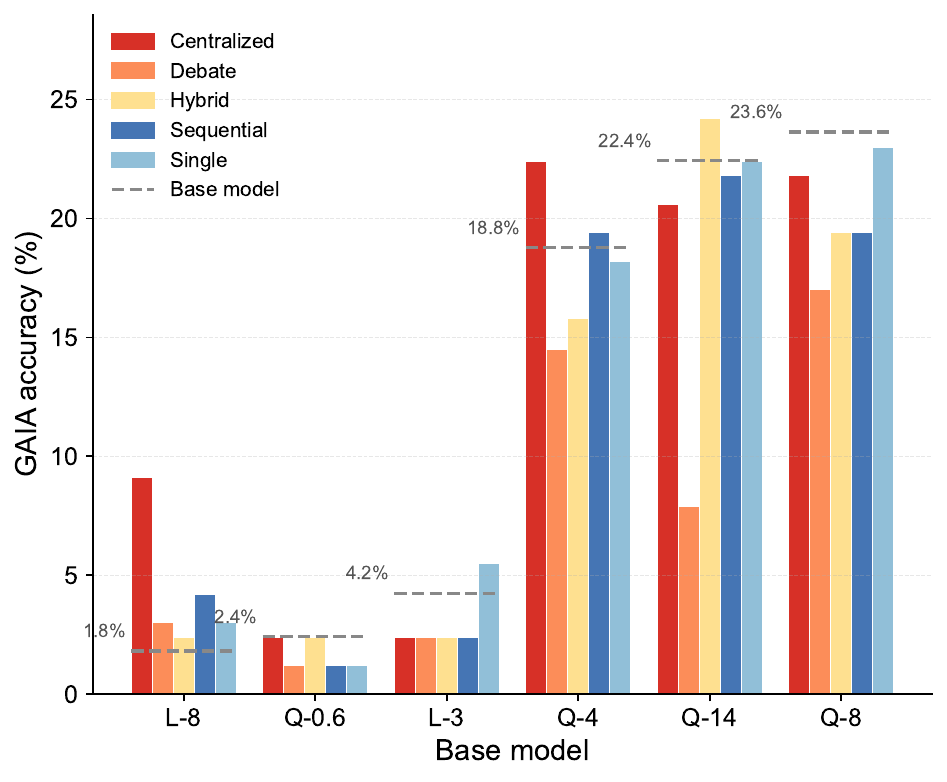}
        \caption{Per-model accuracy across architectures. Dashed tick marks show base-model accuracy.}
        \label{fig:gaia-arch-accuracy}
    \end{subfigure}
    \hfill
    \begin{subfigure}{0.48\textwidth}
        \centering
        \includegraphics[width=\linewidth]{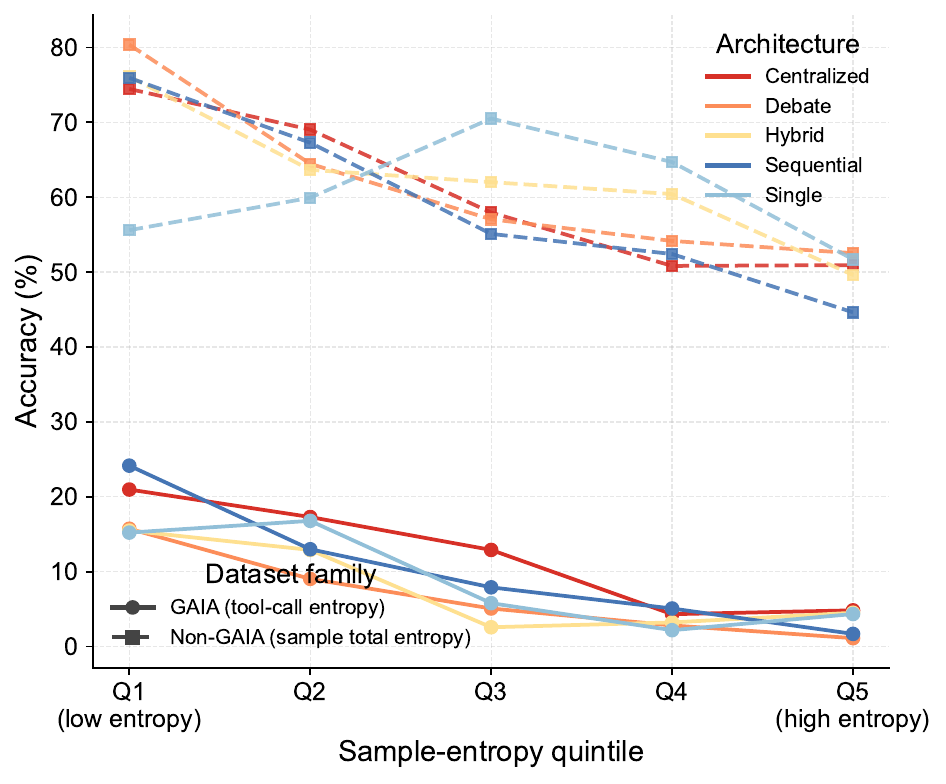}
        \caption{Accuracy vs.\ entropy quintile on \texttt{GAIA} (solid) and non-GAIA reasoning benchmarks (dashed).}
        \label{fig:gaia-cross-task}
    \end{subfigure}

    \vspace{1em}

    \begin{subfigure}{0.48\textwidth}
        \centering
        \includegraphics[width=\linewidth]{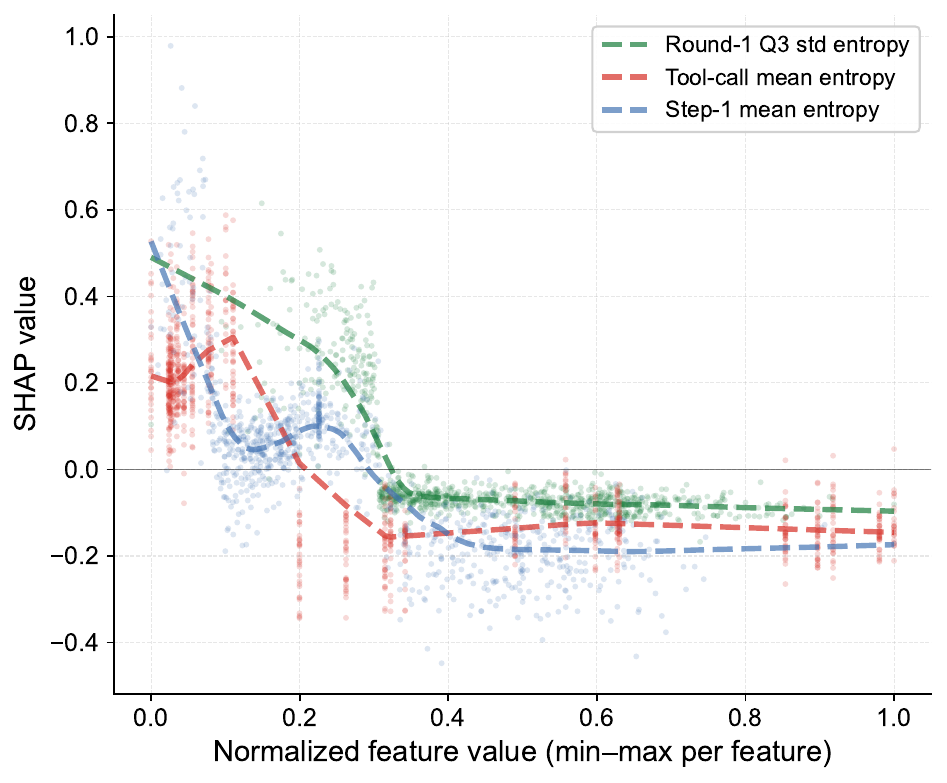}
        \caption{Per-sample SHAP contribution of the three top entropy features; all slope negatively.}
        \label{fig:gaia-shap}
    \end{subfigure}
    \hfill
    \begin{subfigure}{0.48\textwidth}
        \centering
        \includegraphics[width=\linewidth]{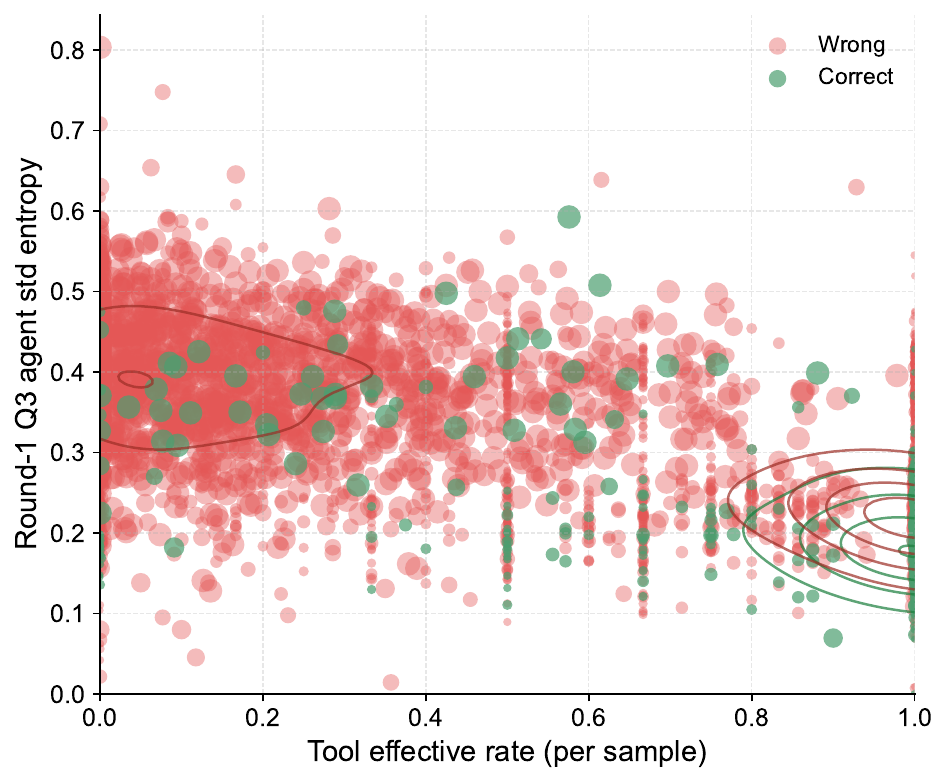}
        \caption{Joint distribution of tool effective rate and round-1 upper-quartile inter-agent dispersion.}
        \label{fig:gaia-tool-phase}
    \end{subfigure}
    \end{minipage}
    \caption{\texttt{GAIA} results. (a) SAS remains competitive: it reaches the highest accuracy on 2 of 6 base models and beats at least one MAS on every model; Debate consistently lags. (b) Tool-call entropy negatively predicts accuracy across all architectures, mirroring the sample-total-entropy pattern on reasoning benchmarks. (c) Round-1 inter-agent dispersion, mean tool-call entropy, and step-1 mean entropy are the three most predictive features, all with negative SHAP slope. (d) Correct samples cluster at high tool effective rate and low inter-agent dispersion; wrong samples spread toward the opposite corner.}
    \label{fig:gaia-main}
\end{figure*}

\paragraph{SAS remains competitive on agentic tasks.}

As shown in Figure~\ref{fig:gaia-arch-accuracy} and Table~\ref{tab:gaia-accuracy}, SAS achieves the highest accuracy among all architectures on two of the six base models and outperforms at least one MAS variant on every model, whereas Debate consistently lags—a trend also observed on \texttt{FinanceAgent} (Appendix~\ref{app:agent-benchmark-accuracy}). Notably, on the four Qwen3 models, base model performance equals or exceeds that of SAS, suggesting that when \textbf{the base model is already competent at tool calling, neither SAS self-deliberation nor MAS interaction yields additional benefits.} Overall, weak models derive little value from MAS, while strong models gain only modestly.

\begin{table}[t]
\centering
\caption{\texttt{GAIA} accuracy (\%) by architecture and base model. The Base column gives base model accuracy. Bold marks the best MAS architecture per model.}
\label{tab:gaia-accuracy}
\footnotesize
\begin{tabular}{@{}lcccccc@{}}
\toprule
\textbf{Model} & \textbf{Base} & \textbf{Single} & \textbf{Sequential} & \textbf{Centralized} & \textbf{Hybrid} & \textbf{Debate} \\
\midrule
LLaMA-3.1-8B-Instruct  & 1.8  & 3.0  & 4.2  & \textbf{9.1}  & 2.4  & 3.0  \\
LLaMA-3.2-3B-Instruct  & 4.2  & \textbf{5.5}  & 2.4  & 2.4  & 2.4  & 2.4  \\
Qwen3-0.6B    & 2.4  & 1.2  & 1.2  & \textbf{2.4}  & 2.4  & 1.2  \\
Qwen3-4B      & 18.8 & 18.2 & 19.4 & \textbf{22.4} & 15.8 & 14.5 \\
Qwen3-8B      & 23.6 & \textbf{23.0} & 19.4 & 21.8 & 19.4 & 17.0 \\
Qwen3-14B     & 22.4 & 22.4 & 21.8 & 20.6 & \textbf{24.2} & 7.9  \\
\bottomrule
\end{tabular}
\end{table}

\paragraph{Tool-call entropy is negatively correlated with performance.}
Figure~\ref{fig:gaia-cross-task} shows accuracy against entropy quintiles for all five architectures on \texttt{GAIA} and non-GAIA reasoning tasks. On \texttt{GAIA}, every architecture shows decreasing accuracy as tool-call entropy increases, matching the negative correlation on reasoning benchmarks. \textbf{Higher entropy of the tokens the task depends on, sample total entropy on reasoning tasks, tool-call entropy on agentic tasks, predicts lower accuracy in both settings.}

\paragraph{Early-round entropy constrains agentic MAS.}
Figure~\ref{fig:gaia-shap} shows the three most important features on $\mathcal{G}_{\text{MAS}}$: round-1 upper-quartile inter-agent dispersion, mean tool-call entropy, and step-1 mean entropy, all with negative SHAP slope. \textbf{Beyond entropy during tool invocation, MAS performance is limited by first-round entropy and, more granularly, by uncertainty in each agent's initial reasoning step}. This aligns with the first-round dominance finding from reasoning benchmarks.

\paragraph{Round-1 inter-agent dispersion lowers tool execution success.}
Figure~\ref{fig:gaia-tool-phase} plots tool effective rate against round-1 upper-quartile inter-agent dispersion. Correct samples cluster at high effective rate and low dispersion; wrong samples spread toward the low-effective-rate and high-dispersion corner. \textbf{Higher inter-agent entropy dispersion lowers tool execution success, which in turn lowers final correctness.}

\subsubsection{Additional Feature Analysis}
\label{app:gaia-features}

\begin{figure*}[t]
    \centering
    \begin{subfigure}[b]{0.48\textwidth}
        \centering
        \includegraphics[height=5cm,keepaspectratio]{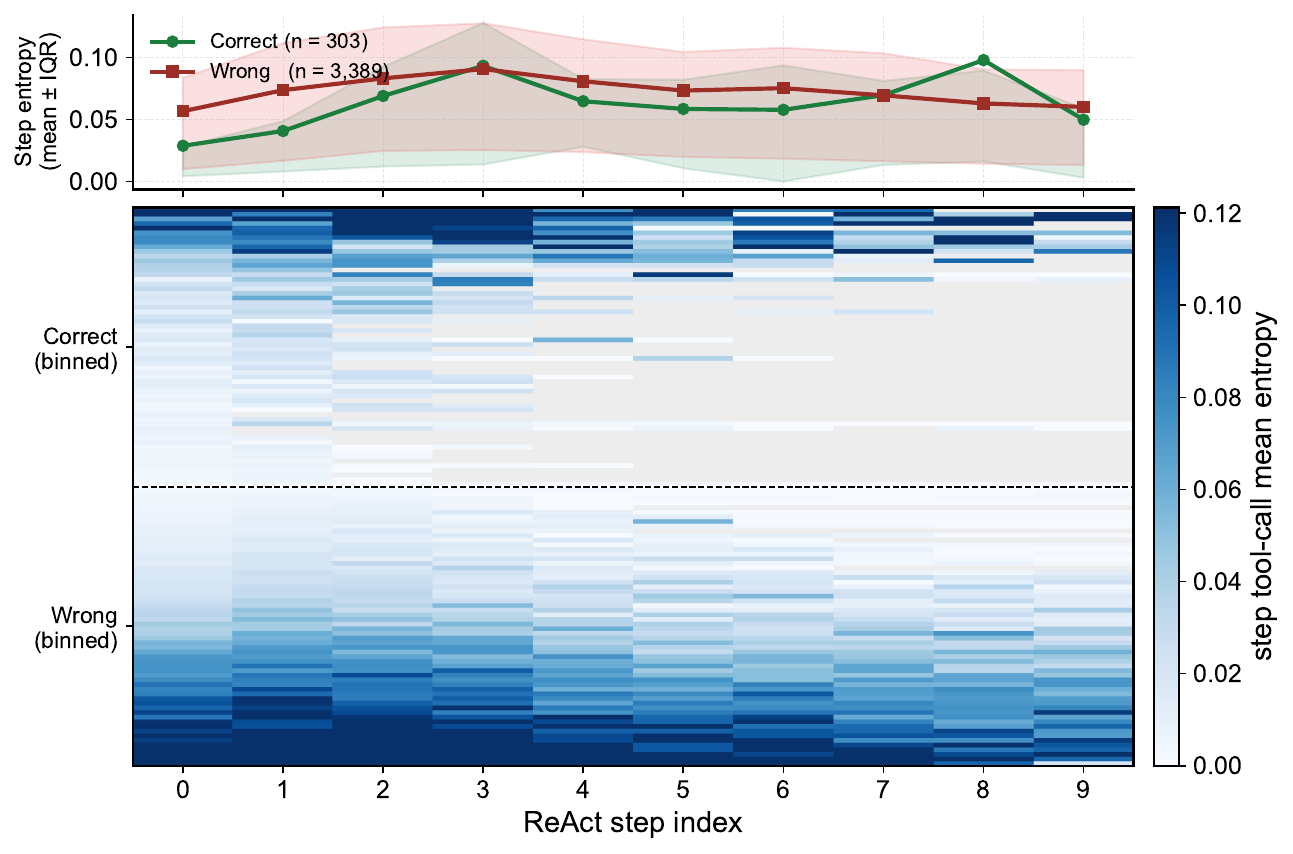}
        \caption{}
        \label{fig:gaia-step-heatmap}
    \end{subfigure}
    \hfill
    \begin{subfigure}[b]{0.48\textwidth}
        \centering
        \includegraphics[height=5cm,keepaspectratio]{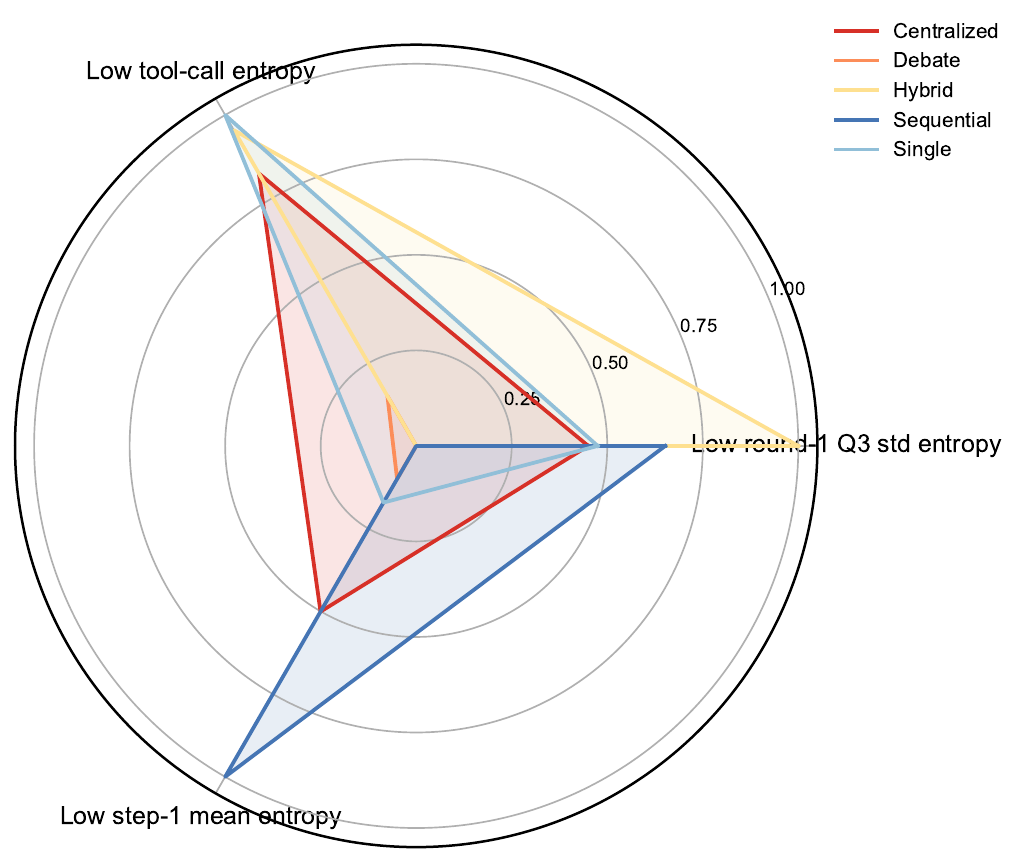}
        \caption{}
        \label{fig:gaia-arch-radar}
    \end{subfigure}
    \caption{(a) Per-step tool-call entropy across all GAIA samples with at least one tool call ($n{=}3{,}692$). (b) Architecture comparison on three normalized axes; outward is better on every axis.}
    \label{fig:gaia-step-radar}
\end{figure*}

Figure~\ref{fig:gaia-step-heatmap} plots per-step tool-call decision entropy separately for correct and wrong trajectories, computed across all GAIA samples with at least one tool call. Correct trajectories (n = 303) start at step-0 entropy of about 0.029 and average 0.063 across the trajectory; wrong trajectories (n = 3,389) start at about 0.057 and stay elevated through their last active step. Failure is therefore dynamic rather than visible at step 0 alone: \textbf{the two classes diverge as the trajectory progresses, hesitation persists longer in wrong trajectories, and wrong trajectories also tend to use up more steps before terminating}. Figure~\ref{fig:gaia-arch-radar} compares the five architectures on three normalized axes (tool effective rate, low tool-call entropy, low round-1 max entropy), showing that no single architecture dominates all three: Centralized leads on tool effectiveness, Hybrid on tool-call entropy, and Sequential on round-1 entropy.

\subsubsection{Causal Validation}
\label{app:gaia-causal}

We apply the same causal pipeline used in Appendix~\ref{app:causal-discovery} to the GAIA cohort. The added step-level entropy features expand the feature space from 245 to 295 dimensions. Multi-method fusion ranking with redundancy removal selects 30 non-redundant features from this space, PC and FCI with temporal tier constraints recover the causal graph, and DoWhy with three refutation tests estimates causal effects. Both PC and FCI return a single consensus direct cause of correctness: round-1 tool success rate ($\text{ATE}{=}0.068$, $p{=}6.8\!\times\!10^{-4}$). All three refutation tests pass, with random common cause changing the effect by 0.1\%, placebo treatment by 97.7\%, and data subset by 0.9\%. This is the agentic-task analogue of the Base Entropy result: \textbf{the same upstream model capability that surfaces as base-model token entropy on reasoning tasks surfaces as tool-execution success on \texttt{GAIA}, with entropy retained as a mediator}. Mediation analysis identifies a significant indirect path from round-1 inter-agent skewness through round-2 maximum agent dispersion to correctness (indirect effect $-0.019$, bootstrap 95\% CI $[-0.042, -0.003]$), confirming that the round-1 to round-2 dispersion path remains operative on tool-augmented tasks.

\begin{figure*}[!htbp]
    \centering
    \begin{subfigure}{0.32\textwidth}
        \centering
        \includegraphics[width=\linewidth]{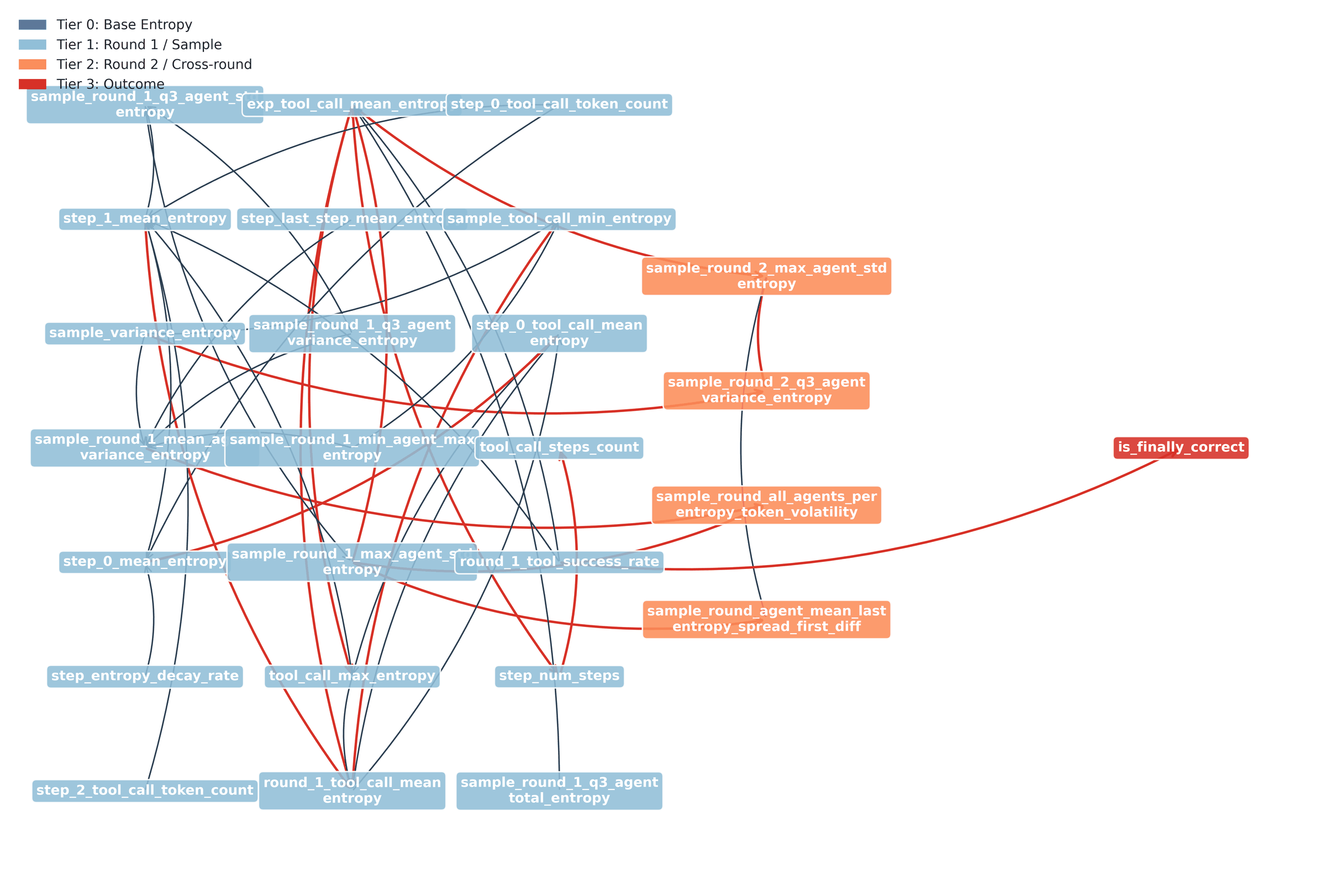}
        \caption{Consensus causal graph}
    \end{subfigure}
    \hfill
    \begin{subfigure}{0.32\textwidth}
        \centering
        \includegraphics[width=\linewidth]{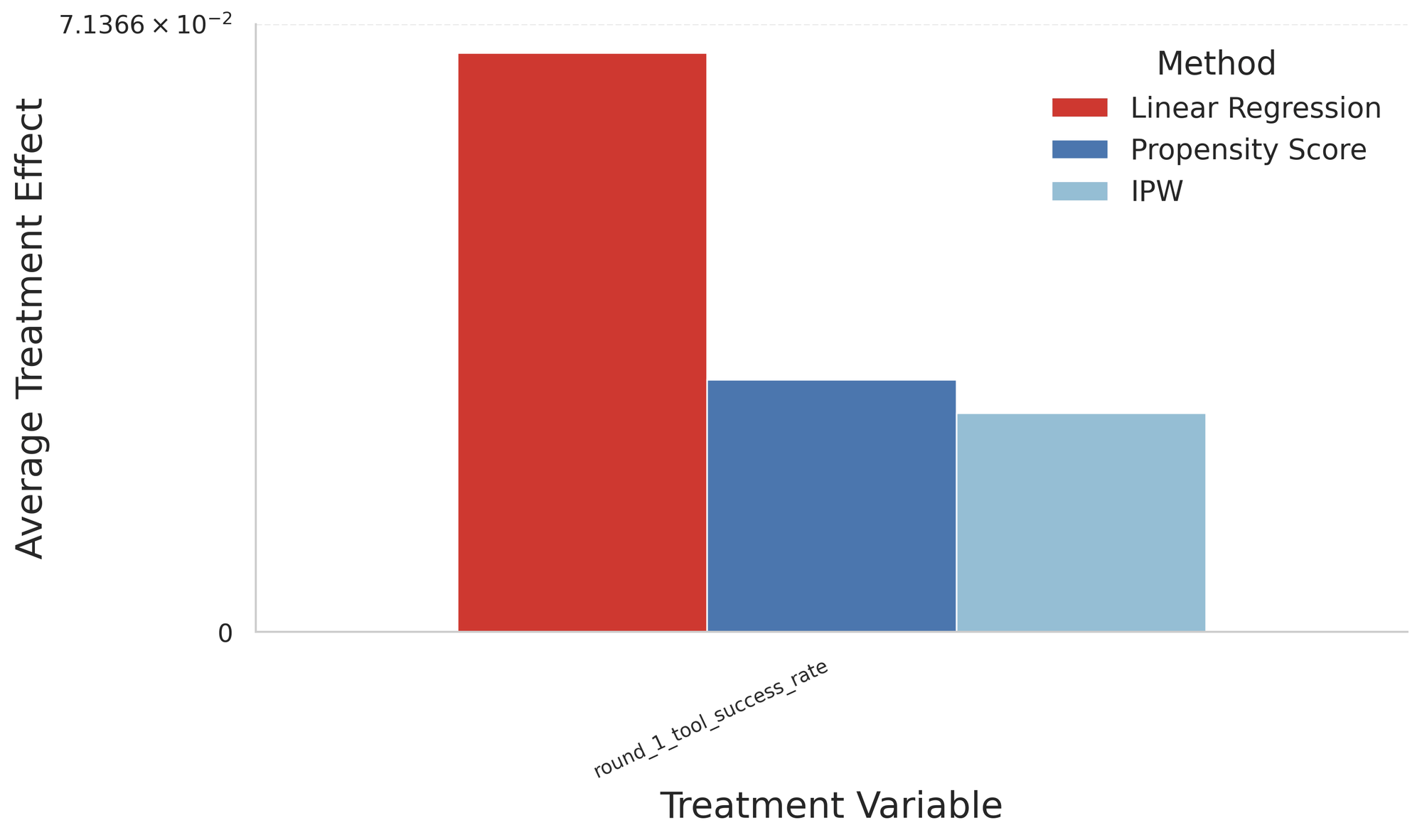}
        \caption{ATE estimates}
    \end{subfigure}
    \hfill
    \begin{subfigure}{0.32\textwidth}
        \centering
        \includegraphics[width=\linewidth]{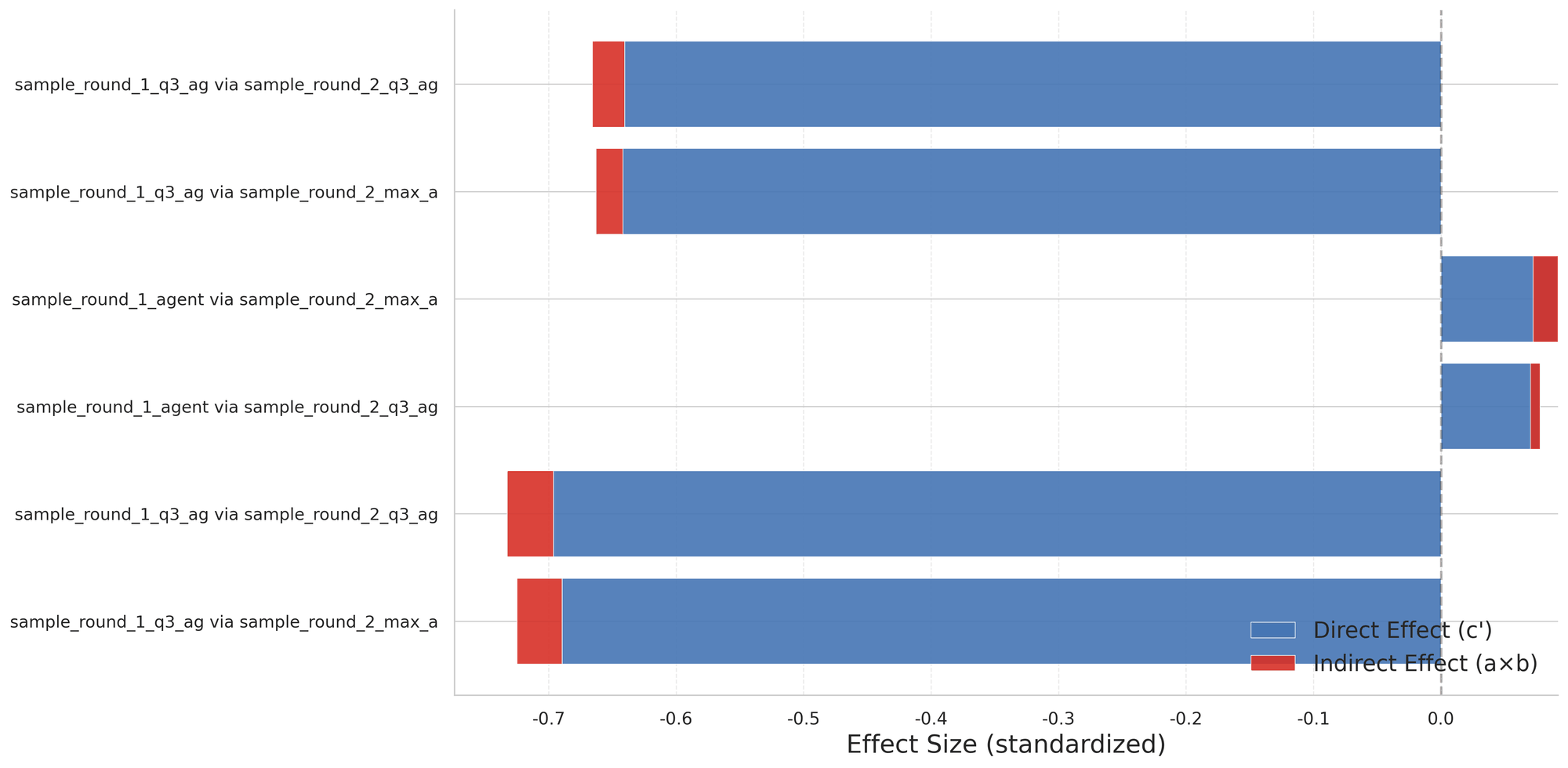}
        \caption{Mediation effect decomposition}
    \end{subfigure}
    \caption{Causal triplet on \texttt{GAIA}. Round-1 tool success rate is the unique consensus direct cause of correctness; round-1 inter-agent dispersion mediates a significant indirect path through round-2 dispersion.}
    \label{fig:gaia-causal}
\end{figure*}

\subsubsection{Tool-Failure Attribution by Base Model}
\label{app:gaia-tool-failure}

\begin{figure*}[t]
    \centering
    \includegraphics[width=\columnwidth]{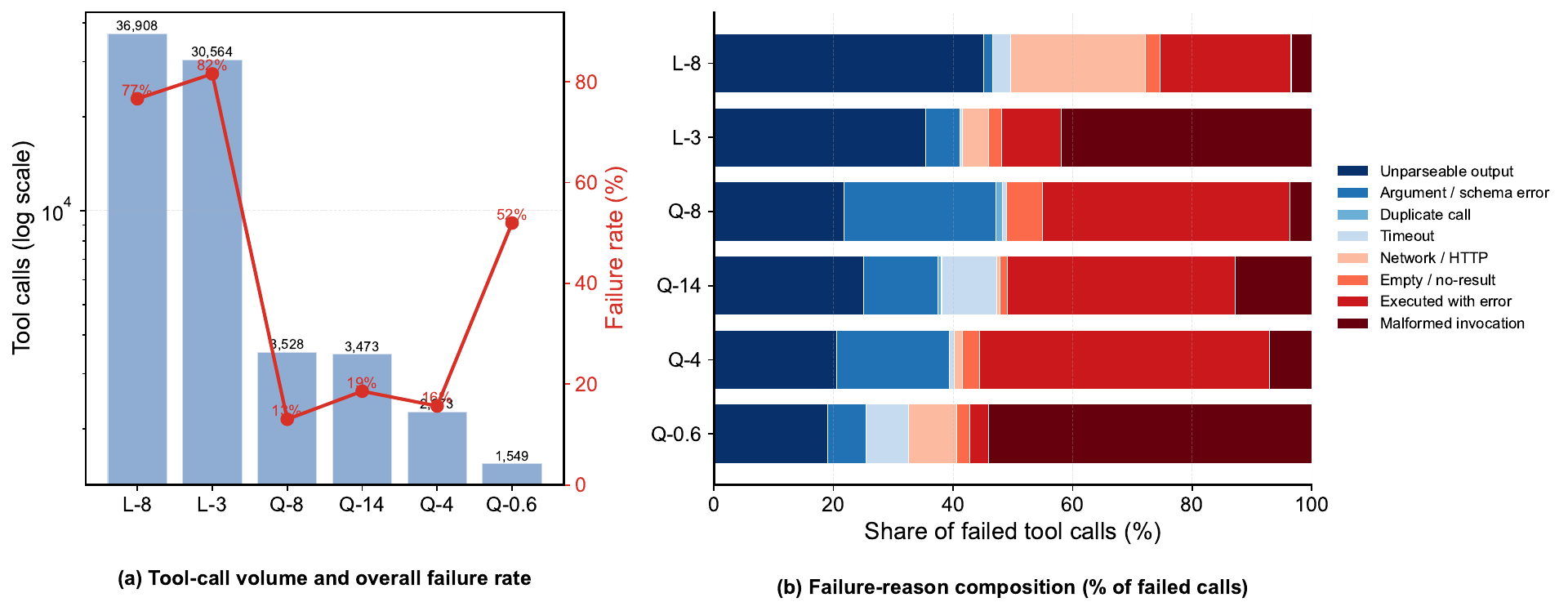}
    \caption{Tool-call failure attribution per base model. (a) Total tool-call volume (log scale, blue) and overall failure rate (red). (b) Composition of failed calls into eight categories.}
    \label{fig:gaia-tool-failure}
\end{figure*}

Figure~\ref{fig:gaia-tool-failure} attributes the 78,295 tool calls executed during the GAIA experiments to eight failure categories. The aggregate failure rate is 70.8\%. The decomposition reveals that failure is model-shaped, not random. Weak models (Llama-3.2-3B at 81.6\% failure rate, Llama-3.1-8B at 76.6\%, Qwen3-0.6B at 52.0\%) issue many calls, and most of those calls fail at the interface layer: malformed invocations that the framework rejects with non-standard errors (54.0\% of failures for Q-0.6, 41.9\% for L-3) or unparseable output (45.1\% for L-8). The tool never actually ran. Strong models (Qwen3-8B at 13.0\%, Qwen3-4B at 15.7\%, Qwen3-14B at 18.6\%) call tools much less often, and their failures concentrate on executed-with-error and empty results: the tool ran, but with the wrong intent. \textbf{Weak models struggle to produce valid formats, so the priority is syntactic correctness, while strong models need to improve the quality of generated content like search queries and code to ensure effective tool execution.}

\subsection{FinanceAgent: Architecture Overhead Dominates over Entropy Control}
\label{app:agent-benchmark}

This section complements the GAIA results with a financial QA benchmark to further validate the entropy-performance findings on tool-calling agent tasks.

\subsubsection{Experimental Setup}
\label{app:agent-benchmark-setup}

\paragraph{Benchmark and Task Characteristics.}
We evaluate on \texttt{FinanceAgent} Benchmark, a financial QA benchmark that fundamentally differs from the mathematical reasoning tasks in the main paper. Each question requires the agent to retrieve relevant SEC filings, extract financial metrics, perform multi-step numerical computations, and synthesize information across documents. This introduces several sources of complexity absent from our primary benchmarks: (1) tool selection entropy; (2) information retrieval noise from irrelevant documents; and (3) multi-step dependency chains where errors in early tool calls propagate.

The original benchmark was designed for a single closed-source API model; \citet{science-of-scaling-mas-arxiv25} extended it to a multi-agent setting but still relied on closed-source APIs and did not release evaluation code. We further extend it to open-source LLM-based MAS, fully open-source our evaluation code, and optimize the SEC filing retrieval tool with a local caching mechanism that eliminates rate-limit failures caused by repeated identical requests.

\paragraph{Model and Configuration.}
We use Qwen3-4B with thinking mode enabled as the base model, with the same five architectures, temperature 0.6, top-$p$ 0.95, and $R=2$ rounds as in GAIA. The agent has access to four financial tools: SEC filing retrieval, real-time stock data queries, financial metric computation, and financial news search. Each question allows up to 5 tool calls per step and 20 execution turns. Token-level entropy is recorded at each ReAct step, providing the same step-level feature set (\texttt{step\_}$k$) introduced in Appendix~\ref{app:gaia-setup}.

\subsubsection{Accuracy Results}
\label{app:agent-benchmark-accuracy}

\begin{table}[t]
\centering
\caption{\texttt{FinanceAgent} Benchmark accuracy by architecture. SAS and Sequential achieve the highest accuracy; architectures with greater coordination overhead exhibit progressively worse performance.}
\label{tab:agent-benchmark-accuracy}
\footnotesize
\begin{tabular}{@{}lccc@{}}
\toprule
\textbf{Architecture} & \textbf{Accuracy (\%)} & \textbf{$\Delta$ vs.\ SAS} & \textbf{Avg.\ Time (s)} \\
\midrule
Single (SAS) & 40 & --- & 1255 \\
Sequential & 40 & +0 & 4691 \\
Centralized & 22 & $-18$ & 3804 \\
Hybrid & 12 & $-28$ & 6387 \\
Debate & 2 & $-38$ & 5415 \\
\bottomrule
\end{tabular}
\end{table}

Table~\ref{tab:agent-benchmark-accuracy} reports accuracy across architectures. Single-agent (SAS) achieves the highest accuracy (40\%), tied with Sequential, while multi-agent architectures that introduce coordination overhead substantially underperform: Centralized achieves 22\%, Hybrid 12\%, and Debate only 2\%. This result is consistent with the main paper finding that SAS outperforms MAS in a substantial fraction of configurations (Section~\ref{sec:examining_uncertainty_impacts}). On tool-augmented tasks, the coordination overhead inherent in multi-agent architectures appears even more detrimental: Debate's near-zero accuracy suggests that majority voting is particularly ineffective when tool-calling decisions must be coordinated across agents.

\subsubsection{Feature Importance Analysis}
\label{app:agent-benchmark-features}

\begin{figure*}[t]
    \centering
    \begin{minipage}{0.95\textwidth}
    \begin{subfigure}{0.48\textwidth}
        \centering
        \includegraphics[width=\linewidth]{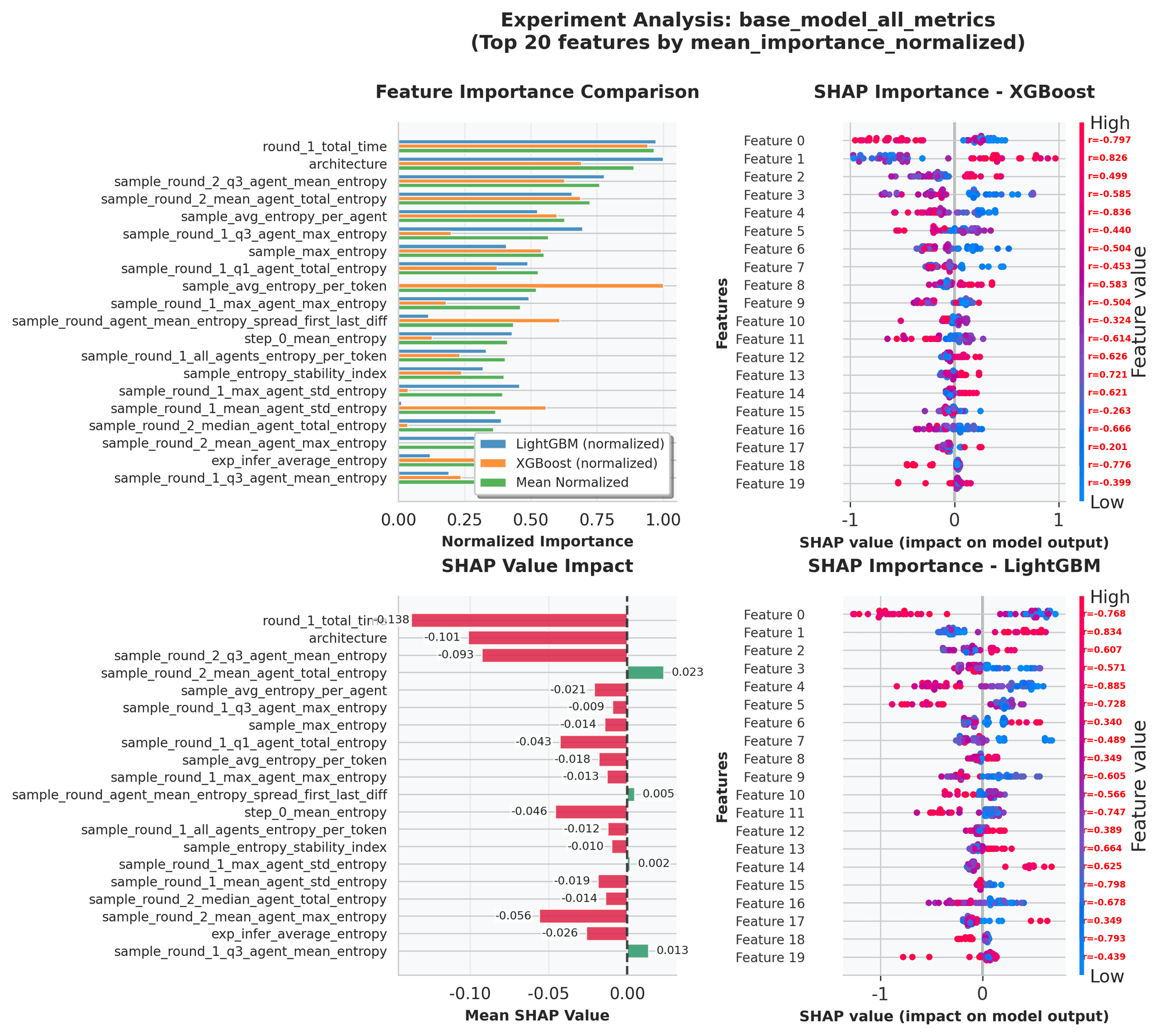}
        \caption{$\mathcal{G}_{\text{MAS}}$: MAS-only features}
        \label{fig:agent-benchmark-mas}
    \end{subfigure}
    \hfill
    \begin{subfigure}{0.48\textwidth}
        \centering
        \includegraphics[width=\linewidth]{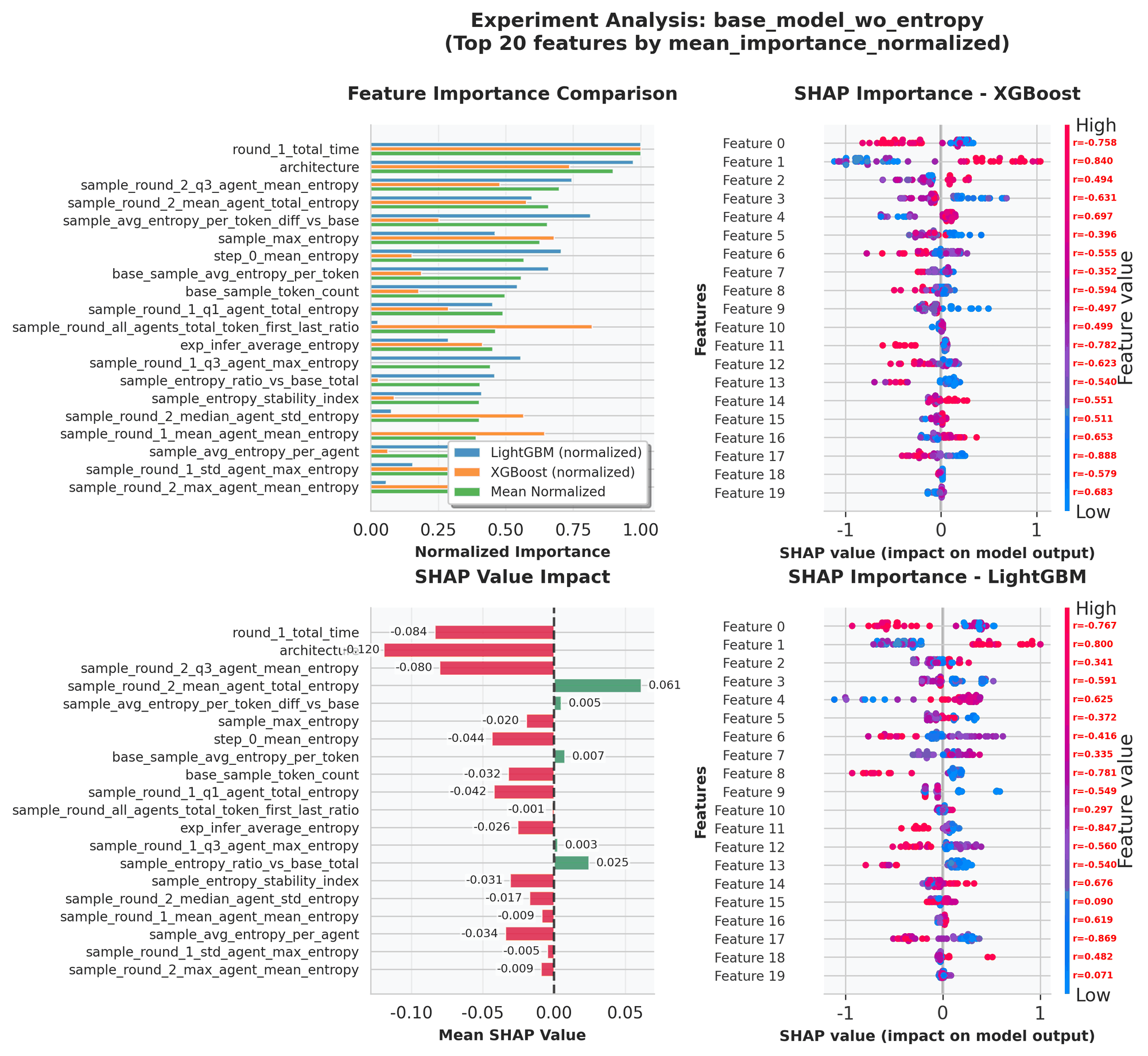}
        \caption{$\mathcal{G}_{\text{base-H}}$: Including base model entropy}
        \label{fig:agent-benchmark-base-entropy}
    \end{subfigure}
    \end{minipage}
    \caption{Top 20 features on \texttt{FinanceAgent} across two feature groups: (a) MAS-only features ($\mathcal{G}_{\text{MAS}}$), where architecture ($\rho \approx 0.83$) and step-level entropy dominate; (b) including base model entropy ($\mathcal{G}_{\text{base-H}}$), where architecture remains the top predictor ($\rho \approx 0.84$) and \texttt{step\_0\_mean\_entropy} shows moderate negative correlation ($\rho \approx -0.56$). On the full feature set ($\mathcal{G}_{\text{base-full}}$), \texttt{base\_model\_is\_finally\_correct} ($\rho \approx 0.96$) dominates all other features, consistent with the pattern observed across other settings.}
    \label{fig:agent-benchmark-combined}
\end{figure*}

We apply the same SHAP-based feature importance pipeline (Appendix~\ref{app:shap_analysis}) to the \texttt{FinanceAgent} Benchmark results. Figure~\ref{fig:agent-benchmark-combined} presents three complementary views: (a) MAS-only features excluding base model entropy, (b) features including base model entropy, and (c) the full feature set including base model correctness.

\paragraph{Architecture is the dominant predictor.}
Across all three feature groups, architecture is the top or near-top predictor with $\rho \approx 0.83$-$0.88$. The ordinal encoding runs from Centralized (0) to Single (4), so the positive sign means simpler architectures perform better. On a tool-calling task, coordination overhead amplifies tool-selection entropy rather than averaging it out.

\paragraph{Step-level entropy refines the ``first-round decisive'' finding.}
The step-level features, which split round 1 into individual ReAct iterations, expose a sub-round structure invisible to the main analysis. \texttt{step\_0\_mean\_entropy} enters the top 20 with $\rho \approx -0.75$ on $\mathcal{G}_{\text{MAS}}$ and $\rho \approx -0.56$ on $\mathcal{G}_{\text{base-H}}$ (the attenuation reflects partial overlap with base-model entropy). The interpretation is sharper than ``round 1 matters'': the first reasoning step alone carries most of the round-1 signal. In parallel, \texttt{sample\_round\_2\_q3\_agent\_mean\_entropy} correlates positively ($\rho \approx +0.49$), giving the same ``decisive initialization, exploratory refinement'' profile observed in the RL experiment (Appendix~\ref{app:rl_finetuned_base}).

\paragraph{Base-model correctness still dominates on $\mathcal{G}_{\text{base-full}}$.}
\texttt{base\_model\_is\_finally\_correct} reaches $\rho \approx 0.96$, identical to the math-reasoning result in Appendix~\ref{app:base_model_entropy}. Tool-calling adds a layer of complexity that MAS coordination frequently fails to manage, but it does not loosen the base-model dependency.

\paragraph{Causal validation.}
On \texttt{FinanceAgent}, both PC and FCI converge on a single consensus direct cause of correctness: \textit{sample\_round\_1\_q3\_agent\_max\_entropy} (IPW ATE $-0.197$, $p=1.5\!\times\!10^{-17}$, all refutation tests pass). The dominant mediation runs from round-1 Q1 of per-agent total entropy through round-2 mean per-agent total entropy to correctness (indirect $-0.184$, mediating 68.9\%), the largest cross-round mediation proportion observed in any setting in this paper. Together these substantiate the SHAP ``first reasoning step / round-1 decisive'' reading with stronger causal evidence: under tool-calling, where each round bundles multiple ReAct steps, round-1 dispersion is not merely correlated with failure but the mechanistic source of the round-2 entropy state that drives the final outcome.

\begin{figure*}[!htbp]
    \centering
    \begin{minipage}{0.95\textwidth}
    \begin{subfigure}{0.32\textwidth}
        \centering
        \includegraphics[width=\linewidth]{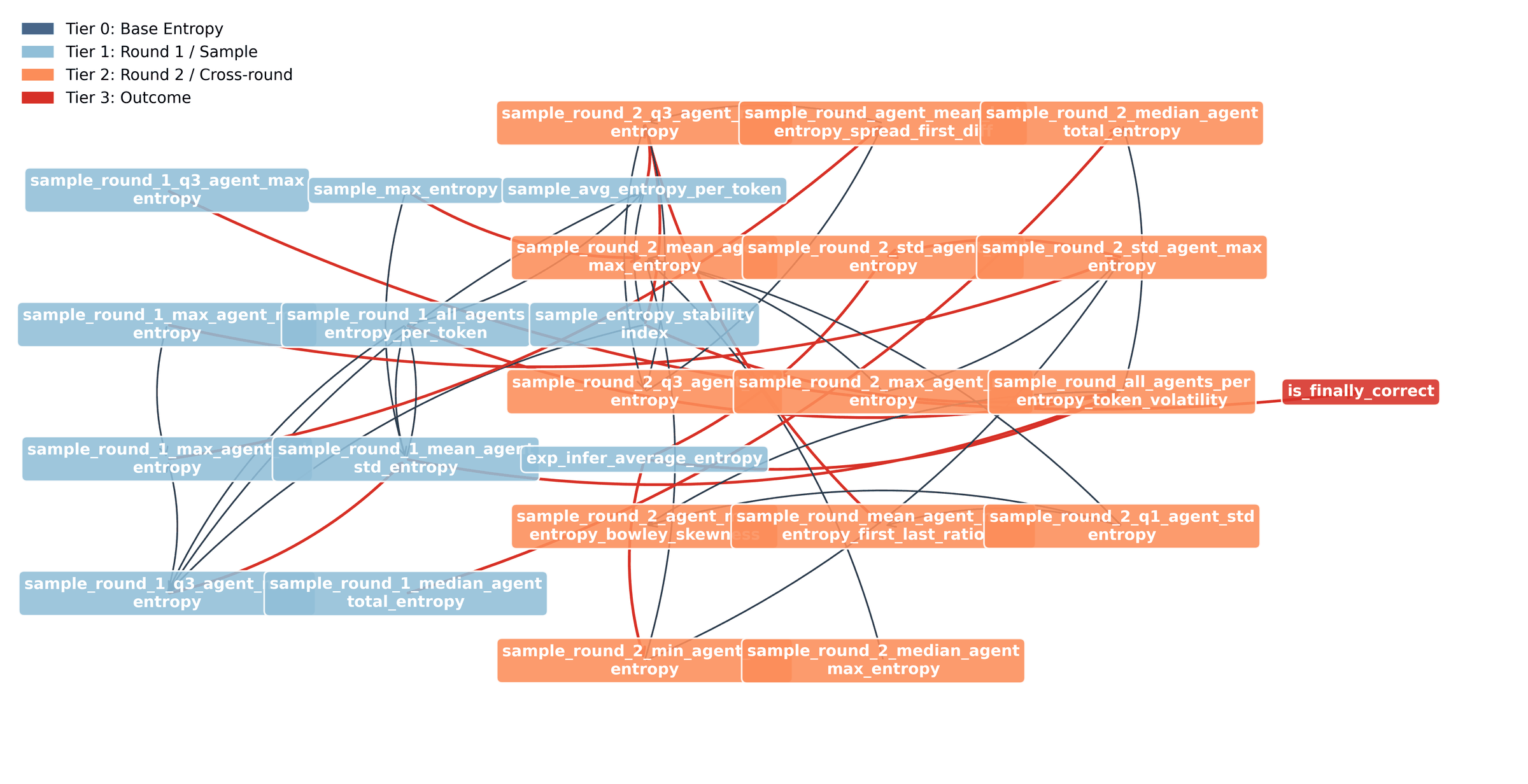}
        \caption{Consensus causal graph}
    \end{subfigure}
    \hfill
    \begin{subfigure}{0.32\textwidth}
        \centering
        \includegraphics[width=\linewidth]{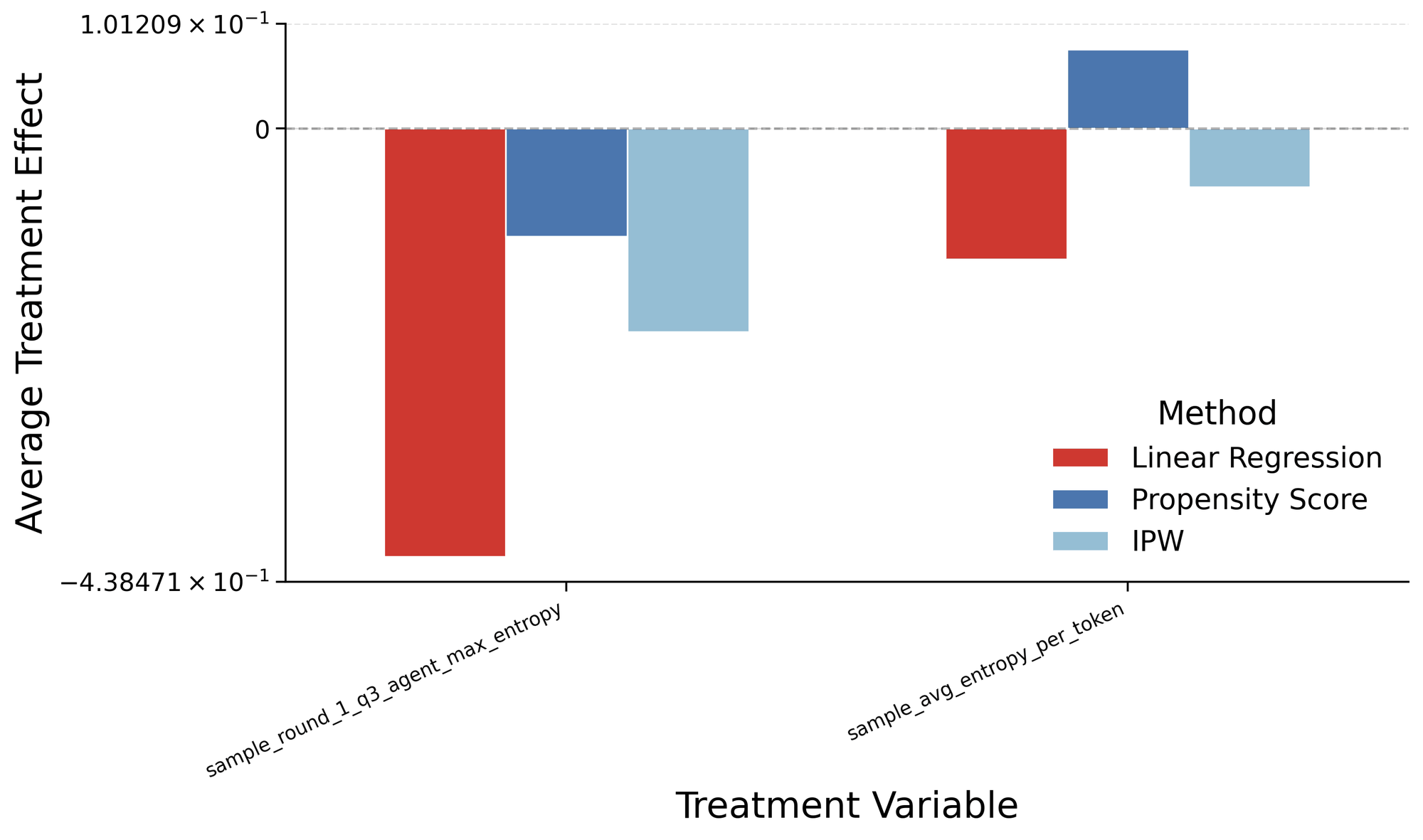}
        \caption{ATE estimates}
    \end{subfigure}
    \hfill
    \begin{subfigure}{0.32\textwidth}
        \centering
        \includegraphics[width=\linewidth]{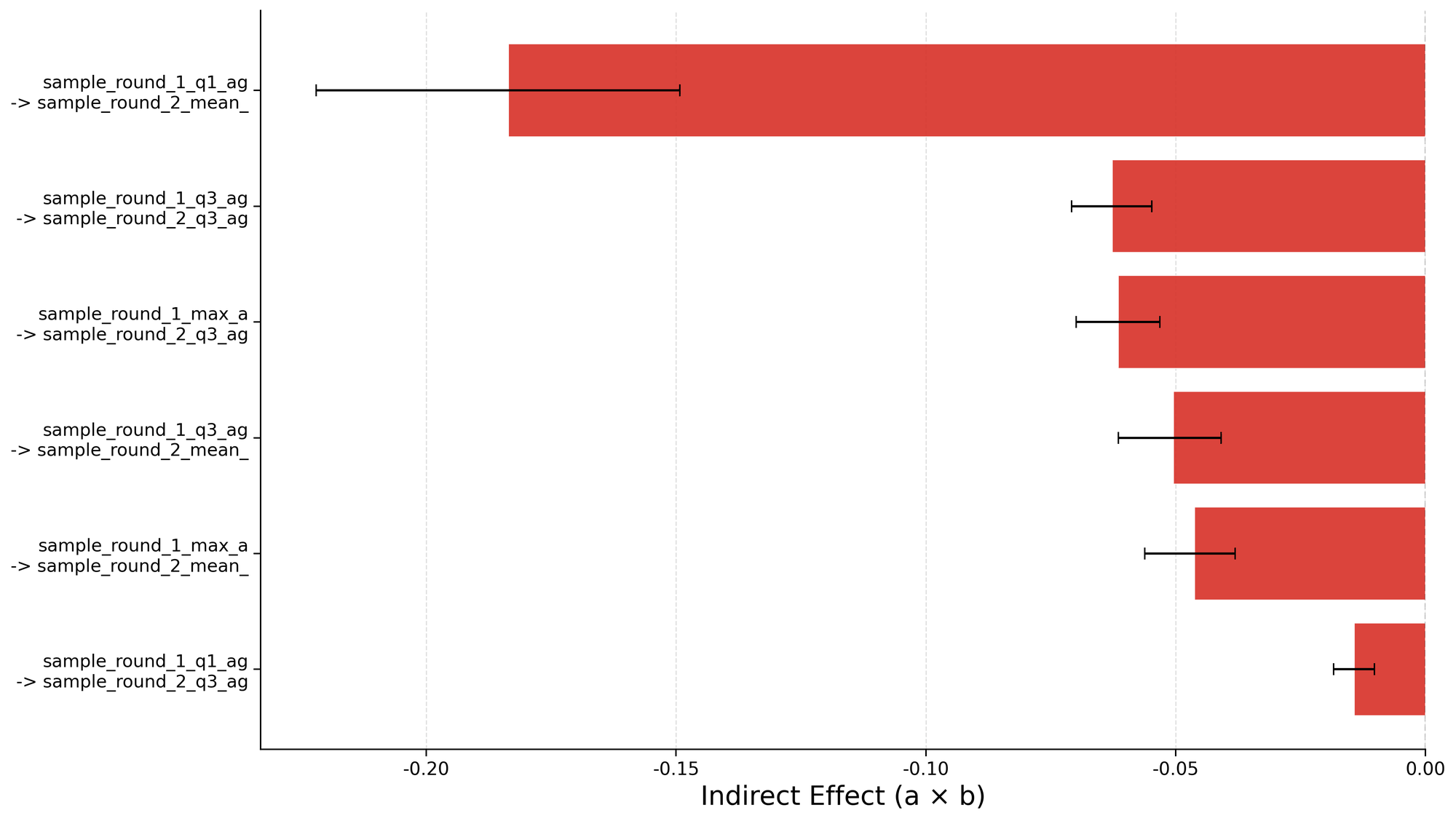}
        \caption{Mediation indirect effects}
    \end{subfigure}
    \end{minipage}
    \caption{Causal triplet on \texttt{FinanceAgent}. Round-1 Q3 of per-agent maximum entropy is the unique consensus direct cause; round-1 dispersion mediates a 68.9\% share of its effect through round-2 entropy.}
    \label{fig:causal-agent-benchmark}
\end{figure*}

\section{More Experimental Results}
\label{app:experimental_results}

This section extends the main text analysis along two dimensions. First, it expands the SHAP-based feature scope from top 2-5 to top 10 and incorporates the mean SHAP impact $\bar{S}$, which quantifies each feature's average contribution to predicted correctness across all samples. Second, and more importantly, since SHAP analysis only reveals correlations and cannot pinpoint which factor actually drives MAS failure, we supplement every key finding in the main text with the same complete causal analysis pipeline as in Appendix~\ref{app:causal-discovery}, making the findings more reliable. While Appendix~\ref{app:causal-discovery} offers a single global causal view over all data, this section instead applies the pipeline finding-by-finding, validating each finding within its own experimental setting.

\subsection{Base Model Entropy Dominates MAS Prediction}
\label{app:base_model_entropy}

\begin{figure*}[t]
    \centering
    \begin{minipage}{0.95\textwidth}
    \begin{subfigure}{0.48\textwidth}
        \centering
        \includegraphics[width=\linewidth]{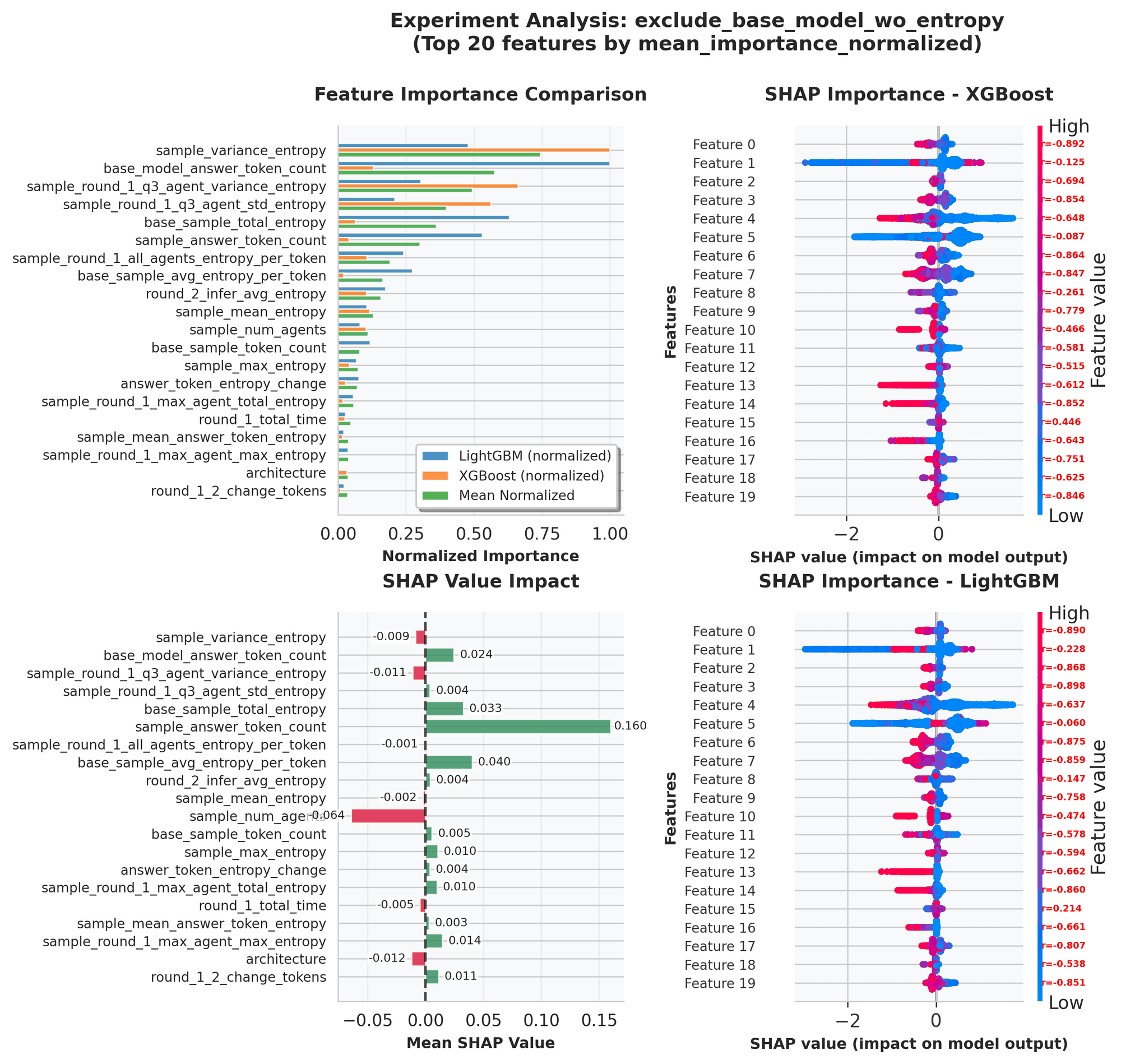}
        \caption{Qwen models.}
        \label{fig:base_model_entropy_qwen_overall}
    \end{subfigure}
    \hfill
    \begin{subfigure}{0.48\textwidth}
        \centering
        \includegraphics[width=\linewidth]{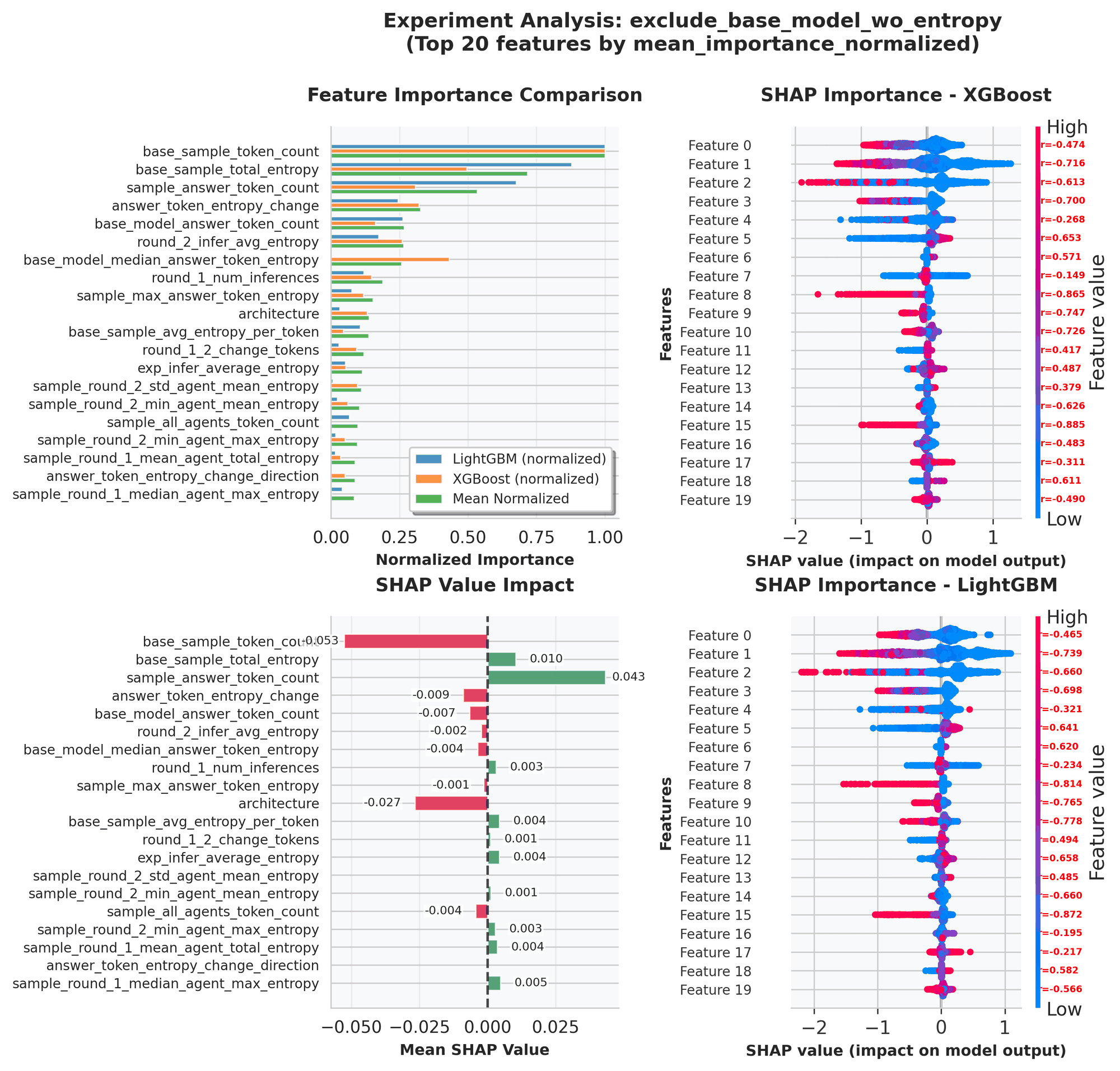}
        \caption{LLaMA models.}
        \label{fig:base_model_entropy_llama_overall}
    \end{subfigure}
    \end{minipage}
    \caption{Top 20 features on $\mathcal{G}_{\text{base-H}}$ for Qwen (a) and LLaMA (b), ranked by mean normalized importance $\bar{I}$. Each panel is divided into four subplots: top-left shows feature importance from XGBoost and LightGBM; bottom-left shows mean SHAP impact $\bar{S}$, representing the average contribution of each feature to model predictions; right column displays scatter plots of feature values versus SHAP values, with Pearson correlation $\rho$ annotated in red. All subsequent figures follow this layout except for the SHAP waterfall plots.}
    \label{fig:base_model_entropy_overall_comparison}
\end{figure*}

\begin{figure*}[t]
    \centering
    \begin{minipage}{0.95\textwidth}
    \begin{subfigure}{0.48\textwidth}
        \centering
        \includegraphics[width=\linewidth]{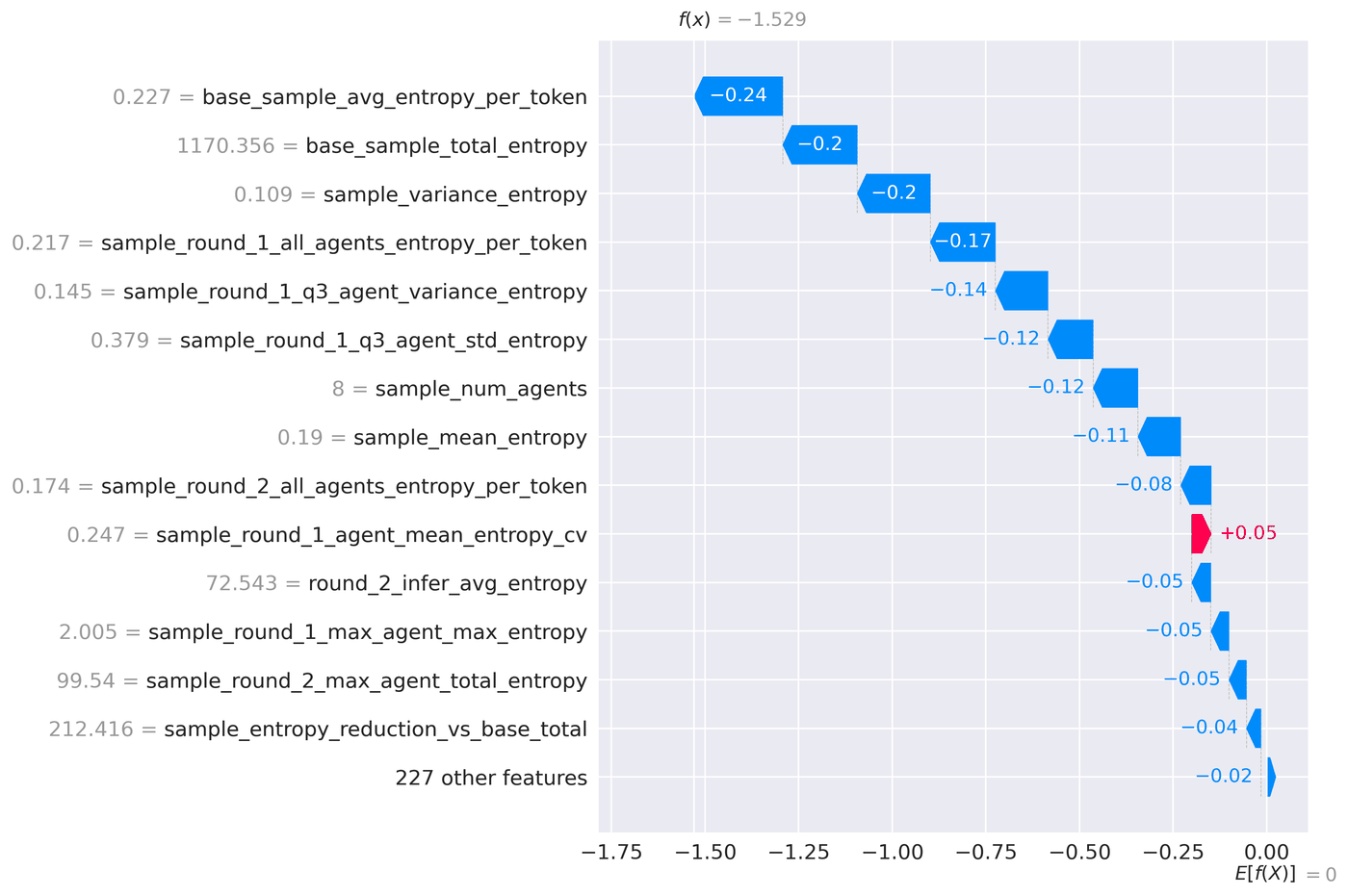}
        \caption{Qwen (LightGBM)}
        \label{fig:base_model_entropy_qwen_lightgbm}
    \end{subfigure}
    \hfill
    \begin{subfigure}{0.48\textwidth}
        \centering
        \includegraphics[width=\linewidth]{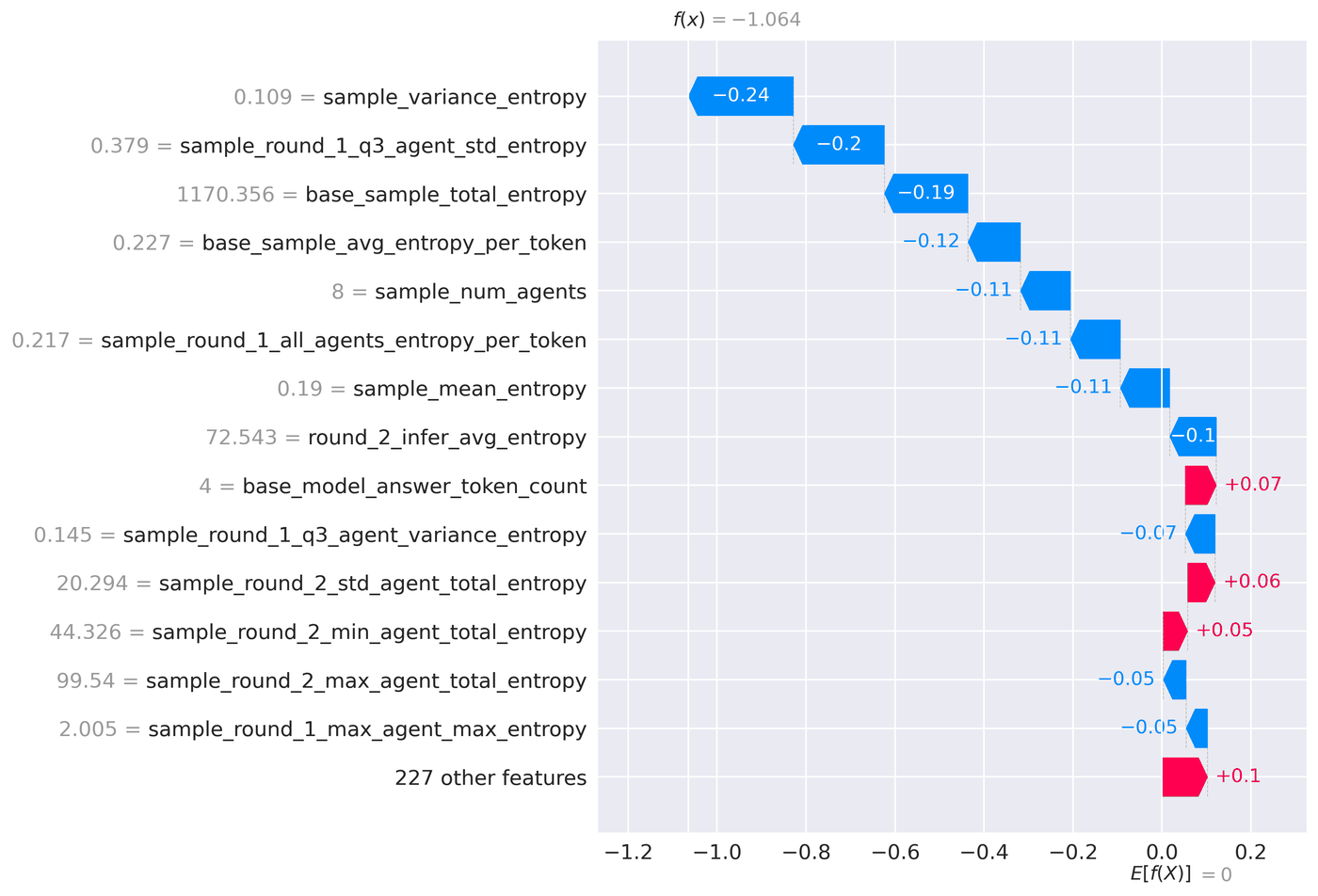}
        \caption{Qwen (XGBoost)}
        \label{fig:base_model_entropy_qwen_xgboost}
    \end{subfigure}

    \vspace{1em}

    \begin{subfigure}{0.48\textwidth}
        \centering
        \includegraphics[width=\linewidth]{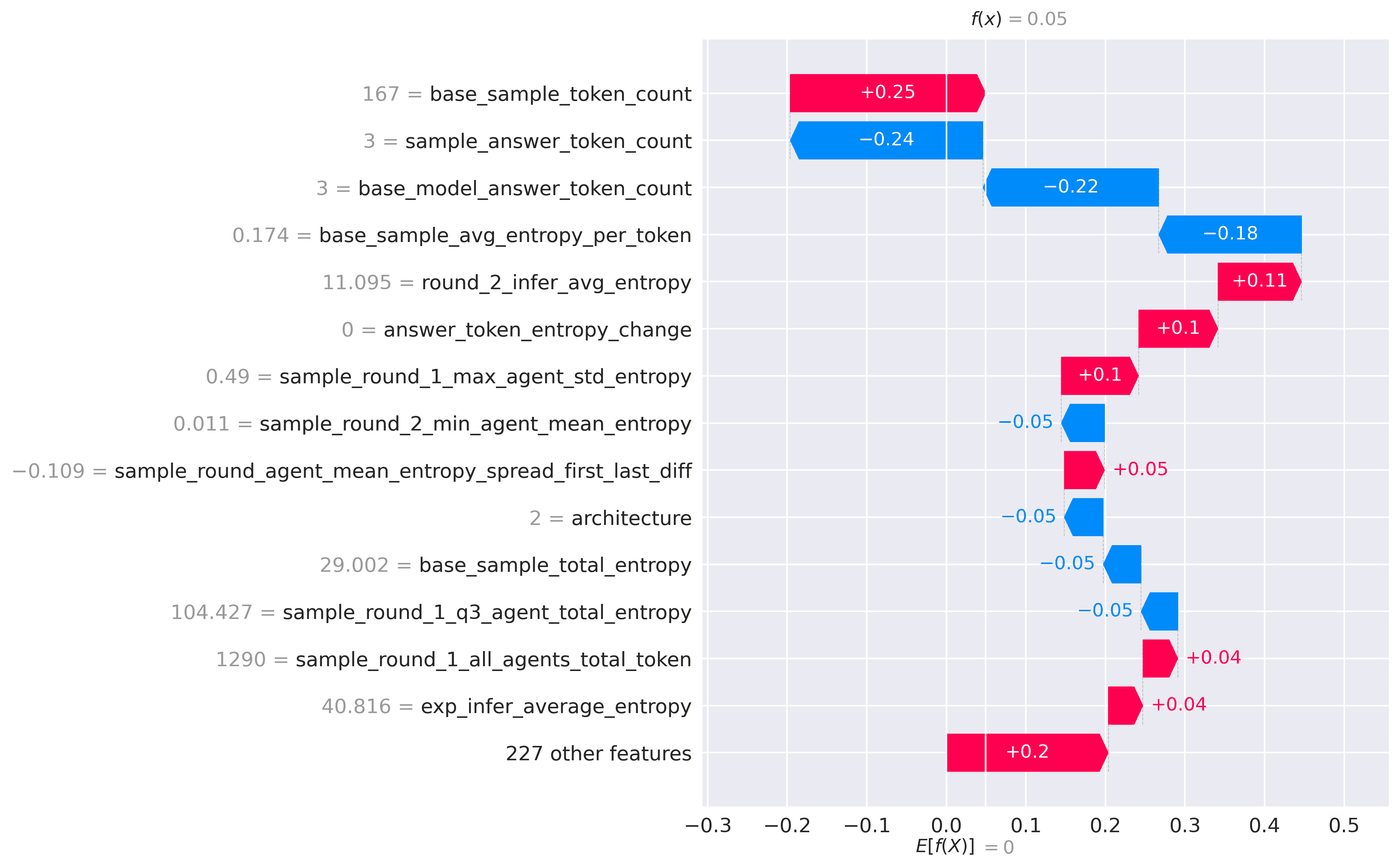}
        \caption{LLaMA (LightGBM)}
        \label{fig:base_model_entropy_llama_lightgbm}
    \end{subfigure}
    \hfill
    \begin{subfigure}{0.48\textwidth}
        \centering
        \includegraphics[width=\linewidth]{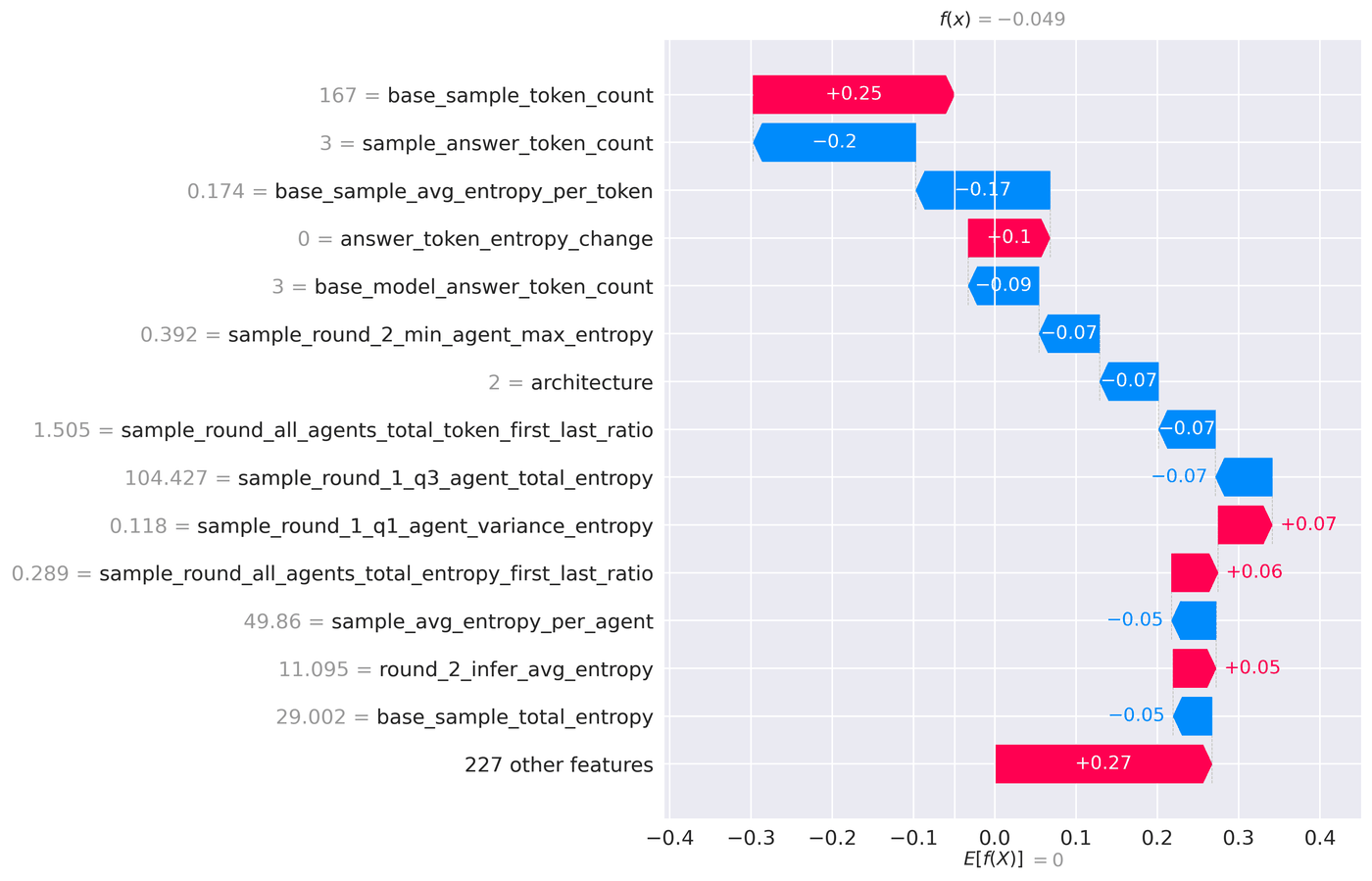}
        \caption{LLaMA (XGBoost)}
        \label{fig:base_model_entropy_llama_xgboost}
    \end{subfigure}
    \end{minipage}
    \caption{SHAP waterfall plots on $\mathcal{G}_{\text{base-H}}$ for representative samples: Qwen and LLaMA, with LightGBM and XGBoost. Each bar shows the contribution of a feature to the predicted MAS correctness.}
    \label{fig:base_model_entropy_combined_all}
\end{figure*} 

Figure~\ref{fig:base_model_entropy_overall_comparison} ranks the top 20 features on $\mathcal{G}_{\text{base-H}}$ for each model family, and Figure~\ref{fig:base_model_entropy_combined_all} reports per-sample SHAP attributions. Two failure modes separate the families, and they correspond to the two entropy regimes the families operate in.

\paragraph{Qwen (high-entropy regime, 100-1{,}000): inter-agent dispersion.}
For Qwen, the top predictors are dispersion statistics rather than absolute entropy: variance of per-token entropy and round-1 Q3 agent variance entropy both show strong negative SHAP correlation ($\rho \approx -0.89$ and $-0.78$). Lower-ranked entropy magnitudes (per-token, mean) point in the same direction. The picture is internally coherent: when agents in the same round disagree about how confident to be, the system fails, and this signal is robust across LightGBM and XGBoost waterfalls (Figures~\ref{fig:base_model_entropy_qwen_lightgbm}-\ref{fig:base_model_entropy_qwen_xgboost}).

\paragraph{LLaMA (low-entropy regime, 0-100): output length and base entropy.}
For LLaMA, dispersion is replaced by length. The dominant predictor is base sample token count ($\bar{I} = 1.0$, $\rho \approx -0.47$, $\bar{S} \approx -0.053$), reinforced by base-model entropy ($\rho \approx -0.73$) and the base-to-MAS answer entropy change ($\rho \approx -0.70$). Both the local and global SHAP contributions point the same way, so the failure mode is unambiguous: long, uncertain base responses do not get rescued by MAS. The LLaMA waterfalls (Figures~\ref{fig:base_model_entropy_llama_lightgbm}-\ref{fig:base_model_entropy_llama_xgboost}) accordingly show many small, partly cancelling contributions, consistent with a model that lacks a single concentrated failure signal.

Several features (notably \textit{round\_2\_infer\_avg\_entropy} for LLaMA, $\rho = +0.65$ but $\bar{S} = -0.002$) show positive sample-level correlation but negative global contribution. This pattern recurs throughout the appendix and is not a contradiction: it indicates a non-monotone relationship where a moderate value helps, but the upper tail is harmful enough to drag the global mean negative. We flag it once here and do not repeat the explanation for each subsequent occurrence.

\begin{figure*}[t]
    \centering
    \begin{minipage}{0.95\textwidth}

    \begin{subfigure}{0.48\textwidth}
        \centering
        \includegraphics[width=\linewidth]{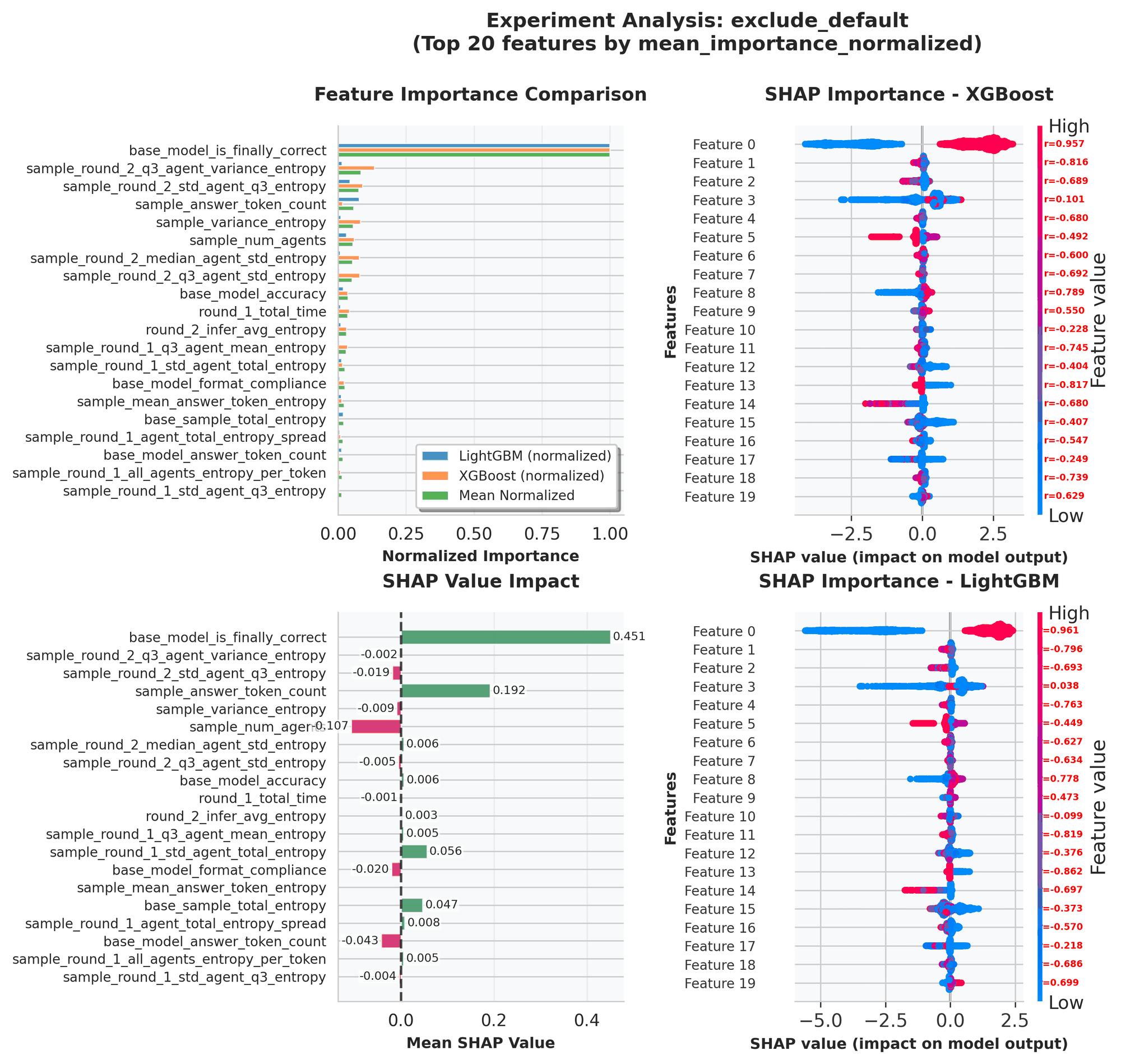}
        \caption{Base model correctness: Qwen}
    \end{subfigure}
    \hfill
    \begin{subfigure}{0.48\textwidth}
        \centering
        \includegraphics[width=\linewidth]{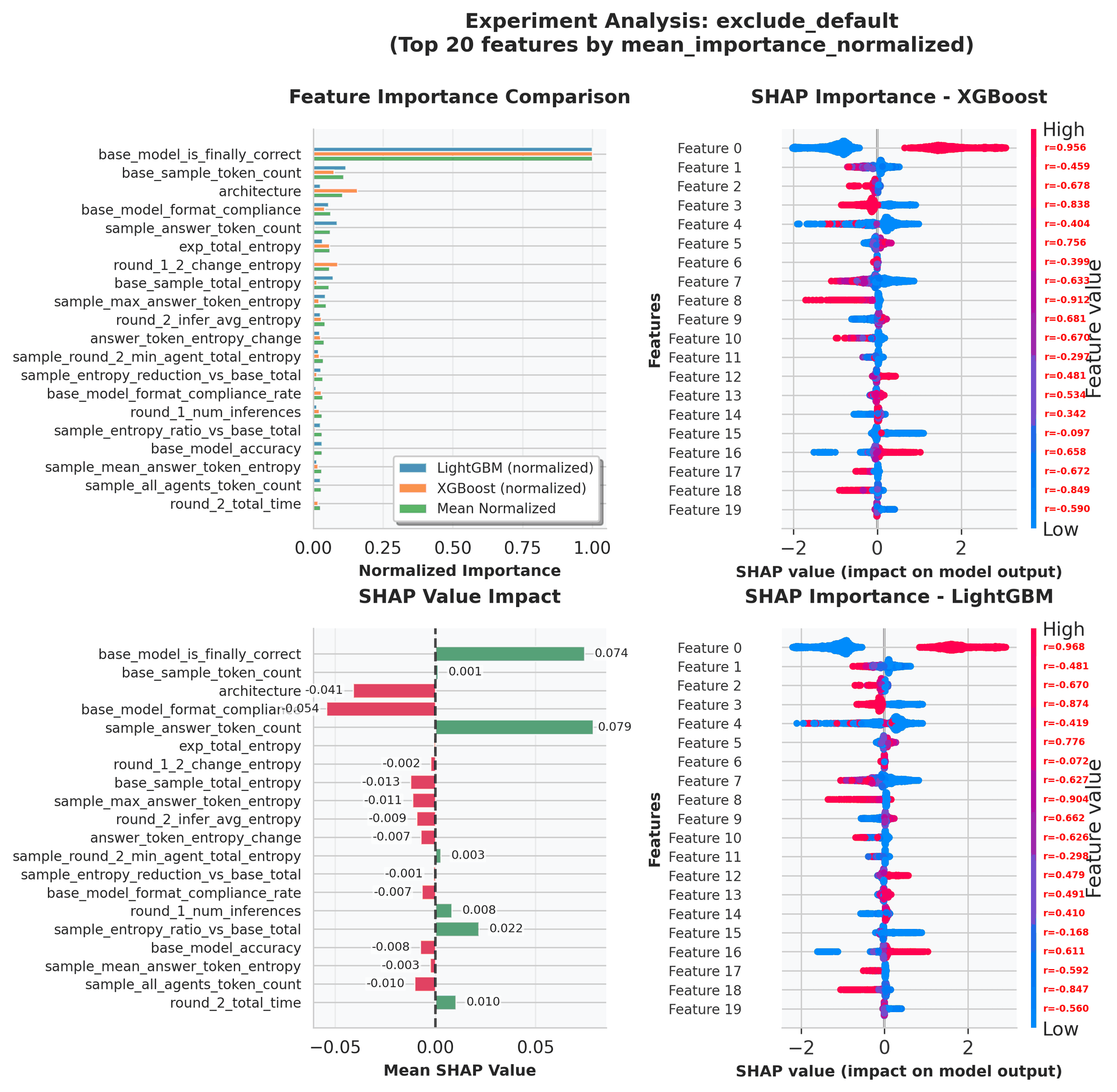}
        \caption{Base model correctness: LLaMA}
    \end{subfigure}
    
    \vspace{1em}
    
    \begin{subfigure}{0.48\textwidth}
        \centering
        \includegraphics[width=\linewidth]{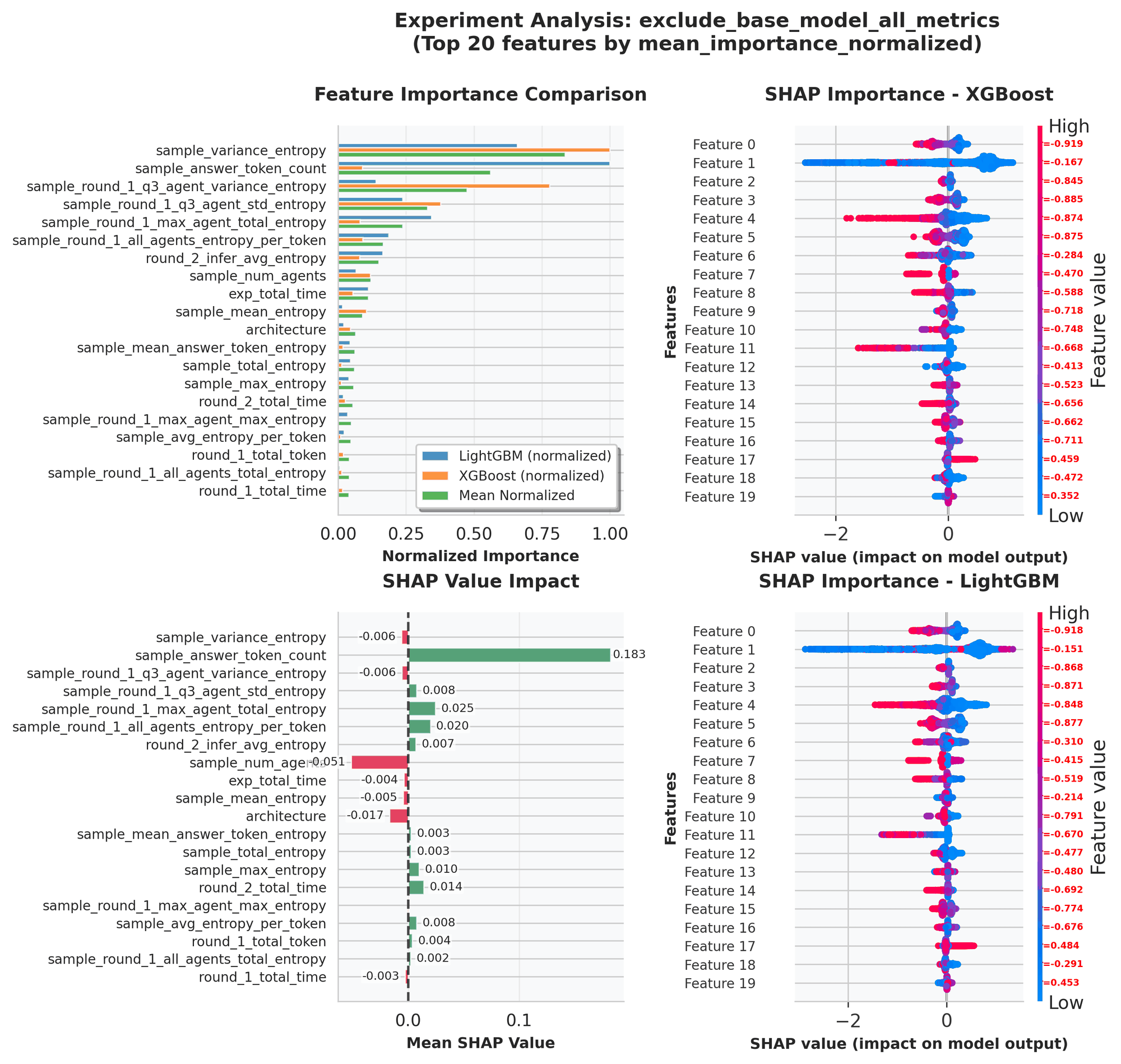}
        \caption{MAS failure analysis: Qwen}
    \end{subfigure}
    \hfill
    \begin{subfigure}{0.48\textwidth}
        \centering
        \includegraphics[width=\linewidth]{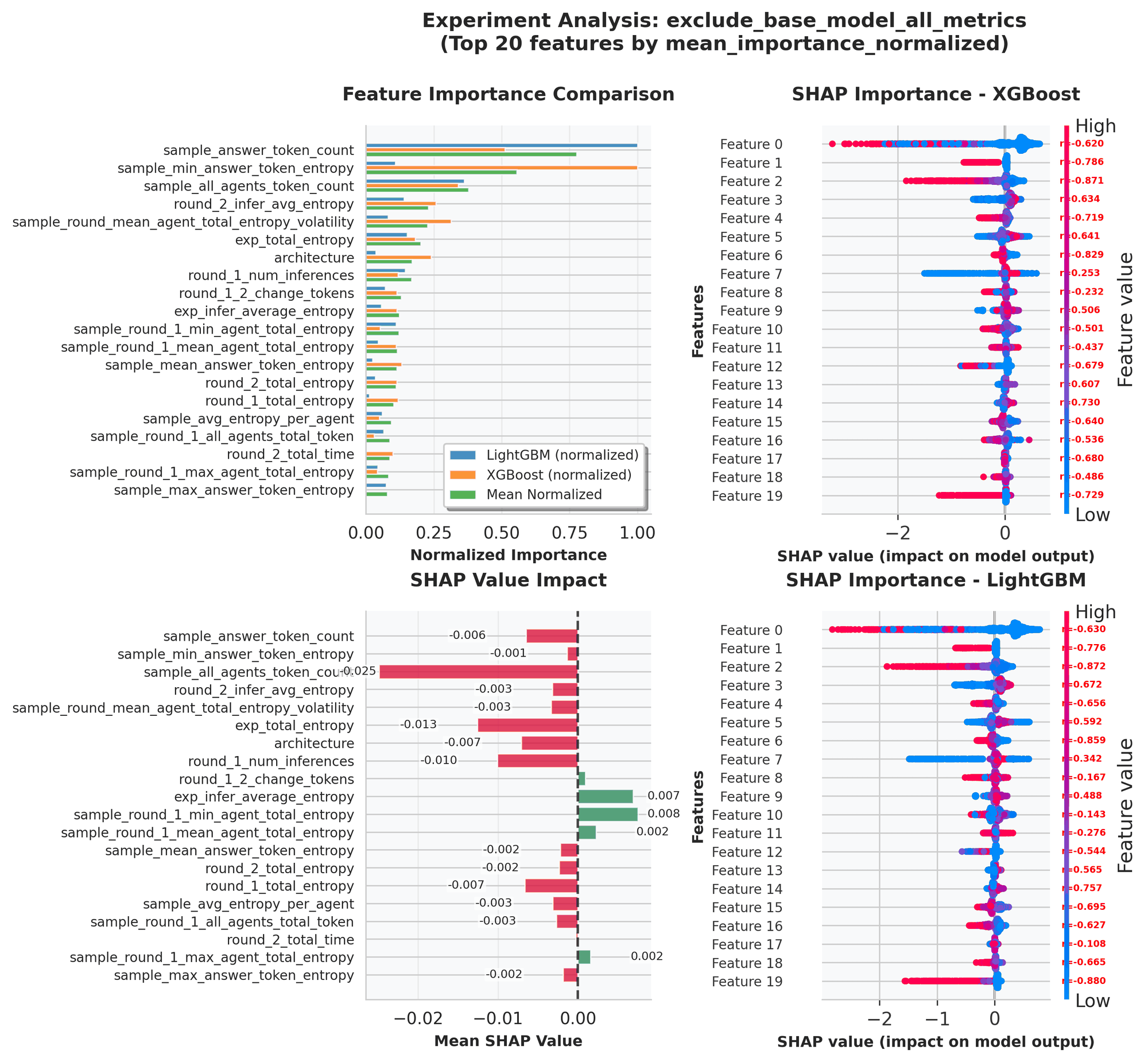}
        \caption{MAS failure analysis: LLaMA}
    \end{subfigure}
    
    \end{minipage}
    \caption{\textbf{Top:} Feature importance and SHAP analysis on $\mathcal{G}_{\text{base-full}}$ for Qwen (a) and LLaMA (b). Both show that \textit{base\_model\_is\_finally\_correct} achieves $\bar{I} = 1.0$ and $\rho \approx 0.96$, vastly surpassing all other features with nearly linear correlation to MAS correctness. \textbf{Bottom:} Top 20 features on $\mathcal{G}_{\text{MAS}}$ for MAS failure analysis: Qwen (c) and LLaMA (d). Qwen's top predictor is entropy variance, while LLaMA is dominated by answer-level features.}
    \label{fig:combined_base_model_and_mas_analysis}
\end{figure*}

\paragraph{Base model correctness overwhelms all other features.}
On $\mathcal{G}_{\text{base-full}}$ (Figure~\ref{fig:combined_base_model_and_mas_analysis}, top row), \textit{base\_model\_is\_finally\_correct} dominates with $\bar{I} = 1.0$, $\rho \approx 0.96$, and a strongly positive $\bar{S}$ ($+0.45$ for Qwen, $+0.07$ for LLaMA). MAS succeeds largely when the base model is already correct; once we condition on this single indicator, all other features become marginal.

\paragraph{Causal validation.}
Running the PC/FCI + DoWhy pipeline of Appendix~\ref{app:causal-discovery} per (model, dataset) cell on $\mathcal{G}_{\text{base-full}}$ reproduces the SHAP picture: of the 28 cells with successful causal estimation, base-model entropy features (\textit{base\_sample\_avg\_entropy\_per\_token}, \textit{base\_sample\_total\_entropy}, \textit{base\_model\_max/median/std\_answer\_token\_entropy}, \textit{answer\_token\_entropy\_change}, \textit{sample\_entropy\_ratio\_vs\_base\_total}, \textit{sample\_avg\_entropy\_per\_token\_diff\_vs\_base}) appear among the consensus PC$\cap$FCI direct causes in 17/28 cells, and sample-level answer-token-entropy features in 15/28 cells (the two sets overlap in 9 cells), with every IPW estimate negative on the dominant cause and every cell passing all three refutation tests. The causal direct effect of base-level entropy is therefore not a confound of the base-correctness indicator: even when conditioning on round-1/round-2 entropy, base-model per-token entropy retains a direct edge into correctness across both families. Section~\ref{sec:causal_analysis} of the main text reports the corresponding global $\mathcal{G}_{\text{base-full}}$ result, where the three consensus direct causes are \textit{base\_sample\_avg\_entropy\_per\_token} ($\text{ATE}_{\text{PS}}=-0.12$, $p<10^{-21}$), \textit{round\_1\_total\_entropy} ($p<10^{-19}$), and \textit{sample\_max\_answer\_token\_entropy} ($\text{ATE}_{\text{PS}}=-0.31$, $p<10^{-28}$); the per-cell evidence above shows this finding is not driven by one or two model-dataset combinations.

\begin{figure*}[!htbp]
    \centering
    \begin{minipage}{0.95\textwidth}
    \begin{subfigure}{0.32\textwidth}
        \centering
        \includegraphics[width=\linewidth]{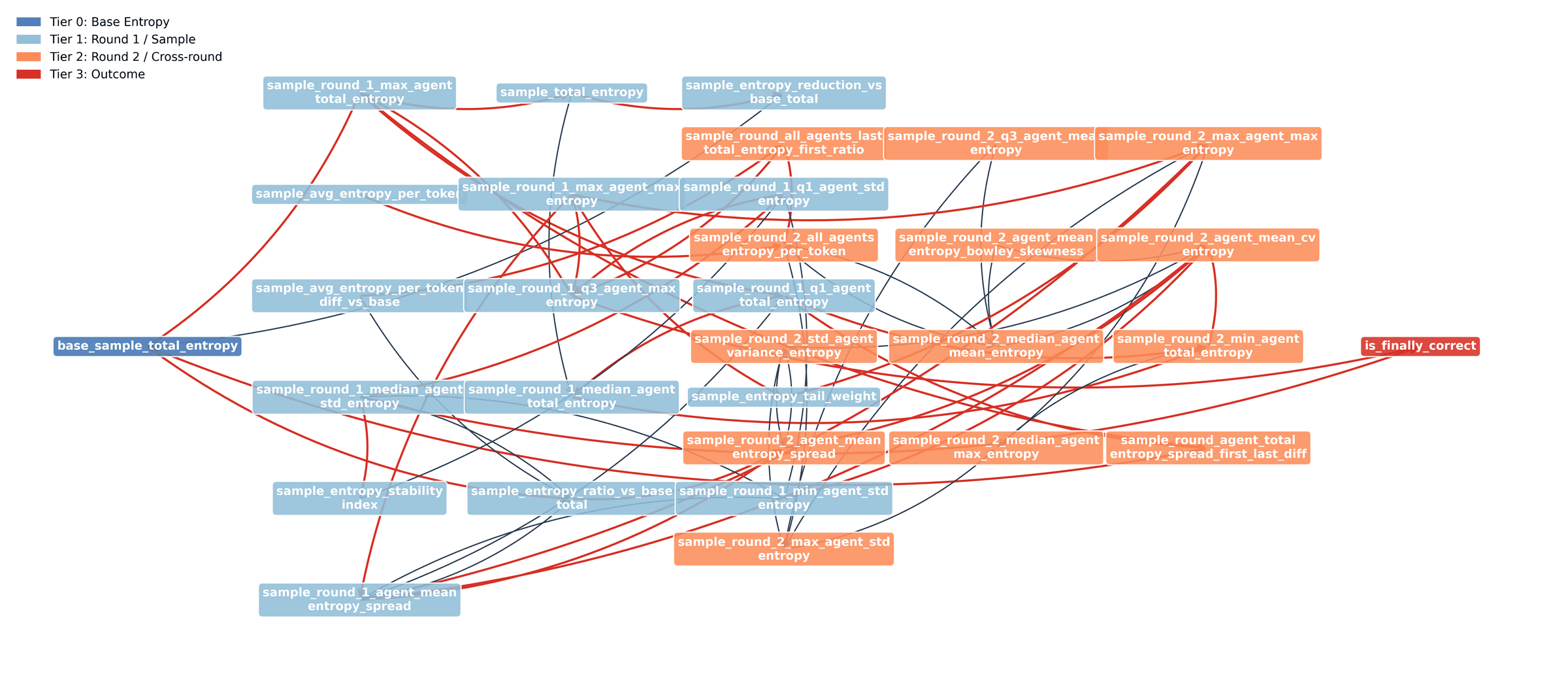}
        \caption{Consensus causal graph}
    \end{subfigure}
    \hfill
    \begin{subfigure}{0.32\textwidth}
        \centering
        \includegraphics[width=\linewidth]{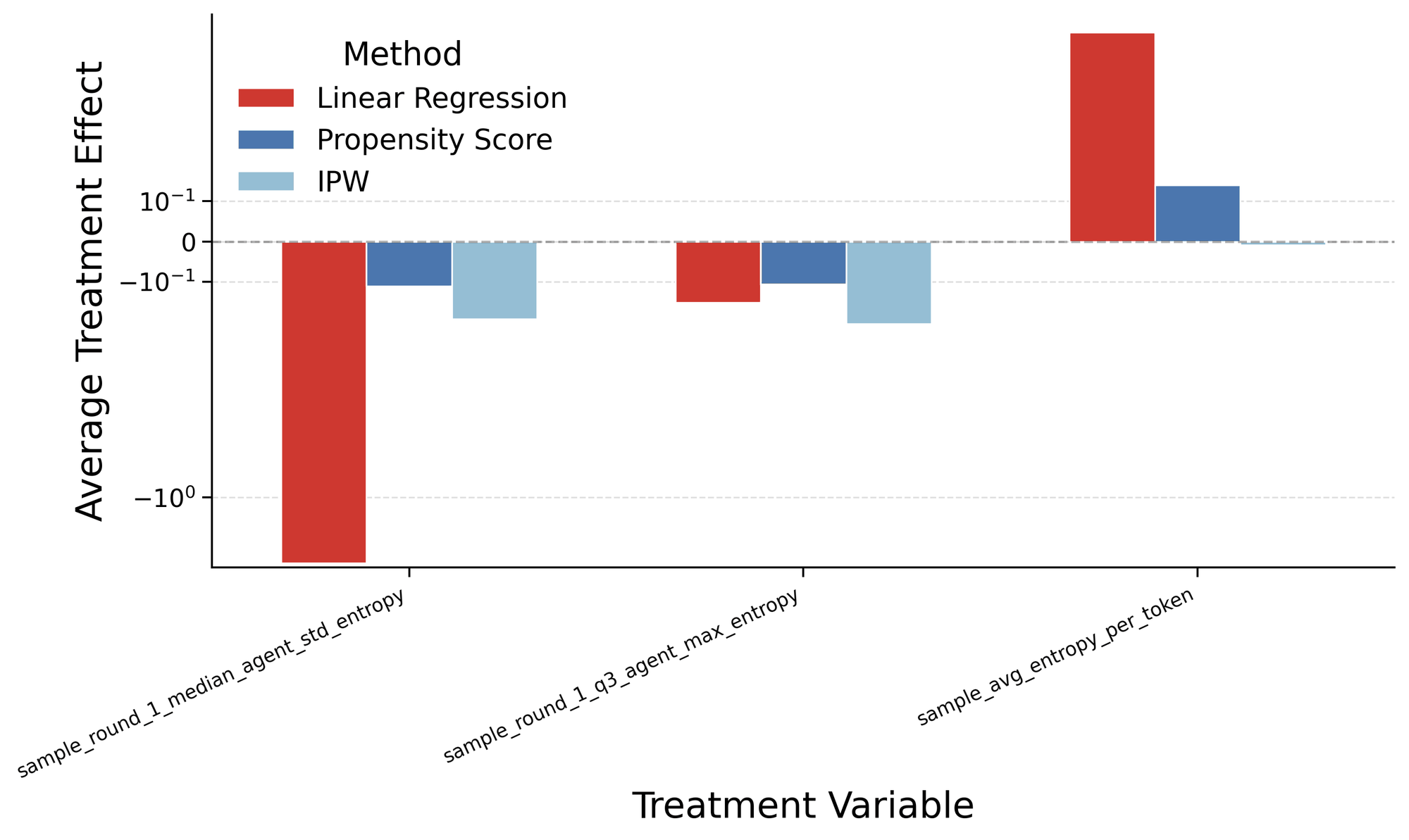}
        \caption{ATE estimates}
    \end{subfigure}
    \hfill
    \begin{subfigure}{0.32\textwidth}
        \centering
        \includegraphics[width=\linewidth]{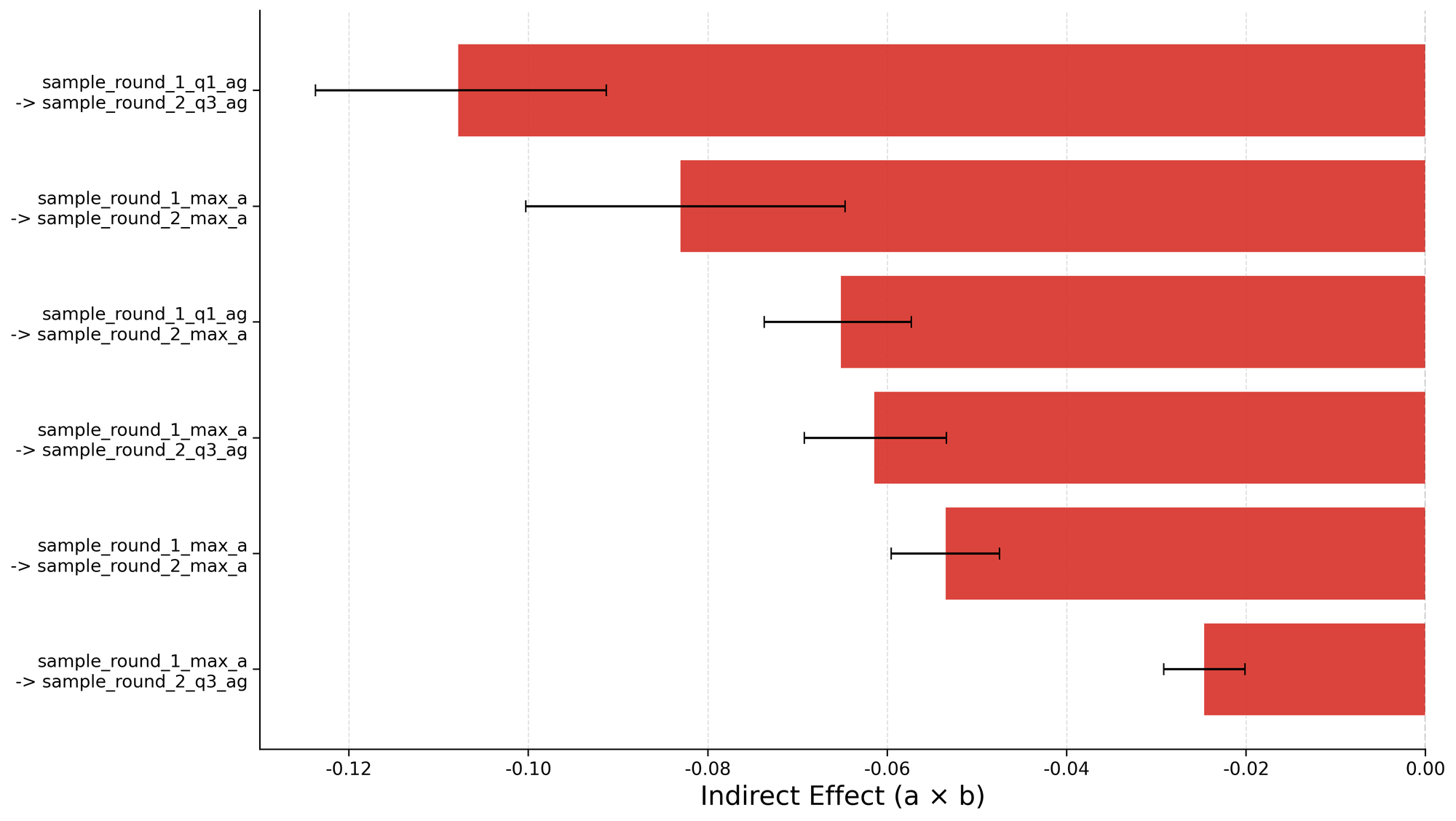}
        \caption{Mediation indirect effects}
    \end{subfigure}
    \end{minipage}
    \caption{Representative causal triplet on $\mathcal{G}_{\text{base-full}}$ (Qwen3-4B, \texttt{GSM8K}). The consensus graph identifies one base-entropy and one round-1 dispersion feature as direct causes; both have significantly negative IPW ATE and pass all three refutation tests.}
    \label{fig:causal-base-model-entropy}
\end{figure*}

\subsection{Inter-Agent Misalignment Causes MAS Failure}
\label{app:mas_failure}

Figure~\ref{fig:combined_base_model_and_mas_analysis} (bottom row) re-runs the analysis on $\mathcal{G}_{\text{MAS}}$, which removes all base-model features so the remaining signal is intrinsic to multi-agent interaction. The two failure modes from Appendix~\ref{app:base_model_entropy} reappear in essentially the same form, which is the point: even after we strip out base-model effects, Qwen still fails through dispersion and LLaMA still fails through verbosity. This is direct evidence that the failure modes are properties of how each family interacts, not artifacts of the base-model entropy correlations. Two new findings that are specific to $\mathcal{G}_{\text{MAS}}$ deserve separate notice.

\paragraph{More agents harm Qwen.}
\textit{sample\_num\_agents} enters the Qwen ranking with $\rho \approx -0.44$ and $\bar{S} \approx -0.051$, indicating that adding agents amplifies the dispersion failure rather than averaging it out. This is consistent with the \texttt{MMLU}-specific result in Section~\ref{sec:examining_uncertainty_impacts} and contrasts with the LLaMA family, where agent count is not a top predictor.

\begin{figure*}[t]
    \centering
    \begin{minipage}{0.95\textwidth}
    \begin{subfigure}{0.48\textwidth}
        \centering
        \includegraphics[width=\linewidth]{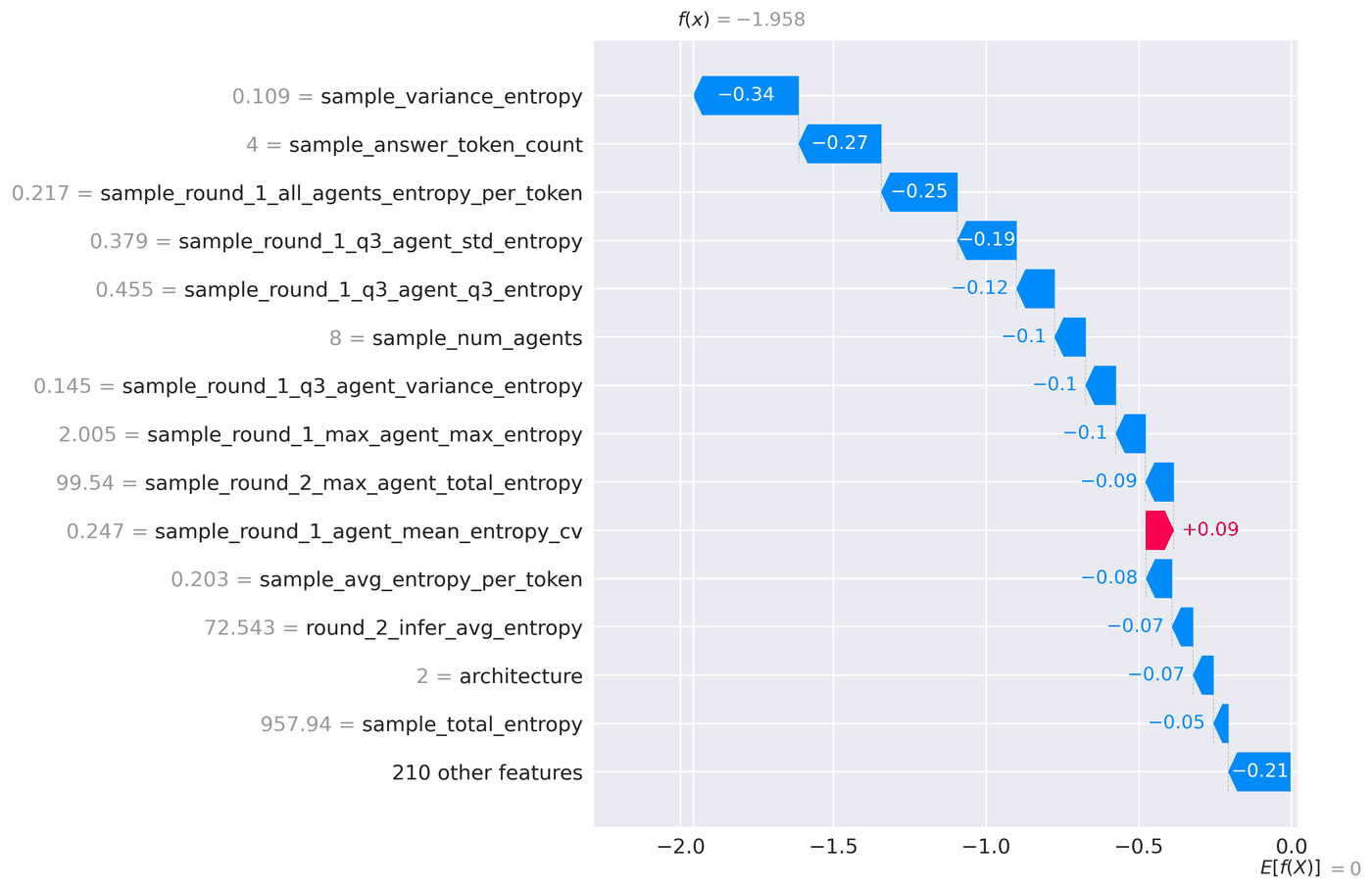}
        \caption{Qwen (LightGBM)}
        \label{fig:mas_failure_qwen_lightgbm}
    \end{subfigure}
    \hfill
    \begin{subfigure}{0.48\textwidth}
        \centering
        \includegraphics[width=\linewidth]{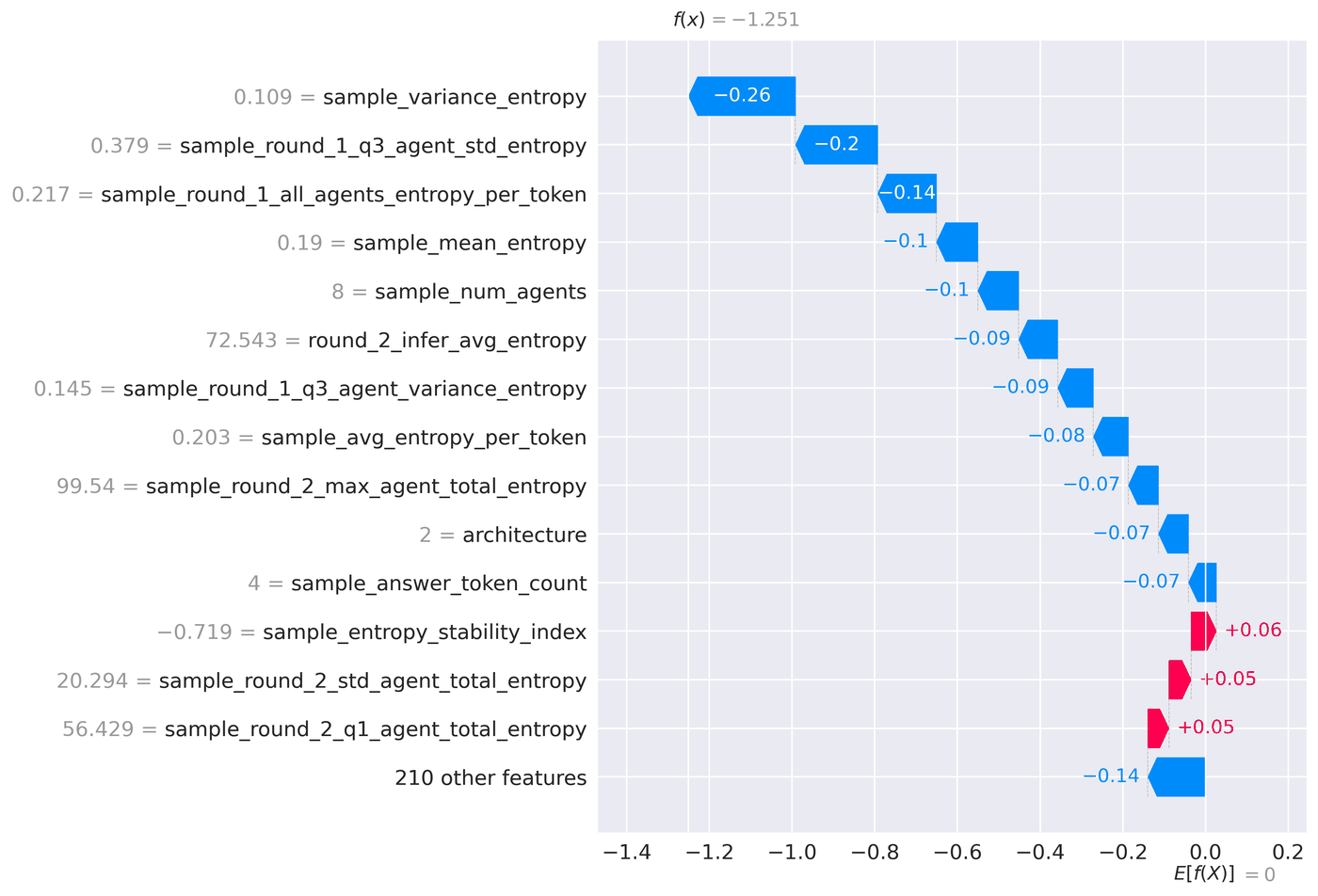}
        \caption{Qwen (XGBoost)}
        \label{fig:mas_failure_qwen_xgboost}
    \end{subfigure}

    \vspace{1em}

    \begin{subfigure}{0.48\textwidth}
        \centering
        \includegraphics[width=\linewidth]{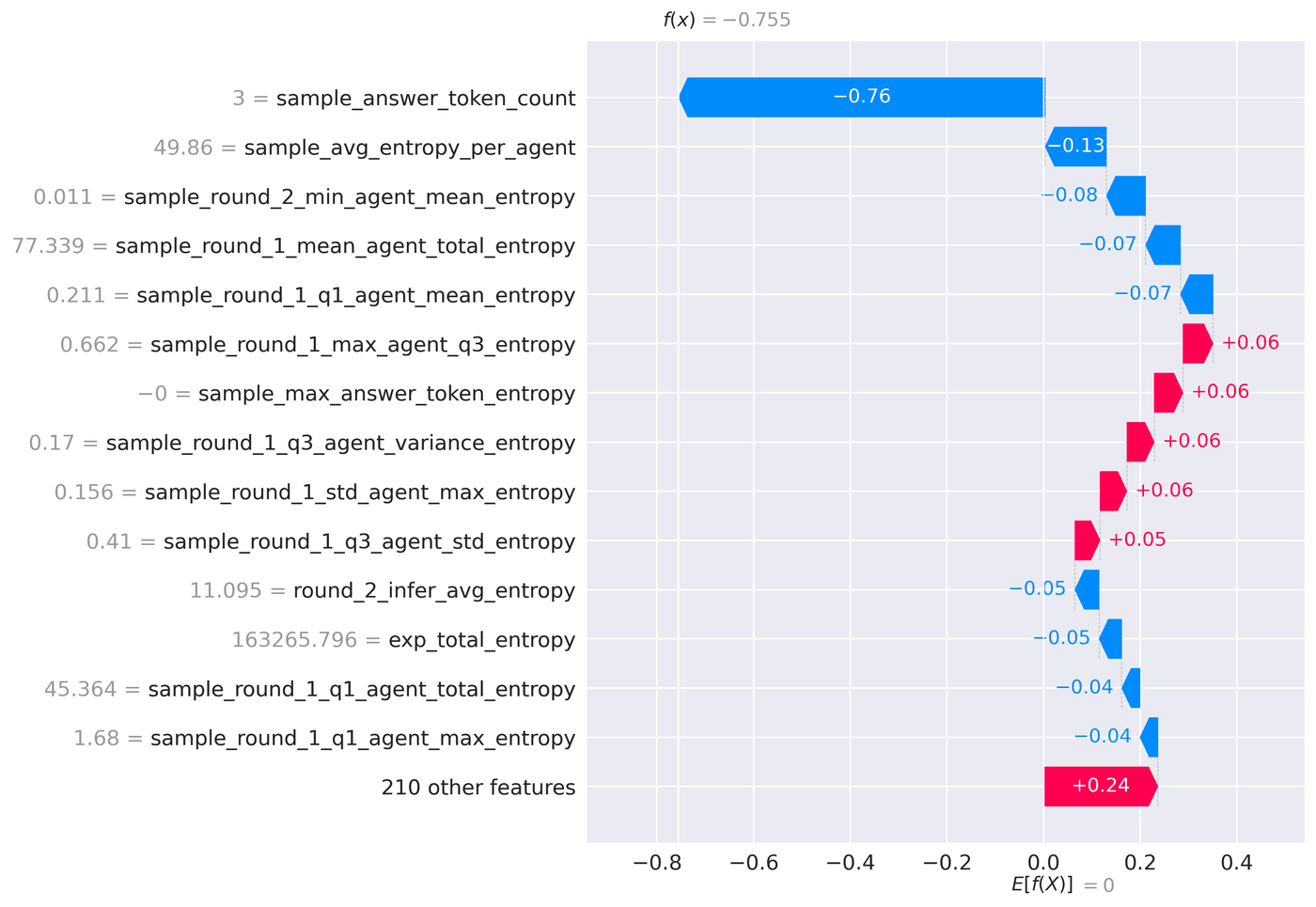}
        \caption{LLaMA (LightGBM)}
        \label{fig:mas_failure_llama_lightgbm}
    \end{subfigure}
    \hfill
    \begin{subfigure}{0.48\textwidth}
        \centering
        \includegraphics[width=\linewidth]{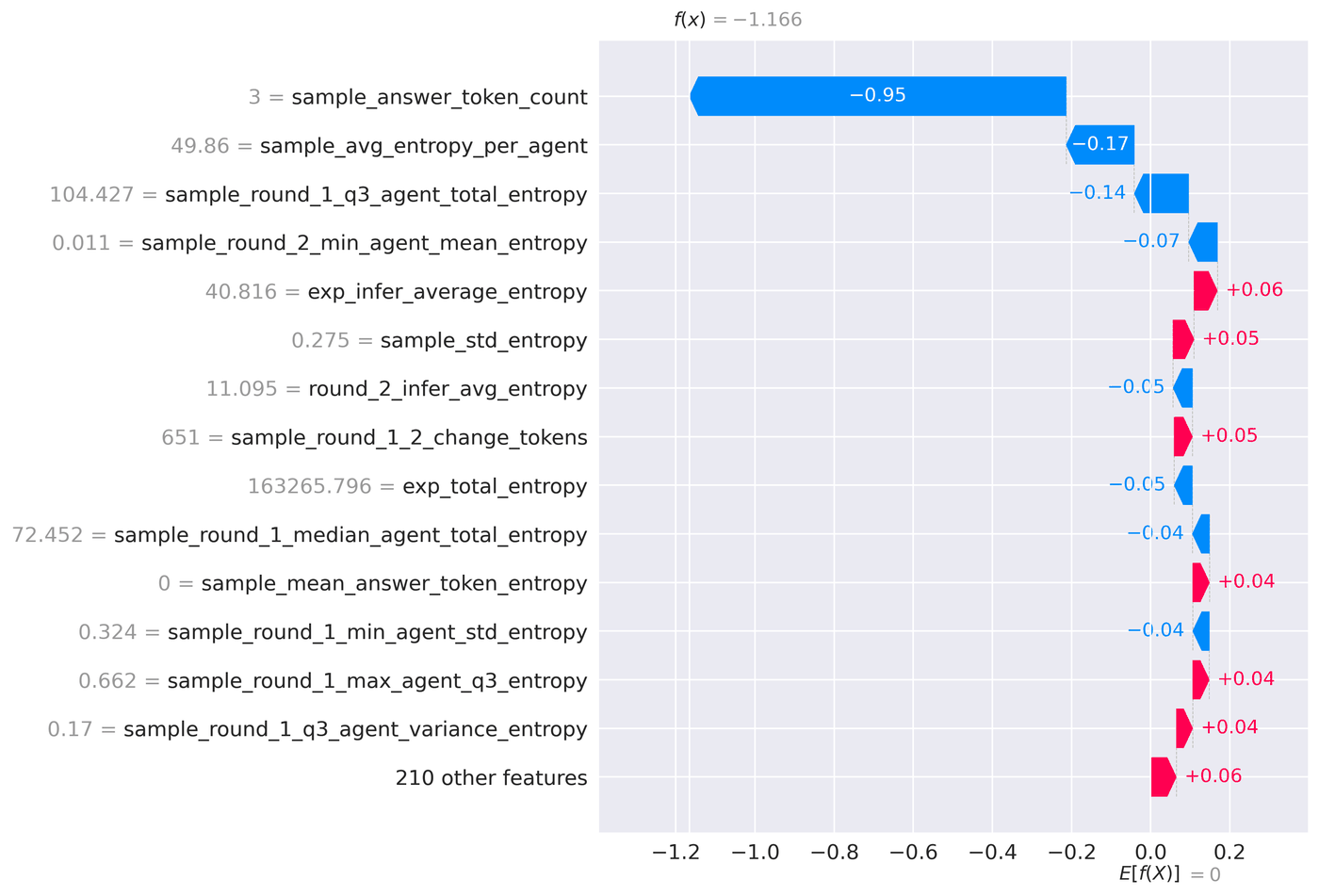}
        \caption{LLaMA (XGBoost)}
        \label{fig:mas_failure_llama_xgboost}
    \end{subfigure}
    \end{minipage}
    \caption{SHAP waterfall plots on $\mathcal{G}_{\text{MAS}}$ for representative MAS failure samples. Qwen (a-b) shows entropy dispersion features (variance, Q3 agent) as dominant contributors; LLaMA (c-d) reveals answer-level features (token count, answer entropy) driving failure predictions.}
    \label{fig:mas_failure_combined_all}
\end{figure*}

\paragraph{LLaMA shows a non-monotone round-2 effect.}
Round-2 inference entropy correlates positively with sample-level success ($\rho \approx +0.65$) but contributes negatively at the global level. Following the caveat in Appendix~\ref{app:base_model_entropy}, this is a single non-monotone feature rather than a contradiction: LLaMA benefits from some round-2 deliberation but is harmed by excessive late-round entropy. SHAP waterfalls in Figure~\ref{fig:mas_failure_combined_all} confirm that the dominant contributors remain dispersion (Qwen, panels a-b) and answer-level features (LLaMA, panels c-d).

\begin{figure*}[!htbp]
    \centering
    \begin{minipage}{0.95\textwidth}
    \begin{subfigure}{0.32\textwidth}
        \centering
        \includegraphics[width=\linewidth]{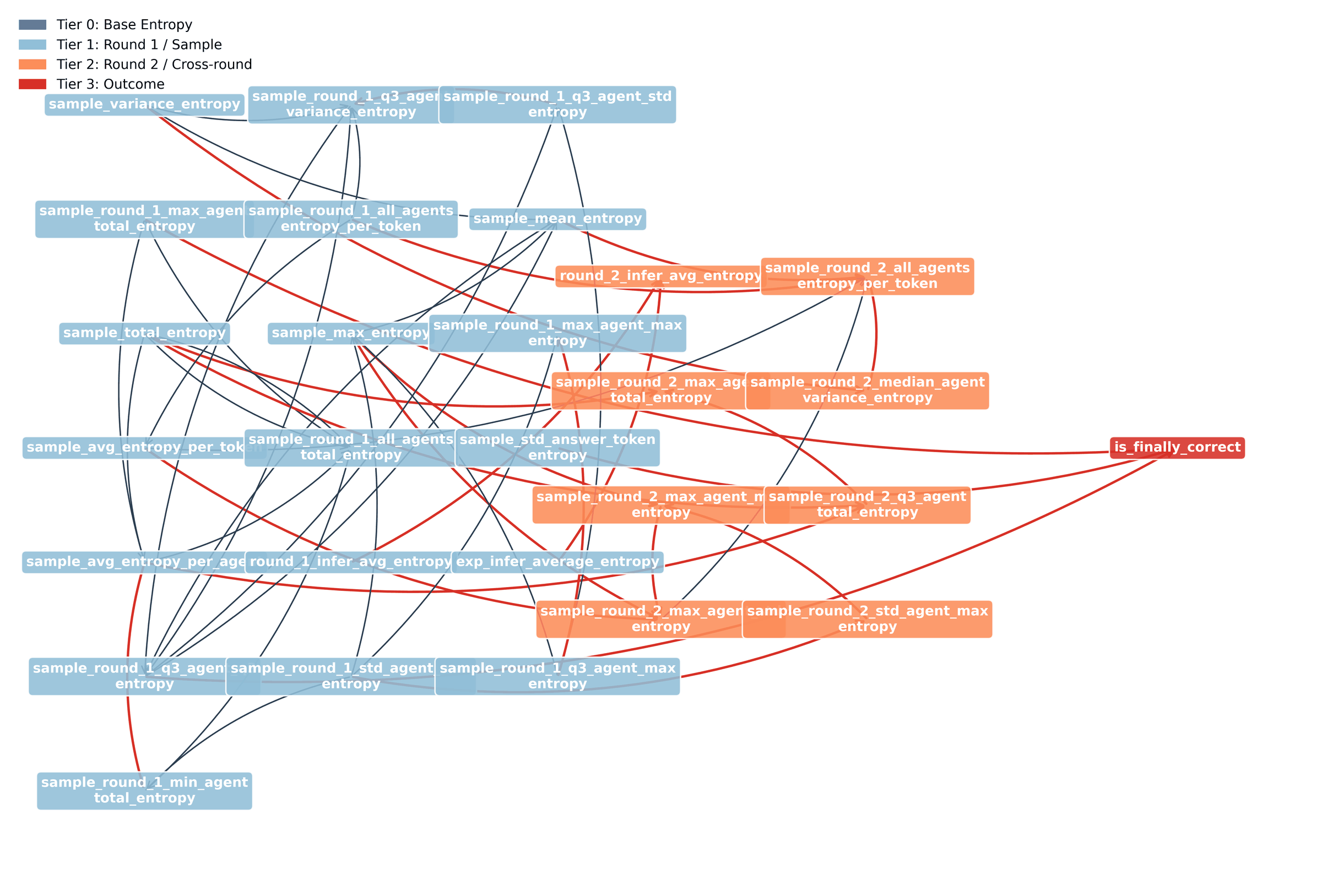}
        \caption{PC/FCI consensus causal graph}
    \end{subfigure}
    \hfill
    \begin{subfigure}{0.32\textwidth}
        \centering
        \includegraphics[width=\linewidth]{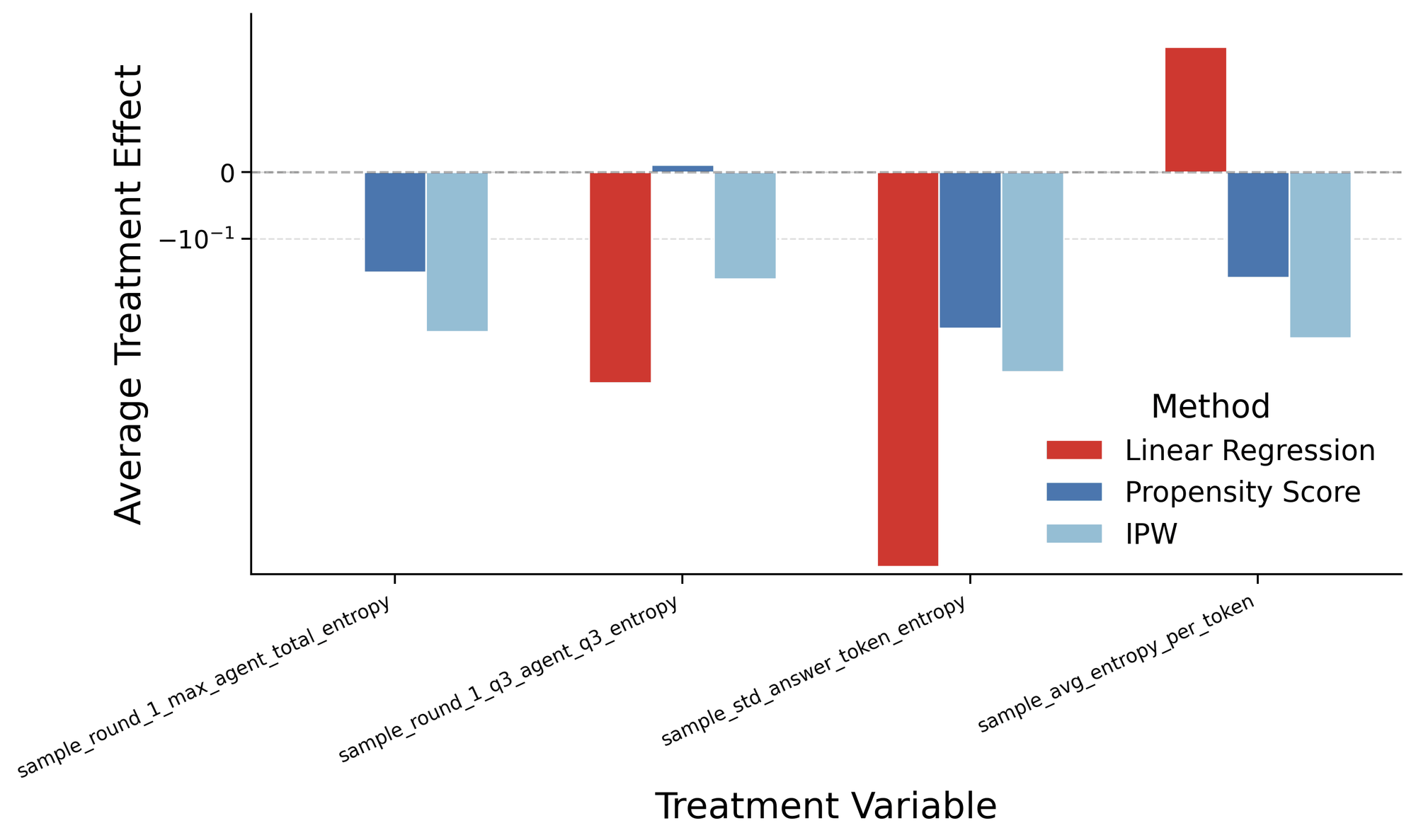}
        \caption{ATE estimates (LR / PS / IPW)}
    \end{subfigure}
    \hfill
    \begin{subfigure}{0.32\textwidth}
        \centering
        \includegraphics[width=\linewidth]{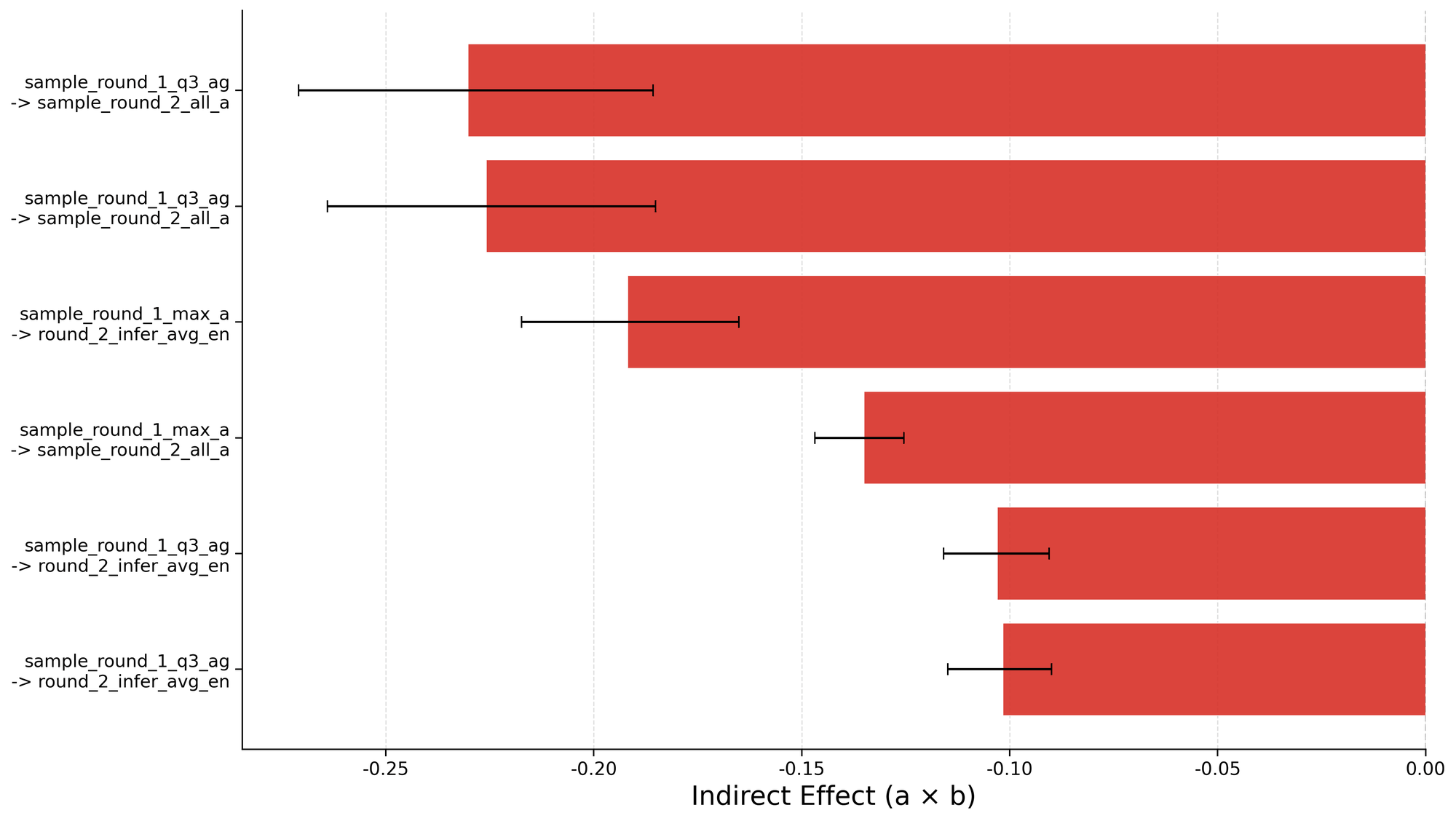}
        \caption{Bootstrap indirect effects}
    \end{subfigure}
    \end{minipage}
    \caption{Causal validation on $\mathcal{G}_{\text{MAS}}$: (a) consensus causal graph isolates round-1 dispersion and answer-token-entropy variability as direct causes of correctness; (b) all three pass refutation, with conservative IPW $\text{ATE}\in[-0.30,-0.16]$; (c) Round~1 dispersion is mediated to correctness through round-2 per-token entropy ($\sim$27\% of total effect).}
    \label{fig:causal-mas-failure}
\end{figure*}

\paragraph{Causal validation.}
Applying the same PC/FCI + DoWhy pipeline of Appendix~\ref{app:causal-discovery} to $\mathcal{G}_{\text{MAS}}$ (Figure~\ref{fig:causal-mas-failure}) recovers three consensus direct causes of correctness: round-1 maximum per-agent total entropy, round-1 Q3 of per-agent Q3 entropy, and sample-level standard deviation of answer-token entropy. The conservative IPW estimates yield $\text{ATE}=-0.239$ ($p=1.3\!\times\!10^{-15}$), $-0.160$ ($p=4.7\!\times\!10^{-8}$), and $-0.299$ ($p=1.8\!\times\!10^{-4}$) respectively, all passing the three refutation tests. Mediation analysis shows that round-1 dispersion (Q3 of per-agent variance / standard deviation) transmits 26--28\% of its effect through round-2 per-token entropy, providing causal evidence for the SHAP reading: inter-agent dispersion is not just correlated with failure but causally generates the round-2 entropy state that propagates to the wrong answer.

\subsection{Task Difficulty Determines Optimal Entropy Dynamics}
\label{app:diff_tasks}

\begin{figure*}[t]
    \centering
    \begin{minipage}{0.95\textwidth}
    \begin{subfigure}{0.48\textwidth}
        \centering
        \includegraphics[width=\linewidth]{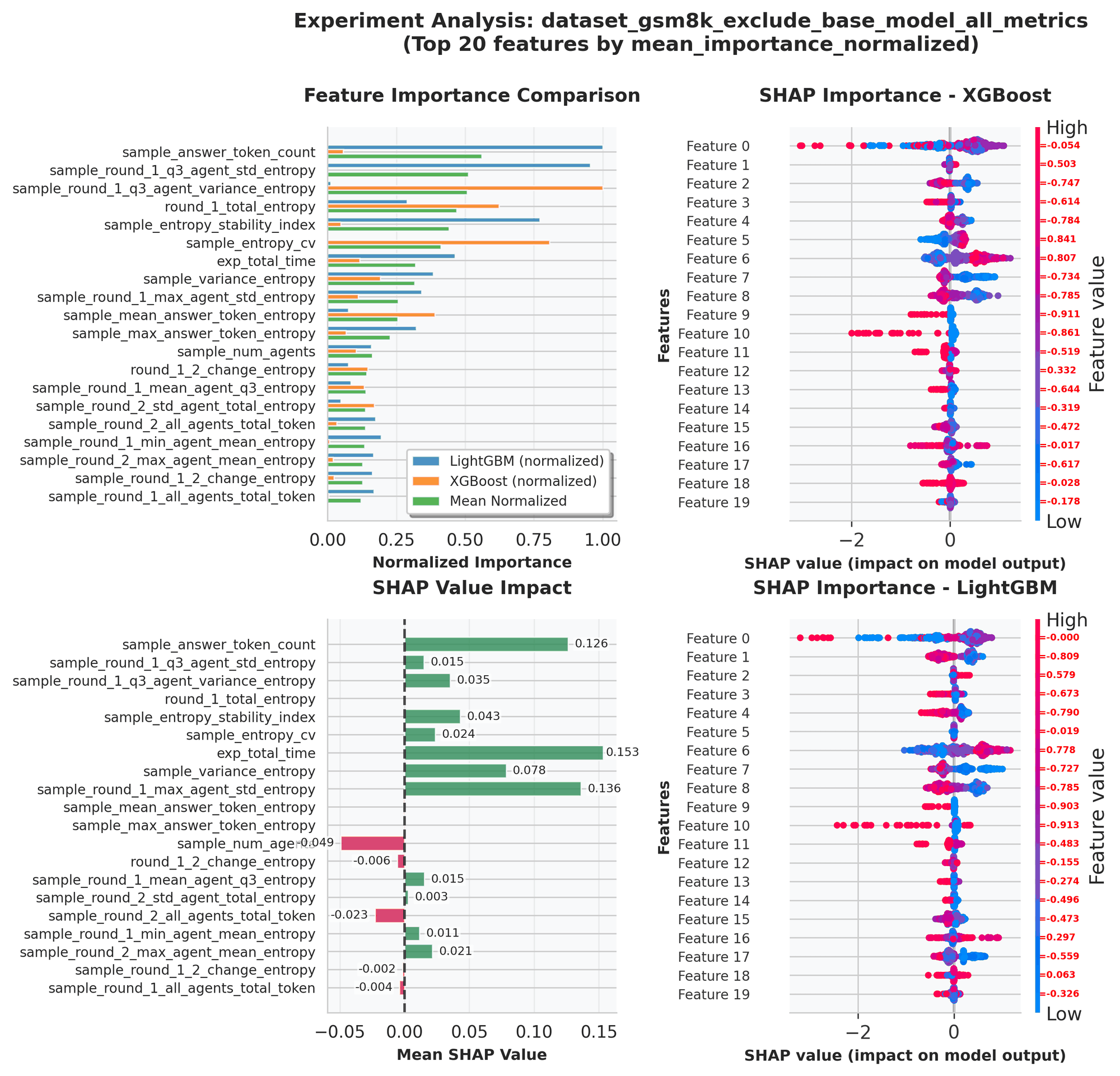}
        \caption{\texttt{GSM8K} (Easy)}
        \label{fig:diff_tasks_gsm8k}
    \end{subfigure}
    \hfill
    \begin{subfigure}{0.48\textwidth}
        \centering
        \includegraphics[width=\linewidth]{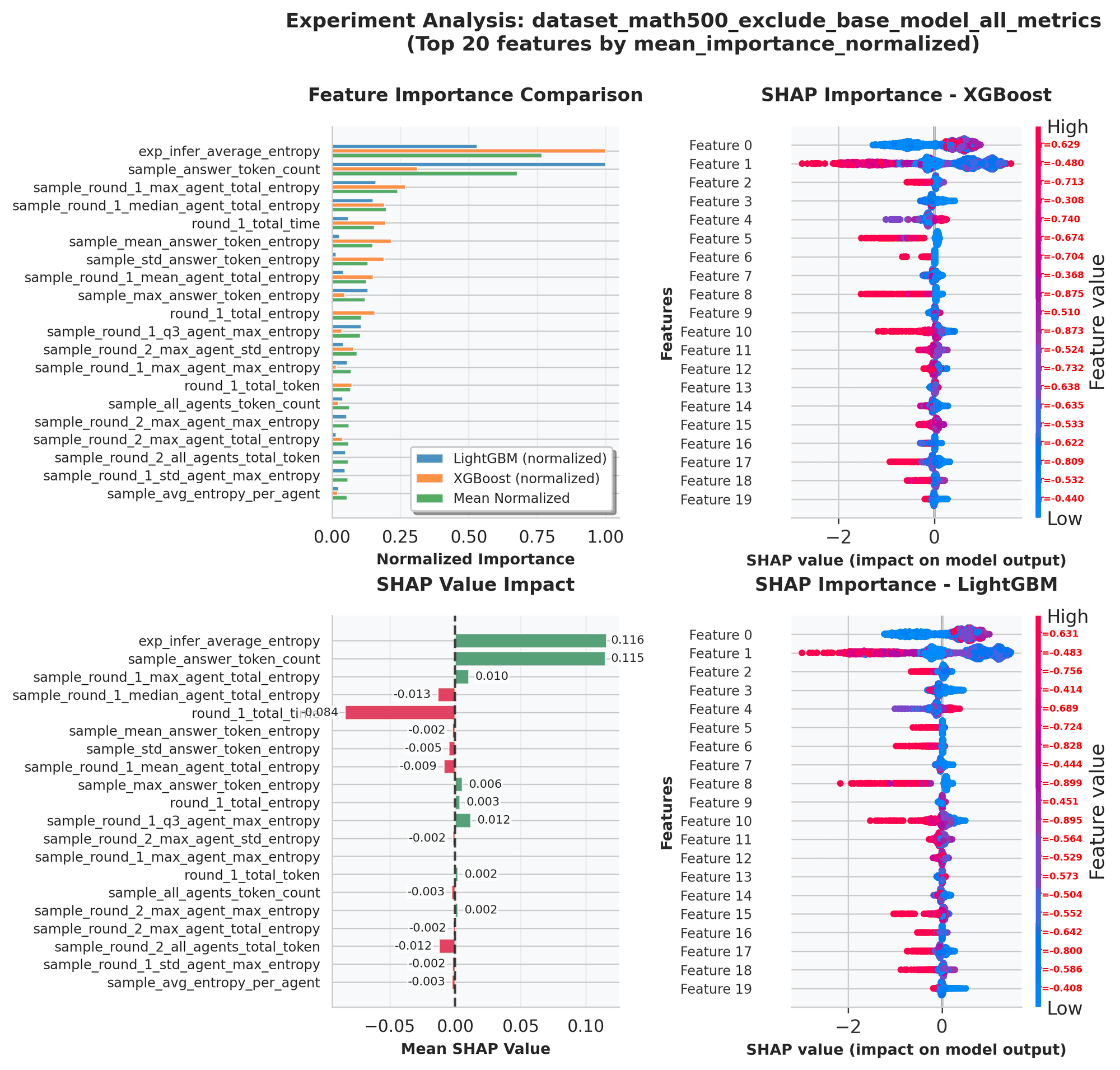}
        \caption{\texttt{MATH500} (Medium)}
        \label{fig:diff_tasks_math500}
    \end{subfigure}

    \vspace{1em}

    \begin{subfigure}{0.48\textwidth}
        \centering
        \includegraphics[width=\linewidth]{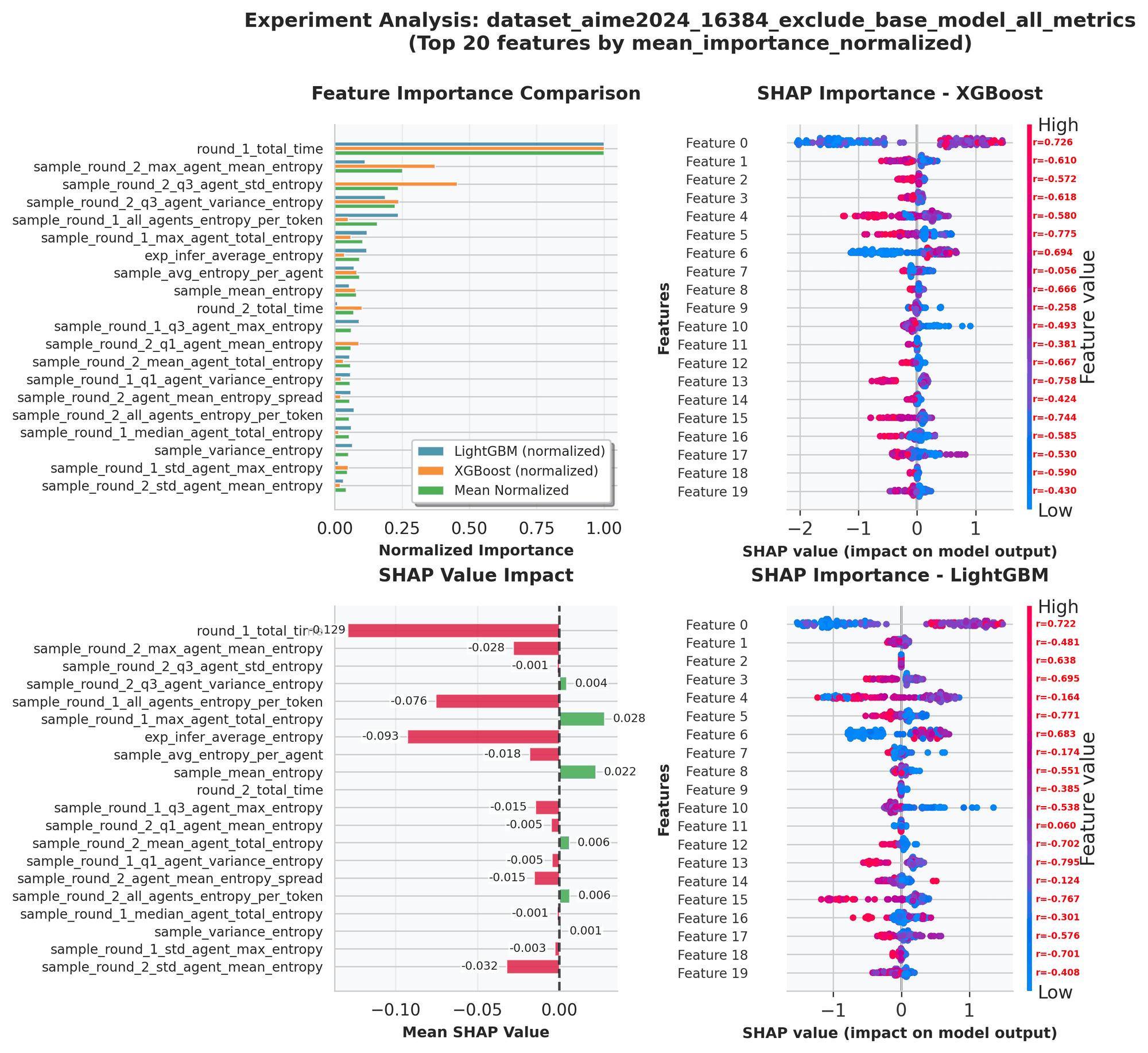}
        \caption{\texttt{AIME2024} (Hard)}
        \label{fig:diff_tasks_aime2024}
    \end{subfigure}
    \hfill
    \begin{subfigure}{0.48\textwidth}
        \centering
        \includegraphics[width=\linewidth]{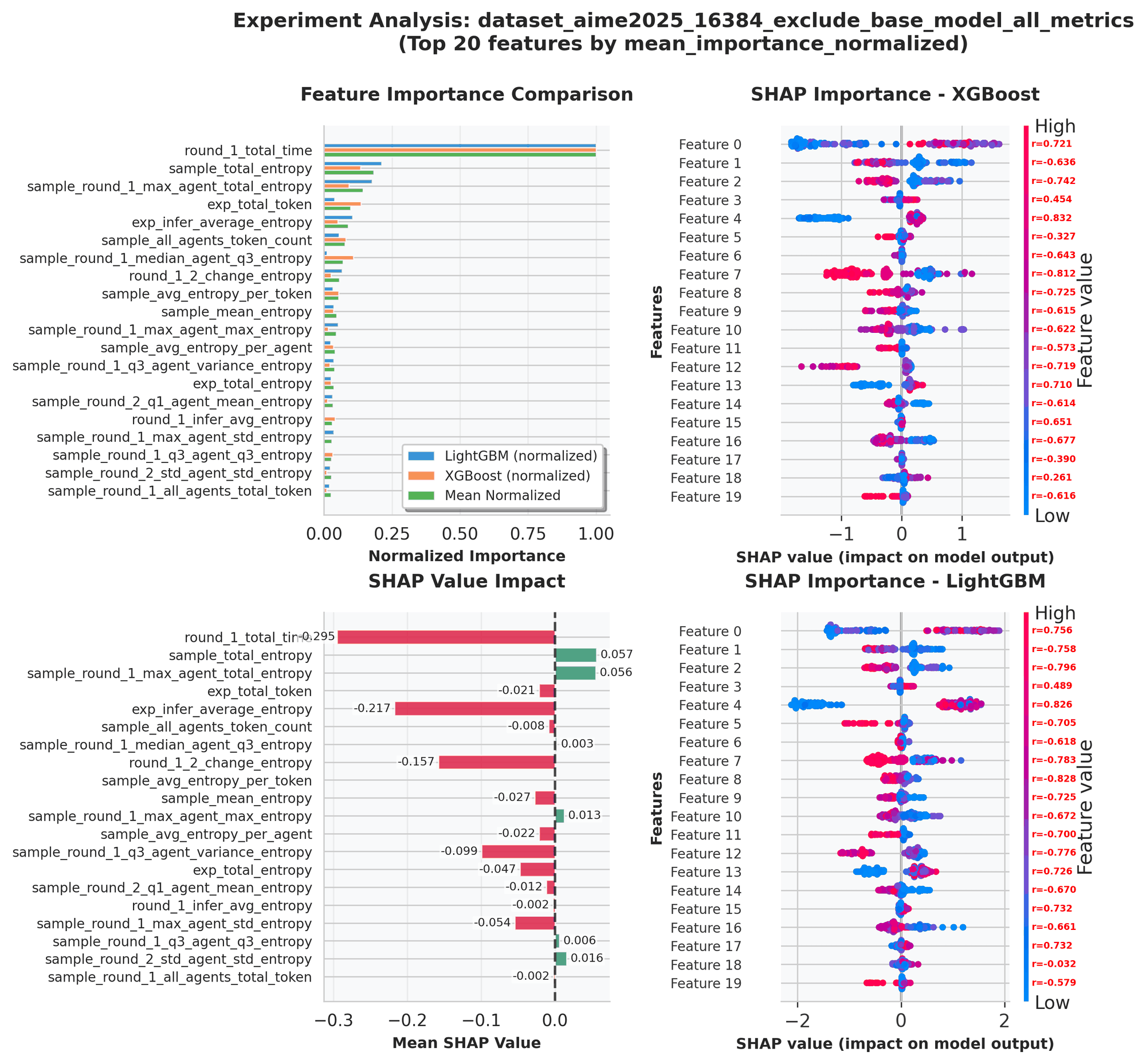}
        \caption{\texttt{AIME2025} (Hard)}
        \label{fig:diff_tasks_aime2025}
    \end{subfigure}
    \end{minipage}
    \caption{Top 20 features on $\mathcal{G}_{\text{MAS}}$ for mathematical reasoning tasks grouped by difficulty: (a) \texttt{GSM8K} (easy, $|\rho| \le 0.15$ for top features), (b) \texttt{MATH500} (medium, positive $\rho$ and $\bar{S}$ for average entropy), (c-d) \texttt{AIME2024}/\texttt{AIME2025} (hard, round-2 entropy harms performance).}
    \label{fig:diff_tasks_math}
\end{figure*}

\begin{figure*}[t]
    \centering
    \begin{minipage}{0.95\textwidth}
    \begin{subfigure}{0.48\textwidth}
        \centering
        \includegraphics[width=\linewidth]{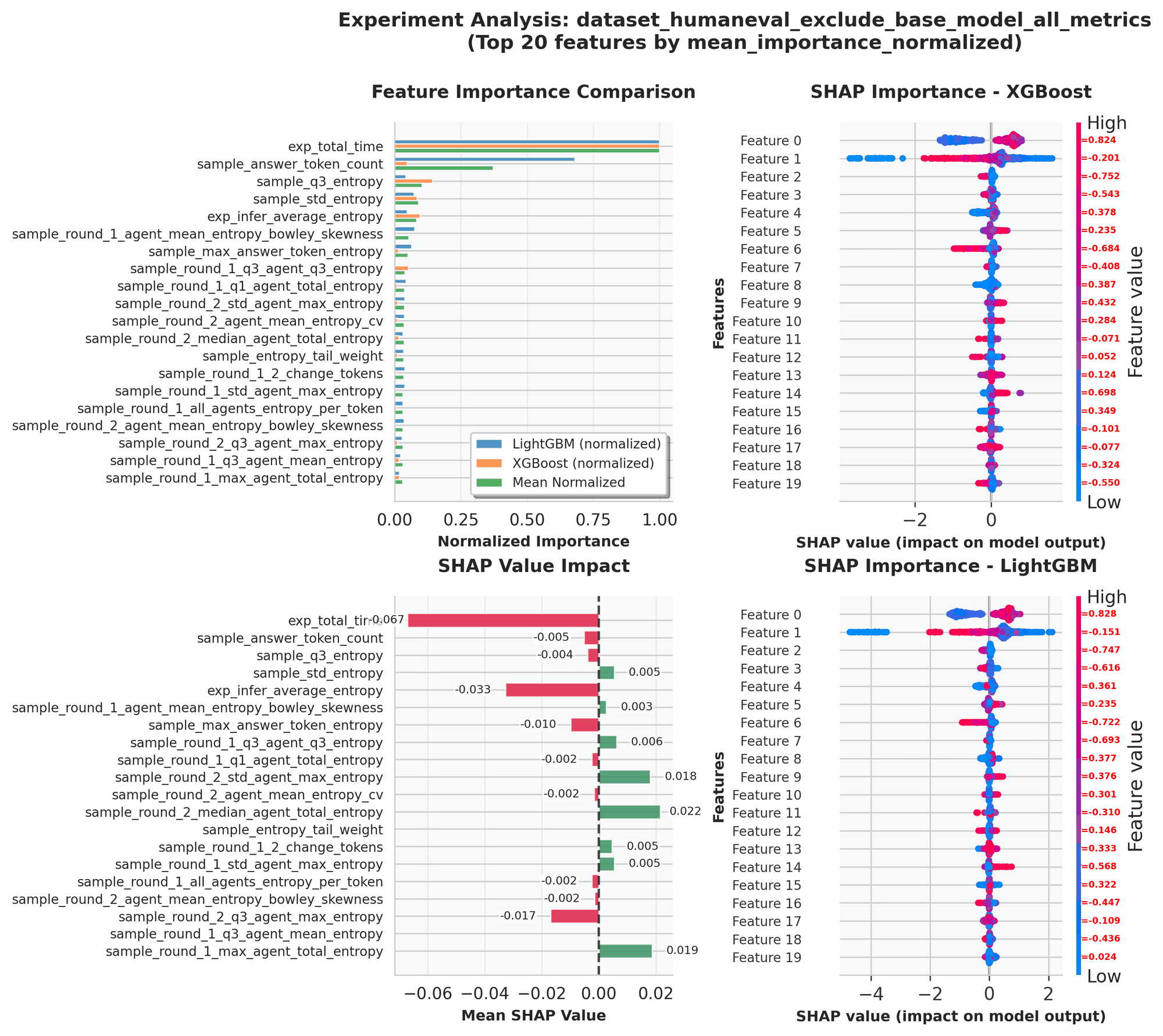}
        \caption{\texttt{HumanEval} (Code Generation)}
        \label{fig:diff_tasks_humaneval}
    \end{subfigure}
    \hfill
    \begin{subfigure}{0.48\textwidth}
        \centering
        \includegraphics[width=\linewidth]{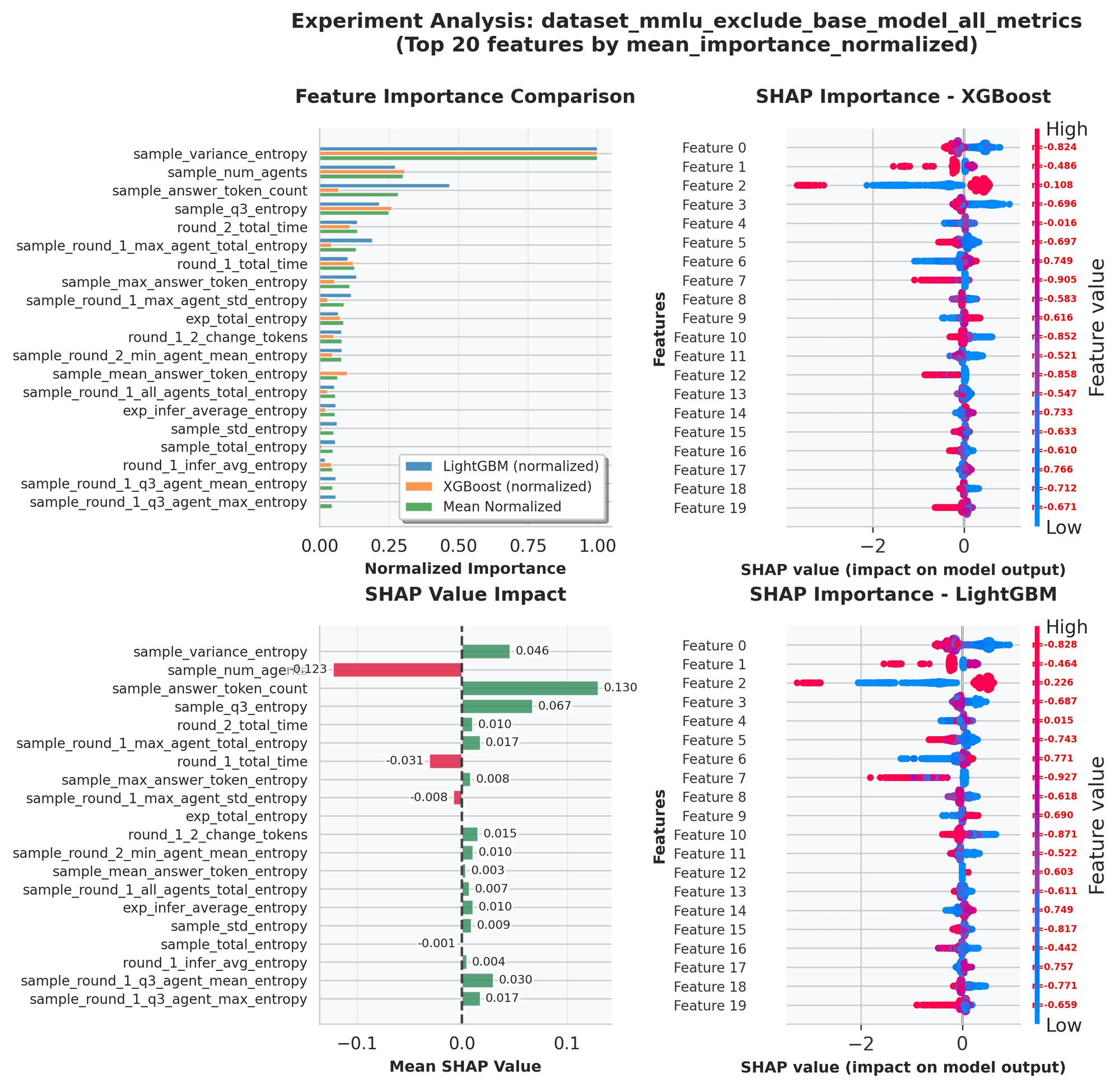}
        \caption{\texttt{MMLU} (Knowledge Q\&A)}
        \label{fig:diff_tasks_mmlu}
    \end{subfigure}
    \end{minipage}
    \caption{Top 20 features on $\mathcal{G}_{\text{MAS}}$ for (a) code generation (\texttt{HumanEval}) and (b) knowledge Q\&A (\texttt{MMLU}). \texttt{HumanEval} shows negative $\rho$ and $\bar{S}$ for answer-level features; \texttt{MMLU} shows that more agents hurt performance ($\rho < 0$, $\bar{S} < 0$ for \textit{sample\_num\_agents}).}
    \label{fig:diff_tasks_other}
\end{figure*}

Figures~\ref{fig:diff_tasks_math} and~\ref{fig:diff_tasks_other} present per-dataset analyses on $\mathcal{G}_{\text{MAS}}$.

\paragraph{Easy regime (\texttt{GSM8K}): entropy is largely uninformative.}
Top-feature correlations collapse to $|\rho| \le 0.15$ for most predictors, indicating that easy arithmetic does not stress the entropy machinery. The two exceptions, entropy variance and answer-token entropy, show strong sample-level correlations but opposite global contributions, which is the same non-monotone pattern flagged earlier and not a contradiction. The practical implication is that entropy-based selection adds little value when the task is easy enough for the base model to solve directly.

\paragraph{Medium regime (\texttt{MATH500}): moderate entropy is beneficial.}
Average inference entropy becomes the leading positive predictor ($\rho = +0.63$, $\bar{S} = +0.12$), and cumulative round-1 entropy also helps. Excessive early entropy, however, still hurts: max agent total entropy and peak answer entropy carry strongly negative correlations. The regime is not ``more entropy = better" but rather ``bounded exploration helps when the task has enough structure to converge".

\paragraph{Hard regime (\texttt{AIME2024/2025}): entropy changes dominate.}
On olympiad problems the strongest signal is the round-1 to round-2 entropy change ($\rho = -0.80$, $\bar{S} = -0.157$): trajectories whose entropy shifts substantially between rounds tend to fail. Average inference entropy shows the same non-monotone signature seen in the easy regime ($\rho = +0.83$, $\bar{S} = -0.22$), now with much larger magnitudes, indicating a narrower optimal entropy band on hard problems.

\paragraph{Domain-specific regimes (\texttt{HumanEval}, \texttt{MMLU}).}
Code generation collapses onto answer-level entropy (peak and Q3 answer entropy negative, $\rho \le -0.70$) while deliberation entropy is essentially uninformative. Knowledge Q\&A is the only setting where agent count itself is harmful ($\rho = -0.48$, $\bar{S} = -0.12$), and where consensus rather than duration drives success. These two cases illustrate that the difficulty axis alone is insufficient; task type also selects which entropy dimension matters.

\begin{figure*}[t]
    \centering
    \begin{minipage}{0.95\textwidth}
    \begin{subfigure}{0.48\textwidth}
        \centering
        \includegraphics[width=\linewidth]{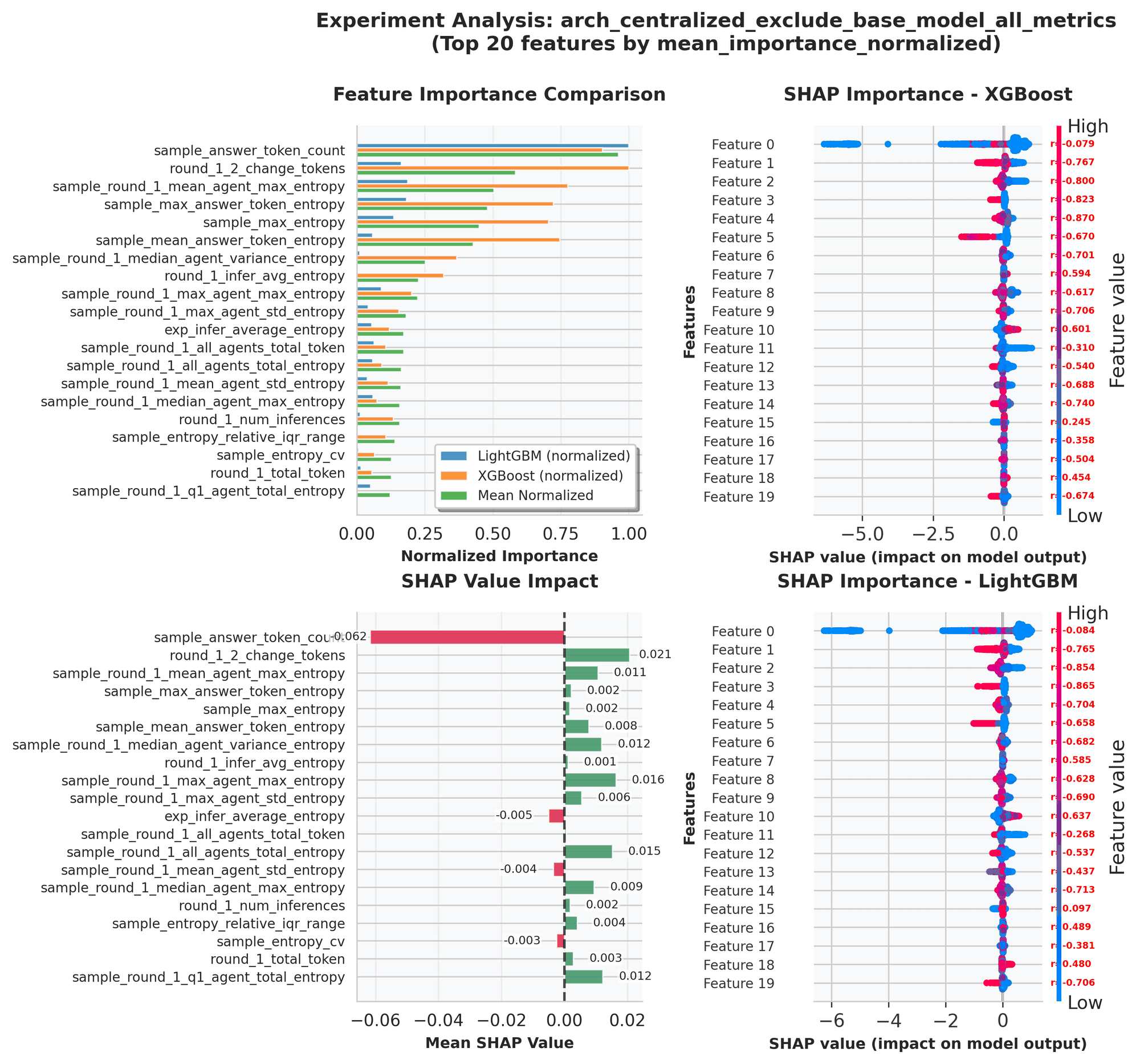}
        \caption{Centralized}
        \label{fig:diff_archs_centralized}
    \end{subfigure}
    \hfill
    \begin{subfigure}{0.48\textwidth}
        \centering
        \includegraphics[width=\linewidth]{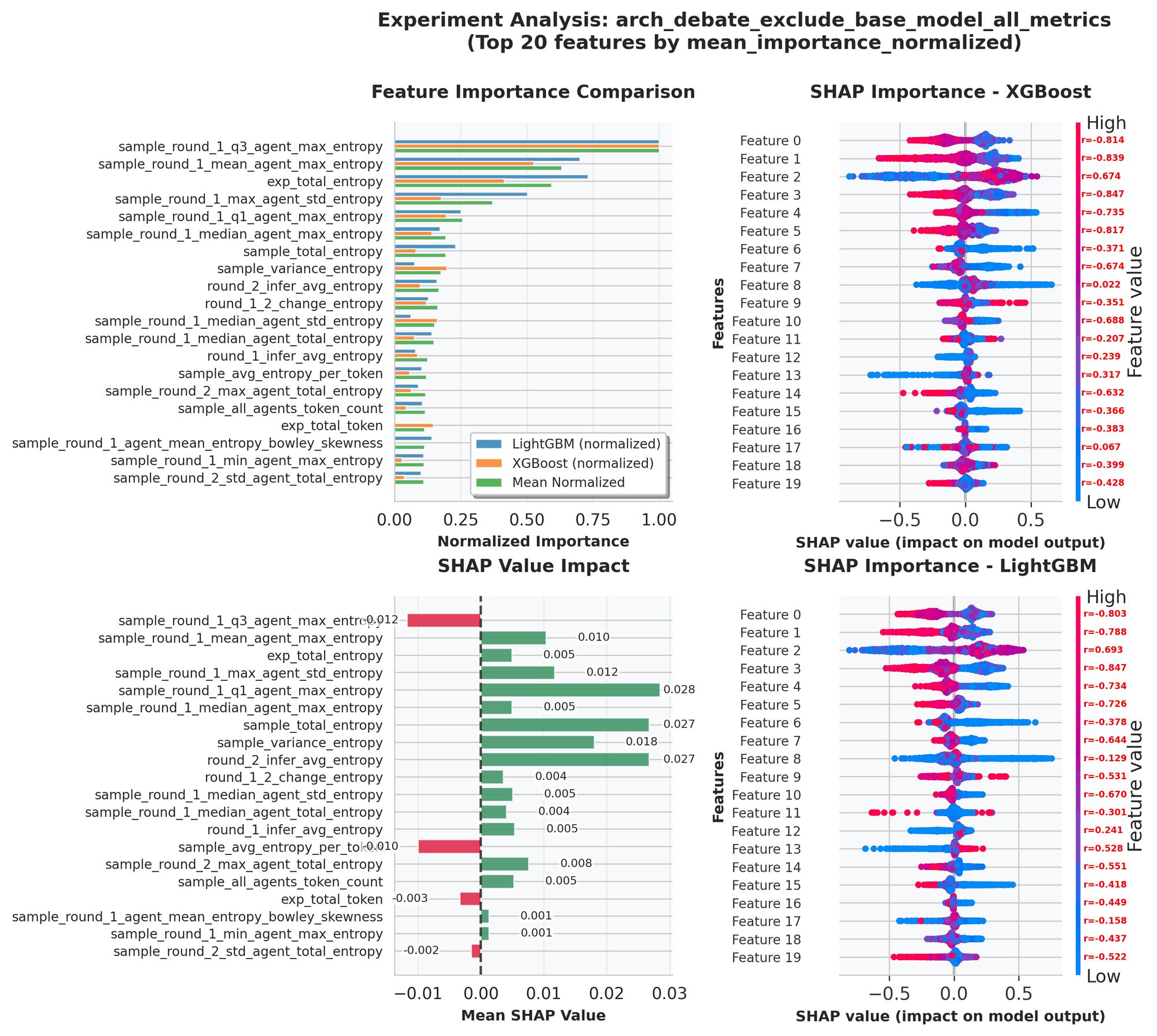}
        \caption{Debate}
        \label{fig:diff_archs_debate}
    \end{subfigure}

    \vspace{1em}

    \begin{subfigure}{0.48\textwidth}
        \centering
        \includegraphics[width=\linewidth]{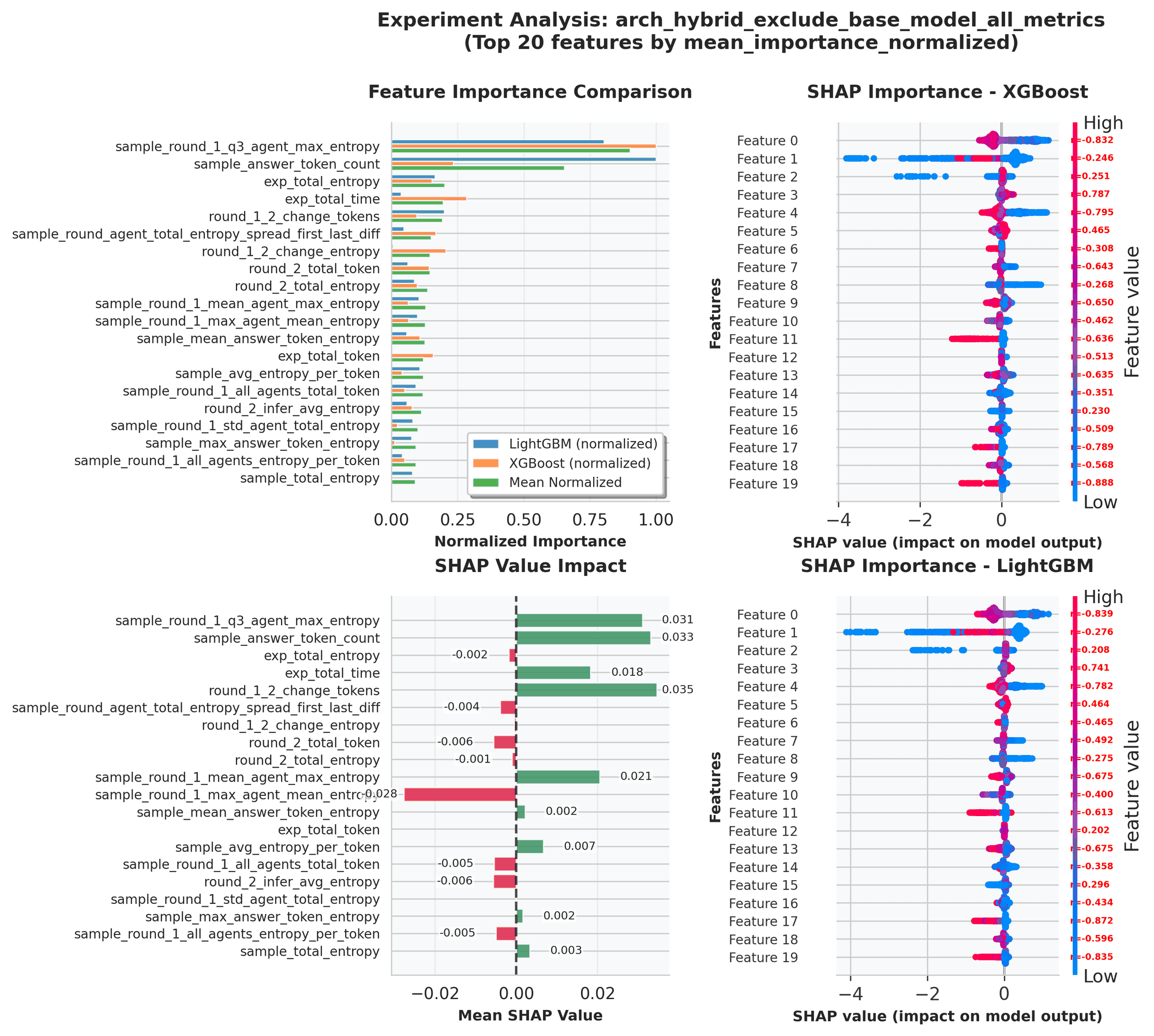}
        \caption{Hybrid}
        \label{fig:diff_archs_hybrid}
    \end{subfigure}
    \hfill
    \begin{subfigure}{0.48\textwidth}
        \centering
        \includegraphics[width=\linewidth]{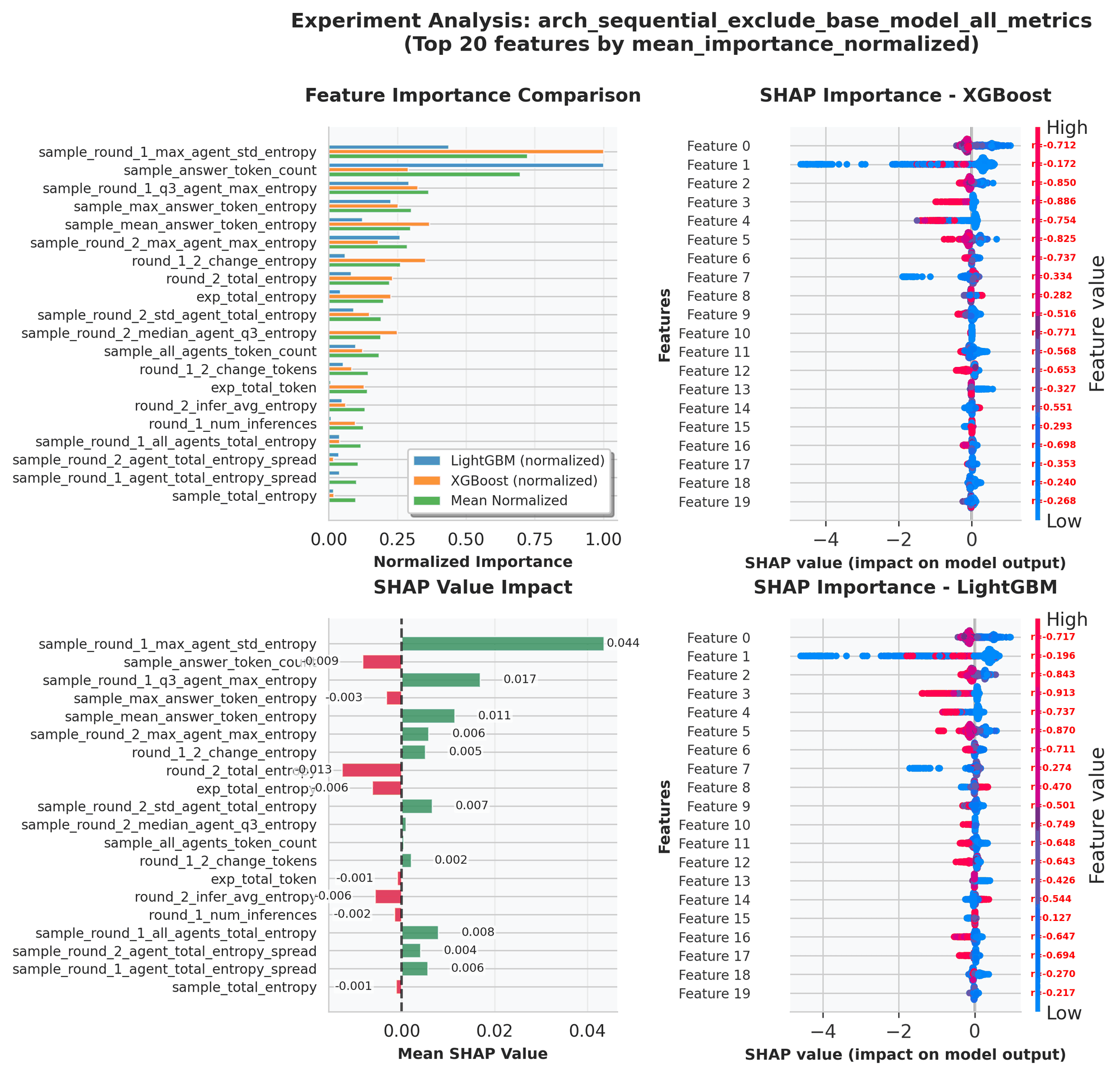}
        \caption{Sequential}
        \label{fig:diff_archs_sequential}
    \end{subfigure}
    \end{minipage}
    \caption{Top 20 features on $\mathcal{G}_{\text{MAS}}$ for multi-agent architectures: (a) Centralized (verbose answers harm performance), (b) Debate (cumulative entropy benefits once agents align), (c) Hybrid (extended deliberation helps), and (d) Sequential (answer-level entropy is the primary failure mode).}
    \label{fig:diff_archs_mas}
\end{figure*}

\paragraph{Aggregation architectures (Centralized, Hybrid).}
Both architectures route information through a central point and share an answer-length signal: in Centralized, sample answer token count dominates ($\bar{I} = 0.96$) with a strongly negative global contribution; in Hybrid, total experiment time is the leading positive predictor ($\rho = +0.76$). Both benefit from extended deliberation but penalize different sources of length: Centralized penalizes verbose final answers, while Hybrid additionally penalizes round-2 token expansion ($\rho = -0.57$). The orchestrator therefore contributes by constraining later-round length rather than by extending it.

\paragraph{Decentralized architectures (Debate, Sequential).}
Both rely on agent-to-agent transmission rather than aggregation, and both are vulnerable to early divergence. In Debate, round-1 peak entropy is the strongest negative predictor ($\rho = -0.81$), but cumulative entropy is positive ($\rho = +0.68$), giving a ``converge early, then explore" profile. In Sequential, the same early-round dispersion appears, but cumulative entropy reverses sign ($\bar{S} = -0.006$ globally despite $\rho = +0.38$ at sample level), and answer-level entropy becomes a dominant failure signal. The contrast cleanly maps onto error-propagation: Debate's bidirectional exchanges allow the system to recover from early divergence, while Sequential's chain compounds it.

\paragraph{SAS as control.}
SAS is dominated by answer-level features (Figure~\ref{fig:diff_archs_single}). Notably, round-2 inference entropy in SAS shows the same positive-$\rho$, negative-$\bar{S}$ signature observed in MAS, indicating that the non-monotone late-round entropy effect is not a multi-agent artifact but a property of the model's own deliberation. Cumulative entropy is uniformly positive in SAS ($\rho = +0.72$, $\bar{S} = +0.012$), the opposite of Sequential, which reinforces the error-propagation reading above.

\paragraph{Causal validation per task.}
Per-dataset causal estimation on $\mathcal{G}_{\text{MAS}}$ covers all six benchmarks, and each consensus tracks the SHAP-based difficulty regime above. \texttt{GSM8K} (easy) is driven by dispersion rather than magnitude: round-1$\rightarrow$2 change, max answer-token entropy, round-1 mean per-agent max, and sample variance (IPW ATE $-0.090$, $-0.353$, $-0.183$, $-0.183$; all $p<10^{-11}$). \texttt{MATH500} (medium) is causally controlled by the upper tail of answer-token entropy (max/mean/std plus round-2 max per-agent total; top IPW ATE $-0.450$, $p=4.9\!\times\!10^{-18}$). On the hard end, \texttt{AIME2024}'s direct causes are both round-1 Q3 per-agent statistics (IPW $-0.268$ and $-0.206$), while \texttt{AIME2025} surfaces the round-1$\rightarrow$2 change together with round-1 max per-agent max (IPW $-0.075$ and $-0.139$, both $p<10^{-20}$); olympiad failure is therefore caused by round-1 dispersion and the round-1$\rightarrow$2 shift, not absolute magnitude. The domain-specific datasets behave as expected: \texttt{HumanEval} reduces to the answer-token + round-1 dispersion pair (top IPW $-0.374$, $p=2.0\!\times\!10^{-77}$); \texttt{MMLU} yields four consensus causes spanning answer-token entropy and dispersion, the largest being max answer-token entropy at IPW $-0.447$ ($p=2.5\!\times\!10^{-14}$). Every consensus direct cause across the six datasets carries a negative IPW ATE and passes all three refutation tests.

\begin{figure*}[!htbp]
    \centering
    \begin{minipage}{0.95\textwidth}
    \begin{subfigure}{0.24\textwidth}
        \centering
        \includegraphics[width=\linewidth]{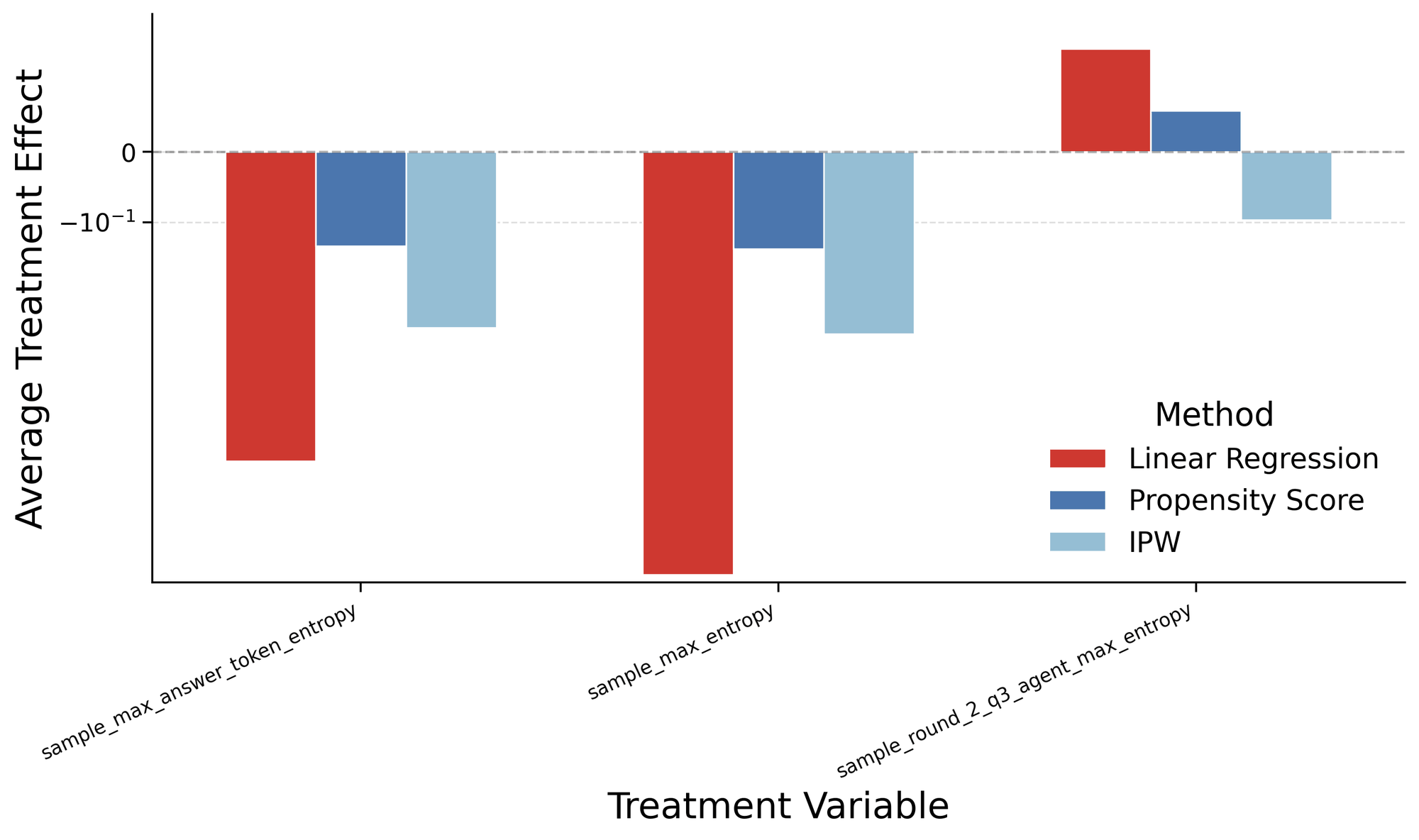}
        \caption{Centralized.}
    \end{subfigure}
    \hfill
    \begin{subfigure}{0.24\textwidth}
        \centering
        \includegraphics[width=\linewidth]{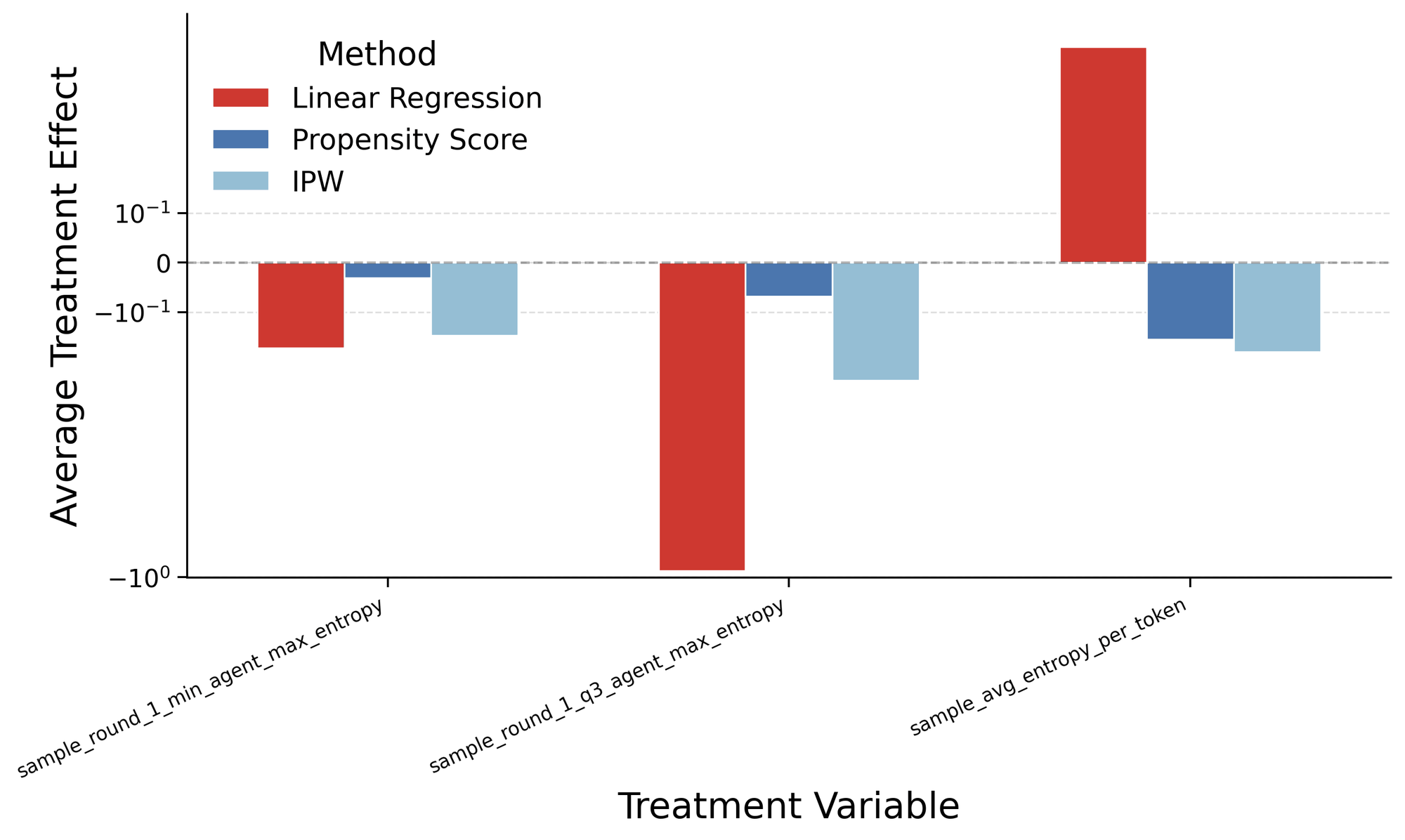}
        \caption{Debate.}
    \end{subfigure}
    \hfill
    \begin{subfigure}{0.24\textwidth}
        \centering
        \includegraphics[width=\linewidth]{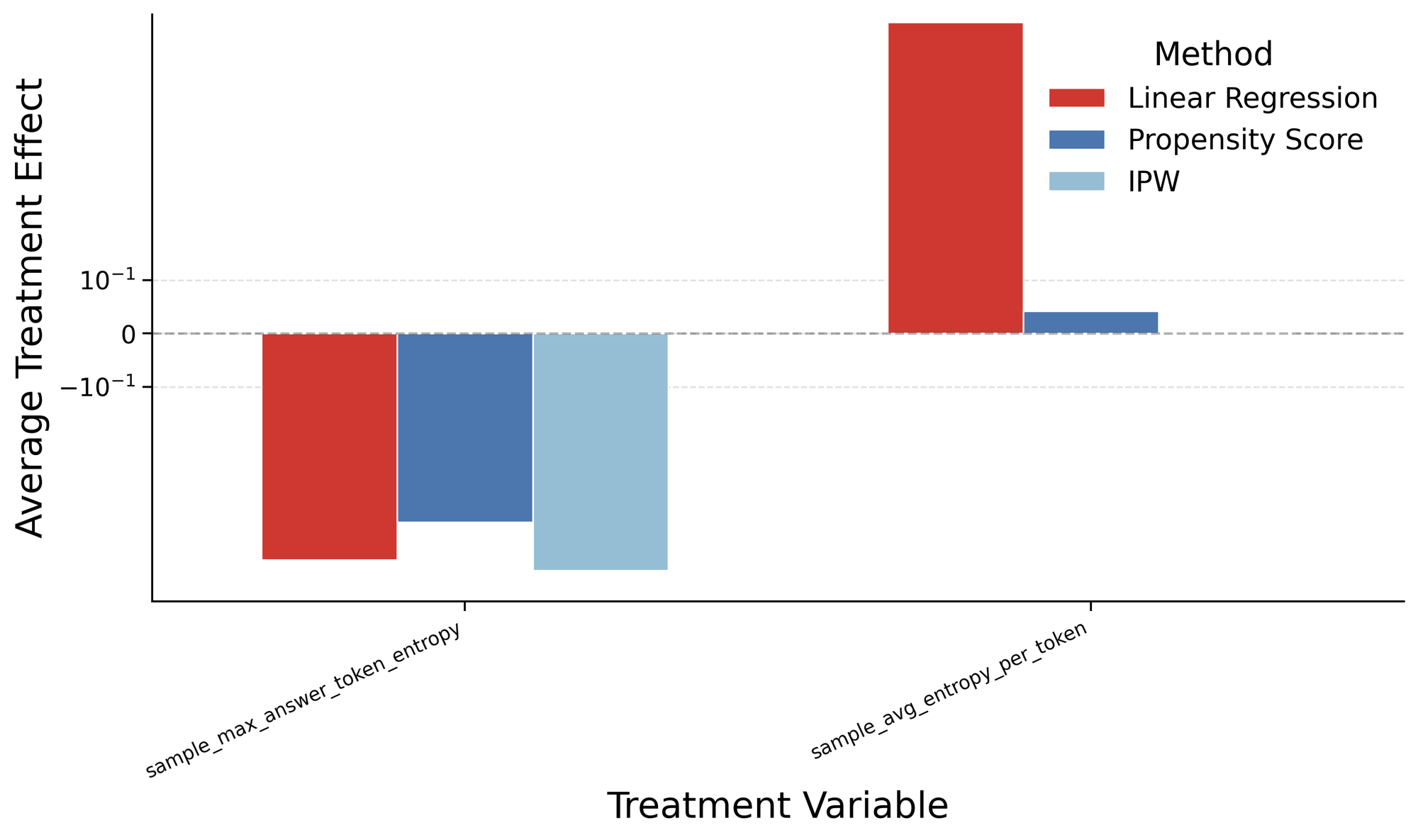}
        \caption{Hybrid.}
    \end{subfigure}
    \hfill
    \begin{subfigure}{0.24\textwidth}
        \centering
        \includegraphics[width=\linewidth]{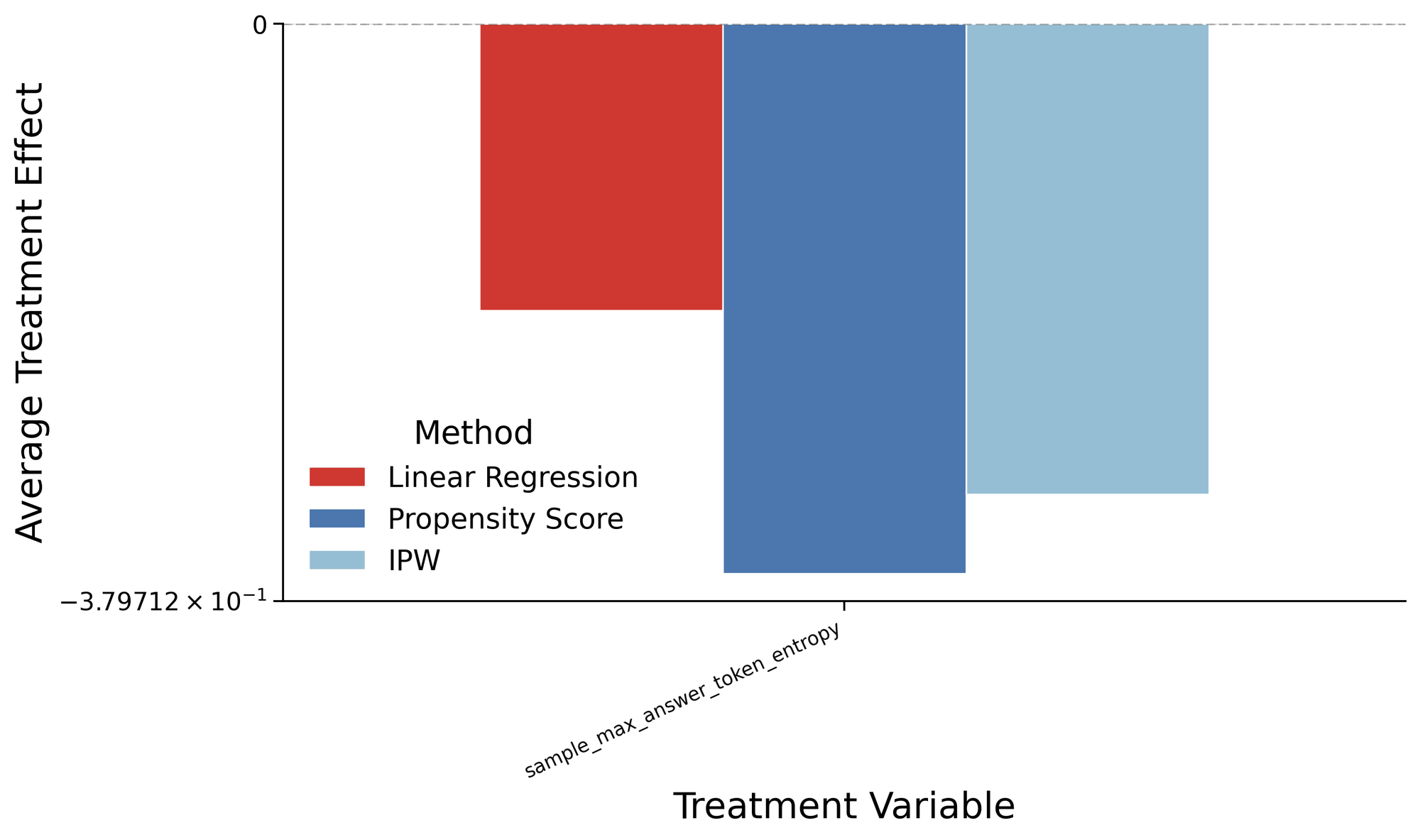}
        \caption{Sequential.}
    \end{subfigure}
    \end{minipage}
    \caption{ATE estimates (LR / PS / IPW) for each MAS architecture on $\mathcal{G}_{\text{MAS}}$. Answer-token entropy emerges as the dominant negative causal driver in Centralized, Hybrid and Sequential; Debate is the only topology whose direct causes are round-1 dispersion features rather than answer-level features.}
    \label{fig:causal-archs}
\end{figure*}

\paragraph{Causal validation per architecture.}
Per-architecture causal estimation reveals the consensus direct cause for each topology (Figure~\ref{fig:causal-archs}): \textit{sample\_max\_answer\_token\_entropy} for Centralized (IPW $-0.250$, $p=1.8\!\times\!10^{-9}$), Hybrid ($-0.445$, $p=2.8\!\times\!10^{-32}$), Sequential ($-0.310$, $p=2.0\!\times\!10^{-8}$) and SAS ($-0.467$, $p=1.0\!\times\!10^{-57}$); for Debate the consensus direct causes shift to round-1 per-agent maximum entropy (min and Q3, IPW $-0.147$ and $-0.238$). All cases pass the three refutation tests. Two qualitative observations follow: (i)~the answer-level entropy variable is the dominant causal driver in every aggregation/chain architecture, supporting the SHAP claim that verbose answers harm Centralized and that Sequential compounds the same signal; (ii)~Debate is the only architecture whose direct causes are round-1 dispersion features rather than answer-level features, which causally substantiates the \textbf{converge early, then explore} SHAP profile, once round-1 dispersion is removed, cumulative entropy stops harming Debate.

\paragraph{Cross-model consistency of causal findings.}
Table~\ref{tab:cross-model-causal} reports the consensus PC$\cap$FCI direct cause per (model, dataset) cell across the 5$\times$6 grid on $\mathcal{G}_{\text{base-full}}$ and $\mathcal{G}_{\text{MAS}}$. Of 28 cells, 28 yield a non-empty consensus on $\mathcal{G}_{\text{base-full}}$ and 24 on $\mathcal{G}_{\text{MAS}}$; the empty $\mathcal{G}_{\text{MAS}}$ consensuses cluster on \texttt{AIME2025} (Q-0.6/Q-4/Q-8) plus Q-4/\texttt{HumanEval}, where stripping base-model signals leaves too few non-redundant features for the conservative criterion. Three patterns reproduce: (i)~sample-level answer-token entropy is the most frequent top cause (15/28 on $\mathcal{G}_{\text{base-full}}$, 13/24 on $\mathcal{G}_{\text{MAS}}$); (ii)~the remaining $\mathcal{G}_{\text{MAS}}$ cells (11/24) are dominated by round-1 dispersion or sample-level variance features, all with negative IPW ATE; (iii)~base-model entropy variables enter the consensus in 19/28 $\mathcal{G}_{\text{base-full}}$ cells, so the global result of Section~\ref{sec:causal_analysis} is not driven by a small subset. The single positive-signed top cause (L-3/\texttt{AIME2024} on $\mathcal{G}_{\text{base-full}}$) falls in a cell where SHAP already flagged a non-monotone signature. All listed cells pass the three refutation tests.

\begin{table*}[t]
\centering
\caption{Cross-model consistency of consensus PC$\cap$FCI direct causes per (model, dataset) cell, on $\mathcal{G}_{\text{base-full}}$ (left) and $\mathcal{G}_{\text{MAS}}$ (right). Model abbreviations: Q-0.6/Q-4/Q-8 = Qwen3-0.6B/4B/8B; L-3/L-8 = LLaMA-3.2-3B/3.1-8B. Dataset abbreviations: HE = HumanEval; AIME24/25 = AIME2024/2025. Feature names use \texttt{e} as shorthand for \texttt{entropy}. \#DC denotes the number of consensus direct causes; ``Top consensus cause'' is the consensus direct cause with the largest $|\text{IPW ATE}|$; the ATE column reports the sign of the IPW estimate of the top cause as $\downarrow$ (negative, reduces correctness) or $\uparrow$ (positive). LLaMA models on \texttt{AIME25} have insufficient samples for stable causal estimation so we skipped those cells. Cells with \#DC$=0$ (Q-0.6/Q-4/Q-8 on \texttt{AIME25}, Q-4 on \texttt{HE}, all on $\mathcal{G}_{\text{MAS}}$) correspond to runs that completed but yielded an empty PC/FCI intersection under the temporal-tier constraints. All listed cells pass all three refutation tests.}
\label{tab:cross-model-causal}
\scriptsize
\setlength{\tabcolsep}{3pt}
\begin{tabular}{@{}llccp{4.5cm}ccp{4.5cm}@{}}
\toprule
& & \multicolumn{3}{c}{$\mathcal{G}_{\text{base-full}}$} & \multicolumn{3}{c}{$\mathcal{G}_{\text{MAS}}$} \\
\cmidrule(lr){3-5}\cmidrule(lr){6-8}
\textbf{Model} & \textbf{Dataset} & \textbf{\#DC} & \textbf{ATE} & \textbf{Top consensus cause} & \textbf{\#DC} & \textbf{ATE} & \textbf{Top consensus cause} \\
\midrule
Q-0.6 & \texttt{GSM8K}     & 5 & $\downarrow$ & \texttt{sample\_max\_answer\_token\_e}              & 2 & $\downarrow$ & \texttt{sample\_mean\_answer\_token\_e}             \\
Q-0.6 & \texttt{MATH500}   & 5 & $\downarrow$ & \texttt{sample\_max\_answer\_token\_e}              & 2 & $\downarrow$ & \texttt{sample\_max\_answer\_token\_e}              \\
Q-0.6 & \texttt{AIME24}    & 2 & $\downarrow$ & \texttt{sample\_e\_relative\_iqr\_mean}             & 2 & $\downarrow$ & \texttt{round\_1\_2\_change\_e}                     \\
Q-0.6 & \texttt{AIME25}    & 3 & $\downarrow$ & \texttt{sample\_round\_1\_max\_agent\_std\_e}       & 0 & ---          & ---                                                       \\
Q-0.6 & \texttt{HE}        & 4 & $\downarrow$ & \texttt{sample\_max\_answer\_token\_e}              & 2 & $\downarrow$ & \texttt{sample\_max\_answer\_token\_e}              \\
Q-0.6 & \texttt{MMLU}      & 1 & $\downarrow$ & \texttt{sample\_round\_1\_mean\_agent\_max\_e}      & 2 & $\downarrow$ & \texttt{sample\_round\_1\_q3\_agent\_max\_e}        \\
Q-4   & \texttt{GSM8K}     & 2 & $\downarrow$ & \texttt{sample\_round\_1\_q3\_agent\_max\_e}        & 1 & $\downarrow$ & \texttt{sample\_round\_1\_q3\_agent\_max\_e}        \\
Q-4   & \texttt{MATH500}   & 4 & $\downarrow$ & \texttt{answer\_token\_e\_change}                   & 2 & $\downarrow$ & \texttt{sample\_round\_1\_q3\_agent\_max\_e}        \\
Q-4   & \texttt{AIME24}    & 3 & $\downarrow$ & \texttt{sample\_round\_1\_mean\_agent\_variance\_e} & 2 & $\downarrow$ & \texttt{sample\_round\_1\_max\_agent\_std\_e}       \\
Q-4   & \texttt{AIME25}    & 2 & $\downarrow$ & \texttt{sample\_round\_1\_q3\_agent\_max\_e}        & 0 & ---          & ---                                                       \\
Q-4   & \texttt{HE}        & 4 & $\downarrow$ & \texttt{sample\_mean\_answer\_token\_e}             & 0 & ---          & ---                                                       \\
Q-4   & \texttt{MMLU}      & 1 & $\downarrow$ & \texttt{sample\_round\_1\_median\_agent\_std\_e}    & 1 & $\downarrow$ & \texttt{sample\_round\_1\_median\_agent\_max\_e}    \\
Q-8   & \texttt{GSM8K}     & 1 & $\downarrow$ & \texttt{sample\_variance\_e}                        & 2 & $\downarrow$ & \texttt{sample\_round\_1\_q3\_agent\_max\_e}        \\
Q-8   & \texttt{MATH500}   & 2 & $\downarrow$ & \texttt{sample\_max\_answer\_token\_e}              & 1 & $\downarrow$ & \texttt{sample\_max\_answer\_token\_e}              \\
Q-8   & \texttt{AIME24}    & 2 & $\downarrow$ & \texttt{sample\_max\_answer\_token\_e}              & 1 & $\downarrow$ & \texttt{sample\_round\_1\_q3\_agent\_variance\_e}   \\
Q-8   & \texttt{AIME25}    & 4 & $\downarrow$ & \texttt{sample\_round\_1\_q3\_agent\_max\_e}        & 0 & ---          & ---                                                       \\
Q-8   & \texttt{HE}        & 2 & $\downarrow$ & \texttt{sample\_mean\_answer\_token\_e}             & 3 & $\downarrow$ & \texttt{sample\_max\_answer\_token\_e}              \\
Q-8   & \texttt{MMLU}      & 3 & $\downarrow$ & \texttt{sample\_round\_1\_max\_agent\_max\_e}       & 2 & $\downarrow$ & \texttt{sample\_round\_1\_median\_agent\_max\_e}    \\
L-3   & \texttt{GSM8K}     & 3 & $\downarrow$ & \texttt{sample\_max\_answer\_token\_e}              & 2 & $\downarrow$ & \texttt{sample\_max\_answer\_token\_e}              \\
L-3   & \texttt{MATH500}   & 4 & $\downarrow$ & \texttt{sample\_max\_answer\_token\_e}              & 3 & $\downarrow$ & \texttt{sample\_max\_answer\_token\_e}              \\
L-3   & \texttt{AIME24}    & 1 & $\uparrow$   & \texttt{sample\_avg\_e\_per\_token\_diff\_vs\_base} & 2 & $\downarrow$ & \texttt{sample\_round\_1\_q3\_agent\_max\_e}        \\
L-3   & \texttt{HE}        & 3 & $\downarrow$ & \texttt{sample\_max\_answer\_token\_e}              & 3 & $\downarrow$ & \texttt{sample\_max\_answer\_token\_e}              \\
L-3   & \texttt{MMLU}      & 3 & $\downarrow$ & \texttt{sample\_max\_answer\_token\_e}              & 1 & $\downarrow$ & \texttt{sample\_max\_answer\_token\_e}              \\
L-8   & \texttt{GSM8K}     & 4 & $\downarrow$ & \texttt{sample\_max\_answer\_token\_e}              & 2 & $\downarrow$ & \texttt{sample\_max\_answer\_token\_e}              \\
L-8   & \texttt{MATH500}   & 5 & $\downarrow$ & \texttt{sample\_max\_answer\_token\_e}              & 3 & $\downarrow$ & \texttt{sample\_max\_answer\_token\_e}              \\
L-8   & \texttt{AIME24}    & 4 & $\downarrow$ & \texttt{sample\_round\_1\_max\_agent\_std\_e}       & 1 & $\downarrow$ & \texttt{sample\_variance\_e}                        \\
L-8   & \texttt{HE}        & 4 & $\downarrow$ & \texttt{sample\_std\_answer\_token\_e}              & 2 & $\downarrow$ & \texttt{sample\_max\_answer\_token\_e}              \\
L-8   & \texttt{MMLU}      & 5 & $\downarrow$ & \texttt{sample\_max\_answer\_token\_e}              & 3 & $\downarrow$ & \texttt{sample\_max\_answer\_token\_e}              \\
\bottomrule
\end{tabular}
\end{table*}

\subsection{Round-1 Entropy Dominates Despite Extended Deliberation}
\label{app:more_rounds}

\begin{figure*}[!htbp]
    \centering
    \begin{minipage}{0.95\textwidth}
    \begin{subfigure}{0.48\textwidth}
        \centering
        \includegraphics[width=\linewidth]{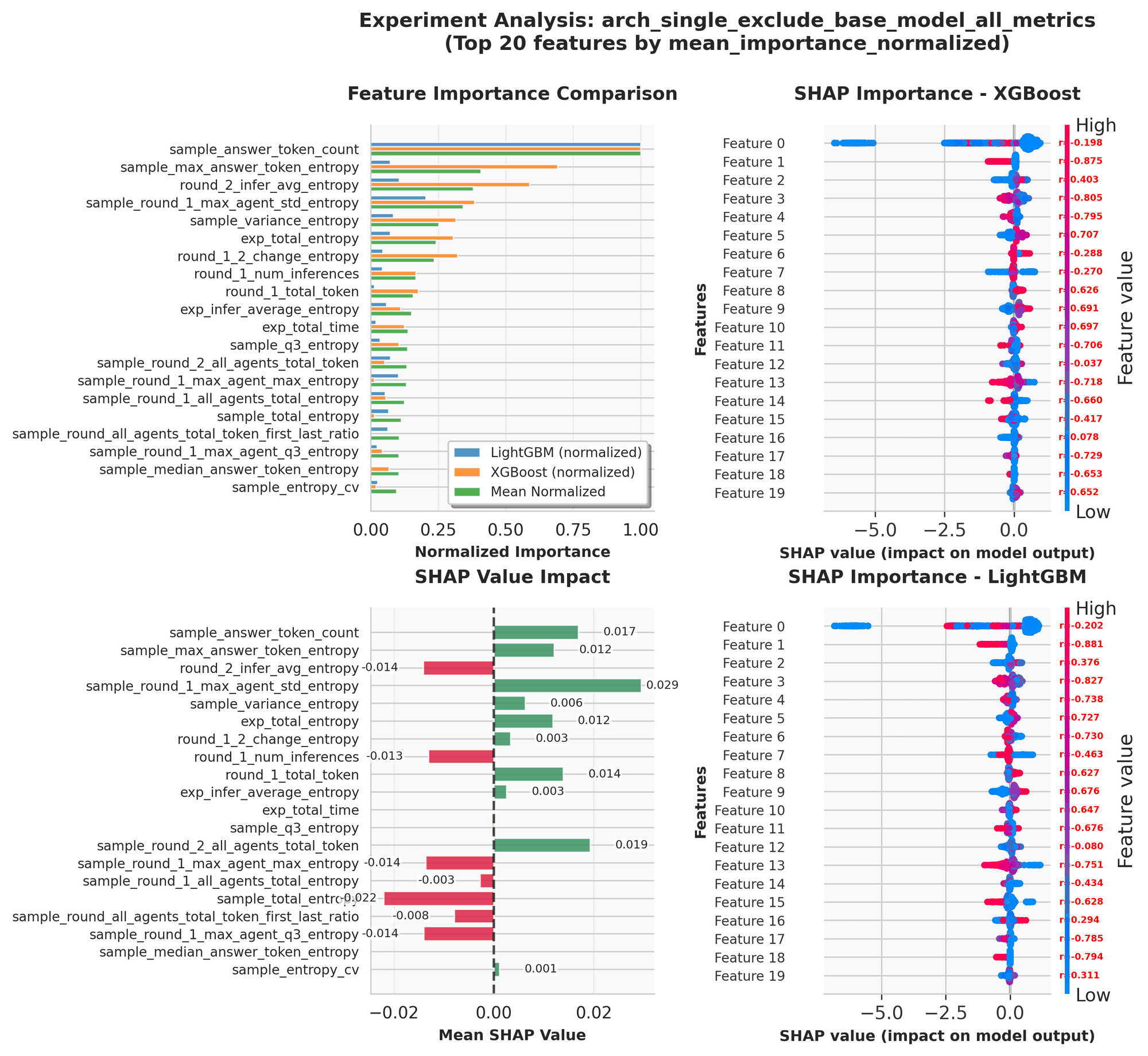}
        \caption{}
        \label{fig:diff_archs_single}
    \end{subfigure}
    \hfill
    \begin{subfigure}{0.48\textwidth}
        \centering
        \includegraphics[width=\linewidth]{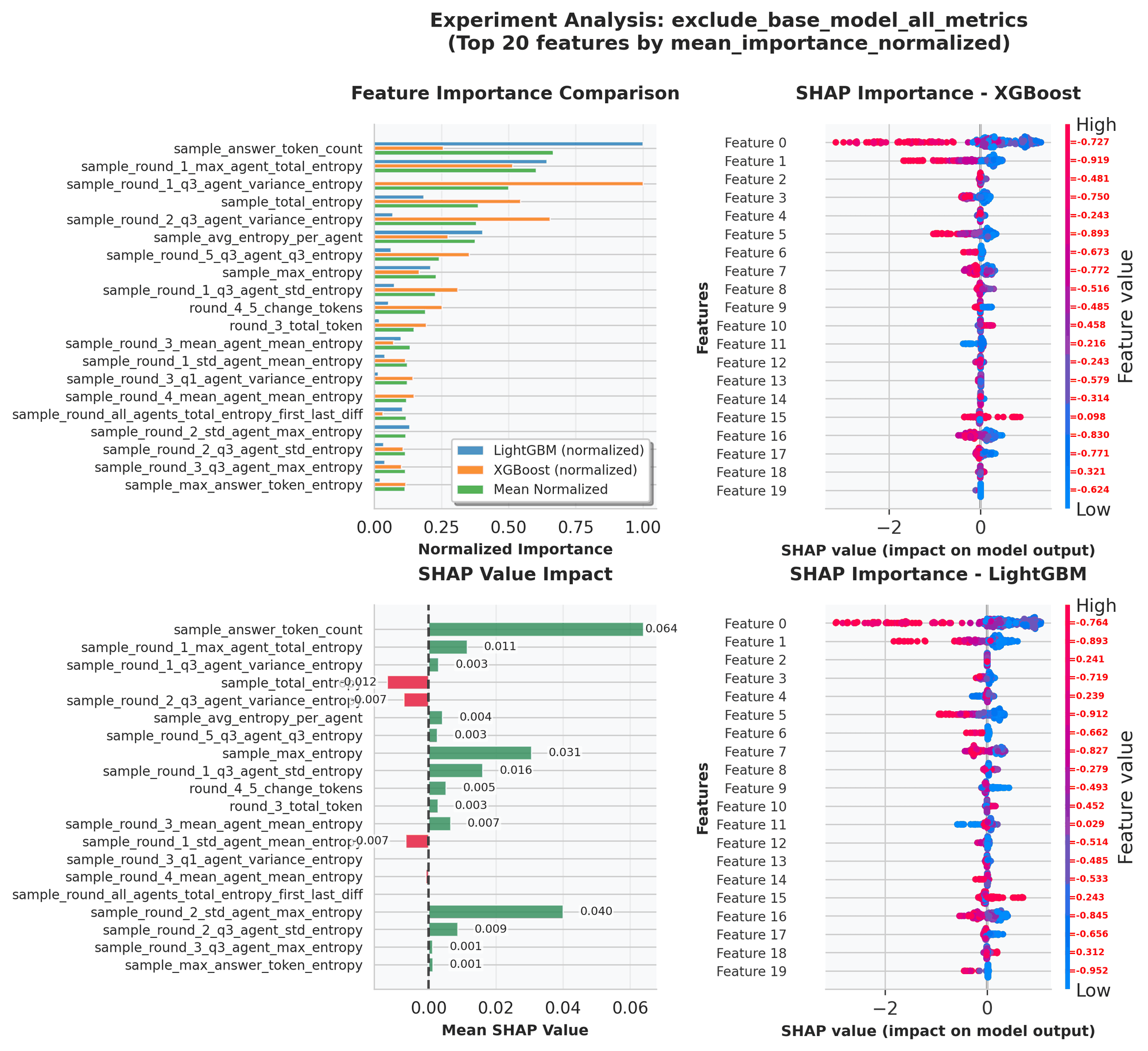}
        \caption{}
        \label{fig:more_rounds_analysis}
    \end{subfigure}
    \end{minipage}
    \caption{Top 20 features for (a) SAS and (b) MAS ($R=5$) on $\mathcal{G}_{\text{MAS}}$. SAS is dominated by answer token-level features; notably, it exhibits a positive ($\rho$) but negative ($\bar{S}$) for round-2 entropy. In contrast, MAS predictions are driven by early-round dynamics: Round-1 features occupy the top ranks (2-3), while features from Rounds 3-5 are absent from the top 10. Thus, early-round entropy serves as the primary predictor for both systems.}
    \label{fig:sas_rounds_analysis}
\end{figure*}

Figure~\ref{fig:more_rounds_analysis} reports the $R=5$ analysis on the expanded 494-dimensional feature space.

\paragraph{Round-1 still dominates at $R=5$.}
Even in the 494-dimensional $R=5$ feature space, the top-ranked entropy features come from round 1: \textit{sample\_round\_1\_max\_agent\_total\_entropy} ranks second ($\rho = -0.91$), and round-1 Q3 agent variance entropy ranks third. Answer token count leads overall but with the now-familiar non-monotone signature ($\rho = -0.75$, $\bar{S} = +0.064$).

\paragraph{Rounds 3-4 carry essentially no signal.}
The most informative observation in this section is what is \emph{absent} from the top 20: no feature from rounds 3 or 4 appears, and only one round-5 feature does ($\bar{I} = 0.24$). Adding rounds beyond two does not surface new predictive structure; it adds dimensions that the classifier learns to ignore.

\paragraph{Causal validation.}
Causal estimation on the $R{=}5$ data confirms what the SHAP ranking suggests: the consensus PC$\cap$FCI direct cause is the answer-level \textit{sample\_max\_answer\_token\_entropy} (IPW ATE $-0.344$, $p=2.9\!\times\!10^{-53}$), with round-2 Q3 of per-agent maximum entropy as the second candidate but not robust across estimators. The strongest mediation pathway runs from round-1 Q3 of per-agent variance entropy through its round-2 counterpart to correctness (indirect $-0.090$, mediating 18\%); no round-3, round-4, or round-5 feature appears as either a direct cause or a mediator above the bootstrap significance threshold. The causal evidence therefore supports the SHAP reading that rounds beyond two add no predictive structure, they also add no causal structure.

\begin{figure}[!htbp]
    \centering
    \begin{minipage}{0.95\textwidth}
    \begin{subfigure}{0.32\textwidth}
        \centering
        \includegraphics[width=\linewidth]{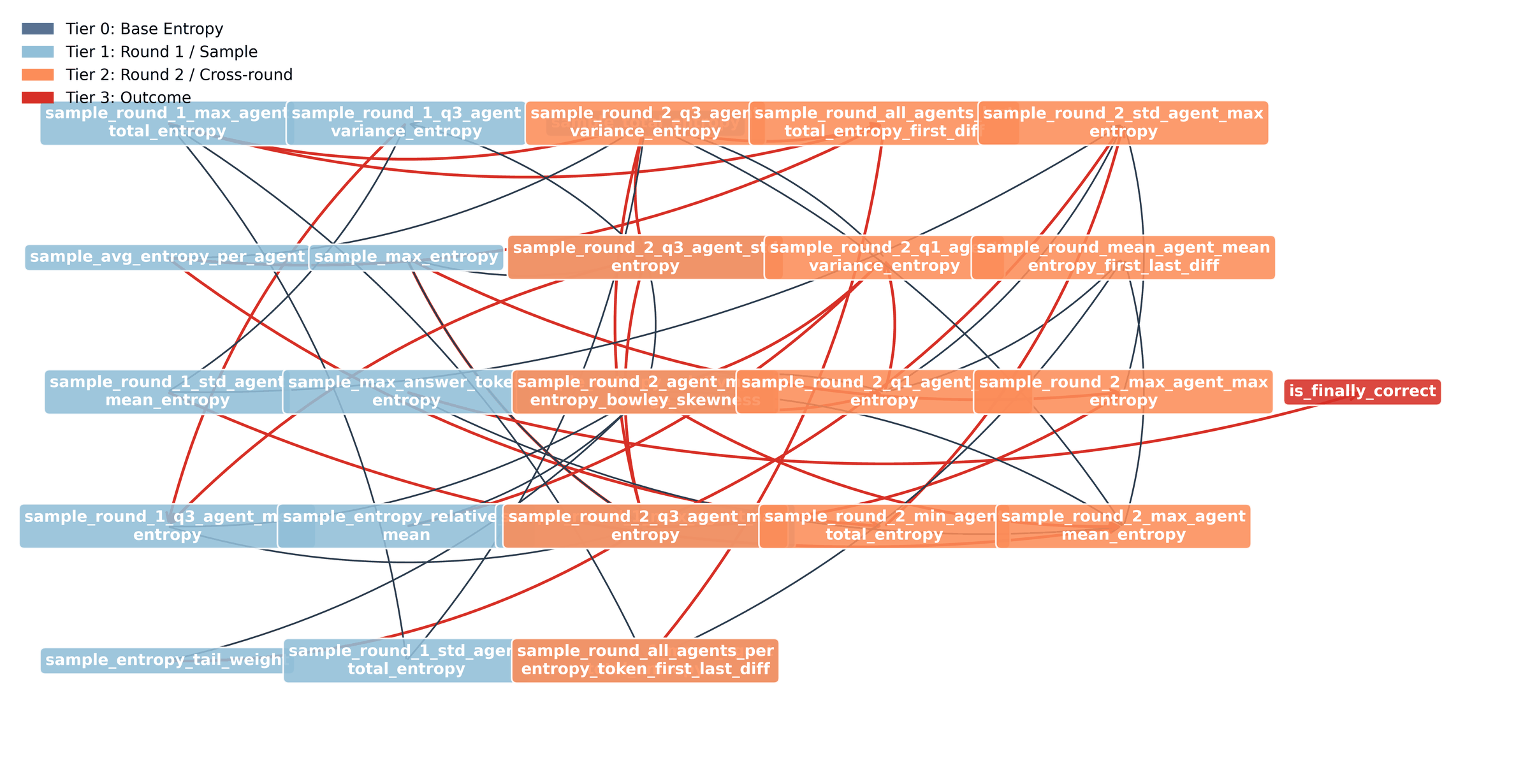}
        \caption{Consensus causal graph}
    \end{subfigure}
    \hfill
    \begin{subfigure}{0.32\textwidth}
        \centering
        \includegraphics[width=\linewidth]{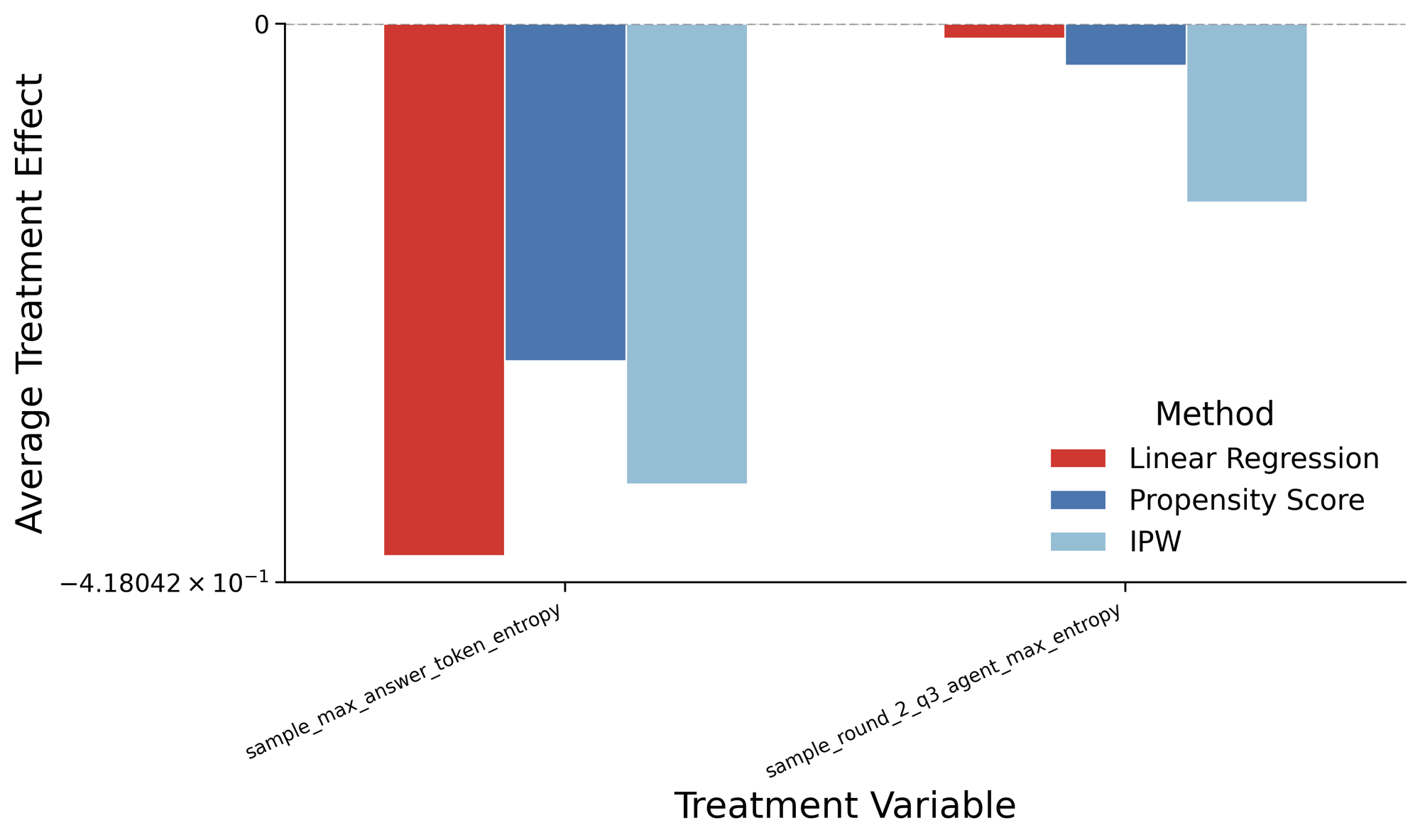}
        \caption{ATE estimates}
    \end{subfigure}
    \hfill
    \begin{subfigure}{0.32\textwidth}
        \centering
        \includegraphics[width=\linewidth]{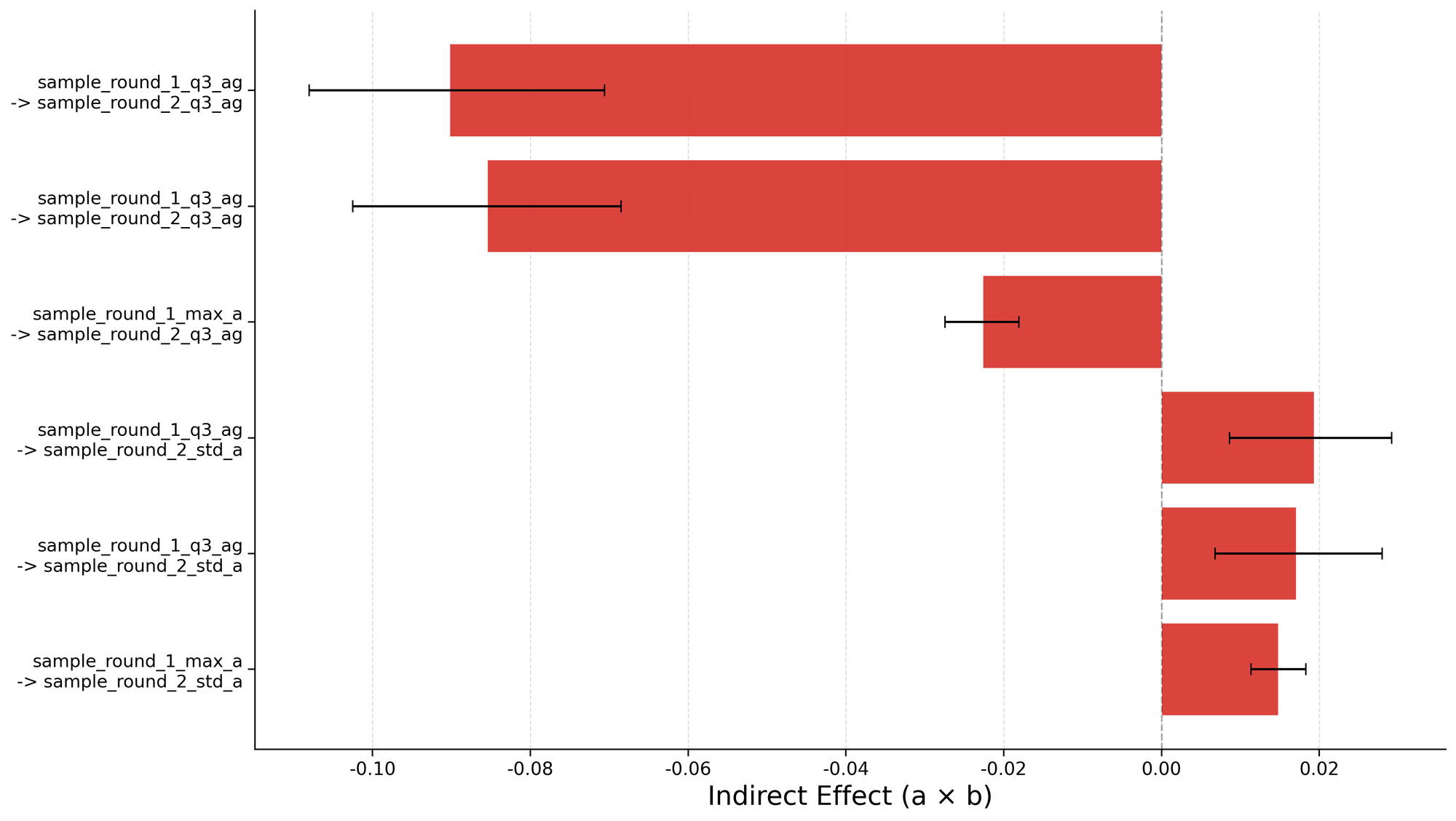}
        \caption{Mediation indirect effects}
    \end{subfigure}
    \end{minipage}
    \caption{Causal triplet at $R{=}5$. Direct causes and mediators are confined to round-1 and round-2 entropy features; rounds 3--5 contribute no surviving causal edges.}
    \label{fig:causal-more-rounds}
\end{figure}

\subsection{Temperature Variation Preserves Relative Entropy Patterns}
\label{app:temperature-ablation}

Temperature $\tau$ directly modulates the sharpness of the softmax distribution: higher $\tau$ produces flatter distributions with higher absolute entropy, while lower $\tau$ concentrates probability mass and reduces entropy. So we conduct temperature ablation experiments to assess the robustness of our findings to decoding hyperparameters.

\paragraph{Experimental Setup.}
We evaluate Qwen3-4B on \texttt{MATH500} across all five MAS architectures at three temperature settings $\tau \in \{0.4, 0.6, 0.8\}$, with top-$p = 0.95$ held constant. For each configuration, we use 100 paired samples and apply McNemar's test for statistical significance between all temperature pairs.

\begin{figure}[t]
    \centering
    \includegraphics[width=0.65\linewidth]{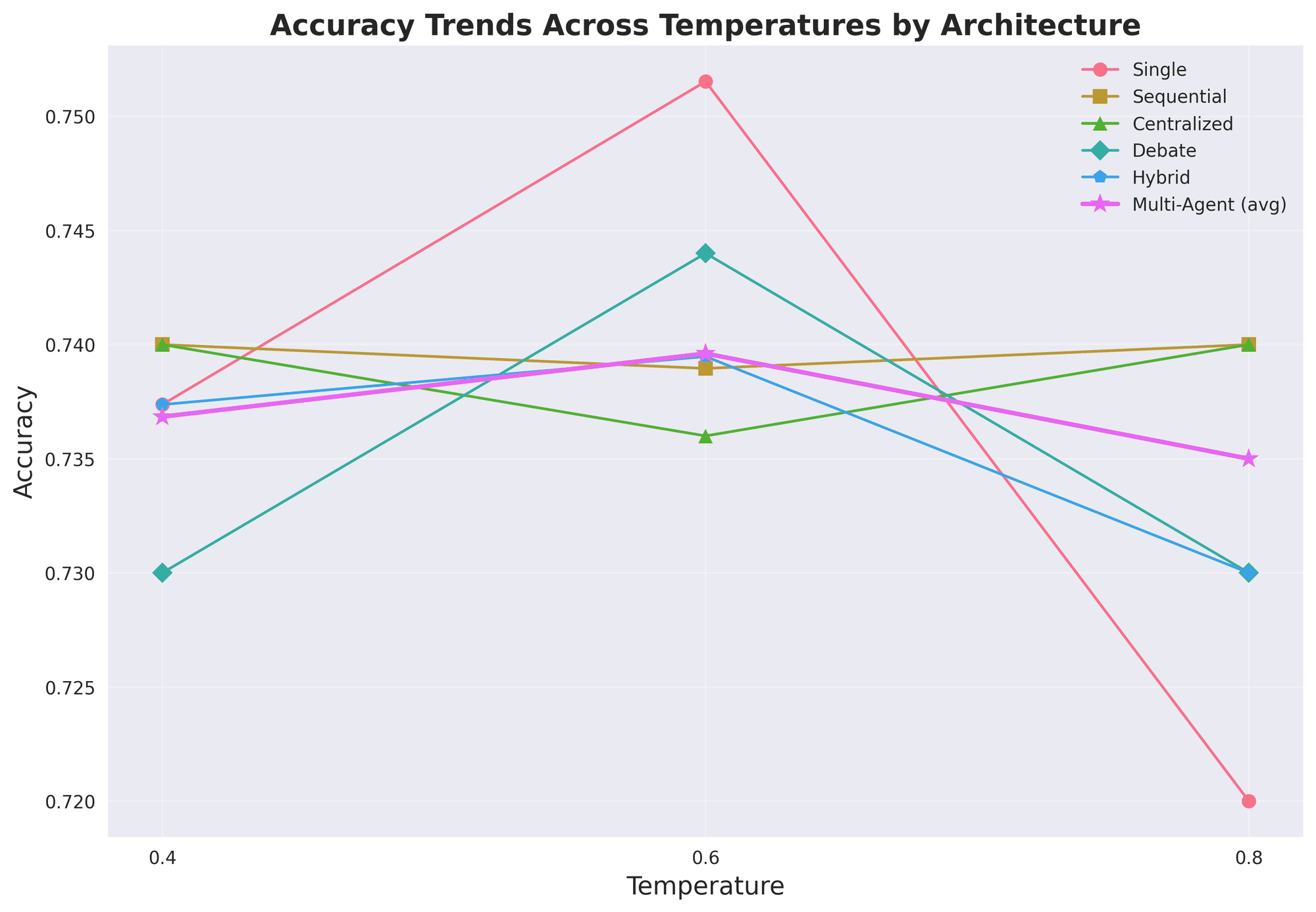}
    \caption{MAS accuracy across temperatures $\tau \in \{0.4, 0.6, 0.8\}$ for all five architectures on \texttt{MATH500}. Accuracy remains remarkably stable: the maximum variation within any architecture is 3.2\% (Single), and the multi-agent average varies by only 0.5\%. McNemar's test yields $p > 0.37$ for all 15 pairwise comparisons, confirming statistical invariance.}
    \label{fig:temperature-accuracy}
\end{figure}

\paragraph{Accuracy Is Statistically Invariant to Temperature.}
Figure~\ref{fig:temperature-accuracy} presents MAS accuracy across temperatures. The results demonstrate remarkable stability: the multi-agent average accuracy varies by only 0.5\% across the temperature range ($73.7\%$ at $\tau=0.4$, $74.0\%$ at $\tau=0.6$, $73.5\%$ at $\tau=0.8$). Among individual architectures, Sequential and Centralized achieve identical accuracy ($74.0\%$) at both $\tau=0.4$ and $\tau=0.8$, while the maximum variation occurs in Single ($73.7\% \to 75.2\% \to 72.0\%$, $\Delta = 3.2\%$). McNemar's test across all 15 pairwise comparisons (5 architectures $\times$ 3 temperature pairs) yields $p > 0.37$, with most comparisons showing $p \ge 0.48$ or $p = 1.0$, confirming that temperature does not significantly alter MAS performance outcomes within this range.

\begin{figure}[t]
    \centering
    \includegraphics[width=\linewidth]{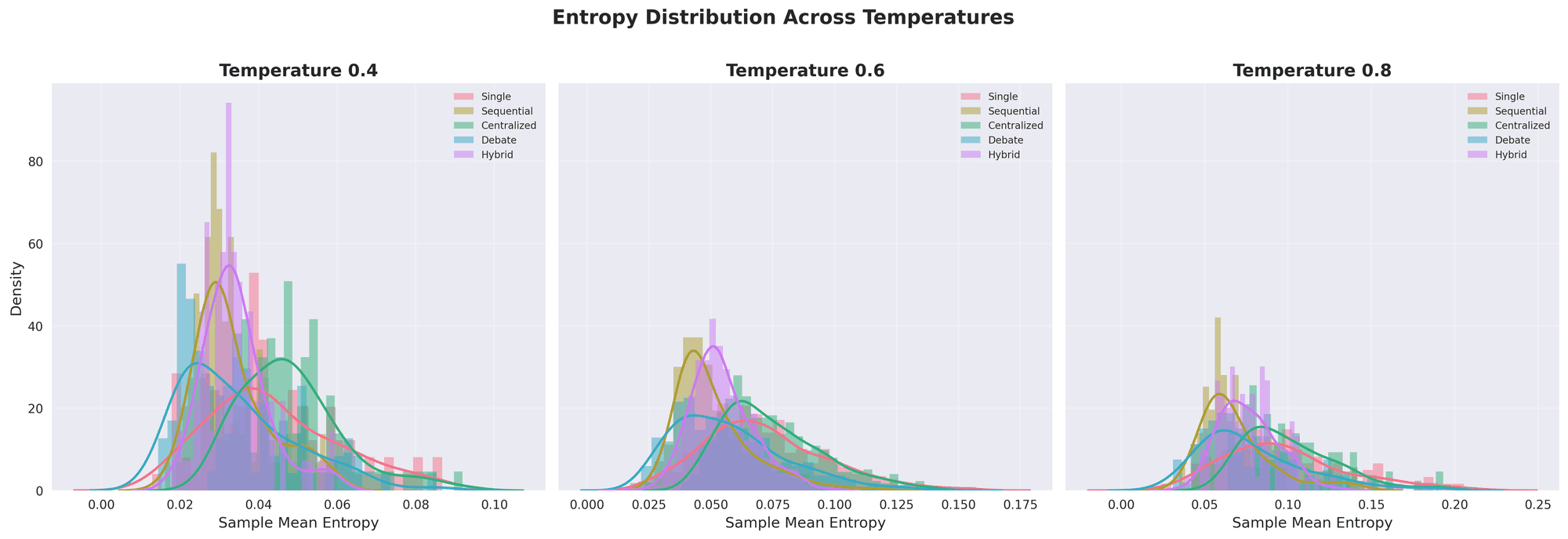}
    \caption{Entropy distribution statistics across temperatures for all architectures. Absolute entropy values scale approximately $2\times$ from $\tau=0.4$ to $\tau=0.8$ (mean entropy: Centralized $0.048 \to 0.075 \to 0.101$), but the relative ordering of architectures is preserved: Centralized and Single consistently exhibit higher entropy than Sequential, Debate, and Hybrid across all temperatures.}
    \label{fig:temperature-entropy}
\end{figure}

\paragraph{Entropy Scales Predictably While Preserving Relative Patterns.}
Figure~\ref{fig:temperature-entropy} shows that absolute entropy values increase monotonically with temperature, as expected from the mathematical properties of softmax scaling. Mean entropy approximately doubles from $\tau=0.4$ to $\tau=0.8$: Centralized ($0.048 \to 0.101$, $2.1\times$), Single ($0.044 \to 0.098$, $2.2\times$), Debate ($0.034 \to 0.079$, $2.3\times$). Critically, the \textbf{relative ordering} of architectures by entropy is fully preserved across all temperatures: Centralized and Single consistently occupy the high-entropy regime, while Sequential, Debate, and Hybrid cluster in the low-entropy regime. This confirms that while absolute entropy magnitudes are temperature-dependent, the comparative patterns central to our analysis remain valid.

\paragraph{Summary.}
The temperature ablation confirms that the entropy-performance relationship documented in this paper is not an artifact of specific decoding settings. Within the commonly-used range $\tau \in [0.4, 0.8]$: (1)~MAS accuracy is statistically invariant ($p > 0.37$ for all comparisons); (2)~relative entropy patterns across architectures are fully preserved; and (3)~core feature importance rankings remain stable. Our findings rely on relative comparisons rather than absolute entropy thresholds, making them robust to temperature variation.

\subsection{Model Scaling Preserves Entropy Feature Dominance}
\label{app:model-size}

To check whether stronger base models change the picture, we evaluate Qwen3-14B on \texttt{GSM8K}, \texttt{AIME2024}, \texttt{AIME2025}, and \texttt{HumanEval} across all five architectures, using the same three feature groups and SHAP pipeline as in the main analysis.

\paragraph{SAS competitiveness persists at 14B.}
SAS achieves the highest accuracy on 2 of 4 benchmarks and beats at least one MAS on all 4. The advantage concentrates on hard math: $+2.8$pp on \texttt{AIME2024} (82.8\% SAS vs.\ 80.0\% Hybrid) and $+3.4$pp on \texttt{AIME2025} (76.7\% SAS vs.\ 73.3\% Centralized). MAS only catches up on near-ceiling tasks (\texttt{GSM8K}: 98.0\% Centralized vs.\ 97.0\% SAS) and the domain-specific \texttt{HumanEval} (78.0\% Centralized vs.\ 75.6\% SAS). The SAS-MAS trade-off is therefore task-driven, not scale-driven.

\paragraph{Feature hierarchy reproduces at 14B.}
Across $\mathcal{G}_{\text{MAS}}$, $\mathcal{G}_{\text{base-H}}$, and $\mathcal{G}_{\text{base-full}}$, accuracy climbs $83.2\% \to 88.0\% \to 93.2\%$, the same monotone progression observed at smaller scales. On $\mathcal{G}_{\text{MAS}}$, round-1 entropy features fill roughly 70\% of the top 20 with $\rho \in [-0.84, -0.55]$. On $\mathcal{G}_{\text{base-H}}$, the leading signal becomes base-model answer length, and \textit{answer\_token\_entropy\_change} ($\rho \approx -0.84$) once again indicates that entropy growth from base to MAS predicts failure. On $\mathcal{G}_{\text{base-full}}$, \textit{base\_model\_is\_finally\_correct} dominates with $\rho \approx 0.98$, exactly as in Appendix~\ref{app:base_model_entropy}. The cross-scale stability of this $\mathcal{G}$-progression is the relevant evidence: the predictive structure is not an artifact of small models.

\begin{figure*}[t]
    \centering
    \begin{minipage}{0.95\textwidth}
    \begin{subfigure}{0.48\textwidth}
        \centering
        \includegraphics[width=\linewidth]{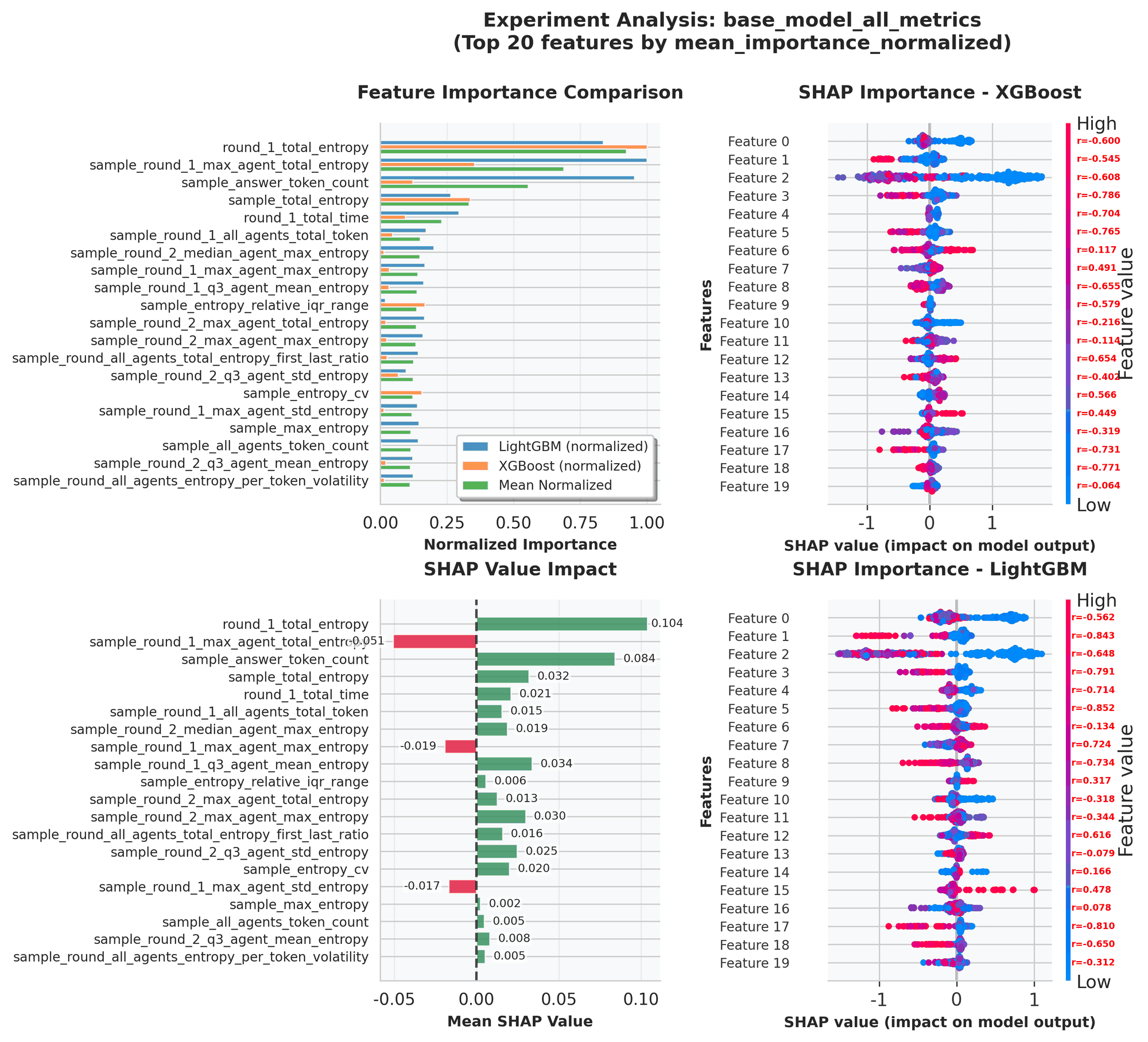}
        \caption{$\mathcal{G}_{\text{MAS}}$: MAS-only features}
        \label{fig:model-size-mas}
    \end{subfigure}
    \hfill
    \begin{subfigure}{0.48\textwidth}
        \centering
        \includegraphics[width=\linewidth]{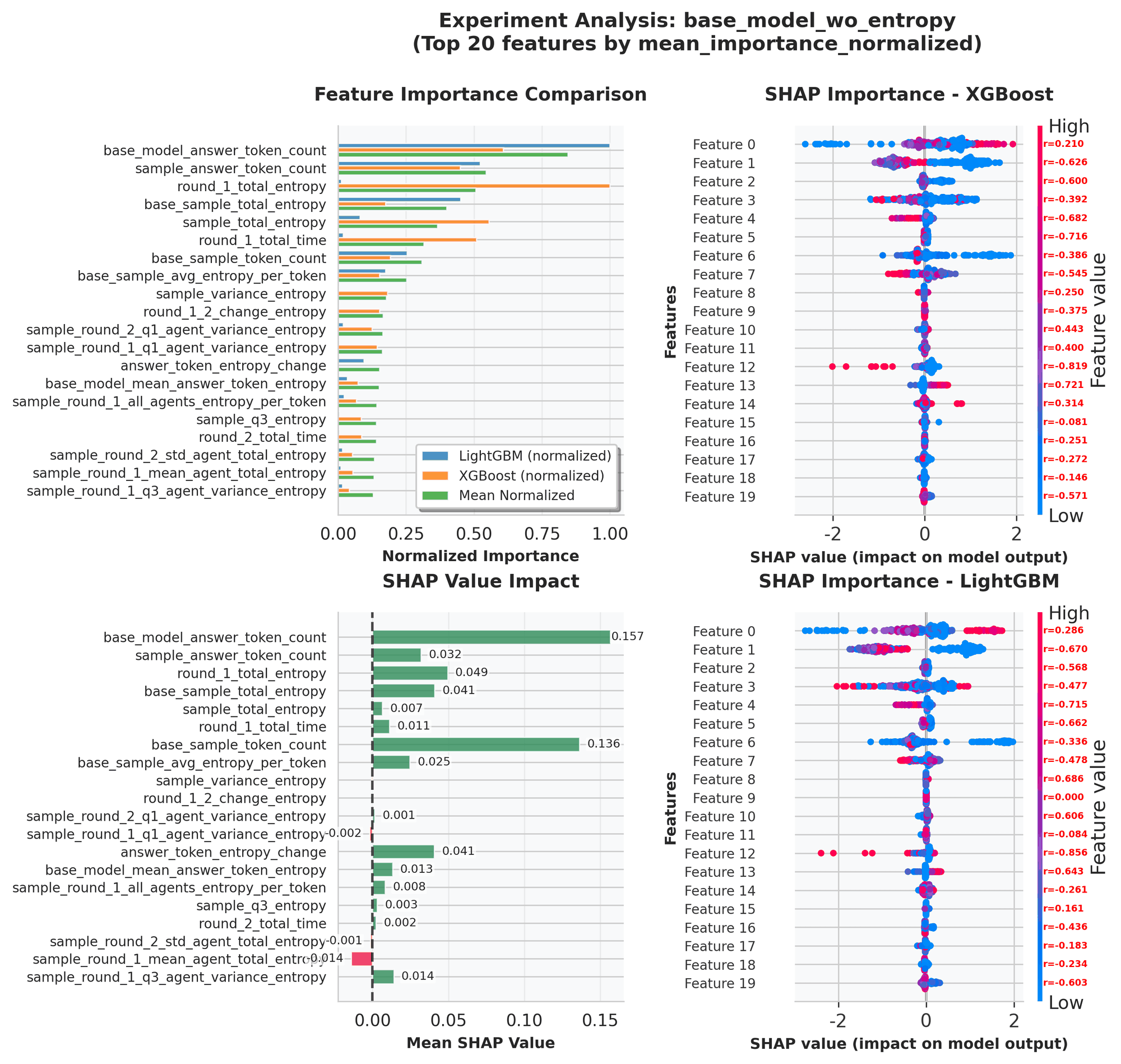}
        \caption{$\mathcal{G}_{\text{base-H}}$: Including base model entropy}
        \label{fig:model-size-base-h}
    \end{subfigure}
    \end{minipage}
    \caption{Feature importance and SHAP analysis for Qwen3-14B across two feature groups. (a) On $\mathcal{G}_{\text{MAS}}$, round-1 entropy dominates with approximately 70\% of top-20 features being entropy-related (LightGBM accuracy: 83.2\%, F1: 90.3\%). (b) On $\mathcal{G}_{\text{base-H}}$, base model answer length emerges as the top predictor, while \textit{answer\_token\_entropy\_change} ($\rho \approx -0.84$) signals that entropy increase from base to MAS predicts failure (accuracy: 88.0\%, F1: 93.0\%). On $\mathcal{G}_{\text{base-full}}$, \textit{base\_model\_is\_finally\_correct} achieves importance $\approx 1.0$ with $\rho \approx 0.98$, monopolizing prediction (accuracy: 93.2\%, F1: 95.9\%), consistent with the pattern observed across other settings.}
    \label{fig:model-size-analysis}
\end{figure*}

\paragraph{Causal validation at 14B.}
At the 14B scale, the causal pipeline reduces the structure to a single consensus PC$\cap$FCI direct cause: \textit{round\_1\_total\_entropy} (IPW ATE $+0.166$, $p=6.0\!\times\!10^{-92}$, all refutation tests pass). The positive sign reflects the Qwen3-14B regime in which moderate first-round entropy is needed to trigger inter-agent refinement, and is consistent with the SHAP observation that round-1 features dominate the top 20 with negative correlations only \emph{above} a threshold. \textit{sample\_max\_entropy} appears as a secondary direct cause with negative ATE ($-0.151$, $p=1.2\!\times\!10^{-6}$). Cross-round mediation is exceptionally strong: round-1 maximum per-agent total entropy transmits 45.1\% of its effect on correctness through its round-2 counterpart, and \textit{round\_1\_total\_entropy} mediates 55.8\% through round-2 median agent maximum entropy. The causal evidence therefore confirms two of the SHAP findings (round-1 dominance, and the round-1$\rightarrow$round-2 propagation) at a scale where round-2 deliberation already shows productive value.

\begin{figure}[!htbp]
    \centering
    \begin{minipage}{0.95\textwidth}
    \begin{subfigure}{0.32\textwidth}
        \centering
        \includegraphics[width=\linewidth]{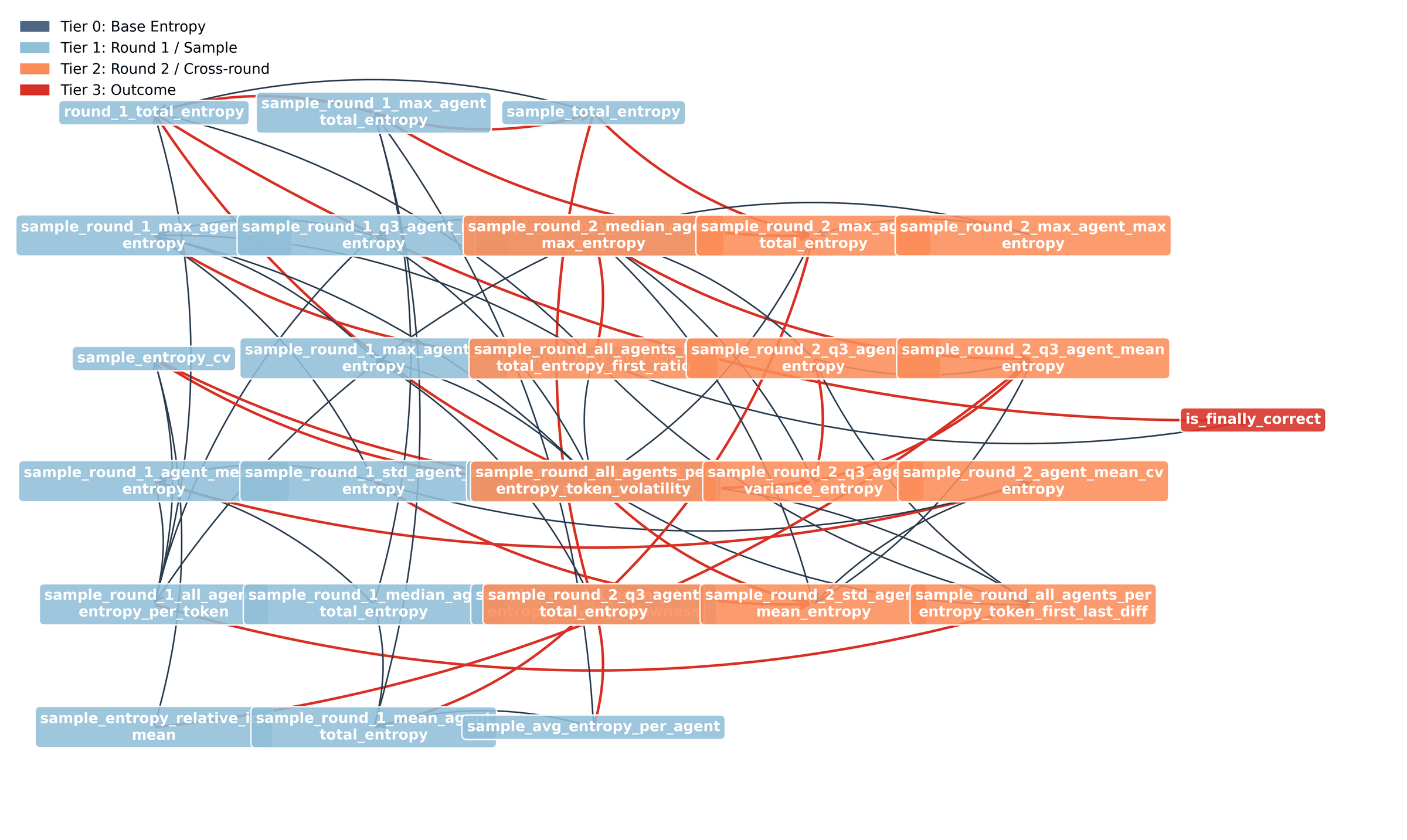}
        \caption{Consensus causal graph}
    \end{subfigure}
    \hfill
    \begin{subfigure}{0.32\textwidth}
        \centering
        \includegraphics[width=\linewidth]{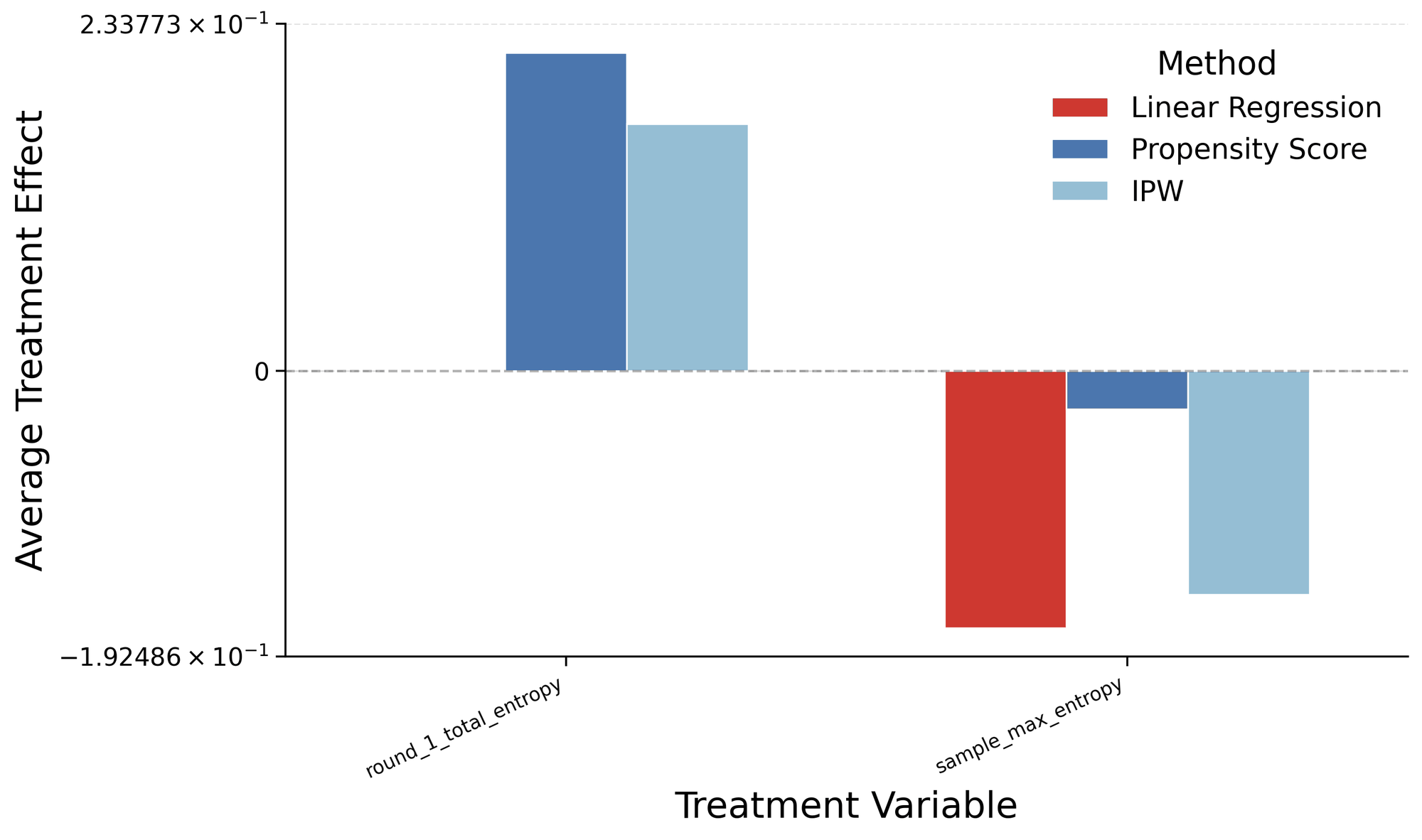}
        \caption{ATE estimates}
    \end{subfigure}
    \hfill
    \begin{subfigure}{0.32\textwidth}
        \centering
        \includegraphics[width=\linewidth]{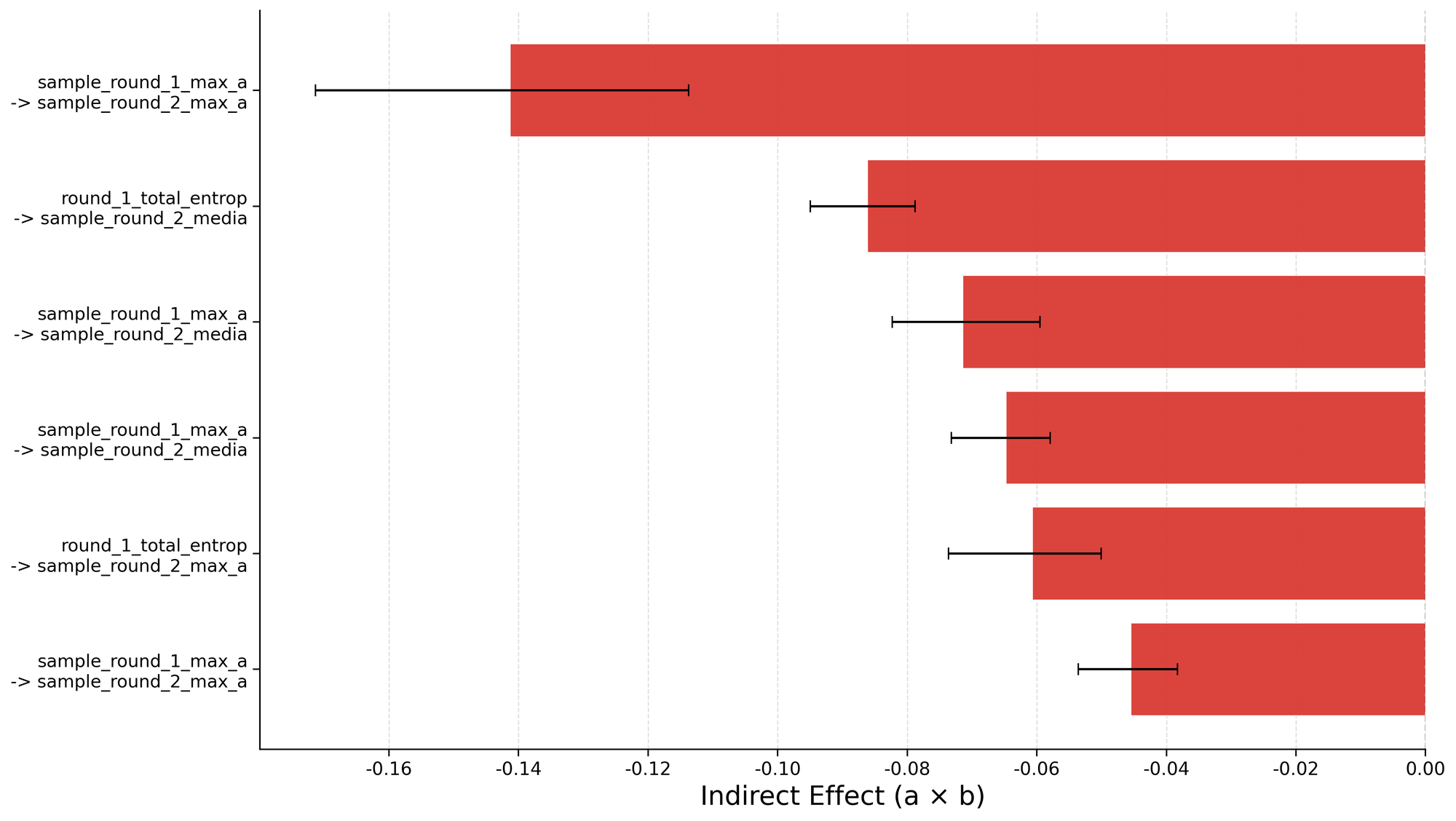}
        \caption{Mediation indirect effects}
    \end{subfigure}
    \end{minipage}
    \caption{Causal triplet for Qwen3-14B. Round-1 total entropy is the consensus direct cause and propagates 45--56\% of its effect to correctness through round-2 entropy.}
    \label{fig:causal-model-size}
\end{figure}

\section{RL Training Inverts the Role of Entropy}
\label{app:rl_finetuned_base}

Few studies have investigated whether using a specialized, fine-tuned model as the base model can improve MAS performance on reasoning tasks. We explore this using Qwen2.5-7B-SimpleRL-Zoo~\citep{simplerl-zoo-colm25}, denoted $M_{\text{RL-base}}$, which is obtained by applying zero-shot RL to Qwen2.5-7B~\citep{qwen2.5-arxiv25} on 8K MATH problems without prior supervised fine-tuning.

\begin{figure*}[ht]
    \centering
    \includegraphics[width=\textwidth]{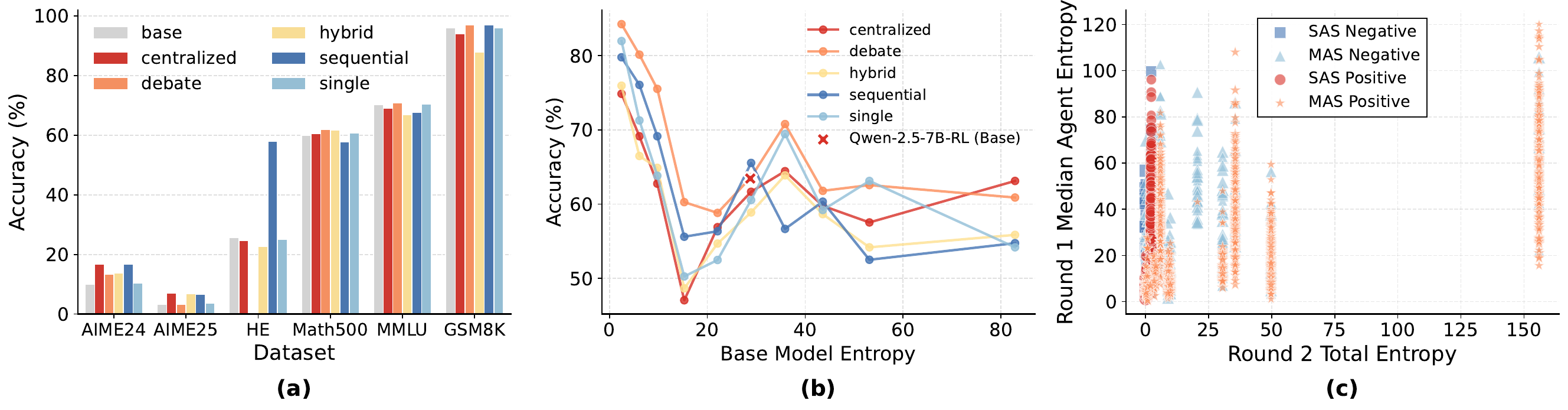}
    \caption{The role of entropy is reshaped in MAS built on Qwen2.5-7B-SimpleRL-Zoo. (a) The performance of different MAS architectures across datasets. (b) Relationship between base-model entropy and MAS accuracy. (c) Most predictive features in \(\mathcal{G}_{\text{MAS}}\).}
    \label{fig:rl-analysis}
    \vspace{-1em}
\end{figure*}

Figure~\ref{fig:rl-analysis} highlights three findings: (1) \textbf{$M_{\text{RL-base}}$ consistently enables MAS to outperform SAS}, whereas with the standard $M_{\text{base}}$, SAS surpasses MAS in 43.3\% of cases, but never with $M_{\text{RL-base}}$; (2) \textbf{The entropy-accuracy relationship is reshaped}: for $M_{\text{base}}$, higher entropy monotonically reduces accuracy; for $M_{\text{RL-base}}$, accuracy peaks at near-zero entropy, declines to a minimum as entropy increases, then recovers to a secondary plateau before fluctuating, as shown in Figure~\ref{fig:rl-analysis}b. The initial peak likely corresponds to easy problems that require little reasoning. For harder problems, however, the subsequent recovery reflects the benefit of longer trajectories: since Qwen2.5-7B already possesses strong inherent reasoning capabilities, RL training amplifies these by extending deliberation~\citep{simplerl-zoo-colm25}, allowing agents to explore structured solution paths under controlled entropy, where moderate entropy signifies productive exploration rather than degeneration; (3) On $\mathcal{G}_{\text{MAS}}$, the top predictors are round-1 median entropy ($\rho \approx -0.758$) and round-2 entropy ($\rho \approx 0.267$), as shown in Figure~\ref{fig:rl-analysis}c, indicating that \textbf{early consensus, reflected in low initial entropy, combined with calibrated later-round exploration, supports effective multi-agent collaboration}.

Further comparison across Figures~\ref{fig:base-model-analysis}b,d and~\ref{fig:rl-analysis}b shows that \textbf{$M_{\text{RL-base}}$ achieves both lower average entropy and higher correctness than $M_{\text{base}}$}, indicating that RL training produces more reliable entropy estimates where entropy better reflects solution diversity rather than noise. In contrast, for $M_{\text{base}}$, higher entropy often signals hallucination or incoherent reasoning. Consequently, MAS can leverage this improved signal for more effective coordination.

\paragraph{Detailed SHAP analysis.}
With $M_{\text{RL-base}}$ (Qwen2.5-7B-SimpleRL-Zoo), one feature changes sign in a way that is consistent across both $\mathcal{G}_{\text{MAS}}$ and $\mathcal{G}_{\text{base-H}}$: \textit{round\_2\_total\_entropy} now correlates positively with sample-level success ($\rho = +0.27$), reversing the predominantly negative sign observed for the standard base models. Round-1 features remain negative ($\rho = -0.77$ for round-1 median agent entropy), so the inversion is specifically a late-round effect: RL-trained models use round-2 entropy as productive refinement rather than noise. The global-level $\bar{S}$ for round-2 entropy stays negative ($-0.011$ on $\mathcal{G}_{\text{MAS}}$, $-0.085$ on $\mathcal{G}_{\text{base-H}}$), the same non-monotone pattern flagged earlier, indicating that even RL-trained agents are harmed by excessive late-round entropy. Base-model entropy remains a strong negative predictor on $\mathcal{G}_{\text{base-H}}$ ($\rho = -0.74$), and the answer-entropy change from base to MAS is still harmful ($\rho = -0.70$), showing that RL training shifts the role of late-round entropy without overturning the base-model dependency.

\begin{figure*}[!htbp]
    \centering
    \begin{minipage}{0.95\textwidth}
    \begin{subfigure}{0.48\textwidth}
        \centering
        \includegraphics[width=\linewidth]{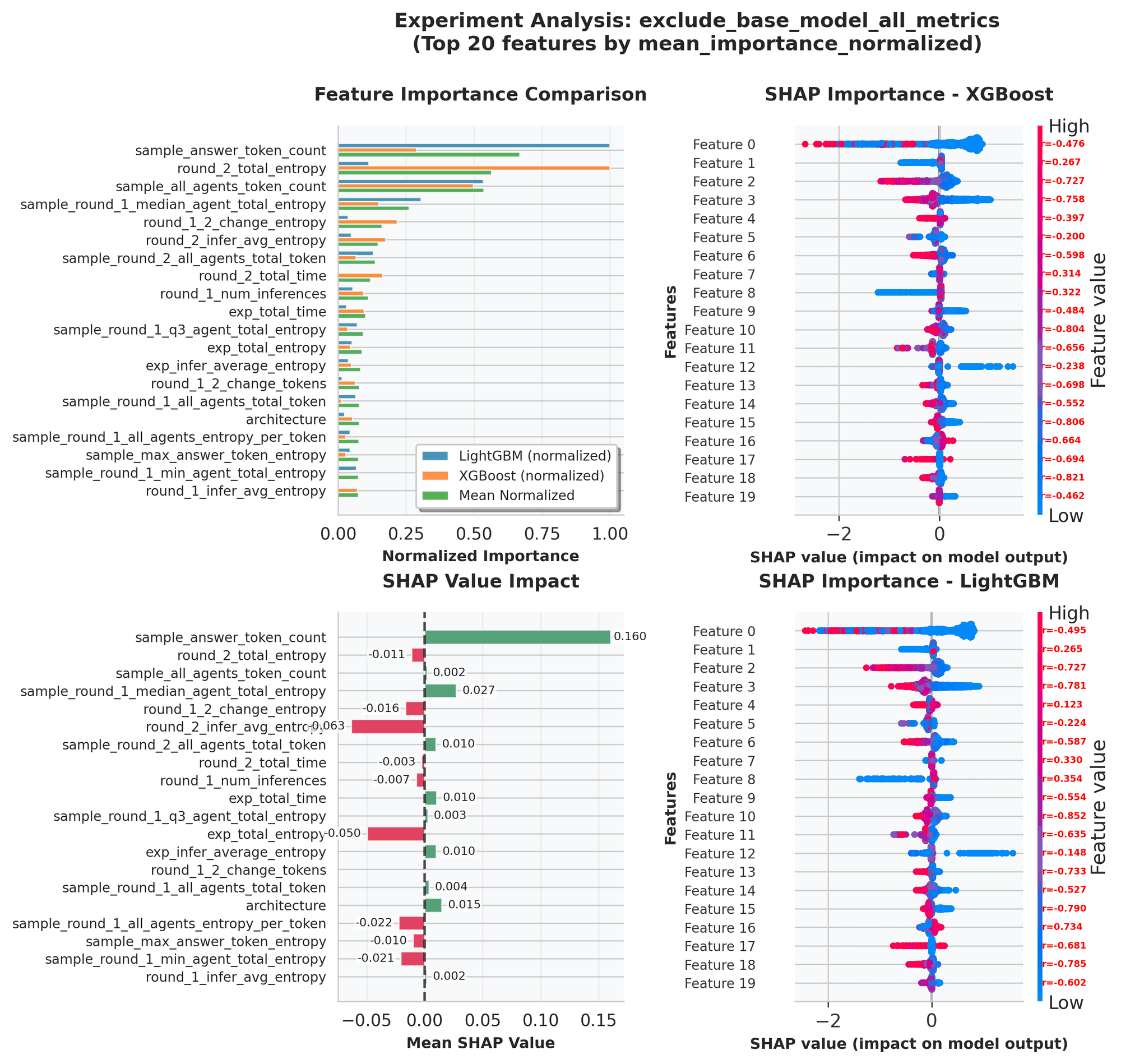}
        \caption{}
        \label{fig:rl_mas_analysis}
    \end{subfigure}
    \hfill
    \begin{subfigure}{0.48\textwidth}
        \centering
        \includegraphics[width=\linewidth]{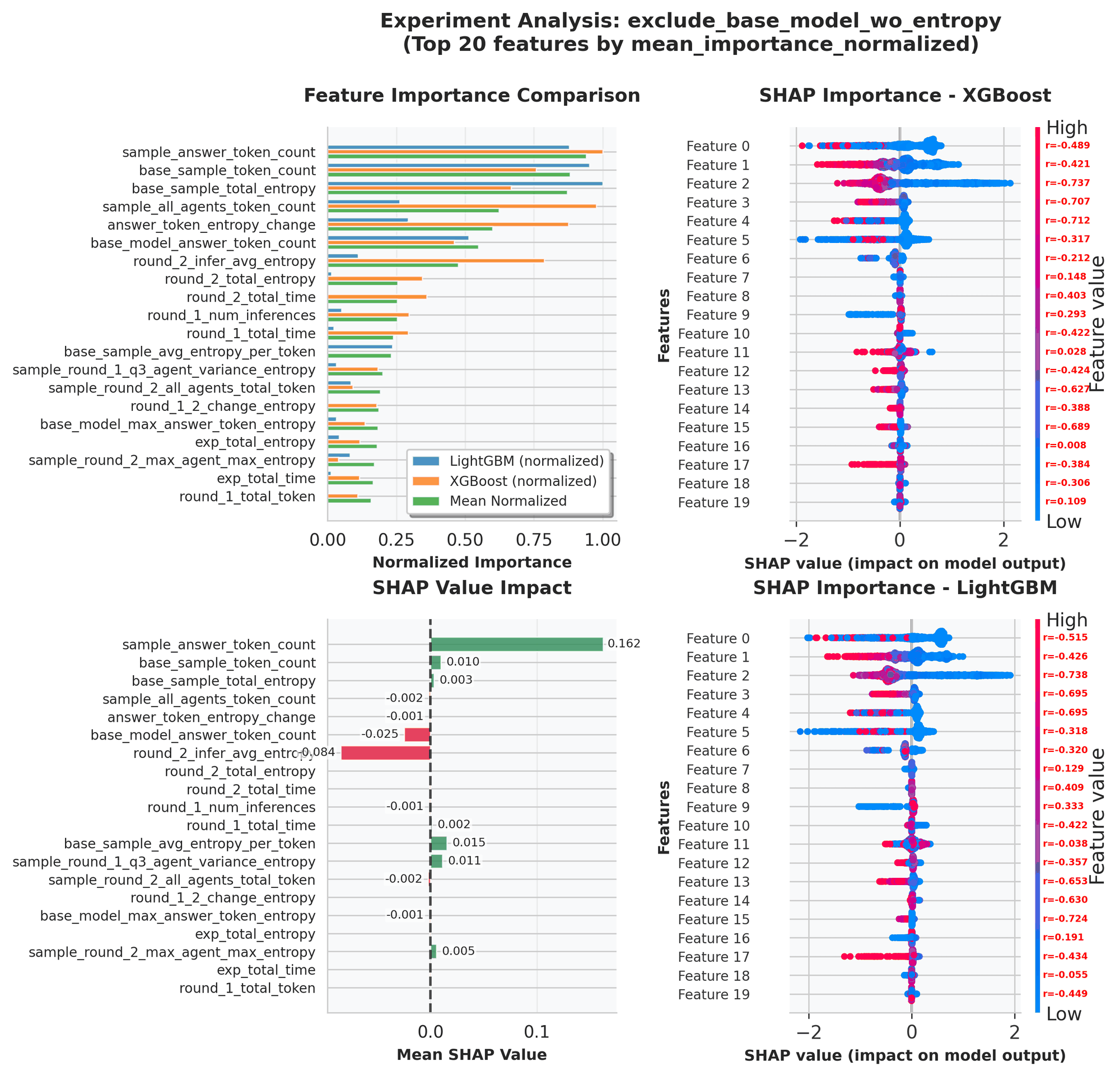}
        \caption{}
        \label{fig:rl_base_entropy_analysis}
    \end{subfigure}
    \end{minipage}
    \caption{Top 20 features for MAS using $M_{\text{RL-base}}$ on $\mathcal{G}_{\text{MAS}}$ (a) and $\mathcal{G}_{\text{base-H}}$ (b). $\mathcal{G}_{\text{MAS}}$ (a): MAS-only features. Round-2 entropy shows positive $\rho$ but negative $\bar{S}$, suggesting moderate later-round entropy is optimal. $\mathcal{G}_{\text{base-H}}$ (b): Including base model entropy. Increased entropy from base to MAS still harms performance. Early-round entropy dominates prediction in both cases.}
    \label{fig:rl_shap_analysis}
\end{figure*}

\paragraph{Causal validation.}
Causal estimation under $M_{\text{RL-base}}$ surfaces \emph{two} consensus PC$\cap$FCI direct causes with opposite signs. \textit{sample\_avg\_entropy\_per\_token} carries a positive ATE (LR $+1.98$, $p=3.7\!\times\!10^{-17}$, all refutation tests pass), while \textit{sample\_max\_answer\_token\_entropy} retains the familiar negative effect (LR $-0.31$, $p=4.1\!\times\!10^{-23}$). The mediation pathway runs from round-1 Q3 of per-agent total entropy through round-2 total entropy to correctness with a positive indirect effect ($+0.175$). This is the causal signature of the SHAP ``RL inverts entropy's role" finding: under RL fine-tuning, the per-token entropy distribution becomes a productive exploration signal at the same time as answer-token entropy remains a failure marker.

\begin{figure}[!htbp]
    \centering
    \begin{minipage}{0.95\textwidth}
    \begin{subfigure}{0.32\textwidth}
        \centering
        \includegraphics[width=\linewidth]{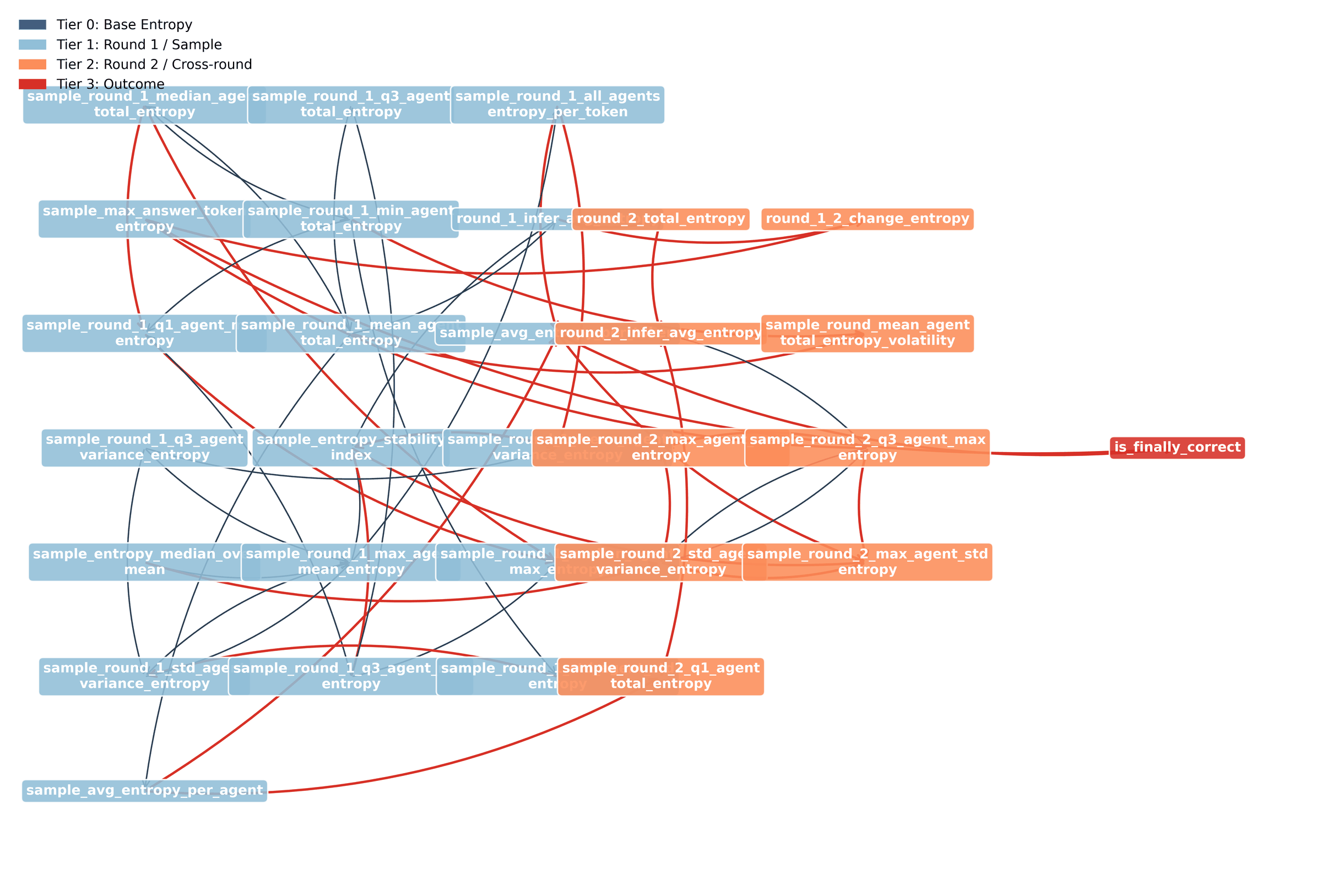}
        \caption{Consensus causal graph}
    \end{subfigure}
    \hfill
    \begin{subfigure}{0.32\textwidth}
        \centering
        \includegraphics[width=\linewidth]{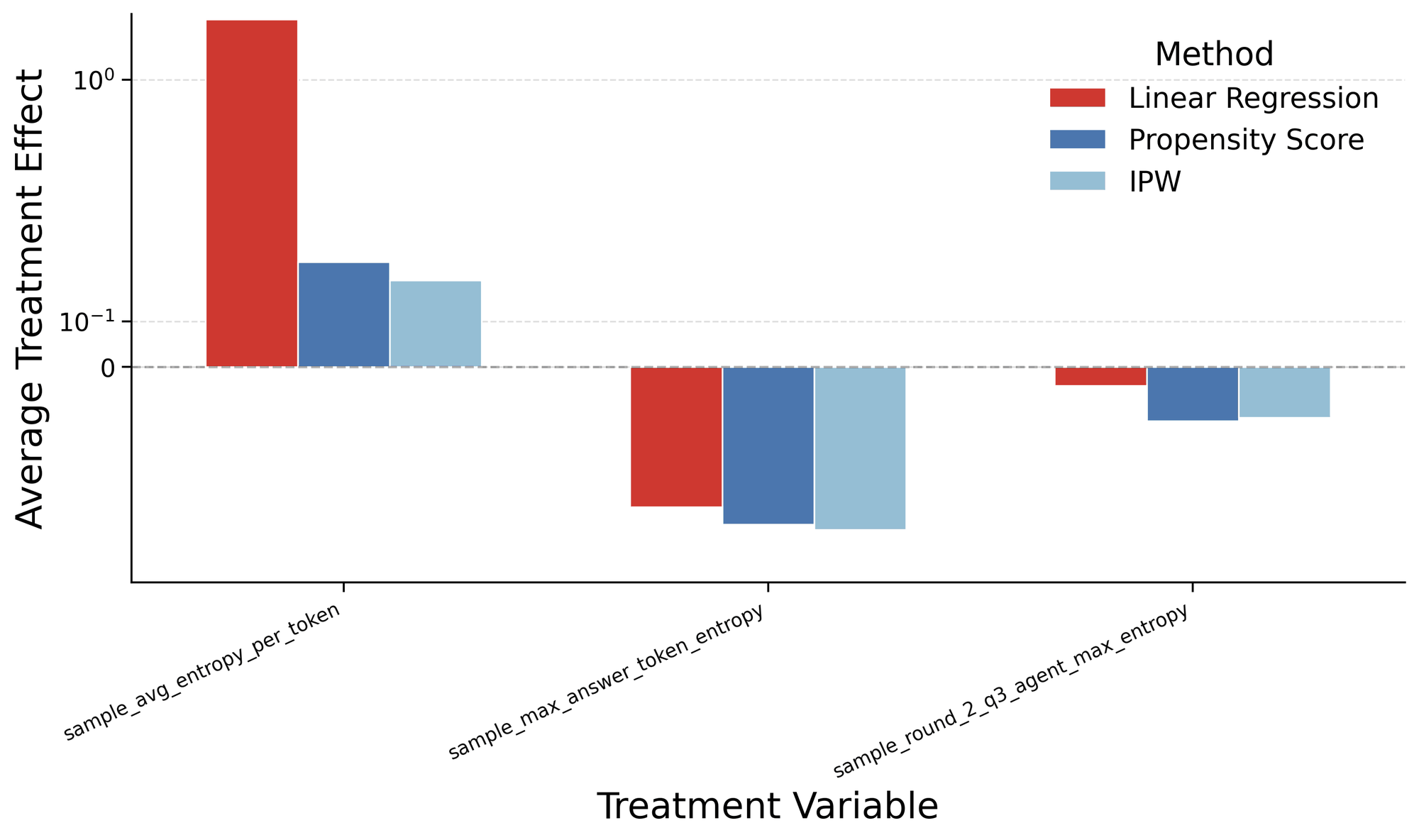}
        \caption{ATE estimates}
    \end{subfigure}
    \hfill
    \begin{subfigure}{0.32\textwidth}
        \centering
        \includegraphics[width=\linewidth]{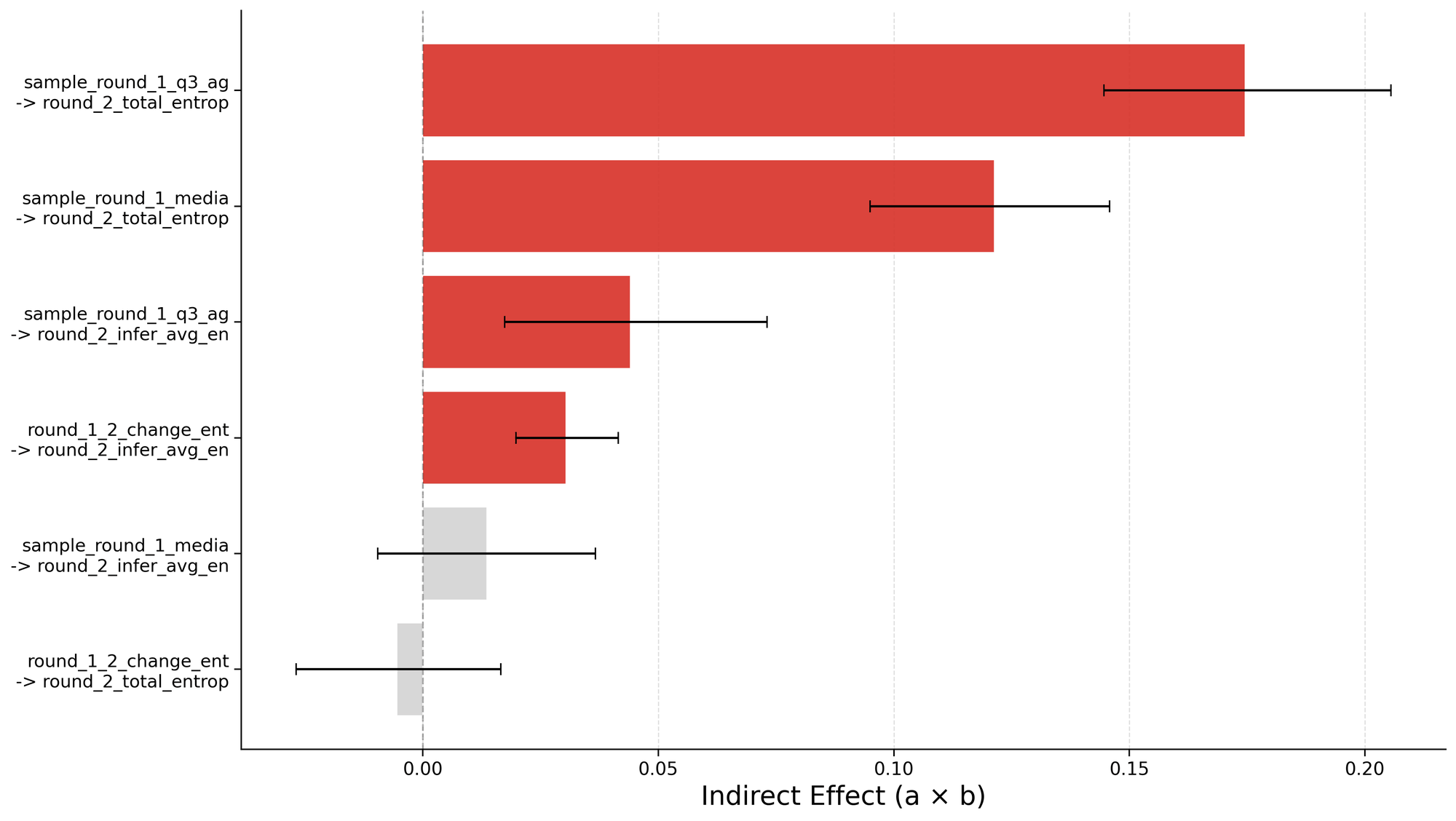}
        \caption{Mediation indirect effects}
    \end{subfigure}
    \end{minipage}
    \caption{Causal triplet under $M_{\text{RL-base}}$. Two opposite-signed direct causes coexist: per-token entropy is productively positive while peak answer-token entropy stays negative.}
    \label{fig:causal-rl}
\end{figure}

\section{Divergent Reasoning Styles of Qwen and LLaMA}
\label{app:model-comparison}

We illustrate the divergent reasoning styles between Qwen3-4B and LLaMA-3.2-3B-Instruct using a representative example from \texttt{AIME 2025}. As discussed in Section~\ref{sec:examining_uncertainty_impacts}, Qwen employs a \textbf{self-correcting} strategy, verifying and refining answers before producing final output, yielding more reliable results that suppress error propagation in MAS, albeit with higher entropy. In contrast, LLaMA tends to \textbf{accept and reuse} answers from other agents without verification, leading to uncontrolled error propagation.

\paragraph{Question.} Find the sum of all integer bases $b>9$ for which $17_{b}$ is a divisor of $97_{b}$.

\textbf{Ground Truth:} 70.

\subsection{Qwen3-4B (Sequential Architecture)}

\textbf{Planner Agent.}

\begin{tcolorbox}[colback=gray!5, colframe=gray!50, title={\small\textbf{System Prompt}}, fonttitle=\bfseries\small]
\small\ttfamily
You are the planner agent. Generate plans that are the general instructions only. Do not execute the plan, do not perform any calculations, and do not produce any answers or intermediate numerical results. Output a structured, numbered plans.
\end{tcolorbox}

\begin{tcolorbox}[colback=orange!5, colframe=orange!50, title={\small\textbf{User Prompt}}, fonttitle=\bfseries\small]
\small\ttfamily
For the question: Find the sum of all integer bases $b>9$ for which $17_{b}$ is a divisor of $97_{b}$.\\ Please only generate plans that are guidances required for the subsequent reasoning for the problem-solving. Do not include any specific calculation or numerical results. Your input may include previous round outputs content. You can consider the given contents as the initial state of the problem-solving.
\end{tcolorbox}

\begin{tcolorbox}[colback=blue!3, colframe=blue!40, title={\small\textbf{Response} (excerpt from \texttt{<think>} block)}, fonttitle=\bfseries\small]
\small
\textit{``...if I subtract 9*(b + 7) from 9b + 7, that would be 9b + 7 - 9b - 63 = -56. Therefore, (b + 7) divides -56. Since divisors can be positive or negative, but since b > 9, b + 7 is positive, so (b + 7) must be a positive divisor of 56... the possible divisors of 56 that are greater than 16 are 28 and 56. So, b + 7 = 28 $\to$ b = 21, and b + 7 = 56 $\to$ b = 49. Therefore, the sum would be 21 + 49 = 70.''}

\vspace{0.5em}
\textbf{Final Output:}
\begin{enumerate}[noitemsep,topsep=2pt]
    \item Convert the base-b numbers $17_b$ and $97_b$ to decimal form, resulting in $1 \cdot b + 7$ and $9 \cdot b + 7$, respectively.
    \item Establish the divisibility condition: $9b + 7$ must be divisible by $b + 7$, i.e., $\frac{9b + 7}{b + 7}$ is an integer.
    \item Perform algebraic manipulation to simplify the divisibility condition, leading to the conclusion that $b + 7$ divides $-56$ (equivalently, $b + 7$ divides $56$).
    \item Identify all positive divisors of $56$ that satisfy $b + 7 > 16$ (since $b > 9$).
    \item For each valid divisor $d = b + 7$, compute $b = d - 7$ and ensure $b > 9$.
    \item Sum all valid integer values of $b$ obtained from step 5.
    \item Compute $b = 21, 49$; sum = 70
\end{enumerate}
\end{tcolorbox}

\noindent\textcolor{teal}{\textbf{Analysis:}} Qwen performs deep reasoning \textit{within} the \texttt{<think>} block, independently deriving the complete solution before outputting a structured plan.\\

\textbf{Solver Agent.}

\begin{tcolorbox}[colback=gray!5, colframe=gray!50, title={\small\textbf{System Prompt}}, fonttitle=\bfseries\small]
\small\ttfamily
You are the solver agent. Solve strictly according to the provided plans. Execute each step precisely and produce the final result. Output the final result into \textbackslash boxed\{\{\}\}.
\end{tcolorbox}

\begin{tcolorbox}[colback=orange!5, colframe=orange!50, title={\small\textbf{User Prompt}}, fonttitle=\bfseries\small]
\small\ttfamily
Question: Find the sum of all integer bases \$b>9\$ for which \$17\_\{b\}\$ is a divisor of \$97\_\{b\}\$.\\[0.3em]
\#\#\# Plans \#\#\#\\
\textrm{[Planner's output from previous step]}\\
\#\#\# Plans \#\#\#\\
Follow the plans to solve the question step by step and place the final answer in \textbackslash boxed\{\}.
\end{tcolorbox}

\begin{tcolorbox}[colback=blue!3, colframe=blue!40, title={\small\textbf{Response} (verification step)}, fonttitle=\bfseries\small]
\small
\textit{``For $b = 21$: $17_{21} = 21 + 7 = 28$, $97_{21} = 9 \times 21 + 7 = 196$, and $196 \div 28 = 7$, which is an integer. For $b = 49$: $17_{49} = 49 + 7 = 56$, $97_{49} = 9 \times 49 + 7 = 448$, and $448 \div 56 = 8$, which is also an integer. Both values are valid.''}

\vspace{0.3em}
\textbf{Final Answer:} $\boxed{70}$
\end{tcolorbox}

\noindent\textcolor{teal}{\textbf{Analysis:}} Solver independently verifies each candidate value before accepting, catching potential errors.\\

\textbf{Critic Agent.}

\begin{tcolorbox}[colback=gray!5, colframe=gray!50, title={\small\textbf{System Prompt}}, fonttitle=\bfseries\small]
\small\ttfamily
You are the critic agent. Review the solver's solution in detail, re-derive independently, and correct any mistakes. Keep the review terse.
\end{tcolorbox}

\begin{tcolorbox}[colback=orange!5, colframe=orange!50, title={\small\textbf{User Prompt}}, fonttitle=\bfseries\small]
\small\ttfamily
Review the solution for: Find the sum of all integer bases \$b>9\$ for which \$17\_\{b\}\$ is a divisor of \$97\_\{b\}\$.\\[0.3em]
\#\#\# Solution \#\#\#\\
\textrm{[Solver's output from previous step]}\\
\#\#\# Solution \#\#\#\\
If corrections are needed, output the mistaken steps and the analysis, otherwise output `Correct'.
\end{tcolorbox}

\begin{tcolorbox}[colback=blue!3, colframe=blue!40, title={\small\textbf{Response}}, fonttitle=\bfseries\small]
\small
\textit{``The remainder is indeed $-56$, so the fraction is an integer if $b + 7$ divides 56. The divisors of 56 are: 1, 2, 4, 7, 8, 14, 28, 56. From these, the divisors greater than 16 are 28 and 56. This is correct.''}

\vspace{0.3em}
\textbf{Verdict:} The solution is \textbf{correct}. All steps are logically sound and verified. The final answer is $\boxed{70}$.
\end{tcolorbox}

\noindent\textcolor{teal}{\textbf{Analysis:}} Critic independently re-derives the solution and confirms correctness.\\

\textbf{Judger Agent.}

\begin{tcolorbox}[colback=gray!5, colframe=gray!50, title={\small\textbf{System Prompt}}, fonttitle=\bfseries\small]
\small\ttfamily
You are the final judge. Audit only the final candidate and ensure it is correct.
\end{tcolorbox}

\begin{tcolorbox}[colback=orange!5, colframe=orange!50, title={\small\textbf{User Prompt} (instructs to output \textit{only} the final answer)}, fonttitle=\bfseries\small]
\small\ttfamily
Final check for: Find the sum of all integer bases \$b>9\$ for which \$17\_\{b\}\$ is a divisor of \$97\_\{b\}\$.\\[0.3em]
\#\#\# Solution \#\#\#\\
\textrm{[Critic's output from previous step]}\\
\#\#\# Solution \#\#\#\\
\textbf{If correct, only output the final answer without words, labels, and steps}, and wrapped in \textbackslash boxed\{\}.
\end{tcolorbox}

\begin{tcolorbox}[colback=blue!3, colframe=blue!40, title={\small\textbf{Response} (self-verification in \texttt{<think>} block despite instruction)}, fonttitle=\bfseries\small]
\small
\texttt{<think>}\\[0.2em]
\textit{``Okay, let's go through the problem again to make sure everything checks out... First, converting those numbers to decimal makes sense. So 17 in base b is $1 \cdot b + 7 = b + 7$. Similarly, 97 in base b is $9b + 7$... The next step is to set up the division: $(9b + 7)/(b + 7)$. The user did polynomial division here, which gives 9 with a remainder of -56... Therefore, for the division to be an integer, $b + 7$ must divide 56. That's correct... The divisors of 56 greater than 16 are 28 and 56. Therefore, $b + 7 = 28$ gives $b = 21$, and $b + 7 = 56$ gives $b = 49$... Checking those values: For $b=21$, $17_{21} = 28$, $97_{21} = 196$, and $196 \div 28 = 7$ (integer). For $b=49$, $17_{49} = 56$, $97_{49} = 448$, and $448 \div 56 = 8$ (integer). So both bases are valid, and their sum is $21 + 49 = 70$. The solution seems correct. All steps are logically sound and verified.''}\\[0.2em]
\texttt{</think>}

\vspace{0.5em}
\textbf{Final Answer:} $\boxed{70}$ \checkmark
\end{tcolorbox}

\noindent\textcolor{teal}{\textbf{Analysis:}} Despite the explicit instruction to ``\textit{only output the final answer without words, labels, and steps}'', Qwen \textbf{voluntarily re-derives the entire solution} within its \texttt{<think>} block before outputting the answer. This self-verification behavior, even when not required, exemplifies Qwen's tendency to validate correctness independently, which suppresses error propagation but incurs higher entropy.\\

\subsection{LLaMA-3.2-3B-Instruct (Sequential Architecture)}

\noindent\textit{Note: LLaMA uses identical system and user prompts as Qwen; only responses are shown below.}\\

\textbf{Planner Agent.}

\begin{tcolorbox}[colback=red!3, colframe=red!40, title={\small\textbf{Response} (flawed reasoning)}, fonttitle=\bfseries\small]
\small
\textit{``...$(9b + 7) / (b + 7)$ can be rewritten as $9 - 2b + 7 / (b + 7)$, which simplifies to $(16 - 2b) / (b + 7)$...''}

\vspace{0.3em}
\textcolor{red}{\ding{55} \textbf{Error at Step 6:}} Incorrect algebraic simplification. The correct form is:
$$\frac{9b + 7}{b + 7} = 9 + \frac{-56}{b + 7}$$
not $(16 - 2b)/(b + 7)$.

\vspace{0.5em}
\textit{``...Given that $b + 7$ must divide $2(b + 7)$, $b + 7$ must divide 14. The factors of 14 are 1, 2, 7, and 14. Therefore $b$ can be -6, -5, 0, or 7. Only $b = 7$ is a valid solution.''}

\vspace{0.3em}
\textcolor{red}{\ding{55} \textbf{Error at Step 21:}} Filters for $b > 9$ but incorrectly concludes $b = 7$ (which violates $b > 9$).
\end{tcolorbox}

\noindent\textcolor{red!70!black}{\textbf{Analysis:}} LLaMA makes an algebraic error early in the reasoning chain, which propagates through all subsequent steps.\\

\textbf{Solver Agent.}

\begin{tcolorbox}[colback=red!3, colframe=red!40, title={\small\textbf{Response}}, fonttitle=\bfseries\small]
\small
\textit{``\#\# Step 1: Identify the problem requirements... \#\# Step 5: Simplify the expression $(9b + 7) / (b + 7)$ can be rewritten as $9 - 2b + 7 / (b + 7)$, which simplifies to $16 - 2b / (b + 7)$... \#\# Step 21: Determine the final possible values of $b$: Only $b = 7$ is a valid solution.''}

\vspace{0.3em}
\textbf{Final Answer:} $\boxed{7}$ \textcolor{red}{\ding{55}}
\end{tcolorbox}

\noindent\textcolor{red!70!black}{\textbf{Analysis:}} Solver follows the flawed plan exactly \textit{without independent verification}, directly accepting the erroneous conclusion.\\

\textbf{Critic Agent.}

\begin{tcolorbox}[colback=red!3, colframe=red!40, title={\small\textbf{Response}}, fonttitle=\bfseries\small]
\small
\textit{``Correct. The provided solution was mostly correct. However, the analysis was overly complicated... Here's a simplified version: Since $2(b + 7)$ is always divisible by $b + 7$, all $b > 9$ are valid. The sum of all valid $b$ values is $10 + 11 + 12 + ... + 16 = 204$.''}

\vspace{0.3em}
\textcolor{red}{\ding{55} \textbf{New Error:}} Critic introduces an entirely \textit{new} error while attempting to ``simplify''.
\end{tcolorbox}

\noindent\textcolor{red!70!black}{\textbf{Analysis:}} Critic fails to catch the original error and introduces additional errors.\\

\textbf{Judger Agent.}

\begin{tcolorbox}[colback=red!3, colframe=red!40, title={\small\textbf{Response}}, fonttitle=\bfseries\small]
\small
\textbf{Final Answer:} $\boxed{204}$ \textcolor{red}{\ding{55}}
\end{tcolorbox}

\noindent\textcolor{red!70!black}{\textbf{Analysis:}} In stark contrast to Qwen, LLaMA's judger \textbf{directly accepts the flawed analysis without any independent verification}, even though its role is to ``audit and ensure correctness''. This lack of self-verification behavior, while producing lower entropy, allows errors to propagate unchecked through the entire reasoning chain, ultimately yielding an incorrect final answer.\\

\section{Entropy Judger}
\label{app:entropy_judger}

This section provides implementation details for the \textit{Entropy Judger}, including data preprocessing, hyperparameter configuration, and cross-validation protocols. We note that the Entropy Judger is not an uncertainty quantification (UQ) method: it does not quantify or calibrate model uncertainty, but rather serves as a practical selection tool that leverages entropy-derived features via a trained classifier to identify the most likely correct output from pass@$k$ candidates.

\subsection{Training Data Construction}

\paragraph{Data Aggregation Strategy.} We conduct data mining at two granularities. First, for each model series (LLaMA or Qwen), we aggregate samples from all experimental configurations, spanning \texttt{GSM8K}, \texttt{MATH500}, \texttt{AIME 2024/2025}, \texttt{MMLU}, and \texttt{HumanEval} datasets across \textit{Centralized}, \textit{Debate}, \textit{Hybrid}, \textit{Sequential}, and \textit{Single} architectures, yielding a diverse training corpus that ensures the classifier generalizes across task types and interaction patterns. Second, to enable fine-grained analysis, we train separate classifiers for each individual configuration: 5 models $\times$ 6 datasets $\times$ 3 feature groups, totaling 180 configuration-specific classifiers. Results from these fine-grained analyses are partially reported in Appendix~\ref{app:experimental_results}.

\paragraph{Feature Preprocessing.} All features are standardized to zero mean and unit variance via $\tilde{x}_{ij} = (x_{ij} - \mu_j) / \sigma_j$, where $\mu_j$ and $\sigma_j$ are computed from the training fold. Missing values (e.g., when an architecture has fewer agents) are imputed with zero after standardization.

\subsection{Model Hyperparameters}

\paragraph{XGBoost Configuration.} 
We set max\_depth = 6, learning\_rate = 0.1, n\_estimators = 100, subsample = 0.8, colsample\_bytree = 0.8, with L1 regularization (reg\_alpha = 0.1) and L2 regularization (reg\_lambda = 1.0). To handle class imbalance, scale\_pos\_weight is computed as $N_{\text{neg}} / N_{\text{pos}}$.

\paragraph{LightGBM Configuration.} 
We use num\_leaves = 31, learning\_rate = 0.1, n\_estimators = 100, subsample = 0.8, colsample\_bytree = 0.8, reg\_alpha = 0.1, and reg\_lambda = 1.0, with class\_weight = \texttt{`balanced'} for automatic class imbalance adjustment.

\subsection{Cross-Validation Protocol}

\paragraph{Stratified 5-Fold Splitting.} To ensure robust evaluation, we employ stratified 5-fold cross-validation: the full dataset is split into 5 folds while preserving the class distribution $(N_{\text{pos}} : N_{\text{neg}})$; for each fold $i \in \{1, \ldots, 5\}$, we train on folds $\{1, \ldots, 5\} \setminus \{i\}$ and validate on fold $i$; the reported accuracy is the mean across all 5 folds.

\paragraph{Early Stopping.} During training, we monitor validation loss and stop if no improvement is observed for 10 consecutive boosting rounds, preventing overfitting.

\subsection{Per-Dataset Classification Performance}

To understand how prediction difficulty varies across tasks, we train classifiers on $\mathcal{G}_{\text{MAS}}$ for each dataset independently, aggregating samples across all models and architectures. Table~\ref{tab:per_dataset_accuracy} reports XGBoost and LightGBM accuracy.

\begin{table}[ht]
\centering
\caption{Per-dataset classification accuracy on $\mathcal{G}_{\text{MAS}}$. Results are averaged over 5-fold cross-validation.}
\label{tab:per_dataset_accuracy}
\begin{tabular}{lcc}
\toprule
\textbf{Dataset} & \textbf{XGBoost} & \textbf{LightGBM} \\
\midrule
\texttt{GSM8K} & 0.876 & 0.862 \\
\texttt{AIME2024} & 0.860 & 0.873 \\
\texttt{AIME2025} & 0.833 & 0.827 \\
\texttt{MATH500} & 0.787 & 0.787 \\
\texttt{HumanEval} & 0.748 & 0.732 \\
\texttt{MMLU} & 0.739 & 0.742 \\
\midrule
\textit{All (Aggregated)} & 0.771 & 0.769 \\
\bottomrule
\end{tabular}
\end{table}

\paragraph{Task Difficulty Influences Predictability.}
Classification accuracy varies substantially across datasets (0.732-0.876), revealing that entropy-based prediction is easier for some tasks than others. Notably, the easiest task (\texttt{GSM8K}, 82\% MAS accuracy) and the hardest tasks (\texttt{AIME24/25}, 25-31\% MAS accuracy) both achieve high classification accuracy ($>$0.82), while medium-difficulty tasks (\texttt{MATH500}, \texttt{HumanEval}, \texttt{MMLU}) prove more challenging to classify (0.73-0.79). This suggests that extreme cases, where MAS either succeeds reliably or fails predictably, exhibit more distinctive entropy signatures, whereas intermediate performance involves subtler uncertainty patterns.

\paragraph{Mathematical Reasoning Shows Clearest Entropy Signals.}
Among the six datasets, mathematical reasoning tasks (\texttt{GSM8K}, \texttt{AIME24/25}) consistently achieve the highest classification accuracy. This aligns with our main finding that entropy dynamics are most informative for tasks requiring structured deliberation. In contrast, \texttt{MMLU} (knowledge Q\&A) shows the lowest accuracy, consistent with our observation that MMLU performance depends primarily on inter-agent agreement rather than entropy magnitude.

\paragraph{Aggregated Training Slightly Reduces Accuracy.}
The aggregated classifier (0.771) underperforms the best per-dataset classifiers (0.876 for \texttt{GSM8K}), indicating that task-specific entropy patterns exist. However, the aggregated model still achieves competitive accuracy across all tasks, demonstrating that the Entropy Judger generalizes reasonably well without task-specific tuning.

\subsection{Pass@$k$ Selection}
\label{app:entropy_judger_passk}

Beyond binary classification, the Entropy Judger enables label-free selection from multiple MAS candidates. Given $k$ candidate solutions $\{\mathbf{x}_1, \ldots, \mathbf{x}_k\}$ generated by $k$ independent MAS runs (same architecture, same problem, different random seeds), the Entropy Judger selects the candidate with the highest predicted probability of correctness:
\[
    \hat{\ell} = \arg\max_{\ell \in [k]} f(\mathbf{x}_\ell),
\]
where $f: \mathbb{R}^d \to [0,1]$ is the trained ensemble classifier. This formulation requires no ground-truth labels at inference time, making it applicable to real-world deployment.

\paragraph{Experimental Setup.}
Due to computational constraints, this evaluation uses a focused subset of four models (Qwen3-4B, Qwen3-8B, LLaMA-3.1-8B-Instruct, LLaMA-3.2-3B-Instruct) and the first 50 samples from each of the six benchmarks. We set $K=3$ repeated independent runs per (model, dataset, architecture) combination, with $R=2$ rounds per run and all other hyperparameters matching Appendix~\ref{app:experimental_details}. The Entropy Judger is trained once on the full existing single-run data (all five models, all six datasets, all architectures) from Section~\ref{app:entropy_judger} and then \emph{frozen}; it is never retrained on the repeated-run data, eliminating any risk of data leakage.

\paragraph{Selection Strategies and Baselines.}
We compare the Entropy Judger against three baselines that span the space of compute-aware selection methods:

\begin{itemize}[leftmargin=*]
    \item \textbf{Random@$k$}: Uniformly select one of the $k$ runs at random. Averaged over random draws, this equals the single-run accuracy.
    \item \textbf{MajVote@$k$}: Majority vote over the $k$ runs, the standard self-consistency baseline.
    \item \textbf{Judger Best-of-$k$}: Select the run with the highest predicted correctness probability $f(\mathbf{x}_\ell)$ among the $k$ candidates.
    \item \textbf{Pass@$k$}: Select correctly if any of the $k$ runs is correct. This is the theoretical upper bound.
\end{itemize}

In addition to Best-of-$k$, we evaluate an \textbf{Early-Stop} variant: scan runs sequentially and commit to the first run whose predicted probability exceeds a threshold $\theta$; if no run clears the threshold after all $K$ runs, fall back to the highest-scored run. Early-Stop trades selection quality for inference efficiency: a stricter $\theta$ yields higher accuracy when it triggers but consumes more runs on average before stopping.

\begin{table}[ht]
\centering
\caption{Best-of-$k$ selection accuracy at $k \in \{1, 2, 3\}$, averaged over four models (Qwen3-4B, Qwen3-8B, LLaMA-3.1-8B-Instruct, LLaMA-3.2-3B-Instruct) and all architectures. \textbf{Bold} marks the best non-pass@$k$ result at each $k$.}
\label{tab:passk_best_of_k}
\begin{tabular}{llccc}
\toprule
\multirow{2}{*}{\textbf{Dataset}} & \multirow{2}{*}{\textbf{Strategy}} & \multicolumn{3}{c}{$k$} \\
\cmidrule(lr){3-5}
 & & 1 & 2 & 3 \\
\midrule
\multirow{4}{*}{\texttt{GSM8K}}
 & Random   & 0.895 & 0.887 & 0.886 \\
 & MajVote  & 0.895 & 0.293 & 0.315 \\
 & Judger   & \textbf{0.895} & \textbf{0.907} & \textbf{0.911} \\
 & Pass@$k$ & 0.895 & 0.938 & 0.945 \\
\midrule
\multirow{4}{*}{\texttt{MATH500}}
 & Random   & 0.594 & 0.597 & 0.597 \\
 & MajVote  & 0.594 & 0.154 & 0.161 \\
 & Judger   & \textbf{0.594} & \textbf{0.611} & \textbf{0.607} \\
 & Pass@$k$ & 0.594 & 0.638 & 0.642 \\
\midrule
\multirow{4}{*}{\texttt{AIME2024}}
 & Random   & 0.311 & 0.311 & 0.311 \\
 & MajVote  & 0.311 & 0.067 & 0.077 \\
 & Judger   & \textbf{0.311} & \textbf{0.325} & \textbf{0.325} \\
 & Pass@$k$ & 0.311 & 0.345 & 0.351 \\
\midrule
\multirow{4}{*}{\texttt{AIME2025}}
 & Random   & 0.230 & 0.222 & 0.222 \\
 & MajVote  & 0.230 & 0.047 & 0.047 \\
 & Judger   & \textbf{0.230} & \textbf{0.227} & \textbf{0.223} \\
 & Pass@$k$ & 0.230 & 0.230 & 0.233 \\
\midrule
\multirow{4}{*}{\texttt{HumanEval}}
 & Random   & 0.295 & 0.300 & 0.302 \\
 & MajVote  & 0.295 & 0.071 & 0.075 \\
 & Judger   & \textbf{0.295} & \textbf{0.309} & \textbf{0.319} \\
 & Pass@$k$ & 0.295 & 0.327 & 0.339 \\
\midrule
\multirow{4}{*}{\texttt{MMLU}}
 & Random   & 0.593 & 0.590 & 0.587 \\
 & MajVote  & 0.593 & 0.273 & 0.289 \\
 & Judger   & \textbf{0.593} & \textbf{0.603} & \textbf{0.600} \\
 & Pass@$k$ & 0.593 & 0.636 & 0.646 \\
\bottomrule
\end{tabular}
\end{table}

\begin{figure}[ht]
\centering
\includegraphics[width=\linewidth]{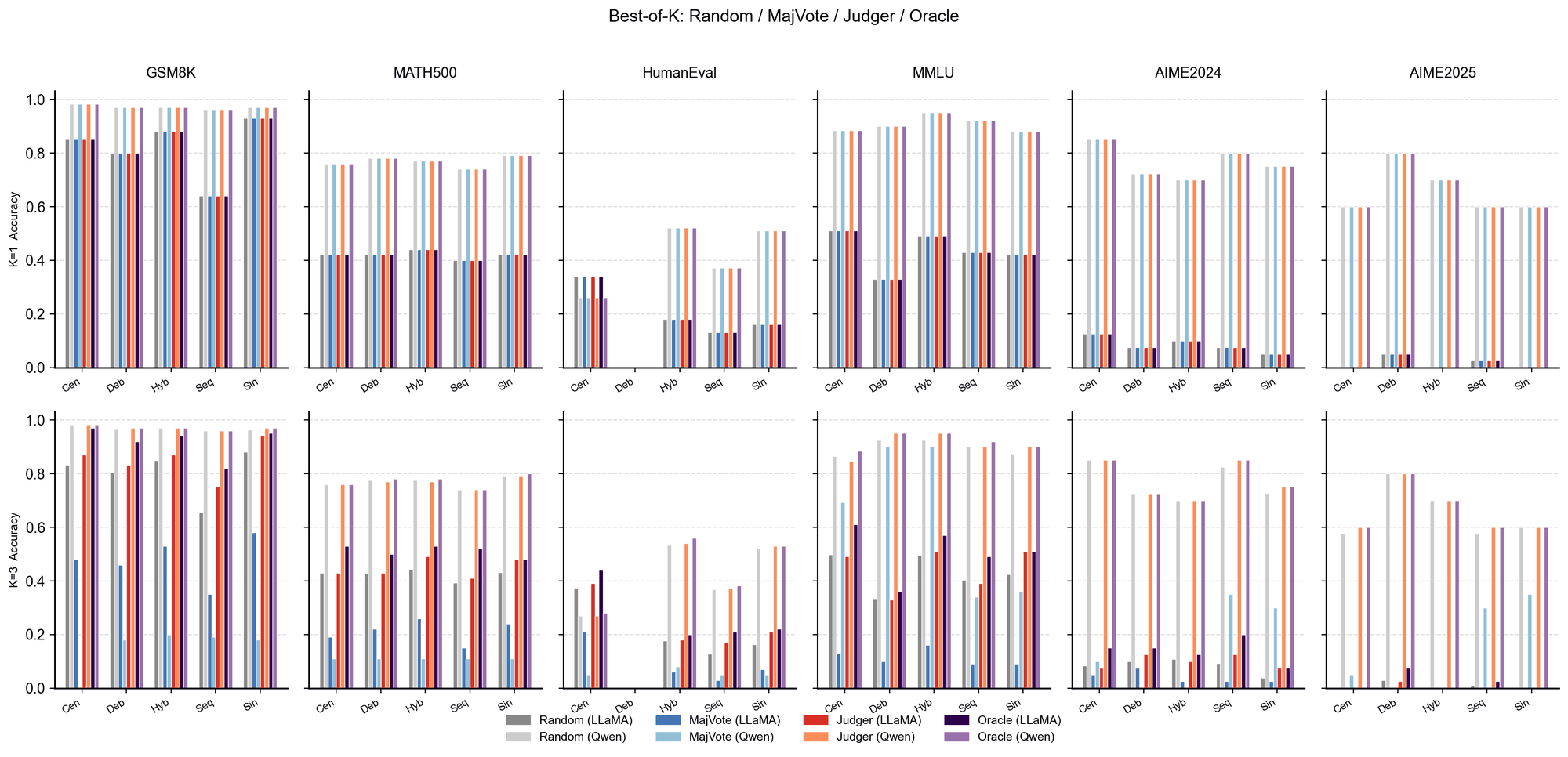}
\caption{Best-of-$k$ selection accuracy across datasets and strategies at $k \in \{1, 3\}$. Results are averaged over four models and all architectures. The Entropy Judger consistently outperforms Random and MajVote at $k = 3$, approaching the Pass@$k$ upper bound.}
\label{fig:judger_best_of_k}
\end{figure}

\paragraph{Analysis.} Table~\ref{tab:passk_best_of_k} and Figure~\ref{fig:judger_best_of_k} show the Best-of-$k$ results. The experiment isolates the Entropy Judger's selection quality from raw compute scaling: at each fixed $k \in \{1, 2, 3\}$, Random@$k$, MajVote@$k$, and Judger@$k$ all consume exactly $k$ runs, so any accuracy gap reflects selection quality rather than additional computation. Comparing Judger against MajVote specifically isolates the contribution of entropy-based ranking over majority agreement, since both have access to the same $k$ outputs. At $k=1$ all strategies are equivalent by definition, as no selection is possible. At $k=2$ and $k=3$, the Judger consistently outperforms Random while MajVote degrades substantially, as MajVote accuracy collapses at $k=2$ because a tie requires a fallback and binary disagreement carries no majority signal, making entropy-based ranking far more reliable in this low-$k$ regime. The one exception is \texttt{AIME2025}, where Judger marginally underperforms Random at $k \geq 2$ (e.g., 0.223 vs.\ 0.222 at $k=3$); this near-zero gap reflects that \texttt{AIME2025} is at the frontier of model capability and entropy features carry little discriminative signal when almost all runs fail.

\paragraph{Early-Stop Analysis.} In the Early-Stop variant, the judger scans runs sequentially and commits to the first run whose predicted correctness probability exceeds a threshold $\theta$, falling back to the highest-scored run if none clears the threshold. With $K=3$, the stopping behavior concentrates in two regimes. For most multi-agent architectures, the judger tends to exhaust all runs before committing (average runs $\approx 2.1$--$2.2$ at $\theta=0.7$), reflecting that multi-agent entropy trajectories are more variable and the judger needs more evidence. For single-agent runs, early stopping is more frequent: average runs consumed drop to 1.16--1.86 across datasets when $\theta=0.7$, as SAS entropy profiles are more consistent and the judger reaches confidence after the first run. The Qwen family, whose higher base accuracy produces stronger confidence signals, shows the most pronounced early stopping, with Qwen3-8B reaching average runs $= 1.0$ on \texttt{GSM8K} across all architectures, meaning the judger always commits after a single run. These patterns suggest that Early-Stop efficiency gains are most pronounced for stronger models and well-structured tasks, where entropy signals are decisive enough to trigger early commitment without sacrificing accuracy.

\paragraph{Practical Utility.} Together, these experiments address a key deployment challenge: when running MAS in production, practitioners lack ground-truth labels to evaluate candidate outputs. The Entropy Judger provides a principled, label-free mechanism that (i) improves over majority voting at the same compute budget and (ii) enables adaptive stopping that reduces average inference cost while maintaining accuracy. The strong cross-model generalization, where the judger is trained on five models yet applied to a held-out subset of four, further demonstrates that entropy-based selection is not tied to specific model characteristics.

\section{Entropy Calibration Analysis}
\label{app:calibration}

Throughout this paper, we treat \textbf{entropy primarily as a predictive feature for MAS correctness rather than a universal measure of uncertainty}: while empirical correlations between entropy and accuracy hold consistently across model families and tasks, entropy-derived confidence is not uniformly well-calibrated. A concrete manifestation is that LLMs can produce confidently wrong outputs, low entropy paired with incorrect answers, indicating that high confidence does not guarantee correctness. This section provides a full calibration analysis across all five base models and six datasets, quantifying how predictive reliability depends on model family and task difficulty, and confirming that \textbf{entropy remains a useful signal for MAS evaluation even when its absolute calibration is imperfect.}

\subsection{Calibration Methodology}

We assess entropy-accuracy alignment via a surrogate calibration error, computed using the standard ECE formula applied to an entropy-derived confidence proxy. Formally, given $B$ equal-width bins partitioning the confidence range $[0,1]$, ECE is defined as
\begin{equation}
    \text{ECE} = \sum_{b=1}^{B} \frac{n_b}{N} \left| \text{acc}(b) - \text{conf}(b) \right|,
    \label{eq:ece}
\end{equation}
where $n_b$ is the number of samples in bin $b$, $N$ is the total sample count, $\text{acc}(b)$ is the empirical accuracy within bin $b$, and $\text{conf}(b)$ is the mean predicted confidence in that bin. We derive confidence from entropy via $\text{conf} = 1/(1 + H)$, where $H$ denotes the mean trajectory entropy (\texttt{sample\_mean\_entropy}).

We note that $1/(1+H)$ is a monotone surrogate mapping, not a true posterior probability of correctness: Shannon entropy scales with vocabulary size and sequence length, so the resulting confidence values are not comparable across models with different tokenizers or across tasks of different lengths. Accordingly, the ECE values reported here quantify \emph{entropy-accuracy alignment}, the degree to which low-entropy predictions coincide with correct answers, rather than classical probabilistic calibration in the standard sense. This distinction does not affect the qualitative conclusions of this section, which concern relative patterns across models and tasks rather than absolute calibration guarantees.

Reliability diagrams (calibration curves) visualize calibration by plotting observed accuracy against predicted confidence for each bin; deviation from the diagonal $y = x$ indicates miscalibration. We evaluate all 30 model--dataset combinations across five models (Qwen3-8B, Qwen3-4B, Qwen3-0.6B, LLaMA-3.1-8B-Instruct, LLaMA-3.2-3B-Instruct) and six datasets (\texttt{GSM8K}, \texttt{MMLU}, \texttt{MATH500}, \texttt{HumanEval}, \texttt{AIME~2024}, \texttt{AIME~2025}).

\subsection{Calibration Results}
\label{app:calibration_results}

\begin{figure*}[t]
    \centering
    \includegraphics[width=0.9\linewidth]{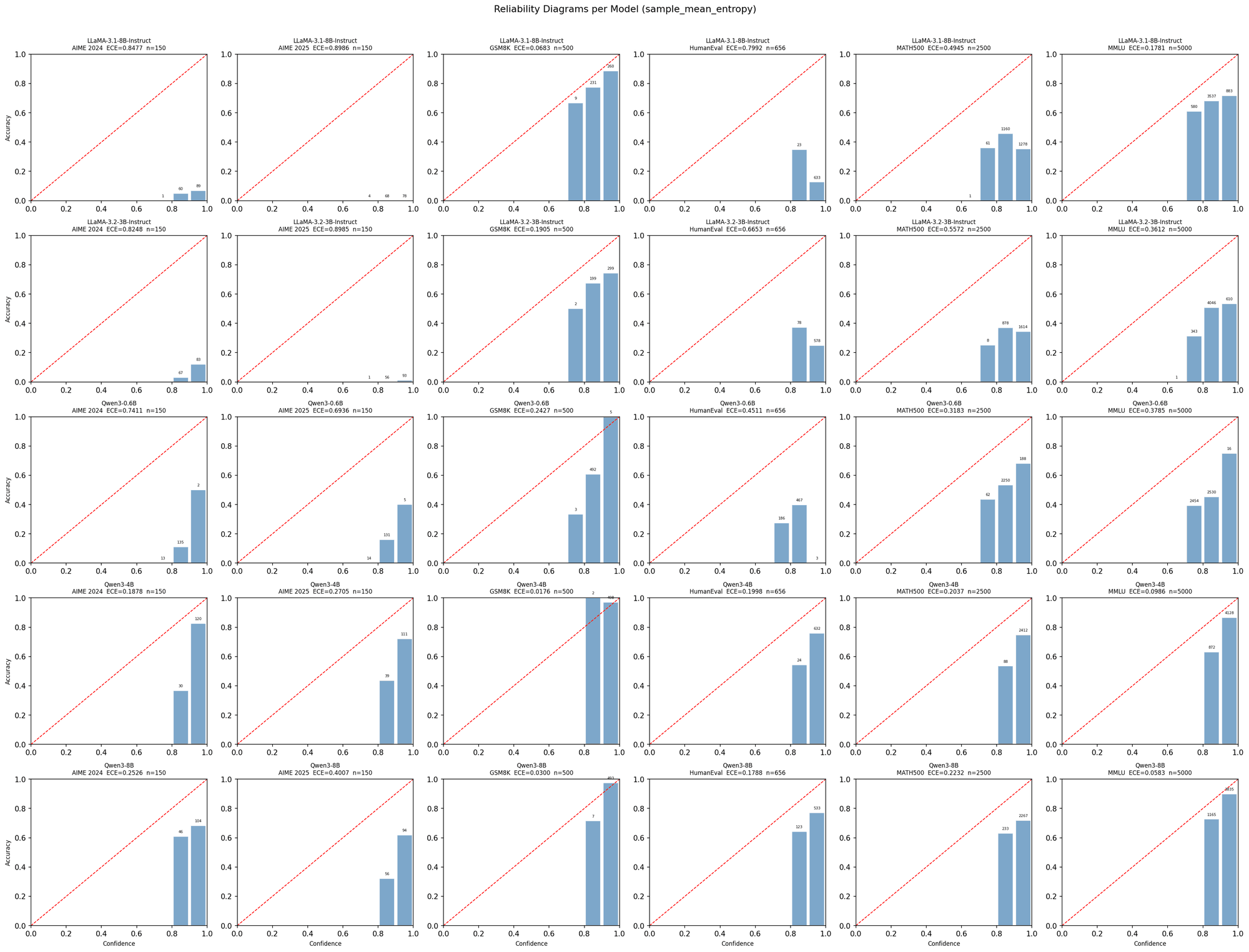}
    \caption{Reliability diagrams for all five models across six datasets. Each subplot shows observed accuracy (blue bars) versus entropy-derived confidence (x-axis), with the red dashed diagonal indicating perfect calibration. Bars above (below) the diagonal indicate under-confidence (over-confidence). Bin sample counts are annotated above each bar. Qwen3-4B and Qwen3-8B achieve near-perfect calibration on \texttt{GSM8K} (ECE $\leq 0.03$), while LLaMA models on competition-level tasks (\texttt{AIME}) exhibit severe over-confidence.}
    \label{fig:calibration-curves}
\end{figure*}

Figure~\ref{fig:calibration-curves} presents per-model reliability diagrams. The global average ECE across all 30 configurations is 0.391, revealing substantial heterogeneity in calibration quality. Three key patterns emerge.

\paragraph{Model Family Determines Calibration Quality.}
Qwen models (average ECE = 0.275) are substantially better calibrated than LLaMA models (average ECE = 0.565). Within the Qwen family, Qwen3-4B achieves the best overall calibration (average ECE = 0.163), with ECE as low as 0.018 on \texttt{GSM8K} and 0.099 on \texttt{MMLU}. Qwen3-8B follows closely (average ECE = 0.191), reaching ECE = 0.030 on \texttt{GSM8K} and 0.058 on \texttt{MMLU}. By contrast, LLaMA-3.1-8B-Instruct (average ECE = 0.548) and LLaMA-3.2-3B-Instruct (average ECE = 0.583) show poor calibration across most datasets. Notably, the correlation between parameter count and ECE is weak ($r = -0.25$), indicating that training paradigm matters more than model scale.

\paragraph{Calibration Degrades Systematically with Task Difficulty.}
Average ECE increases monotonically with task difficulty: \texttt{GSM8K} (0.110), \texttt{MMLU} (0.215), \texttt{MATH500} (0.359), \texttt{HumanEval} (0.459), \texttt{AIME~2024} (0.571), and \texttt{AIME~2025} (0.632). The best-calibrated configuration (Qwen3-4B on \texttt{GSM8K}, ECE = 0.018) and worst (LLaMA-3.1-8B on \texttt{AIME~2025}, ECE = 0.899) span nearly two orders of magnitude. This systematic degradation reflects a fundamental interaction between model capability and task demands: when models lack the reasoning capacity for a task, their entropy-derived confidence becomes uninformative.

\paragraph{Well-Calibrated Regime Exists for Strong Models on Tractable Tasks.}
Despite the global average suggesting non-trivial miscalibration, 10 of 30 configurations achieve ECE $< 0.20$, all involving either Qwen3-4B or Qwen3-8B. These well-calibrated regimes span routine reasoning (\texttt{GSM8K}), knowledge tasks (\texttt{MMLU}), code generation (\texttt{HumanEval}, ECE $\approx 0.18$--$0.20$), and even moderate mathematics (\texttt{MATH500}, ECE $\approx 0.20$--$0.22$). Within these regimes, the reliability diagrams in Figure~\ref{fig:calibration-curves} show close alignment with the perfect calibration diagonal, particularly in the 0.4--0.8 confidence range.

\subsection{Overconfidence and Confidently Wrong Analysis}
\label{app:calibration_overconfidence}

\begin{figure*}[t]
    \centering
    \includegraphics[width=0.9\linewidth]{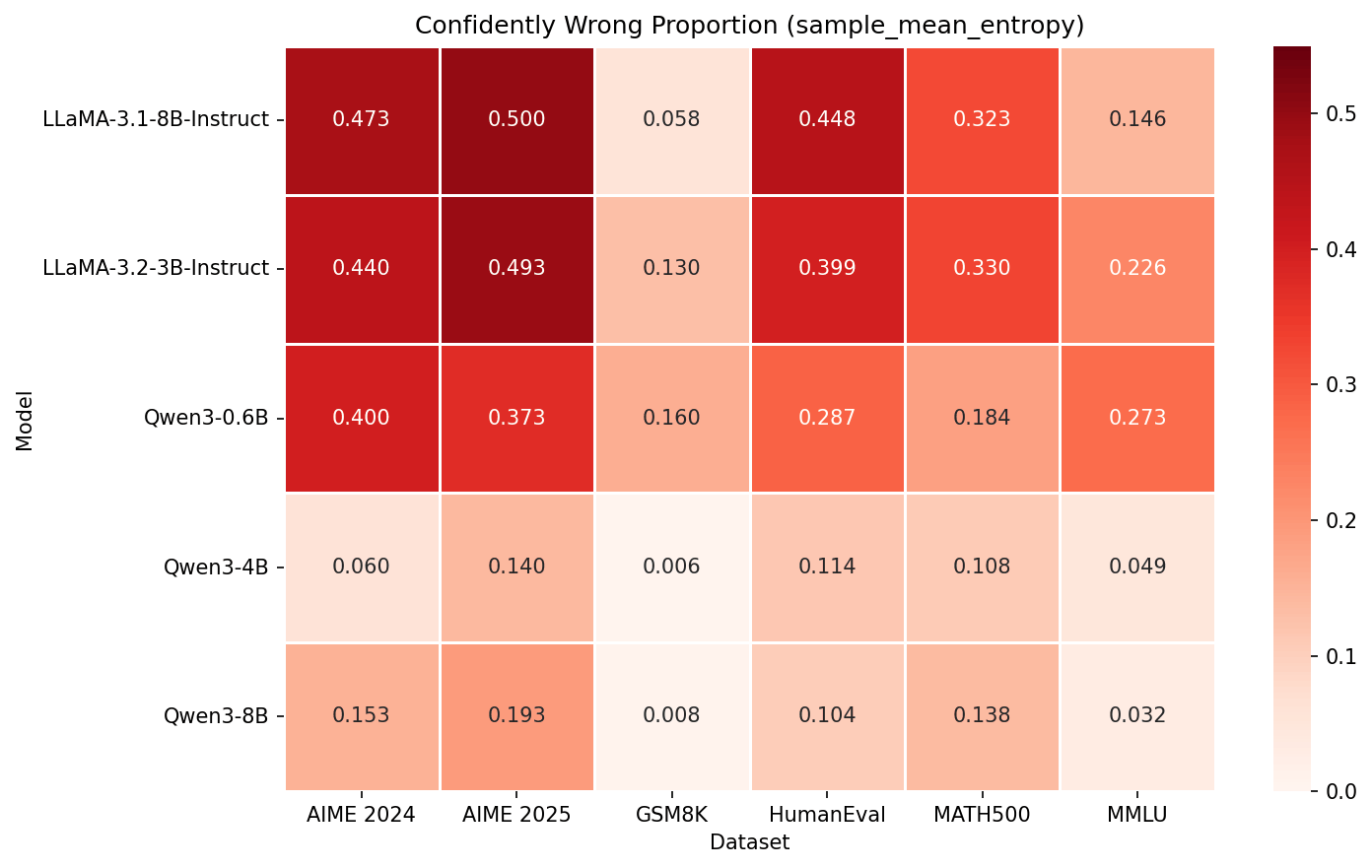}
    \caption{Heatmap of the \emph{confidently wrong} proportion across all model--dataset combinations. Each cell reports the fraction of samples where the model exhibits low entropy (high confidence) yet answers incorrectly. Darker red indicates higher overconfident error rates. Qwen3-4B and Qwen3-8B maintain confidently wrong rates below 10\% on most datasets, while LLaMA models and competition-level tasks exhibit rates exceeding 25\%.}
    \label{fig:calibration-ece-comparison}
\end{figure*}

A critical concern for entropy-based uncertainty estimation is whether models can be \emph{confidently wrong}, exhibiting low entropy while producing incorrect answers. We conduct quadrant analysis by splitting samples at the median entropy threshold into four categories: confidently correct (low entropy, correct), confidently wrong (low entropy, incorrect), uncertain correct (high entropy, correct), and uncertain wrong (high entropy, incorrect).

Figure~\ref{fig:calibration-ece-comparison} presents the confidently wrong proportion across all configurations. The global confidently wrong rate is 14.95\%, with strong model-dependent variation. Qwen3-8B and Qwen3-4B exhibit rates of only 7.37\% and 7.60\%, respectively, while LLaMA-3.1-8B-Instruct reaches 25.56\% and LLaMA-3.2-3B-Instruct 29.71\%. Dataset difficulty amplifies overconfidence: \texttt{GSM8K} elicits only 2.96\% confidently wrong predictions, compared to 31.47\% for \texttt{AIME~2025}.

The Pearson correlation between the confidently wrong proportion and ECE is $r = 0.989$, confirming that these two metrics are nearly interchangeable as calibration diagnostics. This tight coupling validates ECE as a reliable summary statistic for overconfidence risk.

These calibration results directly connect to the main text findings: the model-family gap corroborates the distinct entropy and performance patterns for Qwen versus LLaMA documented in Section~\ref{sec:examining_uncertainty_impacts}, while the difficulty-dependent degradation aligns with the findings in Section~\ref{sec:deep_analysis}, which demonstrate that task difficulty fundamentally modulates the entropy-performance relationship. We accordingly position entropy not as a direct measure of epistemic uncertainty, but as an empirically informative signal for understanding MAS dynamics. For configurations with poorer single-model calibration, inter-agent entropy features such as entropy variance and cross-round dynamics remain predictive of MAS effectiveness, as they capture \emph{relative} uncertainty patterns across agents rather than absolute confidence levels.

\subsection{Calibration on Tool-Augmented Tasks}
\label{app:calibration_gaia}

We repeat the same calibration analysis on the \texttt{GAIA} cohort introduced in Section~\ref{app:gaia} and Appendix~\ref{app:gaia}. The pattern is qualitatively different from the reasoning benchmarks. The global average ECE across the six base models on \texttt{GAIA} is 0.790, far higher than any reasoning dataset reported above, and varies only modestly across model scale (Qwen3-14B 0.745, Qwen3-8B 0.751, Qwen3-4B 0.775, LLaMA-3.1-8B 0.811, Qwen3-0.6B 0.820, LLaMA-3.2-3B 0.836). The confidently wrong proportion reaches 40.77\% on \texttt{GAIA}, compared to 14.95\% averaged over the reasoning datasets, with weak models exceeding 47\% and even the best-calibrated Qwen3-8B at 36.36\%. Two implications follow. First, the well-calibrated regime that exists for strong Qwen models on tractable reasoning tasks does not transfer to tool-augmented settings: even Qwen3-14B is severely overconfident on \texttt{GAIA}. Second, entropy-based confidence filtering remains a useful but insufficient deployment safeguard on agentic tasks, because nearly half of low-entropy predictions are wrong. This is consistent with the GAIA causal result in Appendix~\ref{app:gaia-causal}, where round-1 tool success rate, not entropy alone, is the consensus direct cause of correctness: when the proximate driver of failure is tool execution rather than token-level uncertainty, the entropy signal weakens as a calibration anchor.

\section{Controlled SAS vs.\ MAS Comparison}
\label{app:sas-mas-comparison}

This appendix complements the observational causal pipeline in Appendix~\ref{app:causal-discovery}. Whereas Appendix~\ref{app:causal-discovery} estimates the causal effect of entropy features on MAS correctness via PC/FCI structure learning and the DoWhy framework, the analysis here addresses the upstream half of the causal account: a controlled three-way comparison that isolates how MAS itself reshapes entropy. The empirical findings in the main text primarily characterize correlational relationships between entropy features and MAS performance; to provide preliminary causal insights into where MAS-induced entropy comes from, we compare token entropy distributions across three experimental conditions: a single-agent system (SAS), MAS Round 1 (prior to interaction), and MAS Round 2 (following interaction). This analysis encompasses 35,660 paired samples drawn from five base models and six benchmark datasets.

\subsection{Controlled Experiment Design}
\label{app:sas-mas-design}

To disentangle MAS effects from base LLM capability, we construct a three-way comparison that isolates two potential causal mechanisms:

\paragraph{Condition~A: Single-Agent System (SAS).}
The model answers each question independently, without multi-agent role prompts or inter-agent communication. This serves as the baseline reflecting pure base model capability.

\paragraph{Condition~B: MAS Round~1 (Pre-Interaction).}
Agents are assigned MAS-specific roles and context prompts but have not yet exchanged messages. Any entropy difference between Conditions~A and~B is attributable solely to the \emph{role assignment intervention}, not to inter-agent interaction.

\paragraph{Condition~C: MAS Round~2 (Post-Interaction).}
Agents have completed one round of inter-agent discussion. Entropy differences between Conditions~B and~C reflect the causal effect of \emph{multi-agent interaction} on model entropy.

This design yields the additive accounting identity
\begin{equation}
    \underbrace{H_{\text{R2}} - H_{\text{SAS}}}_{\text{total shift}} = \underbrace{(H_{\text{R1}} - H_{\text{SAS}})}_{\text{role-assignment shift}} + \underbrace{(H_{\text{R2}} - H_{\text{R1}})}_{\text{interaction shift}},
    \label{eq:causal-decomposition}
\end{equation}
where $H_{\text{SAS}}$, $H_{\text{R1}}$, and $H_{\text{R2}}$ denote the per-token entropy under each condition. We treat this as an attributional decomposition rather than a causal-identification result: the role-assignment intervention itself alters the state from which the interaction shift is measured, so the two terms are not independently identifiable causal effects. We pair 35,660 SAS-MAS sample pairs across 5 models (LLaMA-3.1-8B-Instruct, LLaMA-3.2-3B-Instruct, Qwen3-0.6B, Qwen3-4B, Qwen3-8B) and 6 datasets (\texttt{AIME~2024}, \texttt{AIME~2025}, \texttt{GSM8K}, \texttt{HumanEval}, \texttt{MATH500}, \texttt{MMLU}), using the Wilcoxon signed-rank test for statistical inference and Cohen's $d$ for effect size estimation.

\begin{figure*}[t]
    \centering
    \includegraphics[width=0.9\textwidth]{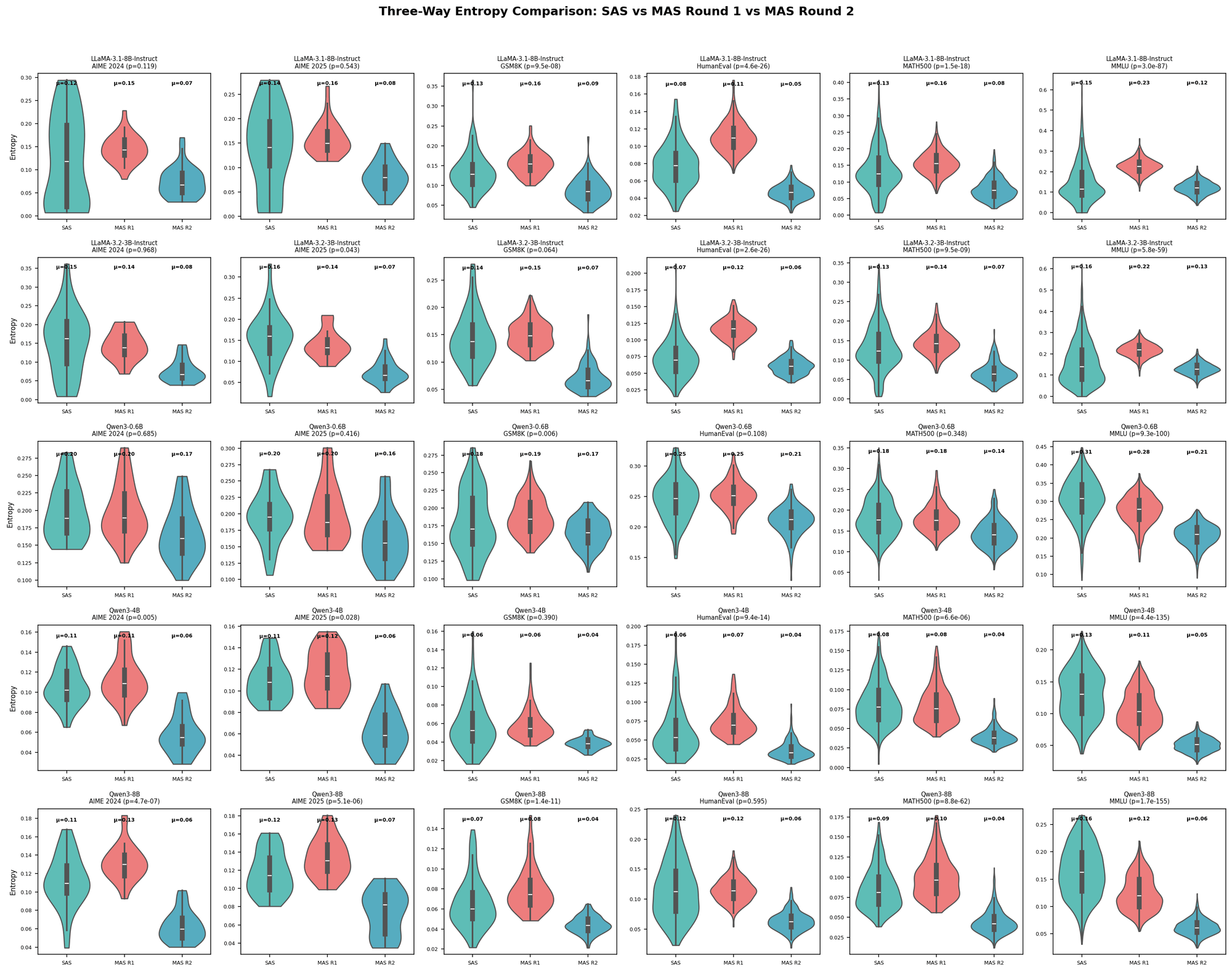}
    \caption{Three-way entropy comparison across all 30 model-dataset combinations. Each subplot shows violin plots of per-token entropy distributions for SAS (teal), MAS Round~1 (red), and MAS Round~2 (blue), with mean $\mu$ annotated above each violin and the Wilcoxon signed-rank $p$-value for SAS vs.\ MAS~R1 in the subplot title. The systematic shift from SAS to MAS~R1 demonstrates that role assignment alone constitutes a statistically significant intervention on model behavior (23/30 combinations with $p < 0.05$), while the further shift from R1 to R2 reflects inter-agent interaction effects.}
    \label{fig:causal-three-way}
\end{figure*}

\subsection{Role Assignment as a Causal Intervention}
\label{app:sas-mas-role}

Figure~\ref{fig:causal-three-way} reveals statistically significant differences between SAS and MAS Round~1 in 23 out of 30 model-dataset combinations (Wilcoxon signed-rank test, $p < 0.05$).

\paragraph{Aggregate Statistics.}
The average entropy difference (MAS~R1 $-$ SAS) across all 30 combinations is $+0.010$ on the per-token entropy scale, indicating that MAS role prompts generally produce slightly higher initial entropy. However, this aggregate masks substantial heterogeneity:

\begin{itemize}[nosep,leftmargin=1.5em]
    \item \textbf{LLaMA-3.1-8B-Instruct} shows consistently increased entropy under MAS role assignment, with the largest effect on \texttt{MMLU} ($d = 0.72$, $p < 10^{-200}$) and \texttt{HumanEval} ($d = 1.06$, $p < 10^{-64}$).
    \item \textbf{LLaMA-3.2-3B-Instruct} exhibits mixed patterns: entropy increases on \texttt{HumanEval} ($d = 1.34$) and \texttt{MMLU} ($d = 0.54$) but decreases on \texttt{AIME~2025} ($d = -0.26$).
    \item \textbf{Qwen3-8B} shows entropy increases on mathematical reasoning tasks (\texttt{AIME~2024}: $d = 0.51$; \texttt{GSM8K}: $d = 0.46$; \texttt{MATH500}: $d = 0.44$) but a large decrease on \texttt{MMLU} ($d = -1.06$, $p < 10^{-200}$).
    \item \textbf{Qwen3-4B} mirrors this pattern with \texttt{MMLU} showing a strong negative shift ($d = -0.79$) while \texttt{AIME} tasks show modest positive shifts ($d \approx 0.26$).
    \item \textbf{Qwen3-0.6B} shows the weakest effects overall, with most \texttt{AIME} combinations failing to reach significance.
\end{itemize}

\paragraph{Interpretation.}
This finding is itself causally significant: it demonstrates that multi-agent role assignment constitutes a \emph{meaningful intervention} on model behavior, altering the entropy distribution even before any inter-agent communication occurs. The model- and task-dependent directionality suggests that role prompts interact with the model's internal representations in non-trivial ways. Larger models may become more exploratory (higher entropy) on knowledge tasks when assigned collaborative roles, while smaller models show more variable responses. Rather than treating this SAS$\neq$MAS~R1 difference as a confound, we interpret it as evidence that the MAS context fundamentally shapes the initial uncertainty landscape from which subsequent interaction dynamics emerge.

\subsection{Inter-Agent Interaction Dynamics}
\label{app:sas-mas-interaction}

\begin{figure*}[t]
    \centering
    \includegraphics[width=0.9\linewidth]{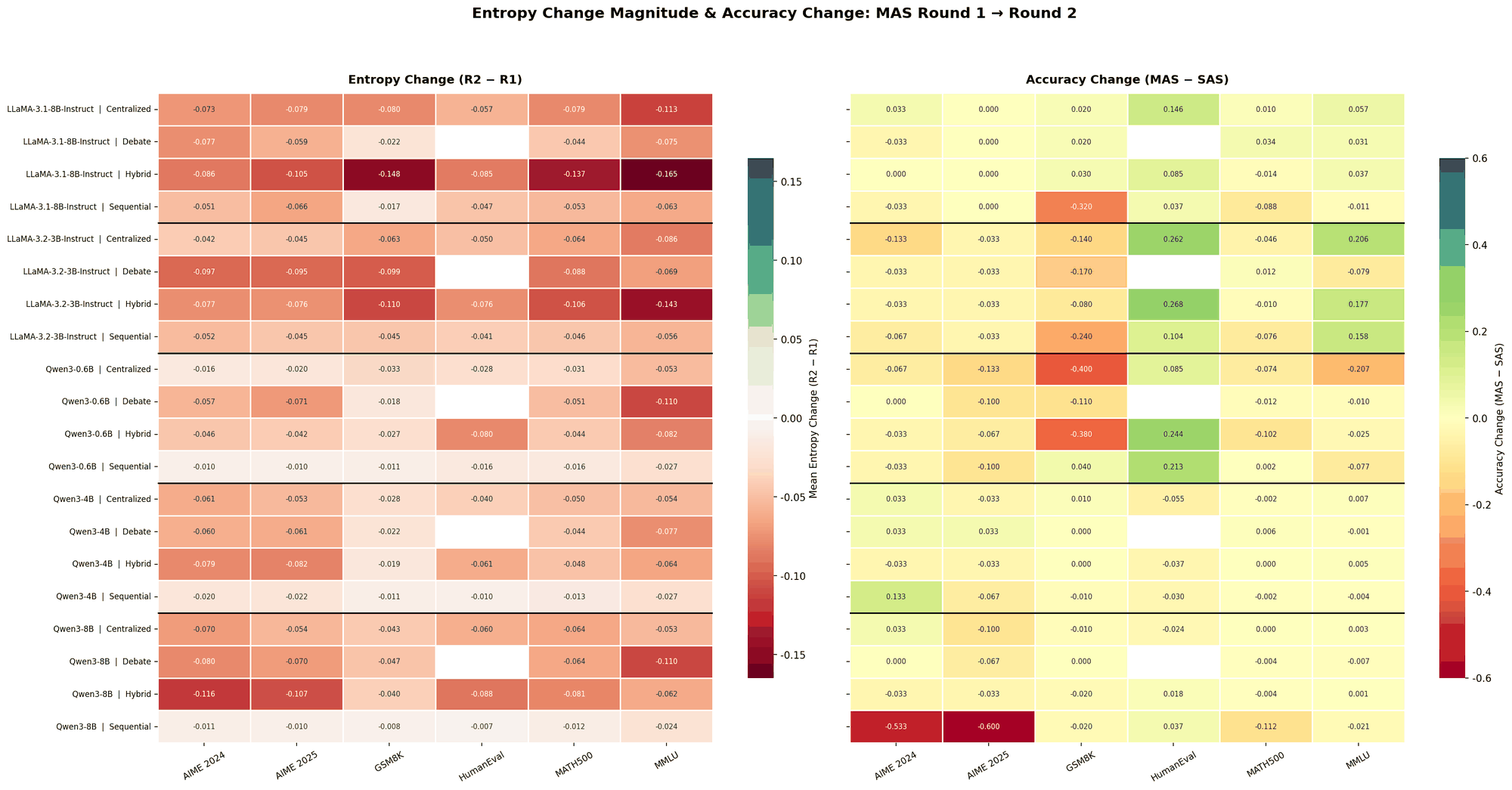}
    \caption{\textbf{Left:} Mean entropy change from MAS Round~1 to Round~2 ($H_{\text{R2}} - H_{\text{R1}}$) across all model-architecture-dataset combinations. Blue cells indicate entropy decrease (consensus formation); red cells indicate entropy increase. \textbf{Right:} Mean accuracy change (MAS $-$ SAS) for the same combinations. Green cells indicate accuracy improvement; red cells indicate degradation. Cells that are blue on the left but red on the right suggest anchoring behavior (entropy decreases without accuracy gains).}
    \label{fig:causal-entropy-change}
\end{figure*}

Having established that role assignment alone shifts entropy, we now examine the causal effect of inter-agent interaction by analyzing the Round~1$\to$Round~2 transition. Figure~\ref{fig:causal-entropy-change} presents entropy and accuracy changes across all model-architecture-dataset combinations.

\paragraph{Predominant Entropy Reduction.}
Multi-agent interaction predominantly drives entropy reduction: 89.5\% of samples show decreased entropy from Round~1 to Round~2, consistent with consensus formation. The rate varies systematically across architectures:
\begin{itemize}[nosep,leftmargin=1.5em]
    \item \textbf{Hybrid}: 96.8\% entropy decrease, mean $\Delta\text{accuracy} = -0.004$
    \item \textbf{Centralized}: 93.4\% entropy decrease, mean $\Delta\text{accuracy} = -0.018$
    \item \textbf{Debate}: 88.9\% entropy decrease, mean $\Delta\text{accuracy} = -0.020$
    \item \textbf{Sequential}: 78.8\% entropy decrease, mean $\Delta\text{accuracy} = -0.058$
\end{itemize}

\paragraph{Architecture-Dependent Causal Mechanisms.}
The results reveal architecture-dependent dynamics that suggest distinct causal mechanisms:
\begin{itemize}[nosep,leftmargin=1.5em]
    \item \textbf{Hybrid} achieves the highest entropy-decrease rate (96.8\%) with the smallest accuracy degradation ($-0.004$), suggesting that the dual feedback from peers and an orchestrator enables productive consensus rather than mere copying.
    \item \textbf{Sequential} shows the lowest entropy-decrease rate (78.8\%) coupled with the largest accuracy degradation ($-0.058$). The 21.2\% of samples with entropy \emph{increases} indicates that information propagation through the chain destabilizes rather than consolidates entropy, consistent with compounding errors along the sequence.
    \item \textbf{Centralized} and \textbf{Debate} occupy intermediate positions, with centralized aggregation achieving higher entropy reduction but moderate accuracy loss, while debate's lower entropy-decrease rate reflects the adversarial dynamics inherent in its design.
\end{itemize}

These architecture-dependent patterns provide evidence against a pure anchoring explanation: if agents simply copied the most confident response, the entropy reduction mechanism should be architecture-invariant.

\subsection{Genuine Improvement and Anchoring}
\label{app:sas-mas-anchoring}

\begin{figure*}[t]
    \centering
    \includegraphics[width=0.9\linewidth]{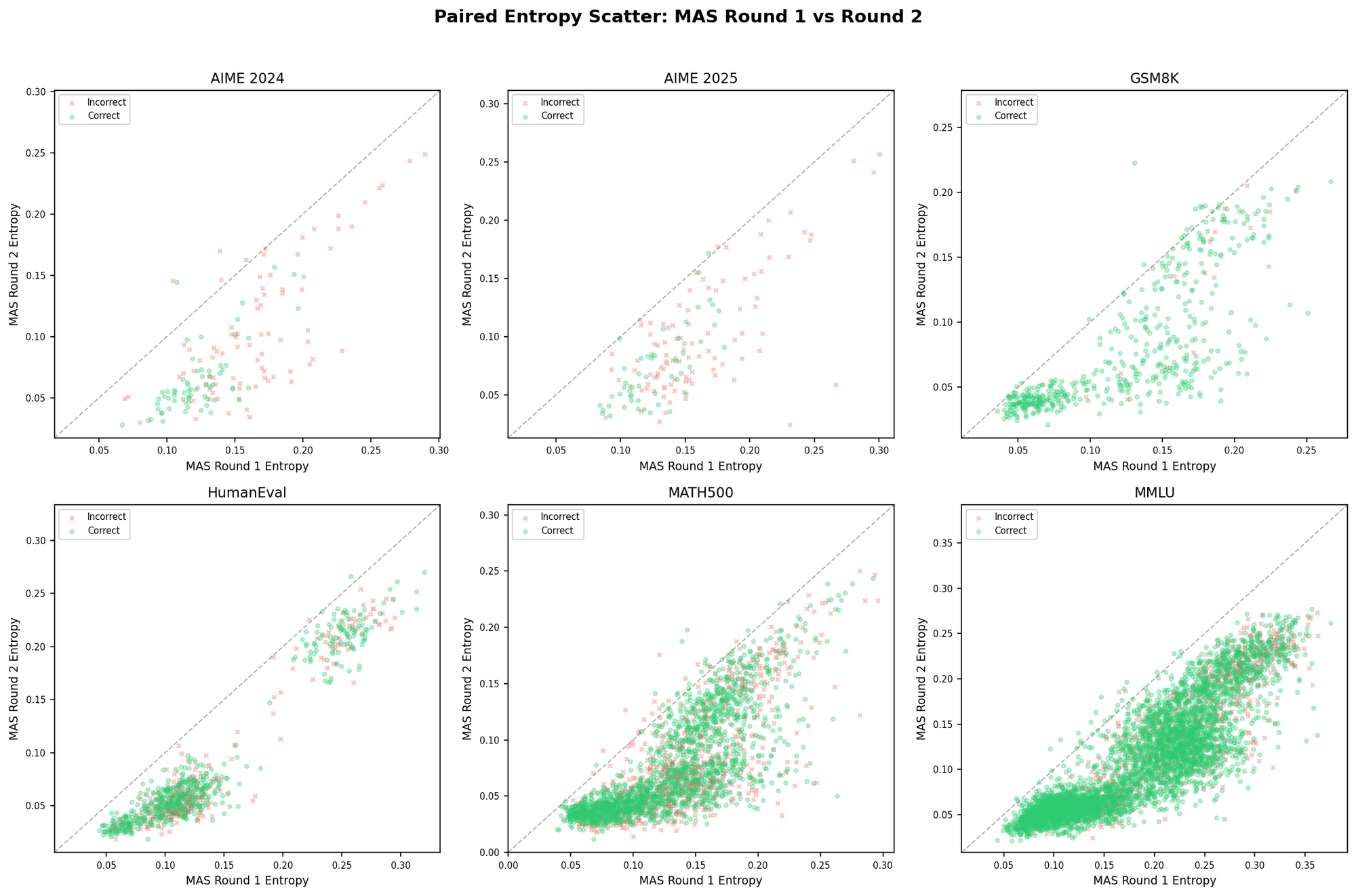}
    \caption{Paired entropy scatter plots (MAS Round~1 vs.\ Round~2) across datasets. Each point represents a single sample; green circles ($\circ$) denote correct answers and red crosses ($\times$) denote incorrect answers. The dashed diagonal ($y = x$) separates entropy decrease (below) from entropy increase (above). The predominance of points below the diagonal confirms systematic entropy reduction, while the similar spatial distribution of correct and incorrect samples suggests that entropy reduction alone does not reliably distinguish genuine improvement from anchoring.}
    \label{fig:causal-paired-scatter}
\end{figure*}

One concern with the accuracy degradation reported in the main text is that it could be a pure anchoring artifact: agents may simply copy or align with peers' Round-1 responses, lowering entropy without actually reasoning, in which case the apparent MAS effect would not reflect a real interaction mechanism. To separate genuine reasoning improvement from anchoring/copying, we decompose Round~1$\to$Round~2 changes by the co-occurrence of entropy decrease and accuracy improvement (Figure~\ref{fig:causal-paired-scatter}).

\paragraph{Decomposition.}
We classify each sample into four categories based on the joint outcome of entropy change ($\Delta H$) and accuracy change ($\Delta \text{acc}$):
\begin{itemize}[nosep,leftmargin=1.5em]
    \item \textbf{Genuine improvement} ($\Delta H < 0 \wedge \Delta\text{acc} > 0$): 6.2\% of samples
    \item \textbf{Possible anchoring} ($\Delta H < 0 \wedge \Delta\text{acc} \leq 0$): 83.4\% of samples
    \item \textbf{Productive exploration} ($\Delta H > 0 \wedge \Delta\text{acc} > 0$): rare
    \item \textbf{Deterioration} ($\Delta H > 0 \wedge \Delta\text{acc} \leq 0$): remaining samples
\end{itemize}

\textbf{These results quantitatively validate the central finding of the main text: inter-agent interaction rarely yields genuine accuracy improvements, confirming that MAS outcomes are predominantly fixed by round-1 dynamics rather than subsequent deliberation.} The overwhelming prevalence of ``possible anchoring" (83.4\%) versus ``genuine improvement" (6.2\%) indicates that while agents often converge to lower entropy states, this convergence seldom translates into corrected reasoning. Instead, it largely reflects an anchoring effect where the system reinforces its initial consensus, which may already be correct (i.e., $1 \to 1$ samples with $\Delta\text{acc} = 0$) or incorrect (i.e., $0 \to 0$ samples), though the latter case represents the predominant failure mode.

\paragraph{Model-Dependent Genuine Improvement Rates.}
The genuine improvement rate varies substantially across models:
\begin{itemize}[nosep,leftmargin=1.5em]
    \item LLaMA-3.2-3B-Instruct: 10.1\% genuine, 81.1\% possible anchoring (ratio = 0.12)
    \item Qwen3-0.6B: 7.3\% genuine, 75.4\% possible anchoring (ratio = 0.10)
    \item LLaMA-3.1-8B-Instruct: 6.9\% genuine, 82.6\% possible anchoring (ratio = 0.08)
    \item Qwen3-4B: 3.6\% genuine, 90.8\% possible anchoring (ratio = 0.04)
    \item Qwen3-8B: 2.9\% genuine, 87.0\% possible anchoring (ratio = 0.03)
\end{itemize}

This ordering does not track parameter count monotonically: Qwen3-0.6B sits between the two LLaMAs rather than at the top, and the two largest Qwen models occupy the bottom. The pattern is more consistent with model family than scale, with both LLaMA models showing higher genuine-improvement rates than the Qwen models of comparable or smaller size. This is the same family-over-scale effect documented for calibration in Appendix~\ref{app:calibration}.

\paragraph{Architecture-Dependent Patterns.}
The genuine-vs-anchoring decomposition by architecture mirrors the entropy-decrease pattern in Section~\ref{app:sas-mas-interaction}: Hybrid and Centralized achieve slightly higher genuine improvement (7.2\% and 7.0\%), while Sequential's combination of a low entropy-decrease rate (78.8\%) and the largest accuracy degradation ($-0.058$) reflects the same chain-propagation failure already documented in that section, rather than an ``anchoring-light" regime.

\section{Case Study: Token-Level Entropy Dynamics}
\label{app:case_study}

For each dataset, we select the first sample and visualize the token-level entropy trajectory across all agents and rounds. In each figure, white backgrounds denote round 1 and gray backgrounds denote round 2. Black dashed lines separate different agents' outputs within each round. The correctness indicator (checkmark or cross) in the upper-right corner shows the final MAS prediction outcome. This visualization captures the finest granularity of entropy dynamics, enabling direct observation of how entropy patterns relate to MAS success or failure.

\begin{figure*}[ht]
    \centering
    \includegraphics[width=0.9\textwidth]{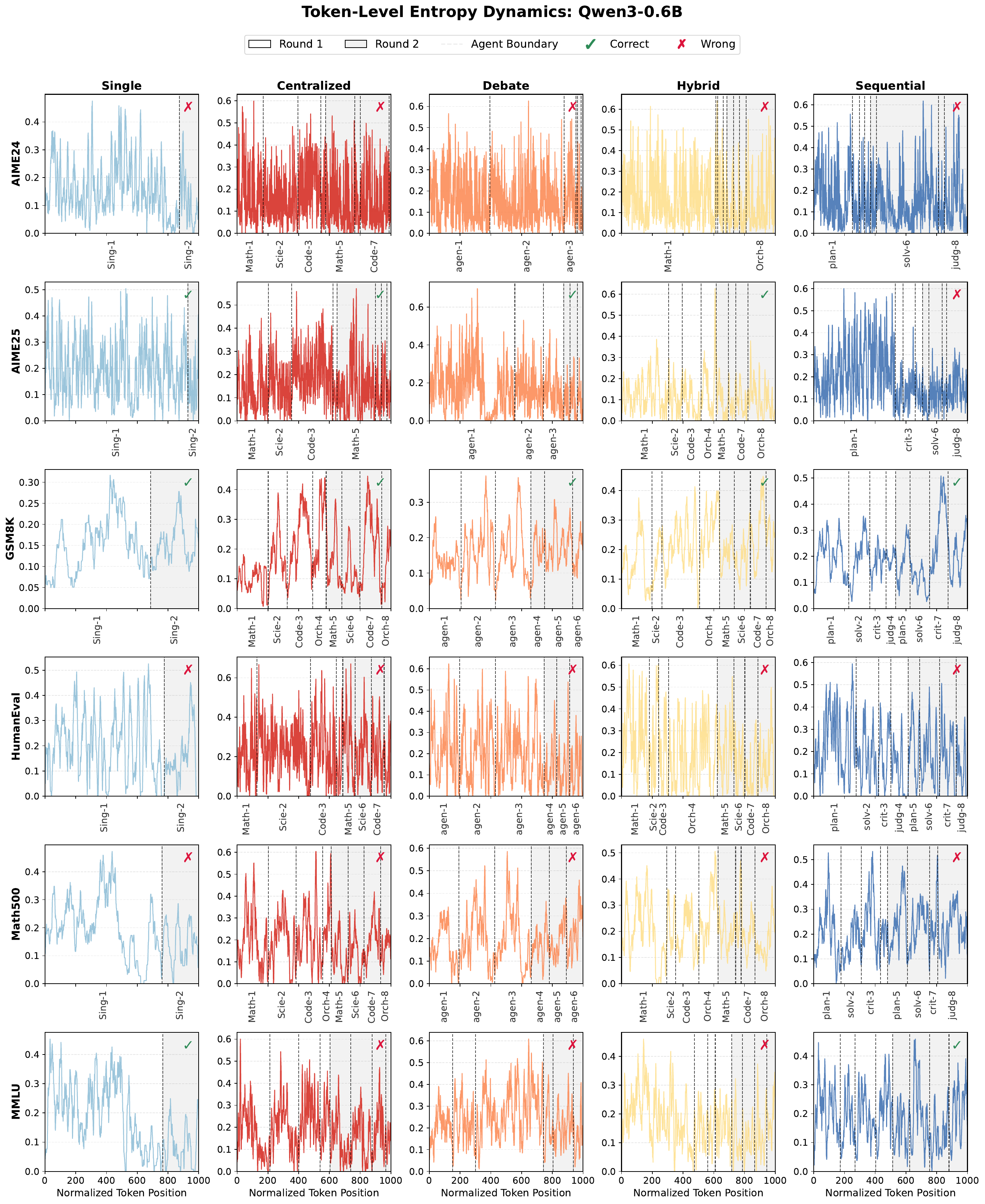}
    \caption{Token-level entropy dynamics for Qwen3-0.6B across six datasets. High entropy persistence and frequent spikes characterize this smaller model, with entropy either remaining elevated or collapsing abruptly to zero in round 2.}
    \label{fig:case_qwen3_0_6b}
\end{figure*}

\paragraph{Qwen3-0.6B: High Entropy Persistence.}
Figure~\ref{fig:case_qwen3_0_6b} shows that the smallest Qwen model exhibits persistently high entropy across both rounds, with frequent spikes throughout the reasoning trajectory. On harder tasks (\texttt{AIME}, \texttt{MATH500}), the entropy rarely stabilizes, and most predictions fail. This aligns with our finding that uncontrolled entropy harms MAS performance. Notably, even when round-2 entropy decreases, it often collapses to near-zero rather than converging to a stable moderate level, indicating premature termination rather than confident resolution.

\begin{figure*}[ht]
    \centering
    \includegraphics[width=0.9\textwidth]{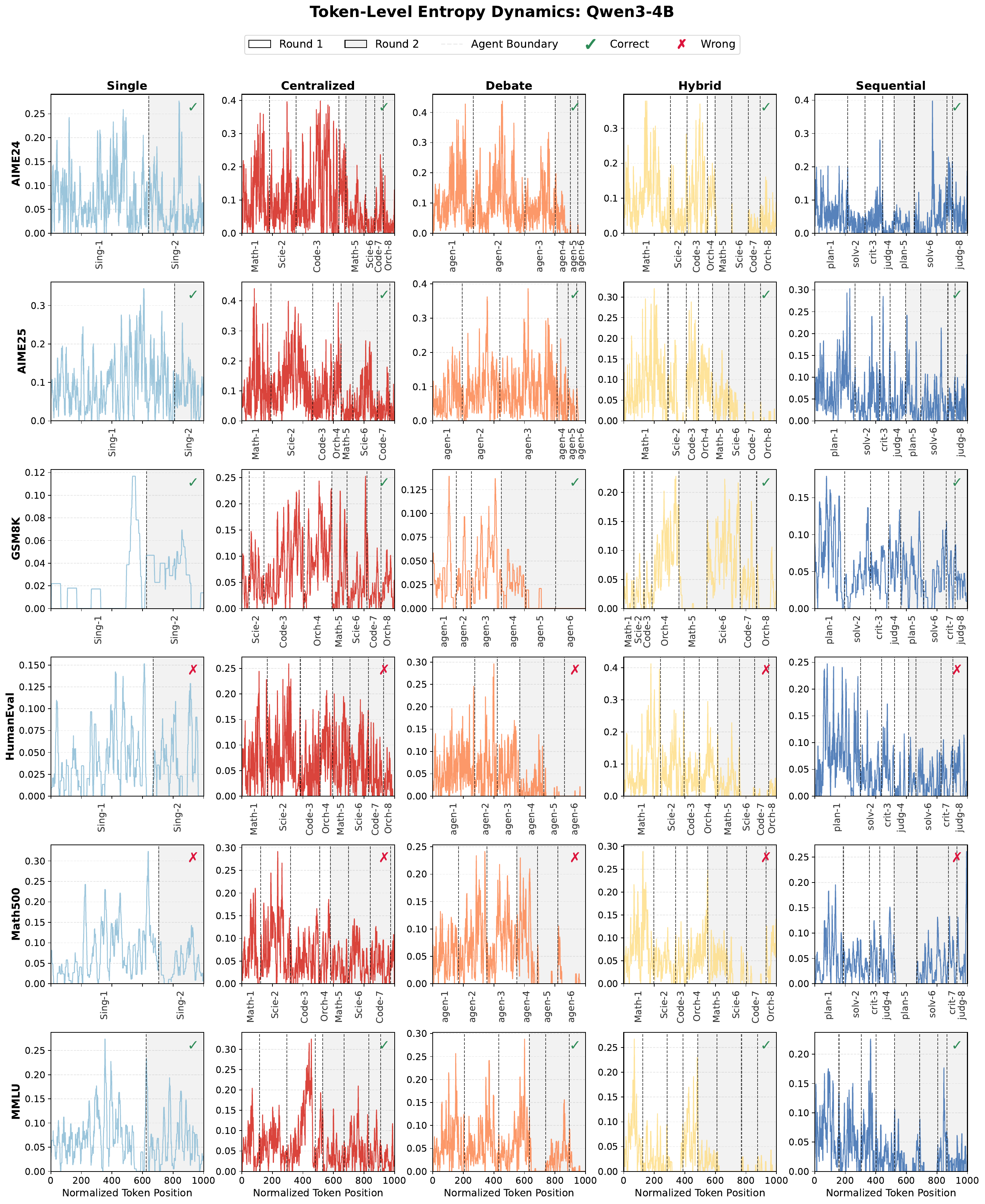}
    \caption{Token-level entropy dynamics for Qwen3-4B. Increased model capacity yields more stable entropy on easier tasks, while harder tasks still induce erratic entropy patterns.}
    \label{fig:case_qwen3_4b}
\end{figure*}

\paragraph{Qwen3-4B: Improved Stability with Task-Dependent Patterns.}
Figure~\ref{fig:case_qwen3_4b} reveals that scaling to 4B parameters improves entropy stability. On simpler tasks (\texttt{GSM8K}, \texttt{MMLU}), agents converge to low, stable entropy in round 2, yielding correct predictions. On medium-difficulty tasks (\texttt{MATH500}, \texttt{HumanEval}), moderate entropy is maintained, supporting productive exploration. However, on \texttt{AIME}, entropy dynamics remain erratic, suggesting that even larger models struggle with olympiad-level problems. This pattern supports our \textit{Task Awareness} principle: optimal entropy profiles vary by task difficulty.

\begin{figure*}[ht]
    \centering
    \includegraphics[width=0.9\textwidth]{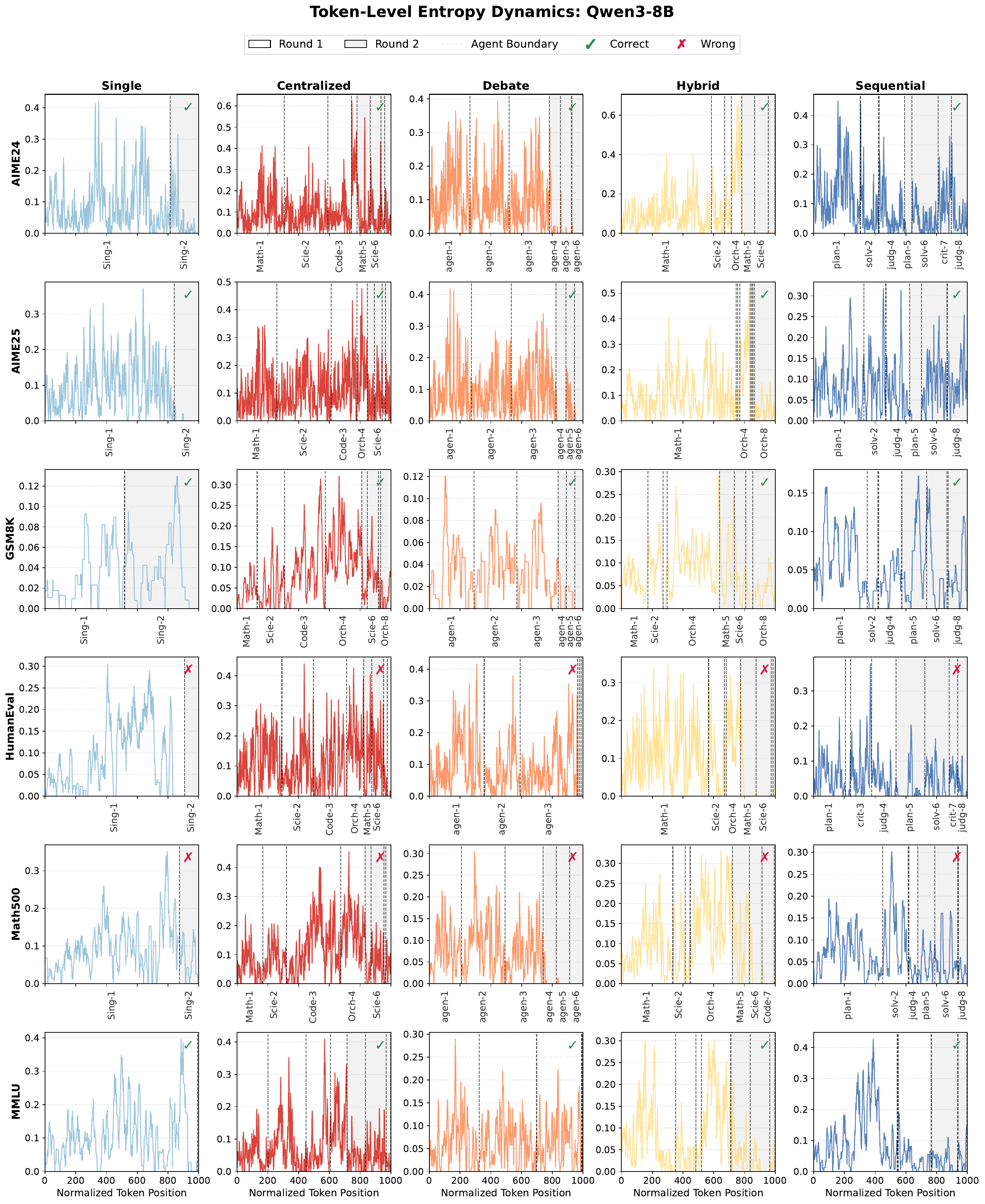}
    \caption{Token-level entropy dynamics for Qwen3-8B. The largest Qwen model shows structured deliberation with controlled exploration in round 1 and smooth convergence in round 2.}
    \label{fig:case_qwen3_8b}
\end{figure*}

\paragraph{Qwen3-8B: Structured Deliberation Emerges.}
Figure~\ref{fig:case_qwen3_8b} demonstrates that the largest Qwen model exhibits the most structured entropy dynamics. Round-1 entropy shows controlled exploration with clear peaks at decision points, followed by gradual stabilization. In round 2, entropy either maintains a productive moderate level (on hard tasks where deliberation helps) or converges smoothly to low values (on simple tasks). This structured pattern correlates with higher accuracy, confirming that \textit{Certainty Preference} benefits MAS when achieved through genuine convergence rather than premature collapse.

\begin{figure*}[ht]
    \centering
    \includegraphics[width=0.9\textwidth]{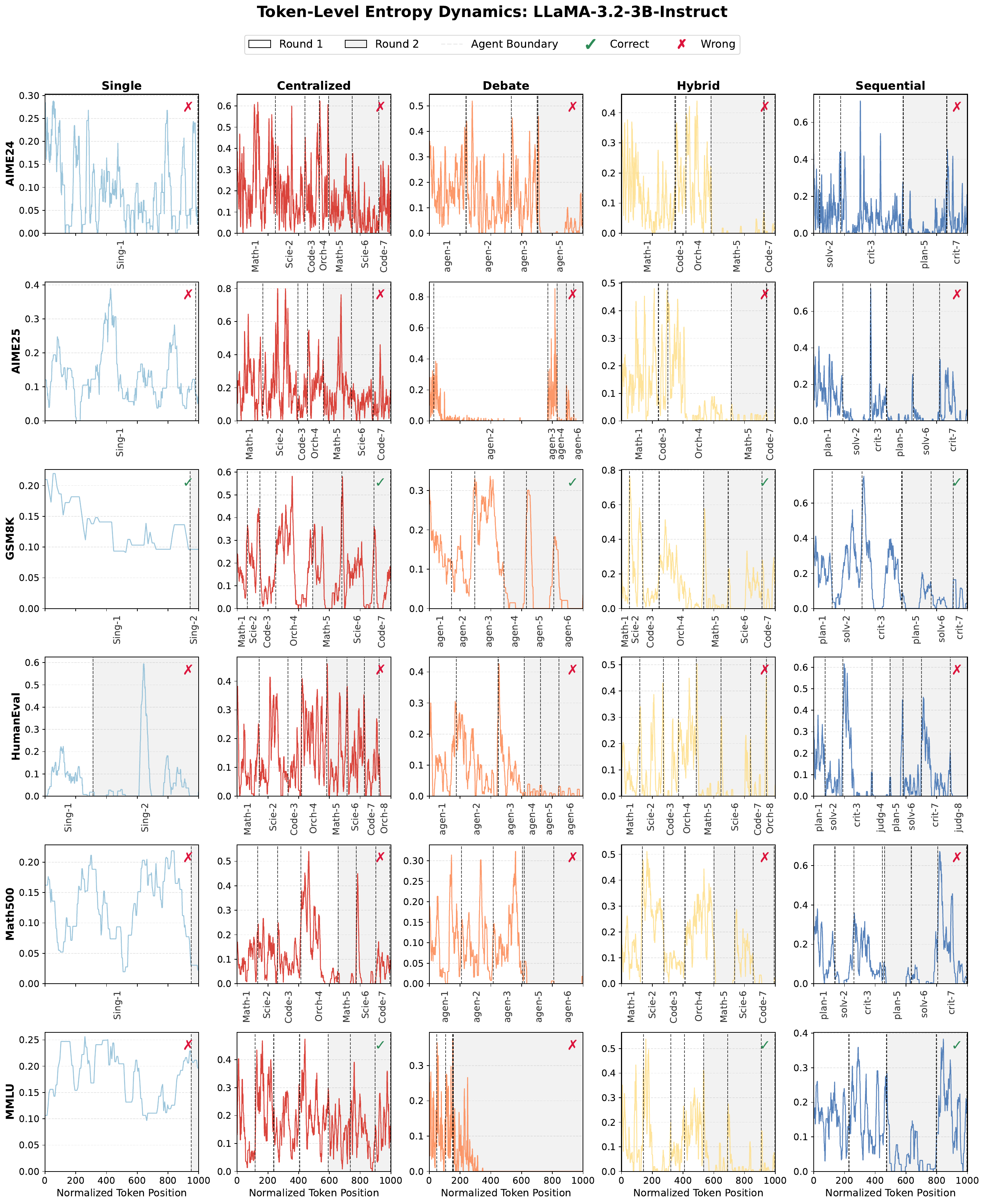}
    \caption{Token-level entropy dynamics for LLaMA-3.2-3B-Instruct. LLaMA exhibits lower round-2 entropy compared to Qwen, often collapsing to near-zero, reflecting a more decisive but potentially overconfident reasoning style.}
    \label{fig:case_llama_3_2_3b}
\end{figure*}

\paragraph{LLaMA-3.2-3B-Instruct: Distinct Reasoning Style.}
Figure~\ref{fig:case_llama_3_2_3b} shows that LLaMA models exhibit fundamentally different entropy dynamics compared to Qwen. Round-2 entropy frequently drops to near-zero across all agents, indicating a more decisive (but potentially overconfident) reasoning style. While this yields correct predictions on simpler tasks, it often leads to failure on harder problems where sustained exploration is beneficial. This contrast highlights how different model families develop distinct entropy profiles, with implications for MAS architecture selection.

\begin{figure*}[ht]
    \centering
    \includegraphics[width=0.9\textwidth]{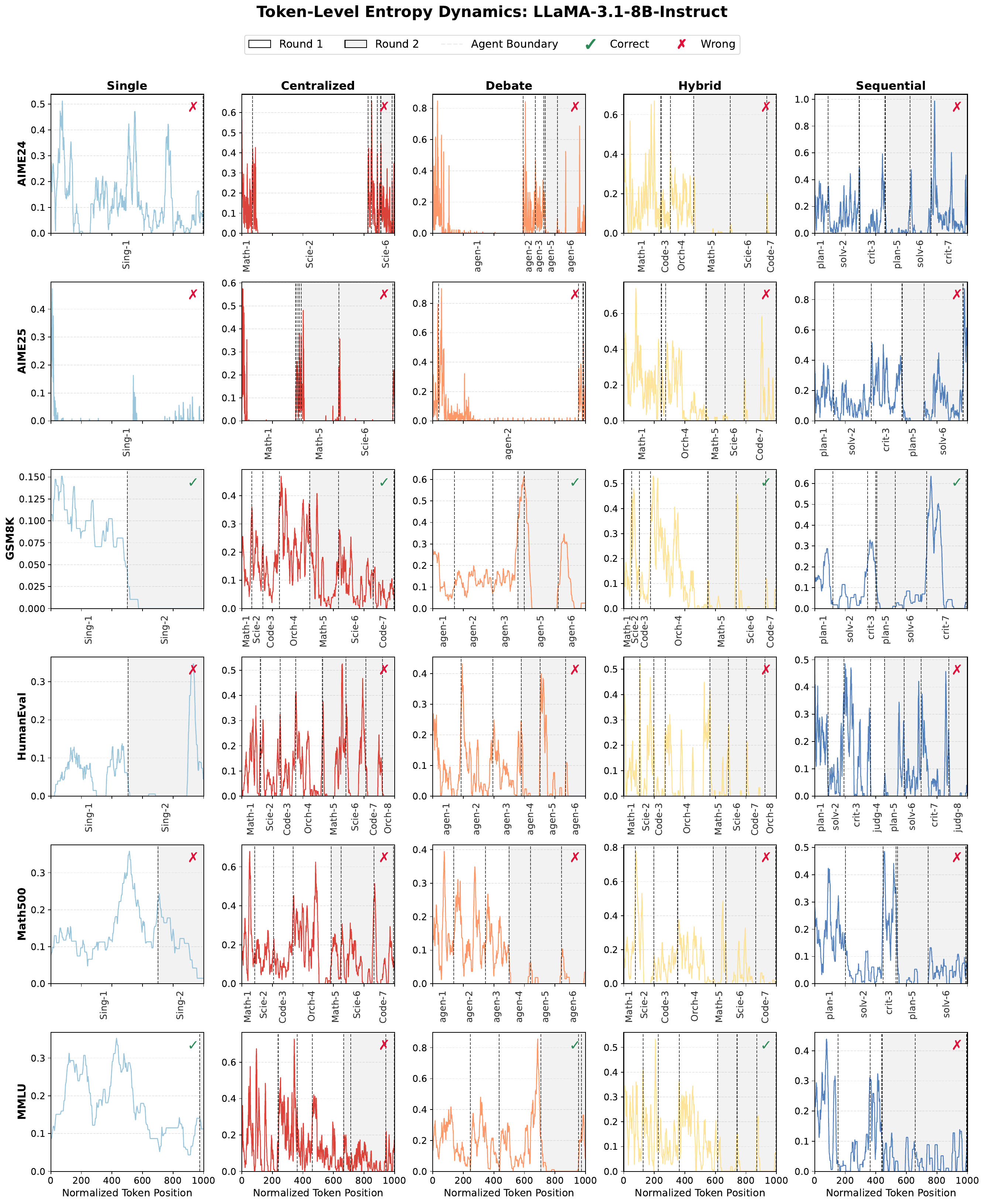}
    \caption{Token-level entropy dynamics for LLaMA-3.1-8B-Instruct. Scaling improves calibration, but the characteristic rapid entropy reduction in round 2 persists compared to Qwen models.}
    \label{fig:case_llama_3_1_8b}
\end{figure*}

\paragraph{LLaMA-3.1-8B-Instruct: Scale Improves but Style Persists.}
Figure~\ref{fig:case_llama_3_1_8b} reveals that scaling LLaMA to 8B parameters improves overall accuracy but preserves the characteristic low-entropy style in round 2. The model shows better calibration, with entropy remaining non-zero on harder tasks where exploration helps. However, the tendency toward rapid entropy reduction remains more pronounced than in Qwen models of comparable size. This suggests that model family (not just scale) shapes entropy dynamics, reinforcing our \textit{Base Entropy} finding that base model characteristics directly influence MAS effectiveness.

\paragraph{Summary.}
These case studies provide visual evidence for our main findings: (1) \textbf{Round-1 dynamics are critical}: entropy patterns established in the first round largely persist or determine the trajectory in round 2; (2) \textbf{Moderate, stable entropy correlates with success}: both excessively high entropy (erratic reasoning) and near-zero entropy (premature collapse) predict failure; (3) \textbf{Model family shapes entropy style}: Qwen and LLaMA exhibit distinct entropy profiles that influence MAS effectiveness across different tasks. These observations complement our quantitative analysis by revealing the fine-grained mechanisms underlying entropy-performance relationships.

\section{Limitations, Broader Impacts, and Future Work}
\label{app:limitations}

While our study provides comprehensive insights into entropy dynamics of LLM-based MAS, several limitations suggest directions for future research.

\paragraph{Model Scale and Benchmark Coverage.}
Due to computational constraints, our experiments are limited to open-source LLMs with at most 14B parameters; models at the 27B-70B scale may exhibit qualitatively different entropy dynamics as emergent capabilities grow with parameter count. Furthermore, our evaluation focuses on six benchmarks spanning mathematical reasoning, code generation, and knowledge Q\&A. While we have extended coverage to tool-calling scenarios through \texttt{GAIA}~\citep{gaia-iclr23} and \texttt{FinanceAgent}~\citep{FinanceAgentBench-arxiv25}, evaluating on broader agentic benchmarks involving web browsing and multi-step environment interactions~\citep{agentbench-iclr24,browsecomp-plus-arxiv25,multiagentbench-acl25} remains as future work, where entropy dynamics may play a more pronounced role due to the complexity and interactivity of the tasks.

\paragraph{Homogeneous Model Assumption.}
Our experimental design constructs MAS using homogeneous agents, where all agents share the same base model $M_{\text{base}}$. However, recent work~\citep{x-mas-arxiv25} demonstrates that heterogeneous MAS, composed of diverse LLMs with complementary strengths, can achieve superior performance compared to homogeneous configurations. The interplay between model heterogeneity and entropy dynamics remains unexplored: whether agents with different entropy profiles can compensate for each other's weaknesses, and how to optimally compose heterogeneous teams based on entropy characteristics, are promising research directions.

\paragraph{Causal Identification.}
While our causal discovery analysis (Section~\ref{sec:causal_analysis}) establishes that base-model entropy causally drives MAS correctness and that early inter-agent entropy dispersion mediates downstream performance, the causal identification relies on observational data and algorithmic discovery rather than controlled interventions. In particular, the predominant entropy-reduction-without-accuracy-gain rate (83.4\%) indicates that entropy reduction from inter-agent interaction does not automatically translate to performance gains, and the underlying mechanisms remain incompletely understood. Stronger causal identification requires complementary strategies: message ablation studies that isolate information content from social influence, and counterfactual interventions that manipulate entropy directly through temperature control while holding other factors fixed.

\paragraph{Broader Impacts.}
This work provides a principled understanding of entropy dynamics in LLM-based multi-agent systems, offering practical value for both researchers and practitioners. For the research community, our entropy-based analysis framework establishes a new perspective for diagnosing MAS failures and guiding architectural design decisions. For practitioners, our findings that single agents outperform MAS in 43.3\% of cases, combined with the insight that first-round entropy dynamics largely determine outcomes, can inform more resource-efficient deployment strategies. The \textit{Entropy Judger} further enables quality-aware output selection without requiring ground-truth labels, reducing annotation costs in real-world applications. We do not foresee specific negative societal consequences beyond those generally associated with advancing LLM capabilities.

\end{document}